\documentclass[a4paper,12pt,notitlepage]{book}
\usepackage[cp1250]{inputenc}
\usepackage{latexsym}
\usepackage[dvips]{graphicx}
\usepackage{graphicx}
\usepackage{enumerate}
\usepackage{amssymb}
\usepackage{fancyhdr}
\fancypagestyle{plain}{%
\fancyhead{}

}

\makeatletter

\newcommand{\linia}{\rule{\linewidth}{0.4mm}}

\renewcommand{\maketitle}{\begin{titlepage}

    \begin{center}

    Institute of Physics \\
		Faculty of Physics, Astronomy and Applied Computer Science\\
		Jagiellonian University\\

    \end{center}

    \vspace{2.5cm}

		\begin{center}

    \small Master Thesis

    \end{center}
    
    \noindent\linia

    \begin{center}

      \LARGE \textsc{\@title}

     \end{center}

     \linia

    \vspace{0.5cm}
		
		\begin{center}
		
		\Large \@author \par

		\end{center}

    \vspace{0.1cm}

		\begin{center}
		
    {\large Supervisor: Prof. dr hab. Michał Prasza\l{}owicz}\\
    
		\end{center}
		
    \vspace*{\stretch{6}}
		
		\vspace{2cm}
    
    \begin{center}

   \large Crackow, 2012

    \end{center}

  \end{titlepage}%

}

\makeatother

\author{Tomasz Stebel}

\title{Quantitative analysis of Geometrical Scaling in
Deep Inelastic Scattering}

\begin{document}

\maketitle

\newpage\thispagestyle{empty}
\mbox{}
\newpage

\thispagestyle{empty}

\begin{center}
\Large \textbf{Abstract} \\
\vspace{1.2cm}
\end{center}

We analyze geometrical scaling in deep inelastic scattering using experimental data from HERA $ep$ collider.
Parallel analyses are performed for energy and Bjorken-$x$ data binnings.
In particular, value of parameter $\lambda$ which governs $x$-dependence of saturation scale is found for both binnings: $\lambda_{\rm{En}}  = 0.352 \: \pm  \: 0.008$ and $\lambda_{\rm{Bj} }  = 0.302 \: \pm  \: 0.004$. We use the following method: $\lambda_{\rm{En}}$ is found as a value for which ratios $\sigma^{W1}_{\gamma^*p}(\tau)/\sigma^{W2}_{\gamma^*p}(\tau)$ are closest to 1 ($\sigma^{Wi}_{\gamma^*p}$ is a photon-proton cross section with definite energy $W_i$ and $\tau = Q^2\: x^{\lambda}$ is a scaling variable); $\lambda_{\rm{Bj} }$ is found analogously using ratios of cross sections with definite Bjorken-$x$ variable. We show also that GS is present for $x<0.2$.

\newpage\thispagestyle{empty}
\mbox{}
\newpage

\tableofcontents

\chapter{Introduction}

It is well known fact that Geometrical Scaling (GS) is present in deep inelastic scattering (DIS) of electrons and protons at small values of Bjorken variable $x$. In this context GS refers to dependence of $\gamma^*$-proton cross section $\sigma_{\gamma^*p}$ only upon one dimensionless variable $\tau$: $\sigma_{\gamma^*p}=\sigma_{\gamma^*p}(\tau)$ while in principle $\sigma_{\gamma^*p}$ may depend on two independent kinematical variables $Q^2$ and $x$: $\sigma_{\gamma^*p}=\sigma_{\gamma^*p}(Q^2,x)$.

This fact was first observed \cite{GBWscal} in the context of Golec-Biernat and W\"{u}sthoff model of $\gamma^*$-proton interaction \cite{GBWmod}. The crucial element of GBW model is the existence of characteristic energy scale $Q_{\rm{sat}}$ (called saturation scale), which depends on $x$. Because of that one can construct dimensionless variable $\tau=Q^2/Q_{\rm{sat}}^2(x)$. Saturation scale is customarily assumed to have power-like dependence on $x$: $Q_{\rm{sat}}^2(x)=Q_{0}^2(x/x_0)^{-\lambda}$ where $\lambda$ is a parameter which must be determined from experimental data. More details about GBW model, in particular on the existence of parton saturation will be given in Chapter 2.

The main aim of this dissertation is thorough analysis of GS using experimental data. We will use combined data \cite{H1ZueusWork} from H1 and ZEUS experiments which were taking data at HERA $ep$ collider.

The full analysis is divided into two pieces: first is based on data arranged into energy bins, second uses data with Bjorken variable binning. Chapter 3 presents method of dealing with data, in particular procedure of binning changing is given.
In this chapter we also present method of finding $\lambda$ exponent. It is based on observation that when GS is present cross section $\sigma_{\gamma^*p}$ plotted in terms of scaling variable $\tau=Q^2x^{\lambda}$ is a universal curve for all experimental points. Constructing ratios $\sigma^{W1}_{\gamma^*p}(\tau)/\sigma^{W2}_{\gamma^*p}(\tau)$ one can find $\lambda_{\rm{min}}(W_1,W_2)$ for which they are closest to 1 (for Bjorken-$x$ binning we can analogously find $\lambda_{\rm{min}}(x_1,x_2)$).   

In Chapters 4 and 5 method from Chapter 3 are applied to $e^{+}p$ and $e^{-}p$ data. In last section of Chapter 4 we search for $Q^2$ dependence of $\lambda$. It was shown \cite{imprGS} that such dependence improve GS in hadronic collisions.

In Appendices some additional analyses are given: we present some variants of method which were checked to give worse results (Appendix A) or are not completely appropriate (Appendices B and C). 

Our results can be shortly summarized as follows:
\begin{itemize}
\item The best $\lambda$ value (different for both binnings):
$$ \lambda_{\rm{Bj} }  = 0.302 \! \pm  \! 0.004 \quad \textrm{and} \quad \lambda_{\rm{En}}  = 0.352 \! \pm  \! 0.008,$$
where $\lambda_{\rm{Bj} }$ is expected to be more reliable result due to better quality of data.
\item Range in $x$ where GS is present (the same for both binnings):
$$ x <0.2 $$.
\end{itemize}

\chapter{Theoretical basics}
\section{DIS - general description}
\subsection{Kinematic variables}

\begin{figure}
\centering
\includegraphics[width=8cm]{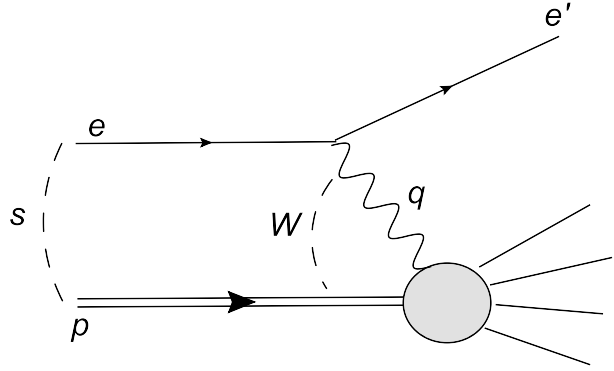}
\caption{Schematic picture of DIS. Some quantities important in problem was denoted.}
\label{zesRys1}
\end{figure}

In Fig. \ref{zesRys1} we shown schematically scattering of electron with initial four-momentum $e$ on proton with four-momentum $p$. Interaction between these two particles is realized by exchange of virtual photon with four-momentum $q$ (in what follows we will not consider cases with weak interaction bosons exchange). As a result of interaction four-momentum of electron changes into $e'$ and new hadrons may be produced. From experimental point of view it is important that detection of $e'$ is sufficient to determine most important quantities which describe DIS (inclusive measurements). We define several kinematic variables:

\begin{itemize}
\item virtuality of exchanged photon \textit{i.e.} square of exchanged four-momentum (with minus sign to make it positive):
\begin{equation}
Q^2=-q^2=-(e-e')^2.
\end{equation}

\item Mandelstam $s$-variable (square of energy in the electron-proton center of mass):
\begin{equation}
s=(e+p)^2.
\end{equation}

\item Energy in the photon-proton center of mass:
\begin{equation}
W=\sqrt{(e+q)^2}.
\end{equation}

\item Bjorken variable $x$:
\begin{equation}
x=\frac{Q^2}{2p\cdot q}=\frac{Q^2}{Q^2+W^2-M_p^2}.
\label{xwrtQ2W2}
\end{equation}
where $M_p$ is a proton mass (in DIS $M_p^2$ often neglected in this formula because it is small comparing to other quantities).  
\end{itemize}

All this quantities are Lorentz invariants \textit{i.e.} they do not depend on reference frame.

\subsection{Cross section and structure functions}

In proton's rest frame we denote: $E$ and $E'$ initial and final energy of electron, respectively; $\nu=E-E'$ energy transfer to proton; $\theta$ scattering angle of electron. Then we can find:
\begin{equation}
x=\frac{Q^2}{2M_p \nu}.
\label{x2Mpni}
\end{equation}

The differential cross section for scattering of an electron on proton can be written as (see for example \cite{DisserKs} or \cite{Atchison}):
\begin{equation}
\frac{\mathrm{d}^2 \sigma}{\mathrm{d} Q^2 \mathrm{d}\nu}= \frac{4\pi \alpha_{\rm{em}}^2}{Q^4} \frac{E'}{E} \left\{ W_2(Q^2,\nu)\cos^2 \frac{\theta}{2}+2 W_1(Q^2,\nu)\sin^2 \frac{\theta}{2} \right\},
\label{crosssecNoninv}
\end{equation}
where $\alpha_{\rm{em}}$ is an electromagnetic coupling constant, $W_2(Q^2,\nu)$ and $W_1(Q^2,\nu)$ are called \textit{structure functions} and contain information about structure of proton. For point-like proton with charge $e_p$ (and spin 1/2) these functions have form:
\begin{equation}
W^{\rm{el}}_1(Q^2,\nu)=e_p^2 \frac{Q^2}{4 M_p^2} \cdot \delta \left(\nu - \frac{Q^2}{2 M_p}\right) \quad \textrm{and} \quad W^{\rm{el}}_2(Q^2,\nu)=e_p^2 \cdot \delta \left(\nu - \frac{Q^2}{2 M_p} \right).
\label{W12el}
\end{equation}

In most cases DIS differential cross section is expressed in terms of $x$ and $Q^2$: 
\begin{equation}
\frac{\mathrm{d}^2 \sigma}{\mathrm{d} x \mathrm{d}Q^2}= \frac{2\pi \alpha_{\rm{em}}^2}{x Q^4} \left\{ (1+(1-y_E)^2)F_2(x,Q^2)- y_E^2 F_L(x,Q^2) \right\},
\label{crosssecInv}
\end{equation}
where $y_E=(p \cdot q) / (p\cdot e)$; $F_2$ and $F_L:=F_2-2x F_1$ are structure functions which can be easy expressed in terms of $W_1$ and $W_2$ (see (\ref{zwWiF})).
\begin{equation}
\end{equation}

\section{Parton model and Bjorken scaling}

In 1968 Bjorken proposed so-called Bjorken scaling hypothesis. It states that in the limit
\begin{equation}
\left\{ \begin{array}{ll}
Q^2 & \rightarrow \infty \\
\nu & \rightarrow \infty \\
\end{array} \right.
\textrm{  and  } x=\frac{Q^2}{2M_p \nu}=\textrm{const.} 
\end{equation}
structure functions $W_i(Q^2,\nu)$ $i=1,2$ are finite and depend only on one variable $x$:
\begin{equation}
\left\{ \begin{array}{llll}
M_p W_1(Q^2,\nu) & = F_1(x,Q^2) & \rightarrow & F_1(x) \\
\nu W_2(Q^2,\nu) & = F_2(x,Q^2) & \rightarrow & F_2(x). \\
\end{array} \right.
\label{zwWiF}
\end{equation}
Experiments confirmed the existence of Bjorken scaling (some distortion was also found). 

\begin{figure}
\centering
\includegraphics[width=7cm,angle=0]{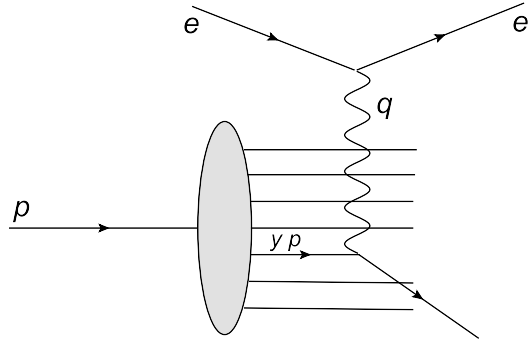}
\caption{Schematic picture of electron-parton interaction.}
\label{zesRys2}
\end{figure}

To explain Bjorken scaling R. P. Feynman proposed so-called parton model. He formulated it in infinite-momentum frame (in such frame proton's momentum is very large): he assumed that proton consists of point-like, spin 1/2, non-interacting particles called \textit{partons}, moreover, in DIS virtual photon interacts only with one of the partons which carries some fraction $y$ of proton's four-momentum (see Fig. \ref{zesRys2}) and transverse momentum of parton is neglected.  

Because partons do not interact with each other one can calculate electron-proton structure functions as sum of weighted electron-parton structure functions:
\begin{equation}
W_i(Q^2,\nu)=\sum_k \int_{0}^{1} \mathrm{d}y_k\: f_k(y_k) w^k_i(y_k,Q^2,\nu) \qquad i=1,2,
\label{WiQ^2}
\end{equation}
where $f_k(y)\mathrm{d}y$ is a probability to find in proton parton $k$ with momentum $y_k p$ (so-called \textit{parton density functions} p.d.f.) and $w^k_i(y_k,Q^2,\nu) \: i=1,2$ are structure functions for scattering of an electron on parton $k$ with mass $y_k M_p$ and charge $y_k e_p$. Because partons are assumed to have spin 1/2 we can use result (\ref{W12el}) replacing $M_p \rightarrow y M_p$ and $e_p \rightarrow y e_p$:
\begin{eqnarray}
\nonumber w^{\rm{k}}_1(y_k,Q^2,\nu) & = & y_k^2 e_p^2 \frac{Q^2}{4 y_k^2 M_p^2} \cdot \delta \left(\nu - \frac{Q^2}{2 y_k M_p}\right), \\
w^{\rm{k}}_2(y_k,Q^2,\nu) &=  & y_k^2 e_p^2 \cdot \delta \left(\nu - \frac{Q^2}{2 y_k M_p} \right) .
\label{w12}
\end{eqnarray}

Inserting (\ref{w12}) to (\ref{WiQ^2}) and integrating over $y_k$ we obtain:
\begin{eqnarray}
M_p W_1(Q^2,\nu) & = & \frac{1}{2} \sum_k e_k^2 f_k(x)\equiv F_1(x), \\
\nu W_2(Q^2,\nu) & = & x \sum_k e_k^2 f_k(x) \equiv F_2(x).
\end{eqnarray}
In particular we can see that:
\begin{equation}
2x F_1(x)= F_2(x)
\end{equation}
which is known as a Callan-Gross relation. It can be also written as $F_L=0$.

We can also give simple interpretation of Bjorken variable $x$. For real particle square of its four-momentum is a square of mass, therefore squares of parton four-momentum before and after interaction are equal:
\begin{equation}
(y p)^2=(y p+q)^2. 
\end{equation}
From this we obtain:
\begin{equation}
y=\frac{Q^2}{2p\cdot q}\equiv x.
\end{equation}
This means that in parton model Bjorken variable $x$ is a fraction of four-momentum carried by struck parton.

\subsubsection{Justification of parton model in QCD}
\label{QCDpartmod}

Results of experiment show that parton model is only an approximation of some more fundamental theory - Quantum Chromodynamics (QCD). The main tool used to study QCD is a perturbation calculus which is based on fundamental QCD property - \textit{asymptotic freedom}: it turns out that strong coupling constant $\alpha_s$ (which is a measure of interactions strength in QCD) decreases with momentum transfer:
\begin{equation}
\alpha_s(Q^2) \sim \frac{1}{\ln (Q^2/\Lambda_{\rm{QCD}}^2)},
\label{alfa_strong}
\end{equation}
where $\Lambda_{\rm{QCD}}\approx 250$ MeV is a scale for which perturbative calculus breaks down, \textit{i.e.} one can use this calculus only for $Q^2 \gg \Lambda_{\rm{QCD}}^2$.

The existence of asymptotic freedom justifies parton model: in the limit $Q^2 \rightarrow \infty$ strong coupling constant is so small that partons (quarks and gluons) do not interact.

\section{Saturation and Geometrical Scaling}

\subsection{DIS at small $x$}

Structure functions $F_1$ and $F_2$ can be related to virtual photon - proton cross section $\sigma_{\gamma^* p}$. It is convenient to separate cross section for photon polarized transversely $\sigma_T$ and longitudinaly $\sigma_L$. It turns out that:
\begin{eqnarray}
\frac{Q^2}{4 \pi^2 \alpha_{\rm{em}}}\sigma_T & = & 2x F_1, \\
\frac{Q^2}{4 \pi^2 \alpha_{\rm{em}}}\sigma_L & = & F_2-2x F_1.
\end{eqnarray}

In particular:
\begin{equation}
\sigma_{\gamma^* p}=\sigma_T+\sigma_L=4 \pi^2 \alpha_{\rm{em}} \frac{F_2}{Q^2}.
\label{sigmF2}
\end{equation}

An alternative and useful way to describe DIS for small $x$ is to work in the reference frame where photon well before the target splits into quark-antiquark pair (dipole) and then interacts with proton (see Fig. \ref{zesRys3}). One can show that $q\bar{q}$ lifetime is about $1/x$ times longer than time of interaction with proton. This means that for small $x$ processes of pair formation and its interaction with proton are separated and one can write cross sections $\sigma_T$ and $\sigma_L$ as a convolution of two factors:

\begin{figure}
\centering
\includegraphics[width=10cm,angle=0]{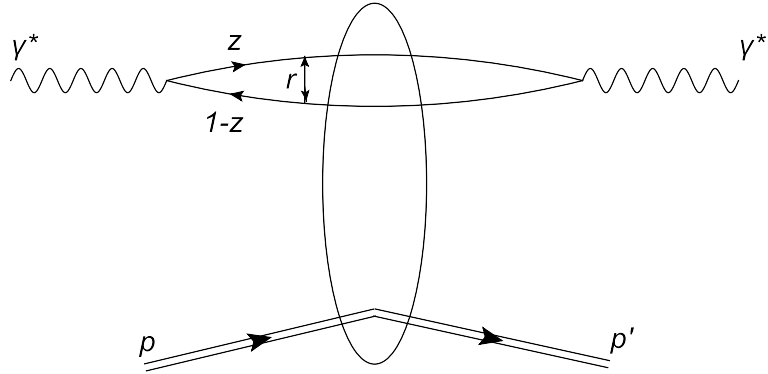}
\caption{Schematic picture of virtual photon - proton interaction.}
\label{zesRys3}
\end{figure}

\begin{equation}
\sigma_{T,L}(x,Q^2)=\int \mathrm{d}^2 \textit{\textbf{r}} \int_{0}^{1} \mathrm{d} z \sum_f | \Psi^f_{T,L}(\textit{\textbf{r}},z,Q^2)|^2 \hat{\sigma}(x,\textit{\textbf{r}}).
\label{factFormula}
\end{equation}

Now we will discuss this relation in more detail:
\begin{itemize}
\item first integration is performed over transverse separation of $q$ and $\bar{q}$ (denoted as $\textit{\textbf{r}}$): this is so-called dipole representation where transverse momentum $\textit{\textbf{k}}_T$ is replaced by its Fourier conjugate variable $\textit{\textbf{r}}$.
\item second integration is performed over photon momentum fraction $z$ carried by quark.
\item sum is performed over quark flavours $f$.
\item \textit{photon wave functions} $\Psi^f_{T,L}(\textit{\textbf{r}},z,Q^2)$ describes spliting of photon with definite polarization into quark-antiquark pair:
\begin{eqnarray}
|\Psi^f_{T}(\textit{\textbf{r}},z,Q^2)|^2 & = & \frac{3 \alpha_{\rm{em}}}{2 \pi^2} e_f^2 \left\{\left[z^2+(1-z)^2\right] \overline{Q}^2 K_1^2(\overline{Q}r)+ m_f^2 K_0^2(\overline{Q}r)\right\}, \nonumber \\
|\Psi^f_{L}(\textit{\textbf{r}},z,Q^2)|^2 & = & \frac{3 \alpha_{\rm{em}}}{2 \pi^2} e_f^2 \left\{4 Q^2 z^2(1-z)^2 K_0^2(\overline{Q}r)\right\}, 
\label{Psiforms}
\end{eqnarray}
where $K_{0,1}$ are the Bessel functions, $\overline{Q}^2=z(z-1)Q^2+m_f^2$ and $m_f$ are quark masses.
\item \textit{dipole cross section} $\hat{\sigma}(x,\textit{\textbf{r}})$ describes the interaction of $q\bar{q}$ pair with proton:
\begin{equation}
\hat{\sigma}(x,\textit{\textbf{r}})=\frac{4\pi}{3}\int_0^\infty \frac{\mathrm{d} l^2}{l^4} \alpha_s f(x,l^2)(1-J_0(lr)),
\label{sigmadipol}
\end{equation}
where $J_0$ is the Bessel function and $f(x,l^2)$ is a so-called \textit{unintegrated gluon distribution}. This name is justified due to relation (valid for large $Q^2$):
\begin{equation}
xg(x,Q^2)=\int_0^{Q^2} \frac{\mathrm{d}^2 \textit{\textbf{l}}}{l^2} f(x,l^2).
\end{equation}
Gluon distribution function $xg(x,Q^2)$ has physical interpretation: in infinite momentum frame it describes a probability to find in proton gluon with momentum $x p$ when one probes it by photon with virtuality $Q^2$.
\end{itemize}

To find form of $\hat{\sigma}(x,r)$ one should calculate unintegrated gluon distribution in the proton. This is not possible using perturbative QCD. One can, however, formulate models of $\hat{\sigma}(x,r)$.

\subsection{GBW model and saturation}

Golec-Biernat and W\"{u}sthoff proposed \cite{GBWmod} model of interaction between $q\bar{q}$ dipole and a proton. They introduced $x$-dependent \textit{saturation radius}:
\begin{equation}
R_0(x)=\frac{1}{Q_0} \left( \frac{x}{x_0} \right)^{\lambda /2}
\label{promsat}
\end{equation}
which sets the scale of dipole size $r$ in dipole cross section:
\begin{equation}
\hat{\sigma}(x,r)=\sigma_0 g\left( \frac{r}{R_0(x)} \right),
\label{sigmodg}
\end{equation}
where $g$ is some function which should have the following asymptotic behavior:
\begin{equation}
\left\{ \begin{array}{llll}
g(\hat{r}) & \sim & \hat{r}^2 & \textrm{for $\hat{r}\rightarrow 0$}\\
g(\hat{r}) & \rightarrow & \textrm{const}. & \textrm{for $\hat{r}\rightarrow \infty$}.
\end{array} \right.
\label{regnag}
\end{equation}
 
In \cite{GBWmod} $g$ was chosen as:
\begin{equation}
g(\hat{r})=1-\exp (-\hat{r}^2/4),
\end{equation}

where $Q_0=1$ GeV sets the set the scale, while constants $x_0$, $\sigma_0$ and $\lambda$ should be determined from fit to experimental data. 

Such formulation has two important properties:

\begin{itemize}

\item $\hat{\sigma}(x,r)=\sigma_0 r^2/ 4R_0^2(x)$ for $r\ll R_0(x)$. Using properties of $|\Psi^f_{T}|^2$ it can be shown that it gives power-like behavior of $F_T\sim Q^2 \sigma_T$ in $x$:
\begin{equation}
F_T\sim x^{-\lambda},
\end{equation}
what agrees with results obtained using so-called BFKL equation. In fact this is a motivation to define $R_0(x)$ with power-like $x$-dependence. However, for very small $x$ power-like rise of structure function would cause problem with unitarity. Some mechanism of taming this rise should be introduced - it is known as \textit{saturation}. Indeed, in GBW model saturation is present:
\item $\hat{\sigma}(x,r)=\sigma_0$ for $r\gg R_0$. As can be shown this leads to logarithmic rise of structure function: $F_T \sim Q^2 \ln (1/x)$ for small $x$ and no problem with unitarity appears.
\end{itemize}

One can interpret saturation radius $R_0(x)$ as a mean distance in transverse plane between partons in proton. Probing proton with $q\bar{q}$ one will "see" dilute state if $r \ll R_0(x)$ and dense if $r \gg R_0(x)$.

Equivalently to $R_0(x)$ one can use \textit{saturation scale} $Q_{\rm{sat}}(x)=1/R_0(x)$.

\subsection{Geometrical Scaling}
\label{GSwypr}

Inserting (\ref{sigmodg}) to (\ref{factFormula}) we get: 
\begin{equation}
\sigma_{T,L}(x,Q^2)=\sigma_0 \int \mathrm{d}^2 \textit{\textbf{r}} \int_{0}^{1} \mathrm{d} z \sum_f | \Psi^f_{T,L}(\textit{\textbf{r}},z,Q^2)|^2 g(r/R_0).
\end{equation}

We can change integration variable $r$ into $\hat{r}=r/R_0$. Using (\ref{Psiforms}) and neglecting quark masses we get:
\begin{eqnarray}
\nonumber \sigma_{T}(x,Q^2)=\sigma_0 \int_0^{\infty} \mathrm{d} \hat{r}^2 \int_{0}^{1} \mathrm{d} z \sum_f \frac{3 \alpha_{\rm{em}}}{\pi} e_f^2 \left[z^2+(1-z)^2\right] \\
\times \quad z(z-1)Q^2 R_0^2 K_1^2\left( \sqrt{z(z-1)Q^2 R_0^2} \: \hat{r} \right) g(\hat{r}).
\end{eqnarray}
Thus we have obtained $\sigma_{T}(x,Q^2)$ which depends only on a dimensionless combination $Q^2 R_0^2\equiv \tau$ (so-called scaling variable). Analogous result can be found for $\sigma_{L}$. Thus we have proved that:
\begin{equation}
\sigma_{\gamma^* p}(x,Q^2)=\sigma_{\gamma^* p}(\tau).
\label{defGS}
\end{equation}

This is so-called \textit{geometrical scaling} (GS). From (\ref{promsat}) we have:
\begin{equation}
\tau=\frac{Q^2}{Q^2_0} \left( \frac{x}{x_0} \right)^{\lambda}.
\label{zmiennaskalowania}
\end{equation}

It should be noted that proving (\ref{defGS}) we have used only assumption that there exists saturation radius $R_0$ which sets the scale of dipole size $r$ in dipole cross section. In particular, function $g$ can be arbitrary and constraints (\ref{regnag}) do not have to be fulfilled. This means that saturation is not necessary to GS existence.

In what follows we shall formulate model-independent way of looking for GS in the HERA data. As it will become clear the only parameter which we will be able to determine is exponent $\lambda$, two other parameters $x_0$ and $\sigma_0$ cannot be determined - one needs a specyfic model for function $g$ to get a handle on these parameters.

\chapter{Method of analysis}

\section{Data}
\label{sectdata}

We use combined data from H1 and ZEUS experiments \cite{H1ZueusWork} 
and consider only neutral current reactions. Data were taken in positron-proton ($e^{+}p$) and electron-proton collisions ($e^{-}p$). We treat them separately.

The following variables are taken from data: 
\begin{description}
\item $Q^{2}$~- virtuality of the exchanged photon,
\item $x$~ - Bjorken variable,
\item $F_{2}$~- value of the structure function,
\item $\Delta F_{2}$~- total uncertainty of the structure function.
\end{description}

To omitt unimportant factors in (\ref{sigmF2}) we define: 
\begin{equation}
\tilde{\sigma}:=\frac{1}{4 \pi^2 \alpha_{\rm{em}}} \sigma_{\gamma^* p}=\frac{F_{2}}{Q^2}.
\label{defsigmtylde}
\end{equation}

As we said in subsection \ref{GSwypr} for GS $\tilde{\sigma}$ depends only upon a scaling variable $\tau$:
\begin{equation}
\tilde{\sigma}(x,Q^2)=\tilde{\sigma}(\tau),
\end{equation}
where:
\begin{equation}
\tau=\frac{Q^2}{Q^2_0} \left( \frac{x}{x_0} \right)^{\lambda}.
\end{equation}
In what follows factor $x_0$ is unimportant so we set it 1. Moreover, measured $Q^2$ values are expressed in GeV units so we also omit $Q^2_0$:       
\begin{equation}
\tau = Q^2 x^{\lambda}. 
\label{deftau} 
\end{equation}.

\begin{figure}
\includegraphics[width=7cm,angle=0]{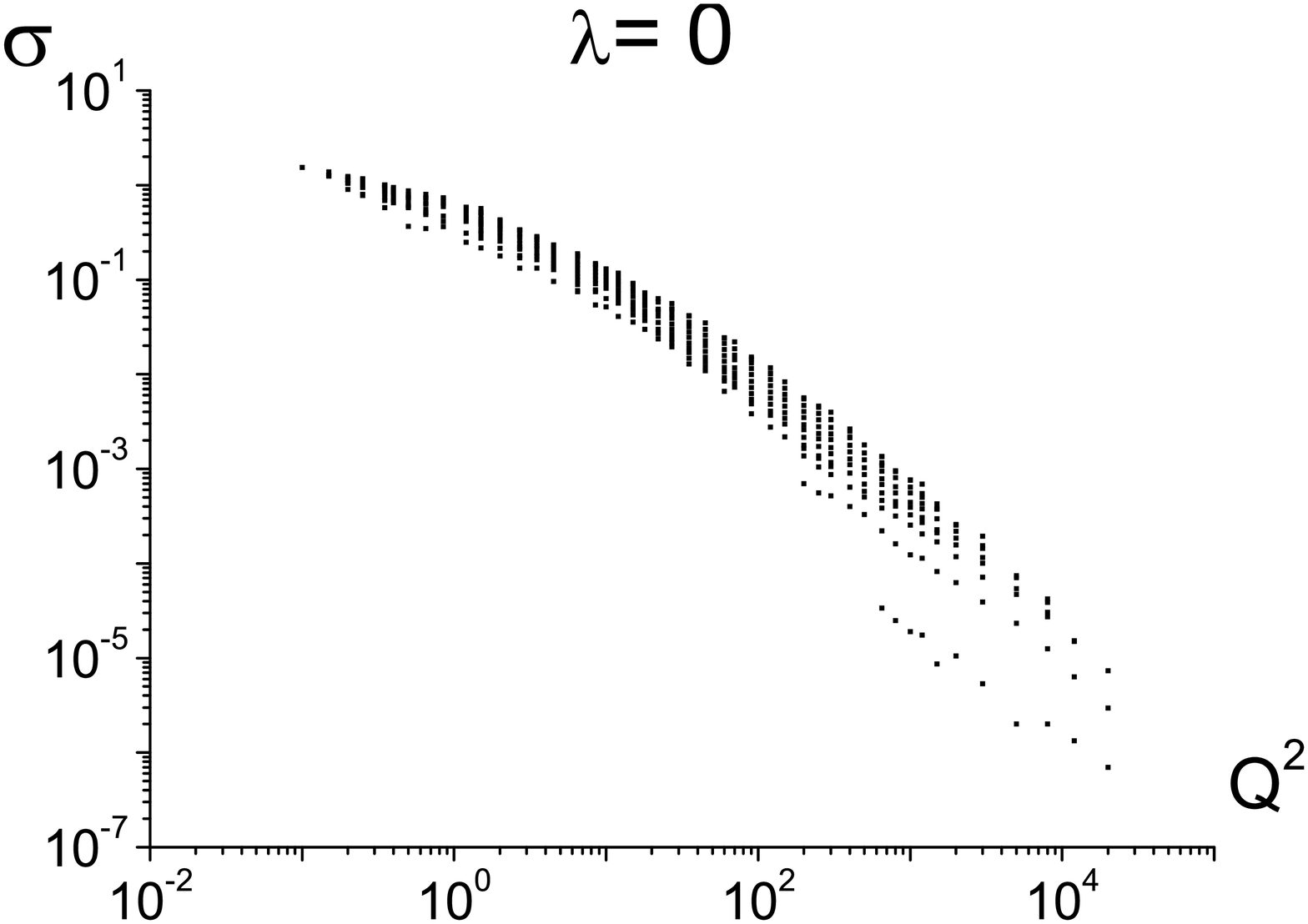}
\includegraphics[width=7cm,angle=0]{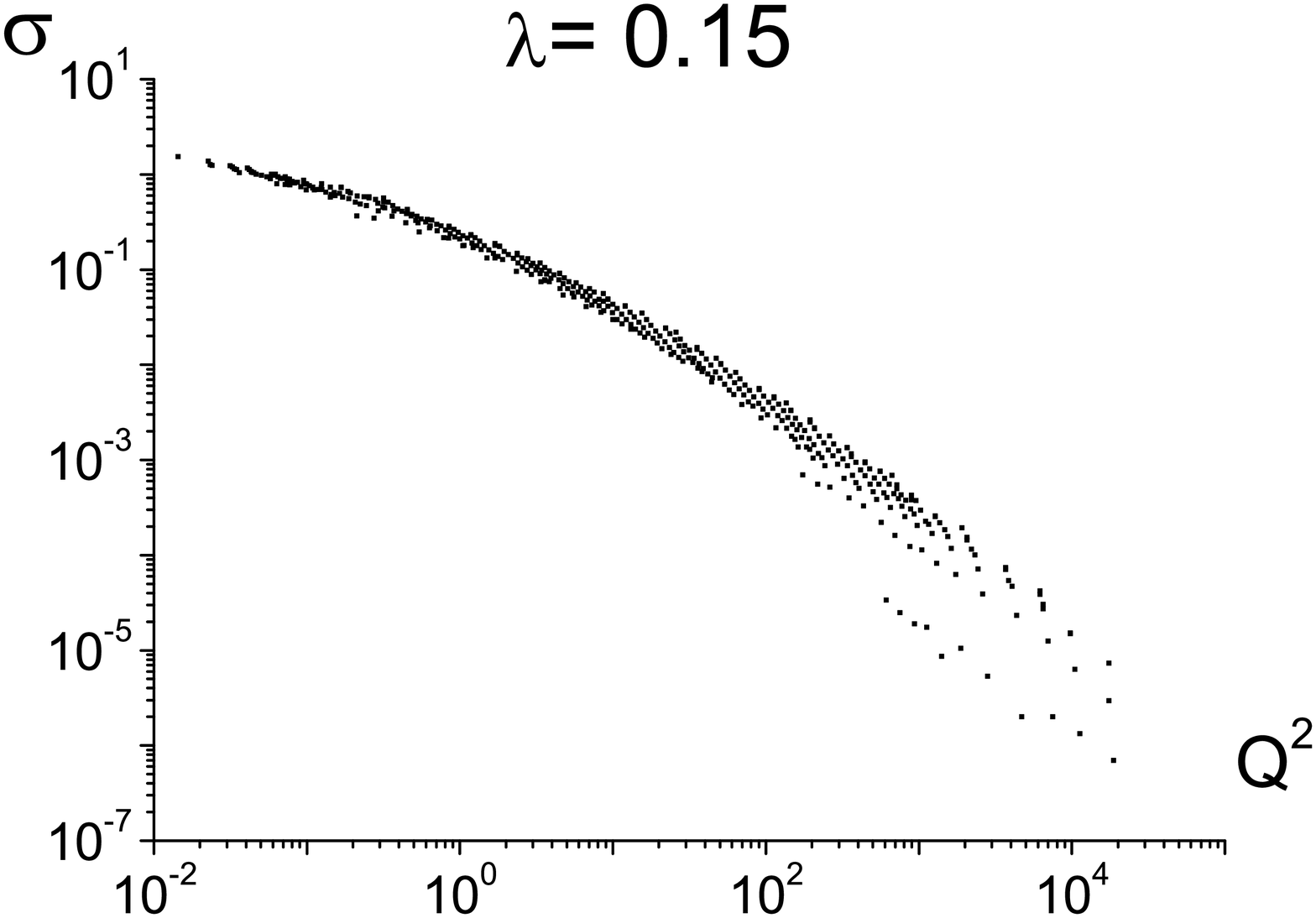}
\includegraphics[width=7cm,angle=0]{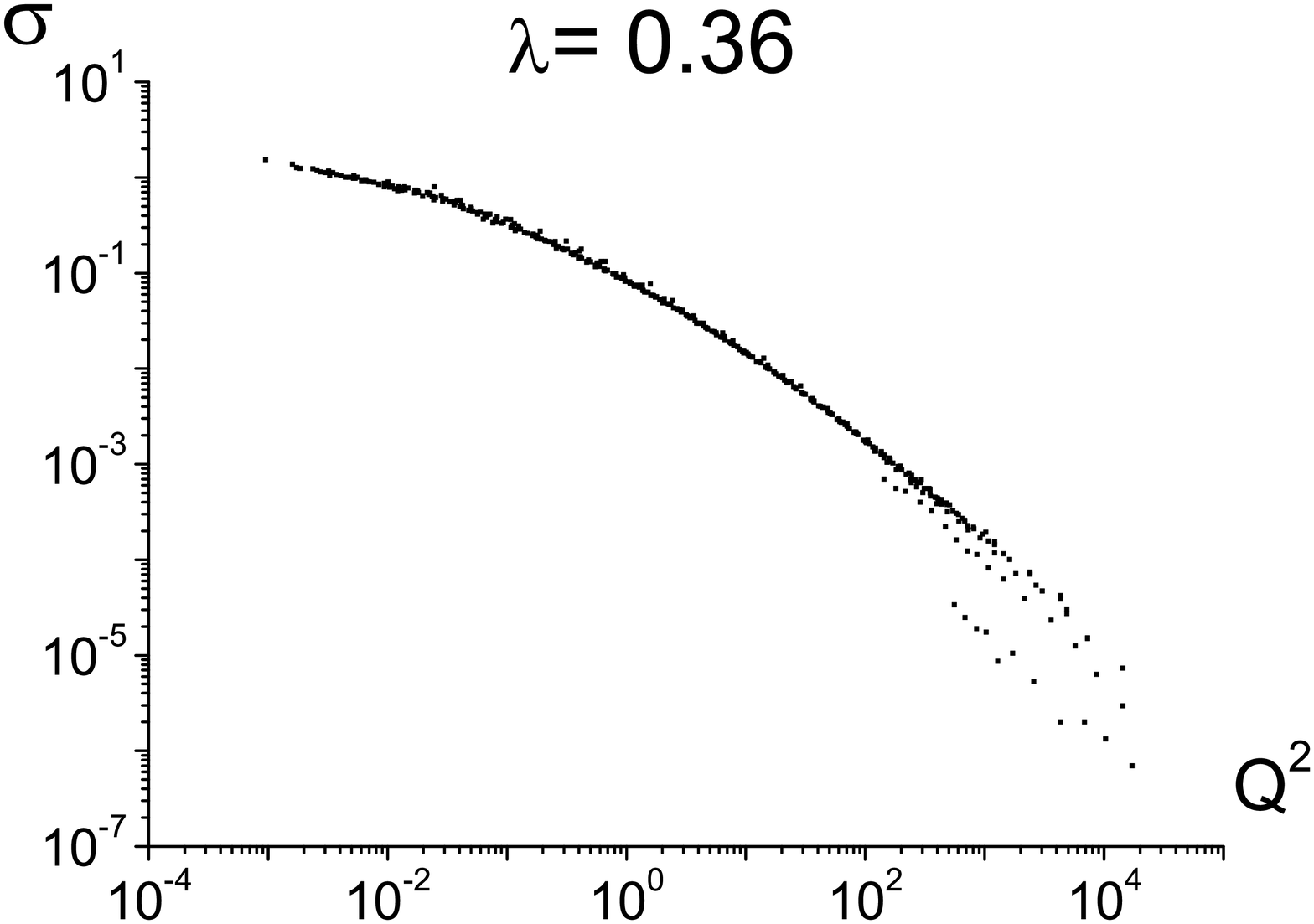}
\includegraphics[width=7cm,angle=0]{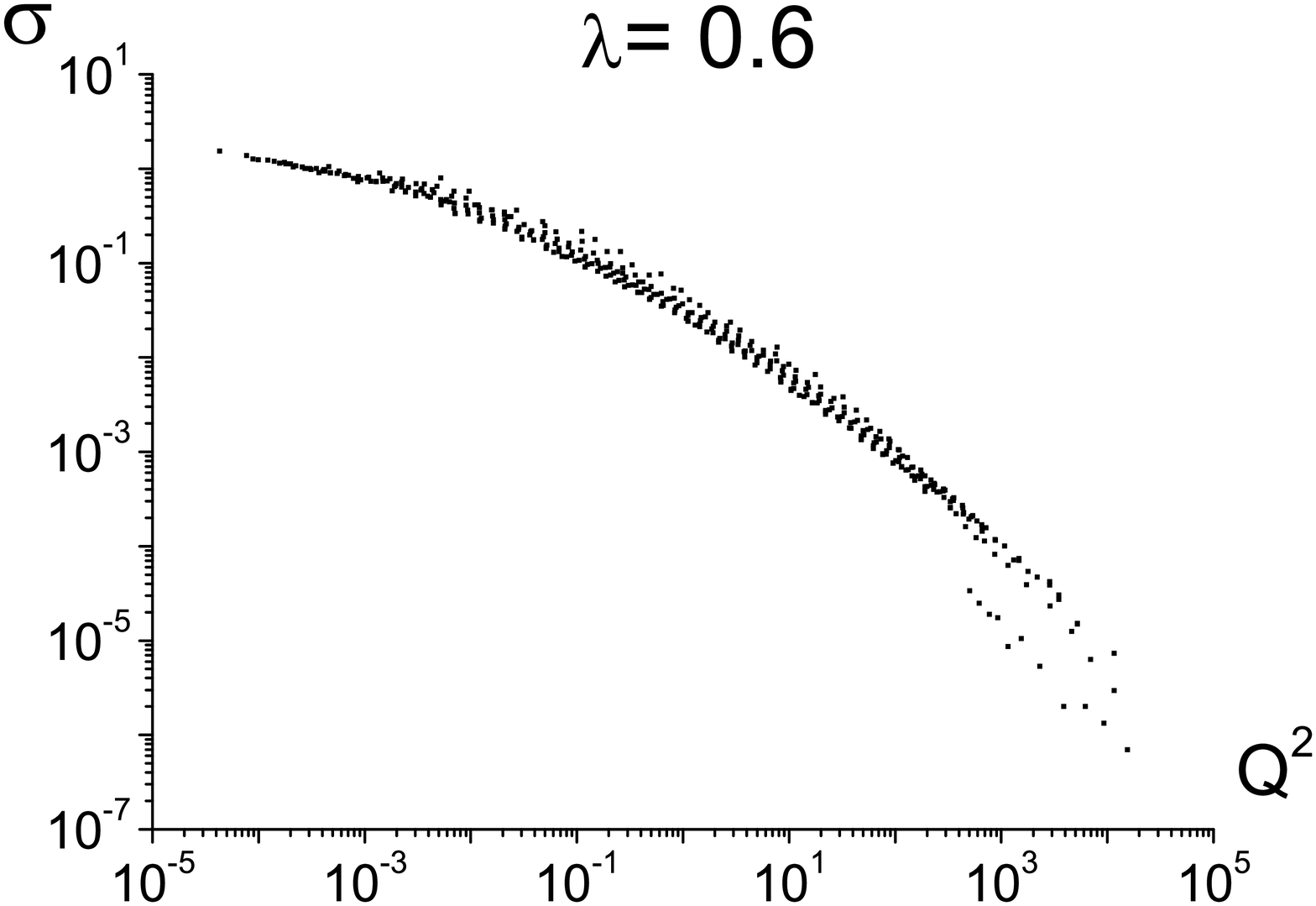}
\caption{$\tilde{\sigma}$ as a function of $\tau$ for $\lambda=0$, $0.15$, $0.36$, $0.6$. For $\lambda=0$ we have $\tau=Q^2$. Uncertainties of $\tilde{\sigma}$ are not shown. $e^{+}p$ data were used.}
\label{zesWyk1}
\end{figure}

In Fig. \ref{zesWyk1} we plot $\tilde{\sigma}$ as a function of $\tau$ for different values of $\lambda$. For $\lambda=0.36$ most of the points form one curve, so one can conclude that GS is present indeed. Some violation of GS can be seen for high values of $\tau$.

\begin{figure}
\includegraphics[width=7cm,angle=0]{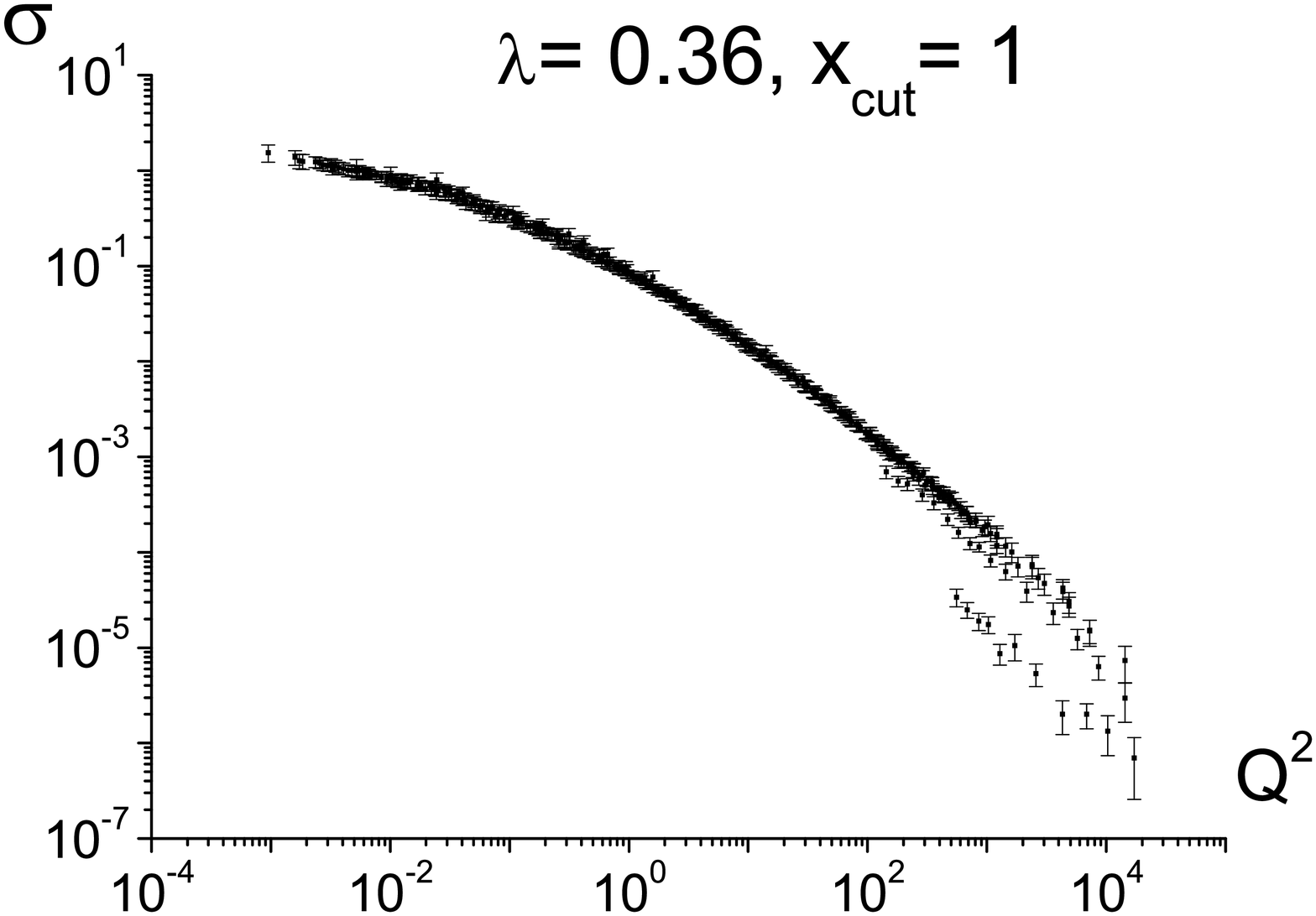}
\includegraphics[width=7cm,angle=0]{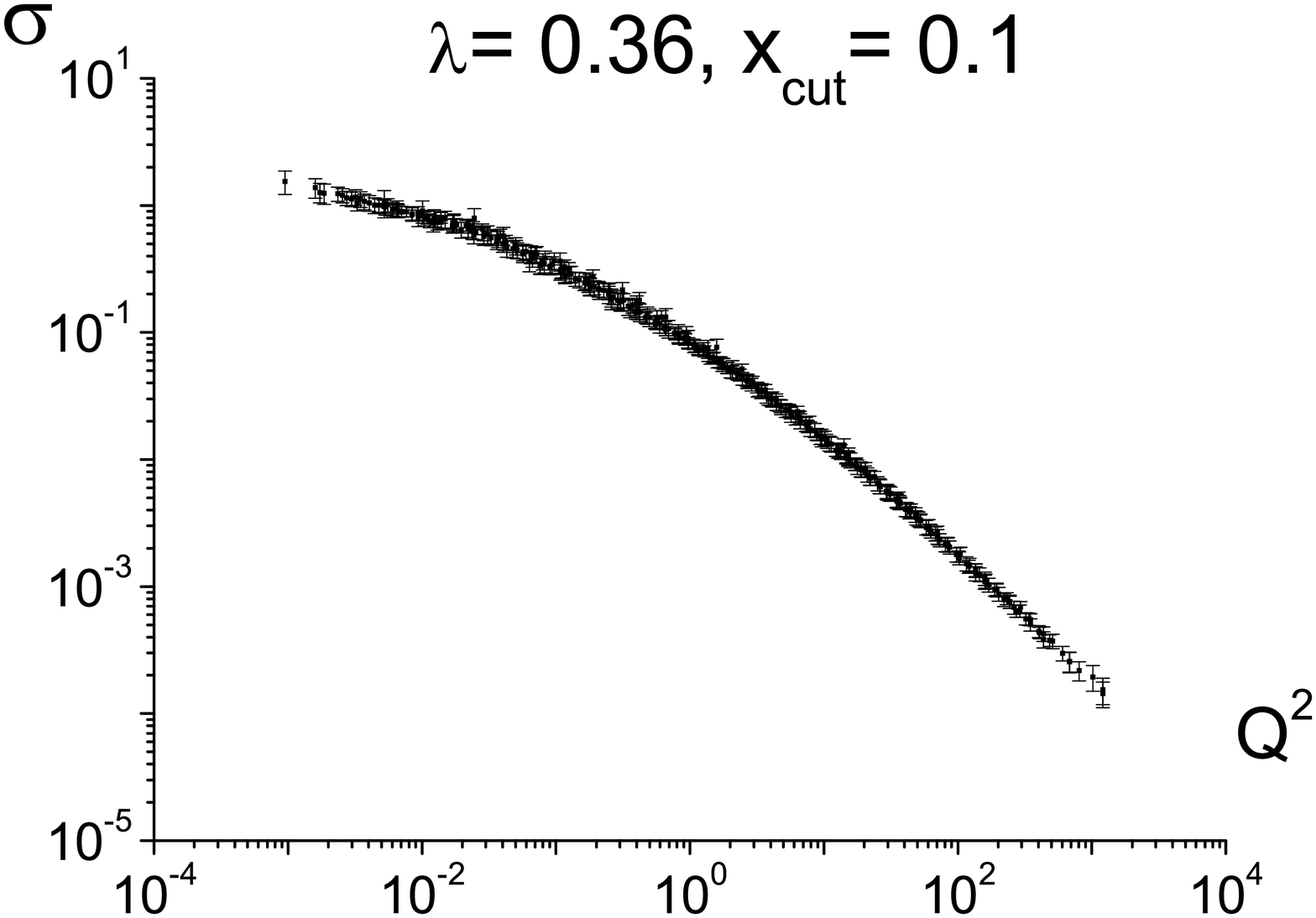}
\caption{$\tilde{\sigma}$ as a function of $\tau$ for $\lambda=0.36$ and $x_{\rm{cut}}=1$, $0.1$. $e^{+}p$ data are used.}
\label{zesWyk2}
\end{figure}

As we can see in Fig. \ref{zesWyk2} if we take only data with low values of $x$ this violation disappears (as we should expect, since GS is present only for small $x$ values). In order to investigate quantitatively the range of $x$ where GS is still present we introduce new parameter $x_{\rm{cut}}$ such that we consider only data with $x\leq x_{\rm{cut}}$. Value $x=1$ is the largest value which is possible due to the kinematics.

\subsubsection{Uncertainties of $\tilde{\sigma}$}

In order to attribute error to the quantity $\tilde{\sigma}=\frac{F_{2}}{Q^2}$ we need to estimate error of $Q^2$ which is not directly provided by HERA experiment. To do this we use values of virtuality $Q^{2}_1,Q^{2}_2,\ldots, Q^{2}_N$ (they are ordered \textit{i.e.} $Q^2_k<Q^{2}_{k+1}$) included in data. They were found using some binning which we do not know. Consider a value of virtuality $Q^2_i$, we assume that it was assigned to bin $(Q^2_i- \Delta_- Q^2_{i} , Q^2_i + \Delta_+ Q^2_{i}$ ).

To find $\Delta_\pm Q^2_{i}$ we use set of $2(N-1)$ equations (two for every $i\in \left\{1,\ldots,N-1\right\}$):
\begin{itemize}
\item $\Delta_+ Q^2_{i} + \Delta_- Q^2_{i+1}=Q^2_{i+1} - Q^2_{i}$ - this condition assures that every value of $Q^2$ is assigned to some bin.
\item $\frac {\Delta_+ Q^2_{i}}{\Delta_-Q^2_{i+1}}= Q^2_{i} / Q^2_{i+1}$ \textit{i.e.} uncertainties are proportional to values of $Q^2_i$. 
\end{itemize}
Solving this equations we get all $\Delta_\pm Q^2_{i}$ except for $\Delta_- Q^2_{1}$ and $\Delta_+ Q^2_{N}$; we define them as: $\Delta_- Q^2_{1}:=\Delta_+ Q^2_{1}$ and $\Delta_+ Q^2_{N}:=\Delta_- Q^2_{N}$.
Total uncertainty of $Q^2_i$ we calculate as a mean value $\Delta Q^2_{i}=\frac{\Delta_- Q^2_{i}+\Delta_+ Q^2_{i}}{2}$. 

Uncertainties of $\tilde{\sigma}$ are calculated from the following formula:
\begin{equation}
\Delta \tilde{\sigma}=\sqrt{\left(\frac{\Delta F_{2}}{Q^2}\right)^2+\left(\frac{F_{2}}{(Q^2)^2} \Delta Q^2\right)^2}.
\label{deltadefinitionF2/Q2}
\end{equation}

In what follows we do not take into account uncertainties of scaling variable $\tau$.

\section{Binning}
\label{sectBinning}

\subsection{Bjorken-$x$ binning}

Data are originally divided into bins with definite value of Bjorken-$x$ variable. We denote those values by $x'$ which means they are determined from Bjorken-$x$ binning (in subsection \ref{sectDivisionofdata} we will use different values denoted by $x$). In what follows we will use only $x'$ which have at least 2 points displayed in Table \ref{tableXbjor+}.
\begin{table}
\begin{tabular}{|c|c|c|c|c|c|} \hline
$x'$ & $5.52\cdot10^{-6}$ & $1.1\cdot10^{-5}$ & $1.58\cdot10^{-5}$ & $2\cdot10^{-5}$ & $3.2\cdot10^{-5}$ \\ \hline
Number of points & 2 & 2 & 2 & 2 & 3 \\ \hline \hline
$x'$ & $3.98\cdot10^{-5}$ & $5\cdot10^{-5}$ & $8\cdot10^{-5}$ & $10^{-4}$ & $1.3\cdot10^{-4}$ \\ \hline
Number of points & 5 & 4 & 5 & 3 & 8 \\ \hline \hline
$x'$ & $2\cdot10^{-4}$ & $2.51\cdot10^{-4}$ & $3.2\cdot10^{-4}$ & $5\cdot10^{-4}$ & $8\cdot10^{-4}$ \\ \hline
Number of points & 9 & 5 & 10 & 14 & 17 \\ \hline \hline
$x'$ & 0.001 & 0.0013 & 0.002 & 0.0032 & 0.005 \\ \hline
Number of points & 2 & 16 & 17 & 20 & 20 \\ \hline \hline
$x'$ & 0.008 & 0.013 & 0.02 & 0.032 & 0.05 \\ \hline 
Number of points & 16 & 19 & 24 & 19 & 18 \\ \hline \hline 
$x'$ & 0.08 & 0.13 & 0.18 & 0.25 & 0.4 \\ \hline 
Number of points & 16 & 14 & 16 & 15 & 16 \\ \hline \hline 
$x'$ & 0.65 & & & &  \\ \hline 
Number of points & 11 & & & &  \\ \hline \hline
\end{tabular}
\caption{$x'$ values used in analysis and numbers of points in bins.}
\label{tableXbjor+}
\end{table}

For a given $x'$ it can happen that several points have the same value of $Q^2$ and differ in $F_{2}$. This results in an ambiguity of $\tilde{\sigma}$. To eliminate this ambiguity we replace those points by one point with the same $Q^2$ and averaged $\tilde{\sigma}$.

\begin{figure}
\centering
\includegraphics[width=12cm,angle=0]{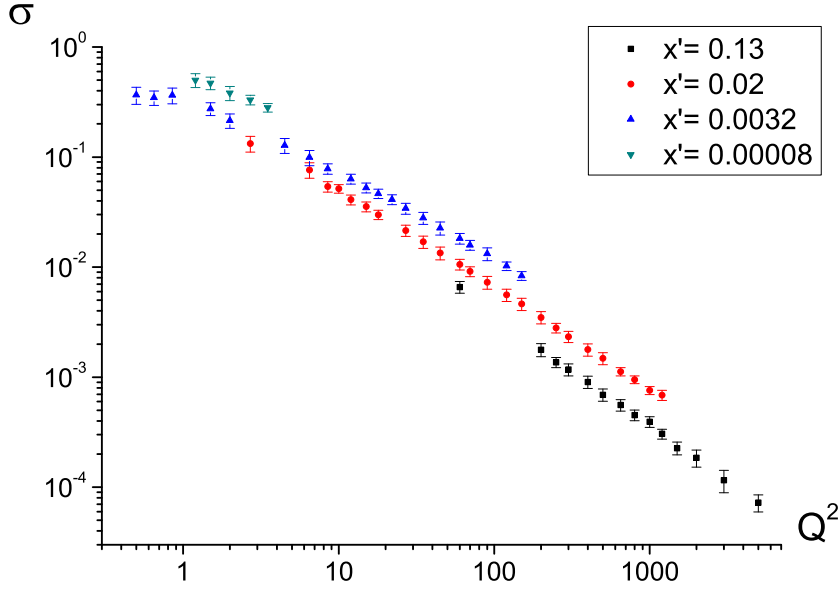}
\caption{Functions $\tilde{\sigma}(Q^2)$ for several values of $x'$. $e^{+}p$ data are used.}
\label{zesWyk28}
\end{figure}

In Fig. \ref{zesWyk28} we show functions $\tilde{\sigma}(Q^2)$ for several values of $x'$.

\subsection{Energy binning}
\label{sectDivisionofdata}

GS can be find not only in Deep Inelastic Scattering. It was argued (\cite{GSinpp} and \cite{GSinhion}) that GS is also present in pp and heavy ion collisions. In hadronic collisions, however, we cannot use Bjorken-$x$ parameter. In the GS analysis for hadronic collisions center-of-mas energy is used to construct scaling variable $\tau$. In order to compare results of GS for DIS and hadronic collisions we shall use energy rather than Bjorken-$x$ binning in the analysis of DIS.  

Energy of $\gamma^* p$ collision $W$ (in the center of mass frame) is not included in data and we have to calculate it using the kinematical relation (\ref{xwrtQ2W2}). We take $M_p=1$ GeV (since all other variables are in GeV units). From (\ref{xwrtQ2W2}) we have:
\begin{equation}
W'=\sqrt{\frac{Q^2}{x'}-\left(Q^2-M_p^2\right)},
\label{w'def}
\end{equation}
where the prime means that value is determinated by Bjorken-$x$ binning (unprimed letters $x$ and $W$ will be reserved for values determined by energy binning).

These values of $W'$ have to be organized in energy bins. To this end we use logarithmic binning with step 1.3 - every consecutive border is 1.3 times greater than preceding one (in appendix \ref{sectEnergybinning} we will consider different binnings). Now we define value of energy $W$ as the mean of two limiting values $W'_{\rm{min}}$ and $W'_{\rm{max}}$ between which it lies (see Table \ref{table1}).

\begin{table}
\begin{tabular}{|c|c|c|c|c|c|c|c|} \hline
$W'_{\rm{min}}\rm{ [GeV]}$ & 10 & 13 & 16.9 & 22 & 28.6 & 37.1 & 48.3 \\ \hline
$W'_{\rm{max}}\rm{ [GeV]}$ & 13 & 16.9 & 22 & 28.6 & 37.1 & 48.3 & 62.7 \\ \hline
$W\rm{ [GeV]}$ & 11.5 & 15 & 19.4 & 25.3 & 32.8 & 42.7 & 55.5 \\ \hline
Number of points & 6 & 3 & 13 & 22 & 22 & 32 & 33 \\ \hline \hline
$W'_{\rm{min}}\rm{ [GeV]}$ & 62.7 & 81.6 & 106 & 137.9 & 179.2 & 233 & \\ \hline
$W'_{\rm{max}}[\rm{ [GeV]}]$ & 81.6 & 106 & 137.9 & 179.2 & 233 & 302.9 & \\ \hline
$W\rm{ [GeV]}$ & 72.2 & 93.8 & 122 & 158.5 & 206.1 & 267.9 & \\ \hline
Number of points & 40 & 40 & 42 & 43 & 44 & 7 & \\ \hline
\end{tabular}
\caption{Energy bins and energies assigned to them. Number of points which are present in different bins are also written (these are values for $e^{+}p$ data).}
\label{table1}
\end{table}

Having $W$ we compute from the formula (\ref{xwrtQ2W2}) new values of Bjorken variable - $x$. They are $0.8-1.3$ times larger than $x'$. In appendix \ref{sectXoryg} we will show how results change when we use $x'$ rather than $x$ in definition of scaling variable $\tau$.

\begin{figure}
\centering
\includegraphics[width=10cm,angle=0]{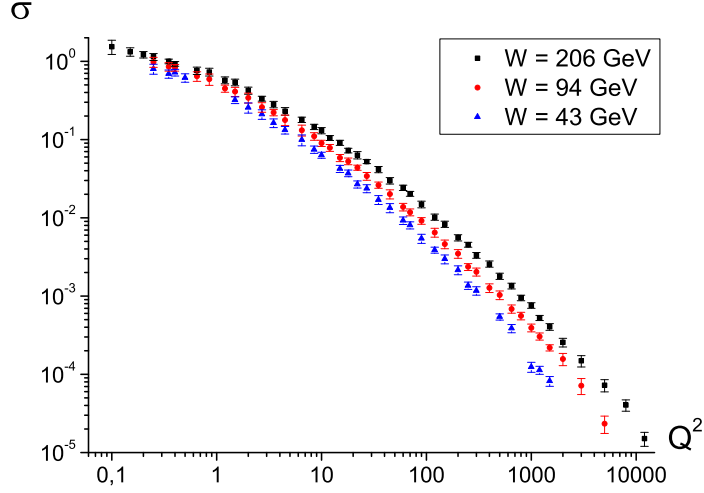}
\caption{Functions $\tilde{\sigma}^W(Q^2)$ (cross section with definite energy $W$) for three energies. $e^{+}p$ data are used.}
\label{zesWyk3}
\end{figure}

For a given energy $W$ it can happen that several points have the same value of $Q^2$ and differ in $F_{2}$ and $x$ values. This results in an ambiguity of $\tilde{\sigma}(Q^2)$. To eliminate this ambiguity we replace those points by one point with the same $Q^2$ and averaged $\tilde{\sigma}$ (see appendix \ref{sectXoryg} where we discuss this subject in more detail). This procedure, however, does not smoothen data completely and quality of data for energy binning is worse than for Bjorken-$x$ (see Fig. \ref{zesWyk41} in appendix \ref{sectXoryg}).

In Fig. \ref{zesWyk3} we can see points for three energies (for clarity we have not displayed points for other values of $W$).

\section{Method of finding $\lambda$}
\label{sectMethodoffinding}

\subsection{Energy binning}

Our aim is to find the value of $\lambda$ for which geometrical scaling is satisfied. We find it for a given $x_{\rm{cut}}$ \textit{i.e.} we search $\lambda$ using only with $x\leq x_{\rm{cut}}$.

When GS is present cross-sections $\tilde{\sigma}^W$ (cross-section with definite energy $W$) for all energies should follow one line if plotted as functions of scaling variable $\tau=Q^2 x^{\lambda}$. In order to quantify GS we choose one energy $W_{\rm{ref}}$ and calculate ratios
\begin{equation}
\label{continiousratio}
\frac{\tilde{\sigma}^{W_{\rm{ref}}}(\tau)}{\tilde{\sigma}^{W}(\tau)}
\end{equation}
for the remaining energies. For ideal GS all ratios should be equal to 1.

In what follows we use $W_{\rm{ref}}=206$ GeV because it gives us the widest range of $\tau$ values and is one of the biggest energies that we have (GS is expected to be present for large energies). In Appendix \ref{sectDiffrentchoiceofW} we will check different choices of $W_{\rm{ref}}$. 

In our analysis we use points from experiment, so ratios (\ref{continiousratio}) should be defined for finite number of points. In general $\tau$ values for $W_{\rm{ref}}$ and $W$ are different so we need to use interpolation: for given $\lambda$ we define a reference curve $f_{\lambda}^{\rm{ref}}(\tau)$ which is made by joining $W_{\rm{ref}}$ points ($\tau_k$ , $\tilde{\sigma}^{W_{\rm{ref}}}_k$) by segments (linear interpolation).

Now for every point with energy $W$ (we assume that $W \neq W_{\rm{ref}}$) we define a ratio ( $i$ labels points with energy $W$):
\begin{equation}
R^W_{i}(\lambda):=\frac{f_{\lambda}^{\rm{ref}}(\tau_i)}{\tilde{\sigma}^{W}_i},
\label{ratiodef} 
\end{equation}
where $\tau_i=Q_i x_i^{\lambda} $ is a value of scaling variable for i'th point of energy $W$ and $\tilde{\sigma}^{W}_i$ is a value of cross section for this point.

To find uncertainty of this ratio we need to estimate uncertainty of $f_{\lambda}^{\rm{ref}}(\tau_i)$. As previously, we should use interpolation because we know uncertainties only for $\tau_k$ values of $W_{\rm{ref}}$ points. We define another curve 
$f_{\lambda}^{\rm{unc}}(\tau)$ which joins points ($\tau_k$ , $\tilde{\sigma}^{W_{\rm{ref}}}_k + \Delta \tilde{\sigma}^{W_{\rm{ref}}}_k$) (we use linear interpolation as previously). Now uncertainty of $f_{\lambda}^{\rm{ref}}(\tau_i)$ (\textit{i.e.} for i'th point of energy $W$) is calculated as:
\begin{equation}
\Delta f_{\lambda}^{\rm{ref}}(\tau_i):=f_{\lambda}^{\rm{unc}}(\tau_i)-f_{\lambda}^{\rm{ref}}(\tau_i).
\end{equation}

In calculations we use only such points of $W$, that theirs arguments $\tau_i$ are within domain of $f_{\lambda}^{\rm{ref}}$ \textit{i.e.} $\tau^{W_{\rm{ref}}}_{\rm{first}} \leq \tau_i \leq \tau^{W_{\rm{ref}}}_{\rm{last}}$, where $\tau^{W_{\rm{ref}}}_{\rm{first}}$, $\tau^{W_{\rm{ref}}}_{\rm{last}}$ are the smallest and the largest values of $\tau$ among points with energy $W_{\rm{ref}}$. 

Uncertainty of ratio $R^W_{i}(\lambda)$ is given by:
\begin{equation}
\Delta R^W_{i}(\lambda)=\sqrt{\left(\frac{ \Delta f_{\lambda}^{\rm{ref}}(\tau_i)}{\tilde{\sigma}^{W}_i}\right)^2+
\left(\frac{ f_{\lambda}^{\rm{ref}}(\tau_i)}{(\tilde{\sigma}^{W}_i)^2}\Delta\tilde{\sigma}^{W}_i \right)^2}.
\label{uncRat}
\end{equation}

\begin{figure}
\centering
\includegraphics[width=10cm,angle=0]{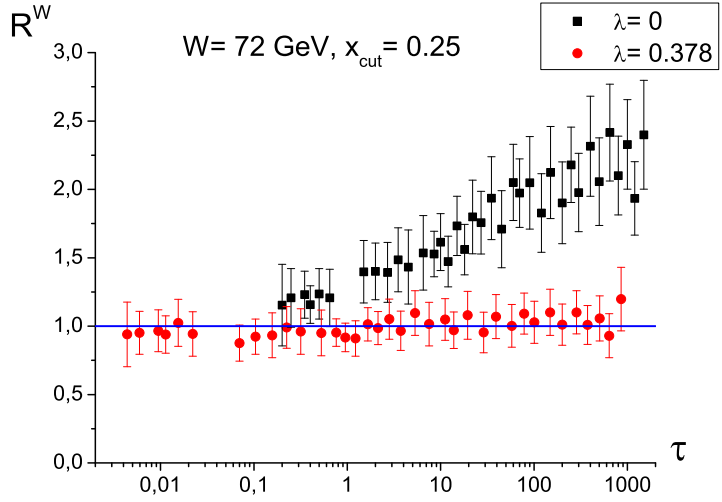}
\caption{Ratios $R^{72}_{i}(\lambda)$ for $\lambda=0$ and $\lambda=\lambda_{\rm{min}}=0.378$ both with $x_{\rm{cut}}=0.25$. $e^{+}p$ data are used.}
\label{zesWyk4}
\end{figure}

In Fig. \ref{zesWyk4} we show as an example two sets of ratios for $W=72$GeV plotted for $\lambda=0$ and $\lambda=0.378$. We see that some ratios decrease several times when we use scaling variable rather than $Q^2$.

Our aim is to find such $\lambda$ for given energy $W\neq W_{\rm{ref}}$ that deviations $R^W_{i}(\lambda)-1$ are minimal. Taking into account uncertainties $\Delta R^W_{i}(\lambda)$ we define $\chi^2$ function:

\begin{equation}
\chi^2(W,x_{\rm{cut}};\lambda):=\sum\limits_{i\in W; x \leq x_{\rm{cut}}} \frac{(R^W_{i}(\lambda)-1)^2}{(\Delta R^W_{i}(\lambda))^2},
\label{defchi2}
\end{equation}
where $i\in W; x \leq x_{\rm{cut}}$ means that we sum over points corresponding to given energy $W$ and values of $x$ are not larger than $x_{\rm{cut}}$. If for some $\lambda_{\ast}$ we have $\chi_W^2(\lambda_{\ast})=0$ this means that all points with $W$ lie exactly on $f_{\lambda}^{ref}(\tau)$.

We will search $\lambda_{\rm{min}}\left(W,x_{\rm{cut}}\right)$ which minimizes $\chi^2$ for given $W$ and $x_{\rm{cut}}$. We estimate uncertainty of $\lambda_{\rm{min}}$ requiring that: 
\begin{equation}
\chi^2(\lambda_{\rm{min}}\pm \Delta^\pm \lambda_{\rm{min}})-\chi^2(\lambda_{\rm{min}})=1,
\label{defdeltalamb}
\end{equation}
where we have omitted arguments $W$, $x_{\rm{cut}}$ to simplify notation. Uncertainty $\Delta^+ \lambda_{\rm{min}}$ (found for $\lambda > \lambda_{\rm{min}}$) is in general different than $\Delta^- \lambda_{\rm{min}}$ (found for $\lambda < \lambda_{\rm{min}}$).

\subsection{Bjorken-$x$ binning}
\label{metanalysB-x}

We will use the same method like for energy binning. All relevant quantities will be denoted with primes to be distinguished from the energy binning case. To find $\lambda'_{\rm{min}}$ we need to choose some $x'_{\rm{ref}}$ and its points will define reference curve $f^{\rm{ref'}}_{\lambda}$. It turns out, that it is not possible to choose one $x'_{\rm{ref}}$ such that domain of $f^{\rm{ref'}}_{\lambda}$ covers range of $\tau$ for all other $x'$ (see Fig. \ref{zesWyk28}). We can, however, choose some $x'_{\rm{ref}}$ and find $\lambda'_{\rm{min}}$ for all $x'$ which have at least points within domain of $f^{\rm{ref'}}_{\lambda}$. For given $x'_{\rm{ref}}$ we will use only such $x'$ that: 1) it has at least two points within $f^{\rm{ref'}}_{\lambda}$ domain and 2) $x'<x'_{\rm{ref}}$ (see below for justification). 

Having $f^{\rm{ref'}}_{\lambda}$ we can define ratios $R'_{i}$:
\begin{equation}
R'_{i}(\lambda):=\frac{\tilde{\sigma}^{x'}_i}{f_{\lambda}^{\rm{ref'}}(\tau'_i)},
\label{ratioXdef} 
\end{equation}
where $\tau'_i=Q_i x_i'^{\lambda} $ is a value of scaling variable for i'th point of $x'$ and $\tilde{\sigma}^{x'}_i$ is a value of cross section for this point.

Note that these ratios are defined as inverses of ratios for energy binning (see formula (\ref{ratiodef})). This can be explained by two facts:
\begin{itemize}
\item one should define ratios such that they are greater than 1 for $\lambda=0$ (this is because in definition of $\chi^2$ we have terms of the form $(R'-1)^2$ and for $R'<1$ they behaves differently than for $R'>1$ when we change $\lambda$).
\item number of points rises with values of $x'$ (see table \ref{tableXbjor+}). This means that we should require $x' < x'_{\rm{ref}}$ to ensure that $f_{\lambda}^{\rm{ref'}}$ can be constructed and has wide range of $\tau'$ values. In Fig. \ref{zesWyk28} one can see that if $x' < x'_{\rm{ref}}$ then $\tilde{\sigma}^{x'}_i > f_{\lambda}^{\rm{ref'}}(\tau'_i)$ \textit{i.e.} ratios $R'$ are greater than 1 if they are defined as in formula (\ref{ratioXdef}).
\end{itemize}

Uncertainties of ratios are given by:
\begin{equation}
\Delta R'_{i}(\lambda)=\sqrt{\left(\frac{\Delta \tilde{\sigma}^{x'}_i}{f_{\lambda}^{\rm{ref'}}(\tau_i)}\right)^2+
\left(\frac{\tilde{\sigma}^{x'}_i}{(f_{\lambda}^{\rm{ref'}}(\tau'_i))^2}\Delta f_{\lambda}^{\rm{ref'}}(\tau'_i) \right)^2}.
\label{uncRatX}
\end{equation}

We define also $\chi'^2$ function:
\begin{equation}
\chi'^2(x'_{\rm{ref}},x';\lambda):=\sum\limits_{i \in x'} \frac{(R'_{i}(\lambda)-1)^2}{(\Delta R'_{i}(\lambda))^2},
\end{equation}
where we sum over all $x'$ points which are in $f^{\rm{ref'}}_{\lambda}$ domain. $\lambda'_{\rm{min}}(x'_{\rm{ref}},x')$ is a minimum of $\chi'^2$, and its uncertainty is defined by formula (\ref{defdeltalamb}).

\chapter{Results for $e^{+}p$ data}

In this chapter we present results for $\lambda$ obtained when we method presented in section \ref{sectMethodoffinding} have applied to $e^{+}p$ data. 
Data are taken from Tables 6-13 from \cite{H1ZueusWork}.

\section{Energy binning}
\label{ePlresEB}

\subsection{$\lambda_{\rm{min}}(x_{\rm{cut}})$ for given $W$}
\label{deplammin}
\begin{figure}
\includegraphics[width=4.5cm,angle=0]{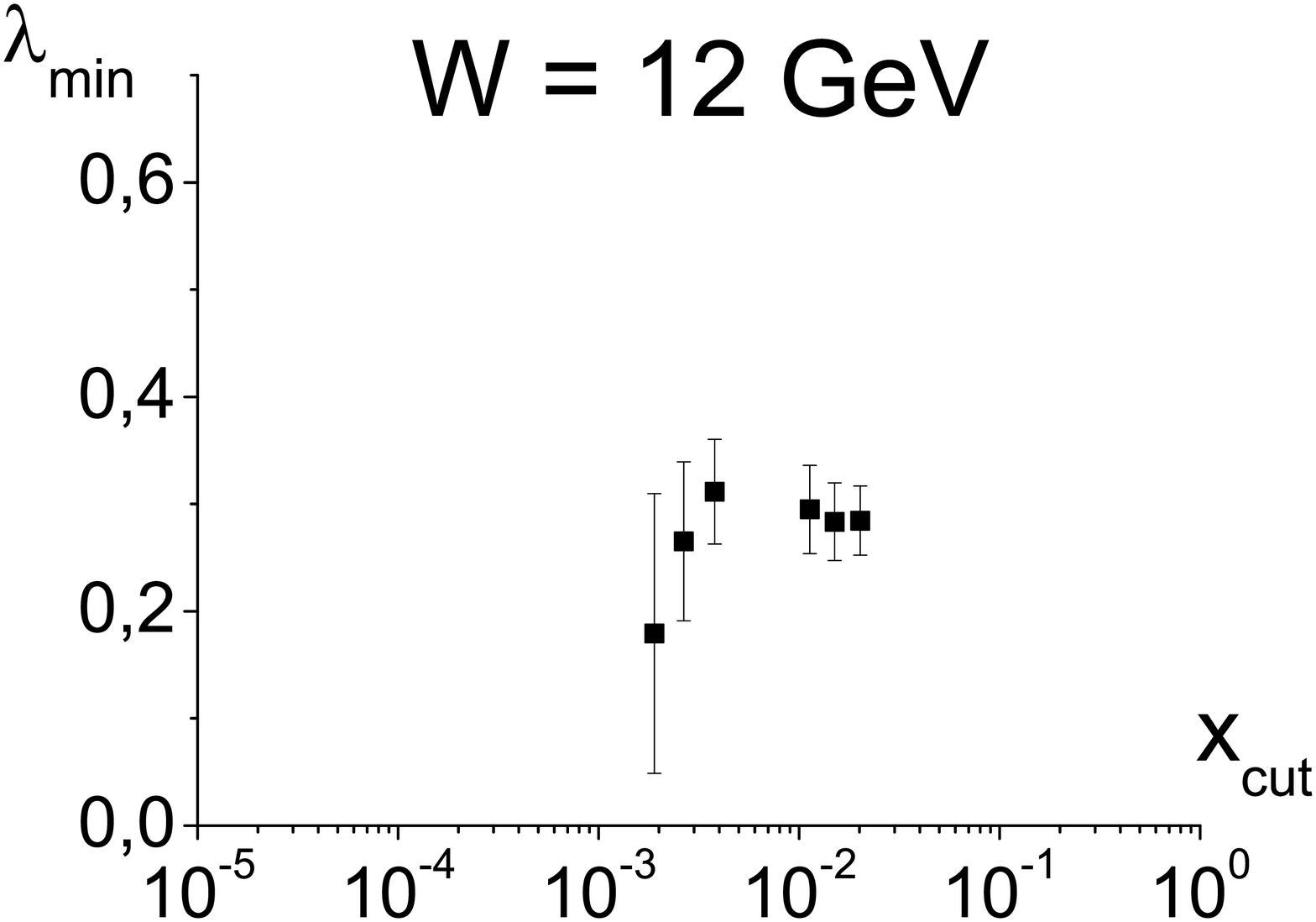}
\includegraphics[width=4.5cm,angle=0]{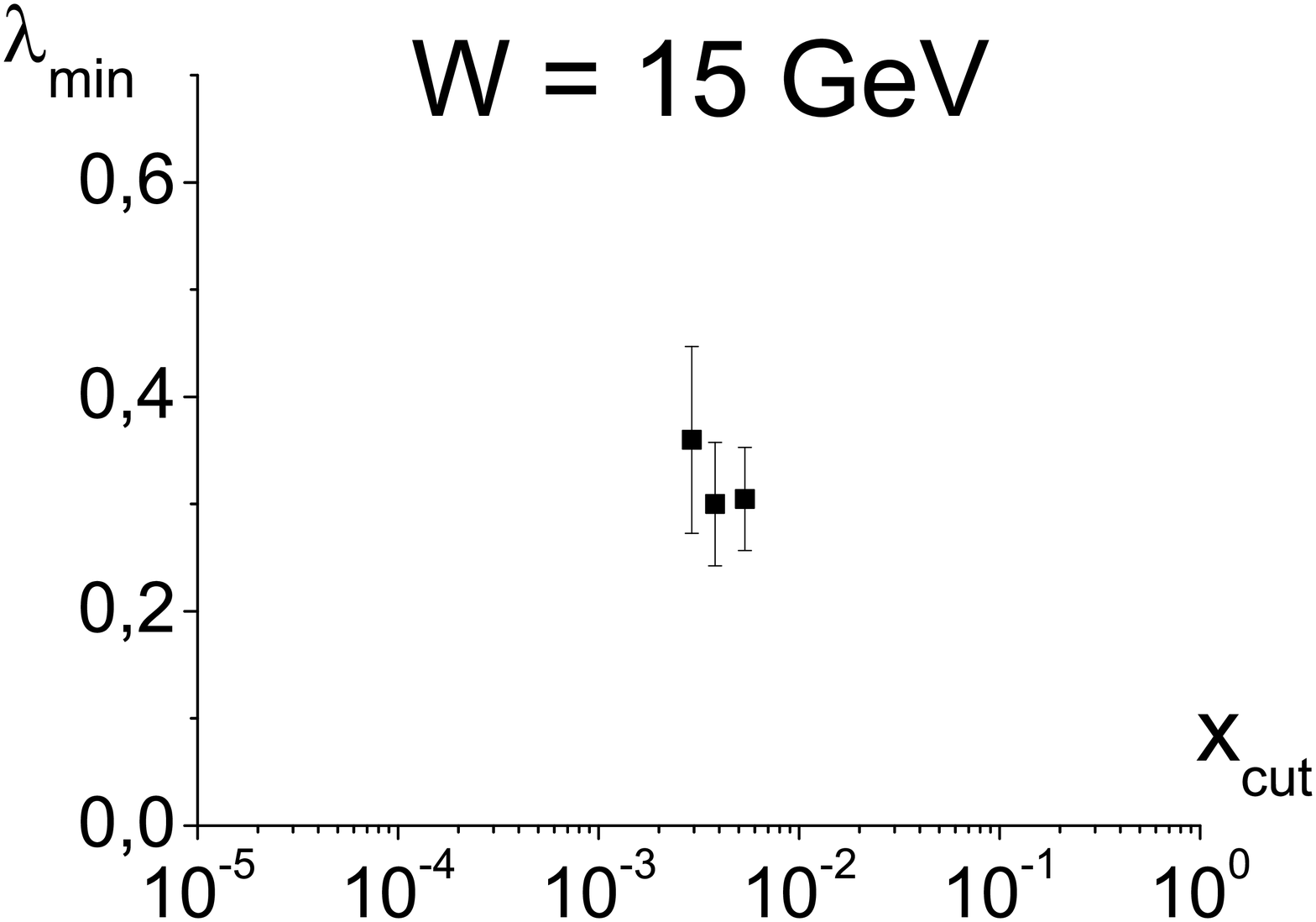}
\includegraphics[width=4.5cm,angle=0]{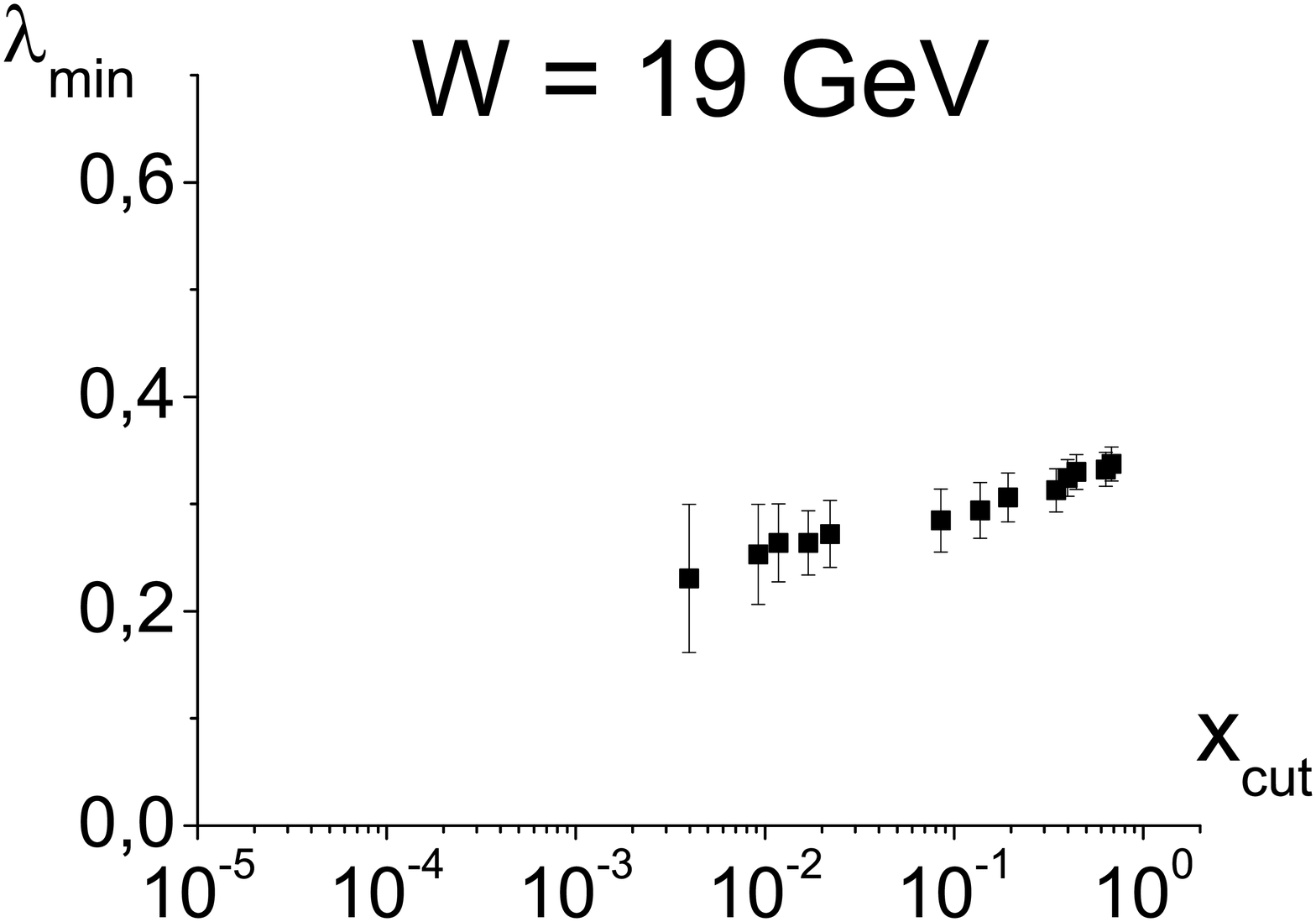}
\includegraphics[width=4.5cm,angle=0]{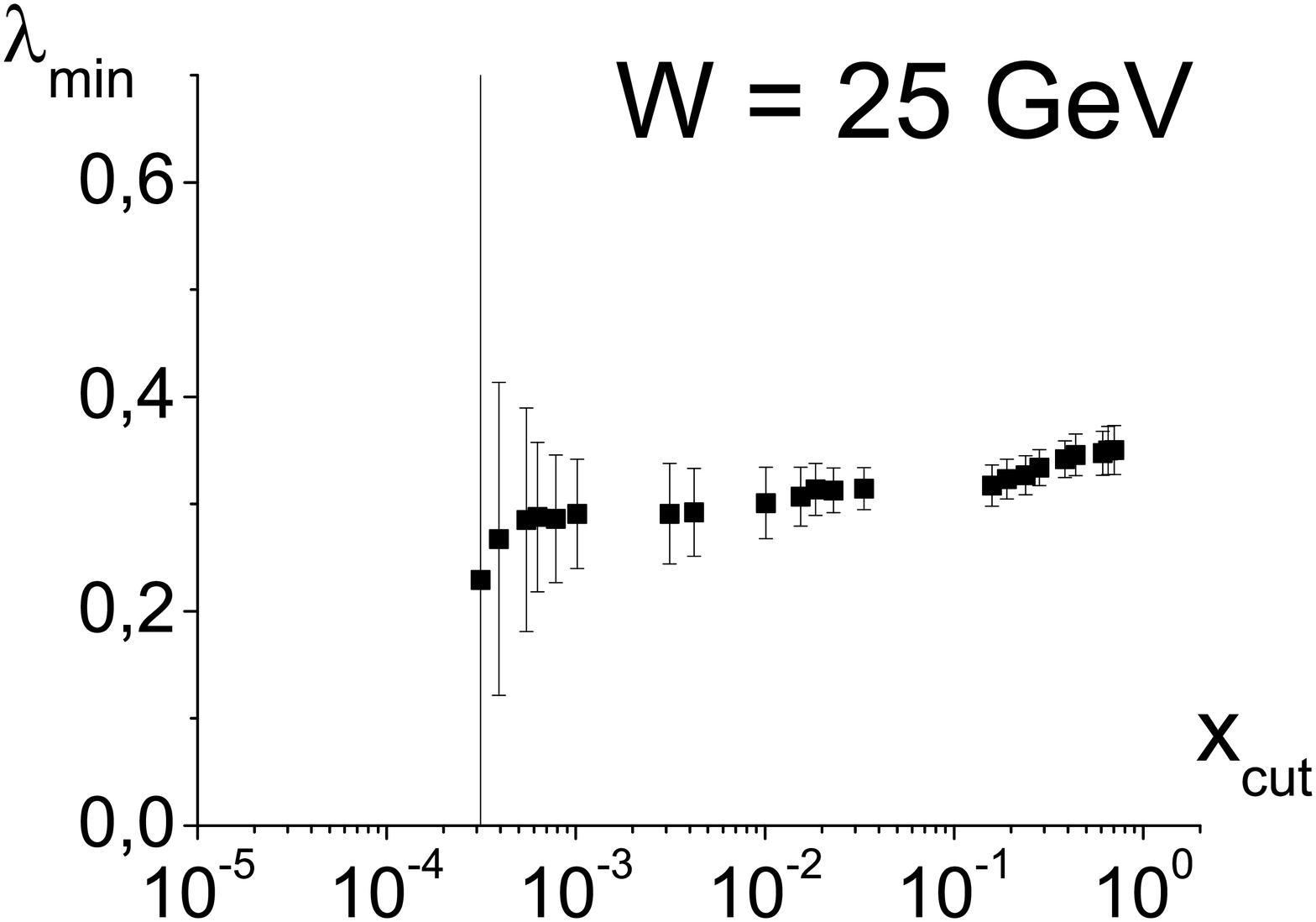}
\includegraphics[width=4.5cm,angle=0]{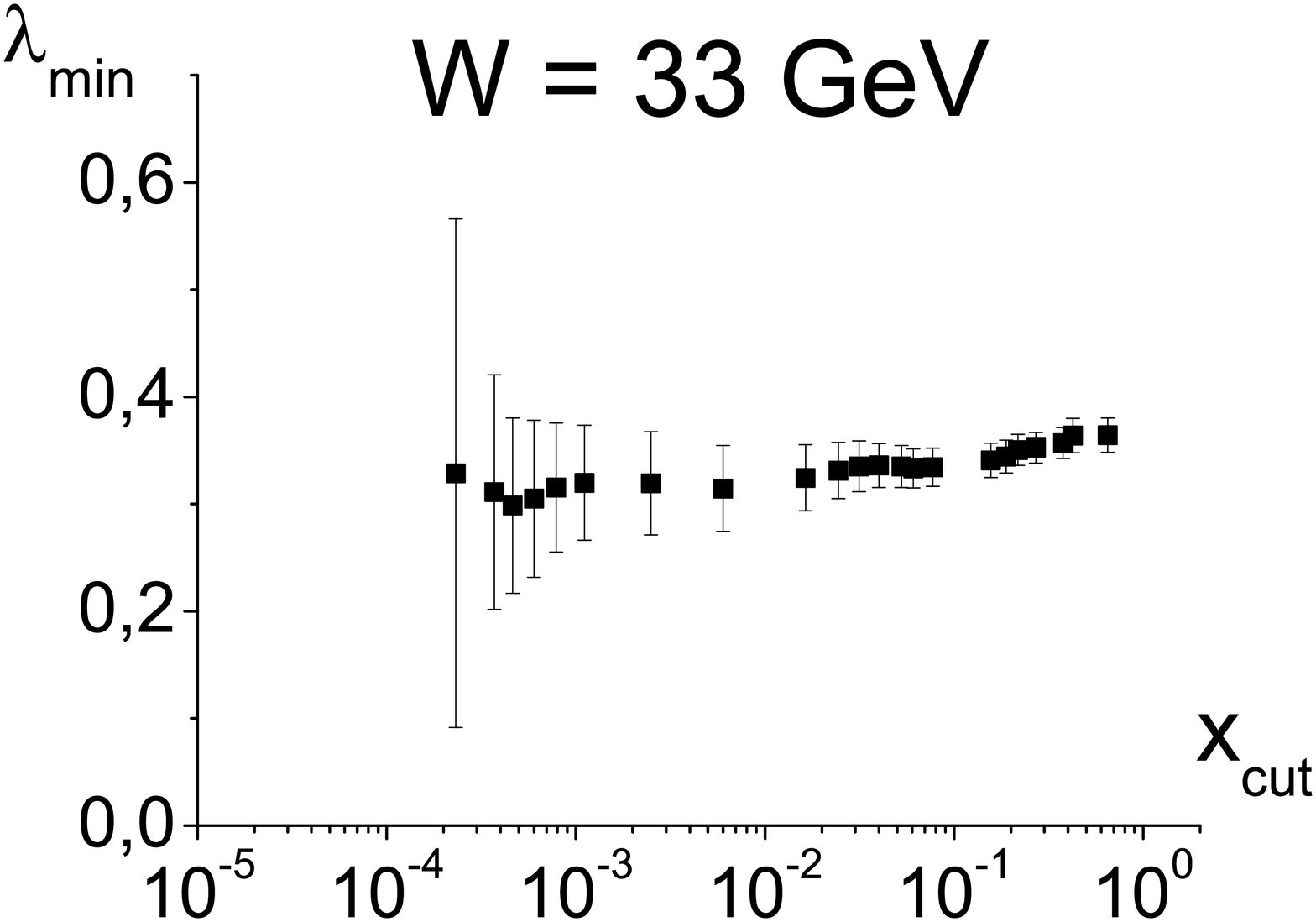}
\includegraphics[width=4.5cm,angle=0]{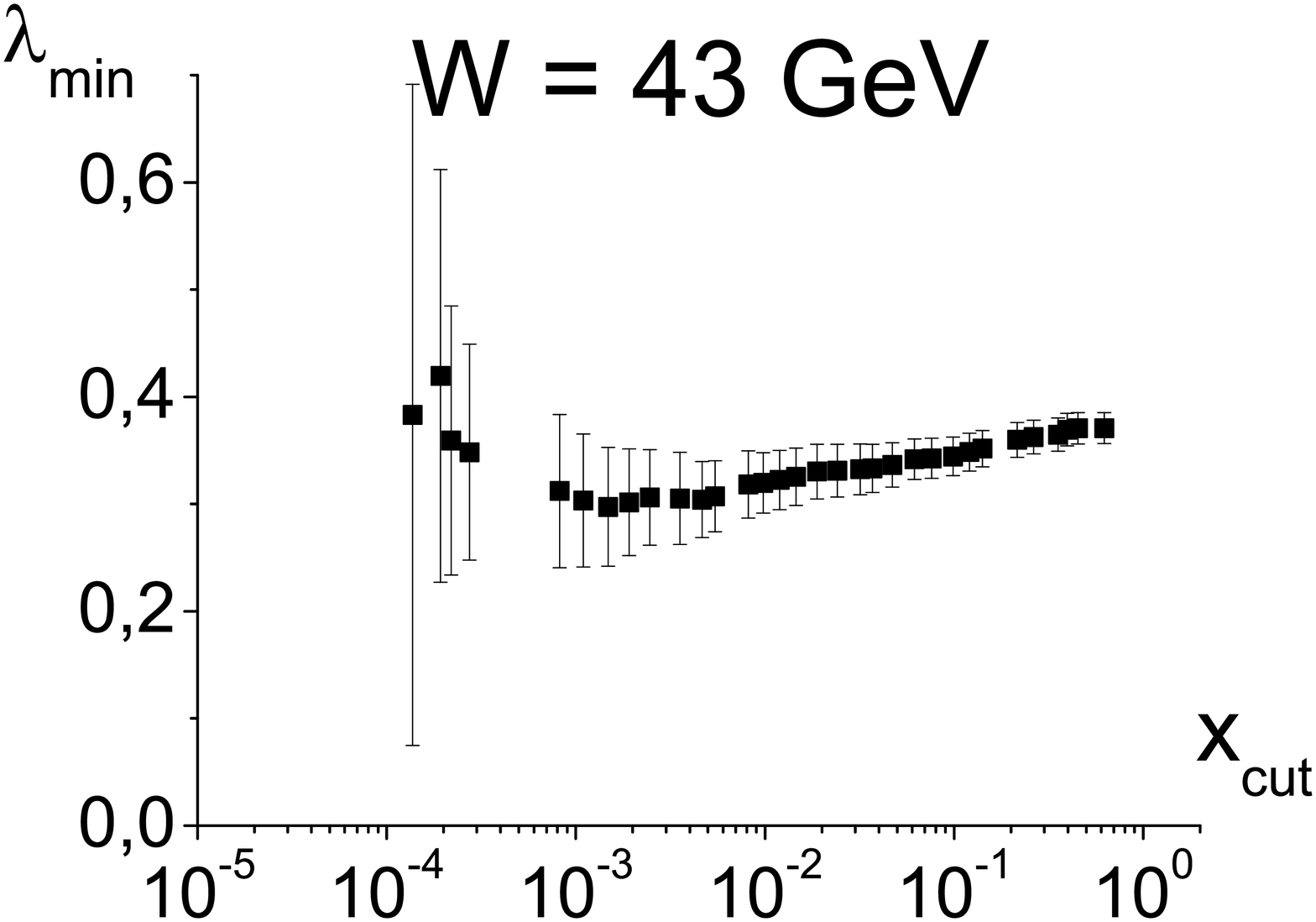}
\includegraphics[width=4.5cm,angle=0]{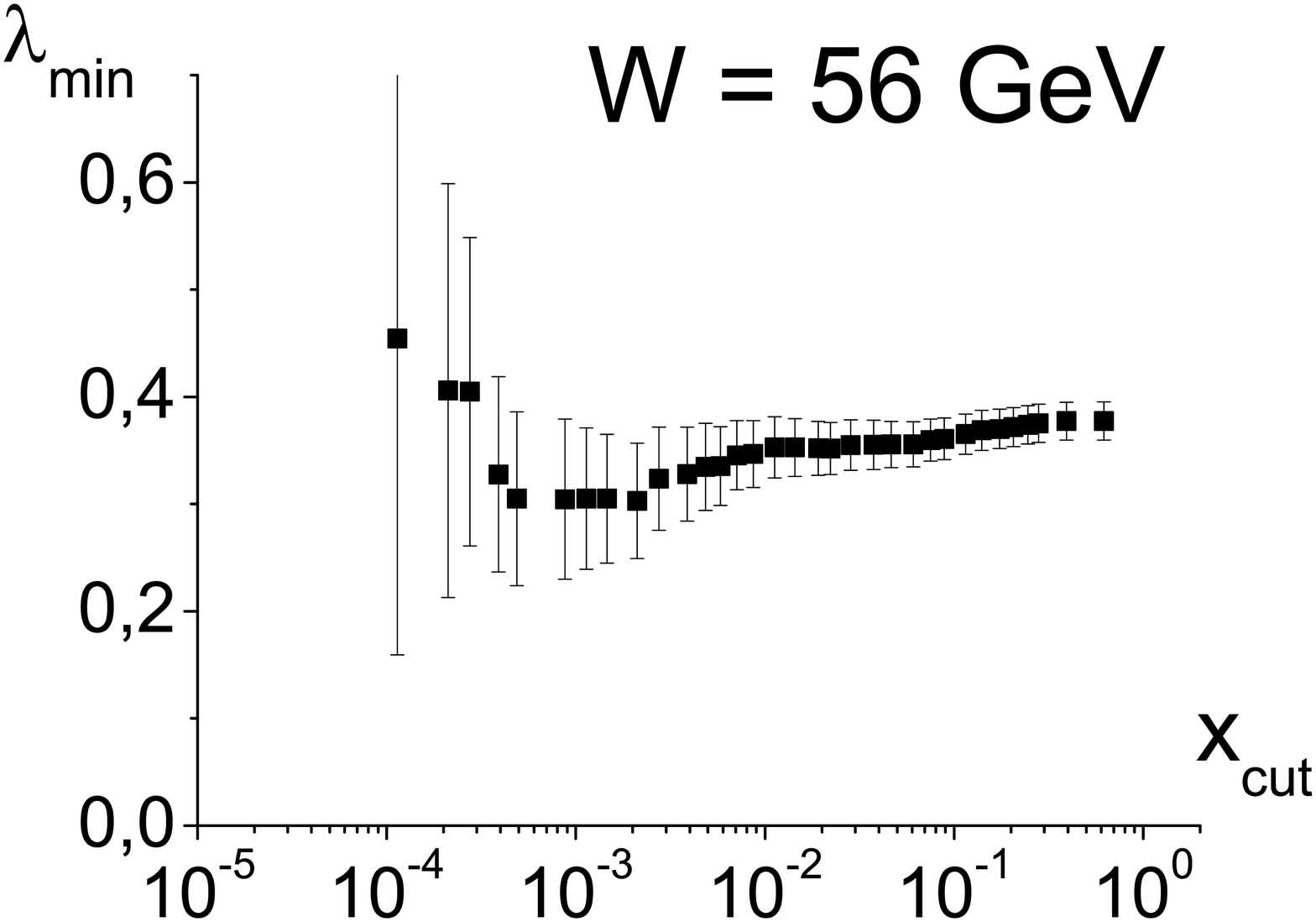}
\includegraphics[width=4.5cm,angle=0]{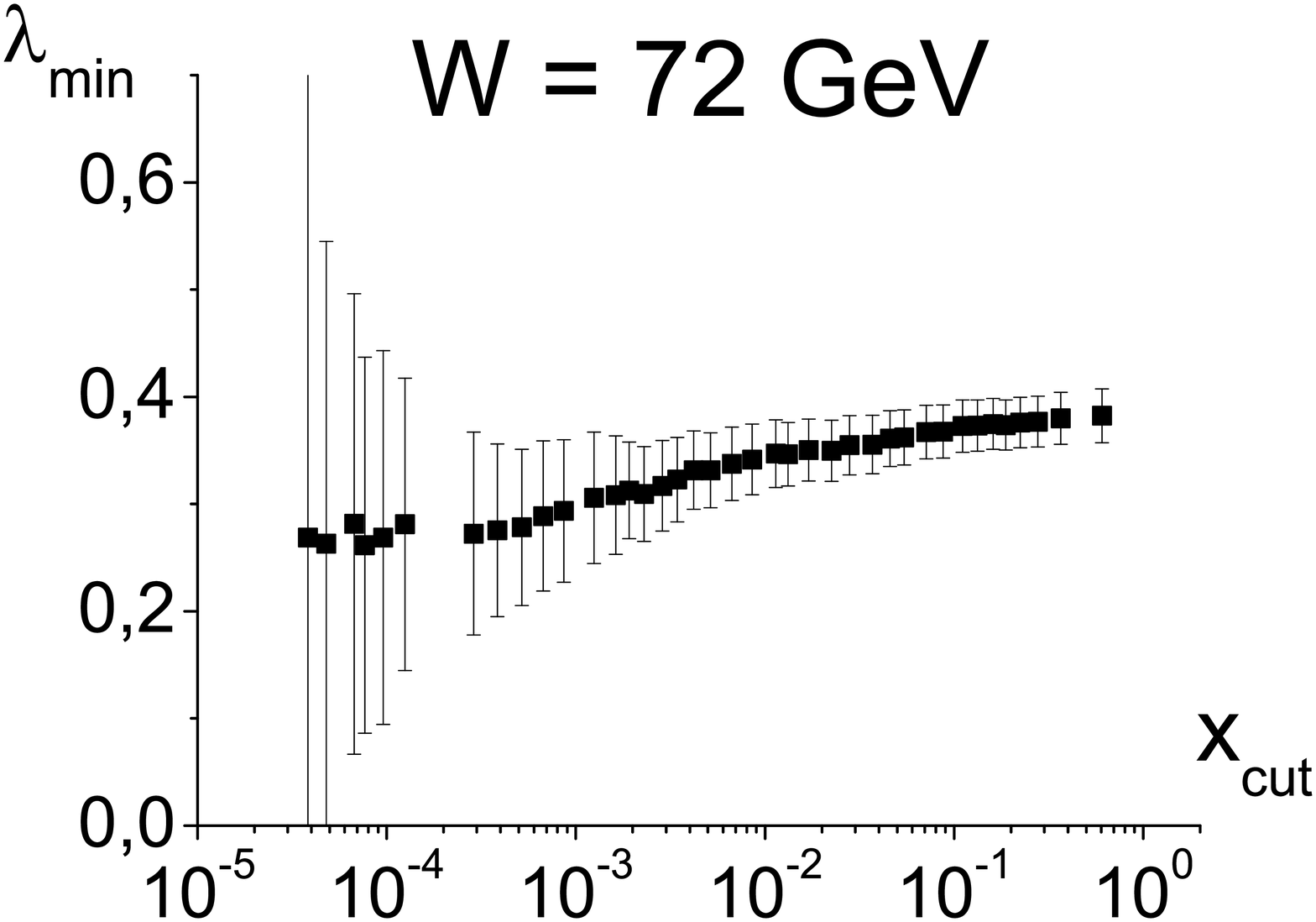}
\includegraphics[width=4.5cm,angle=0]{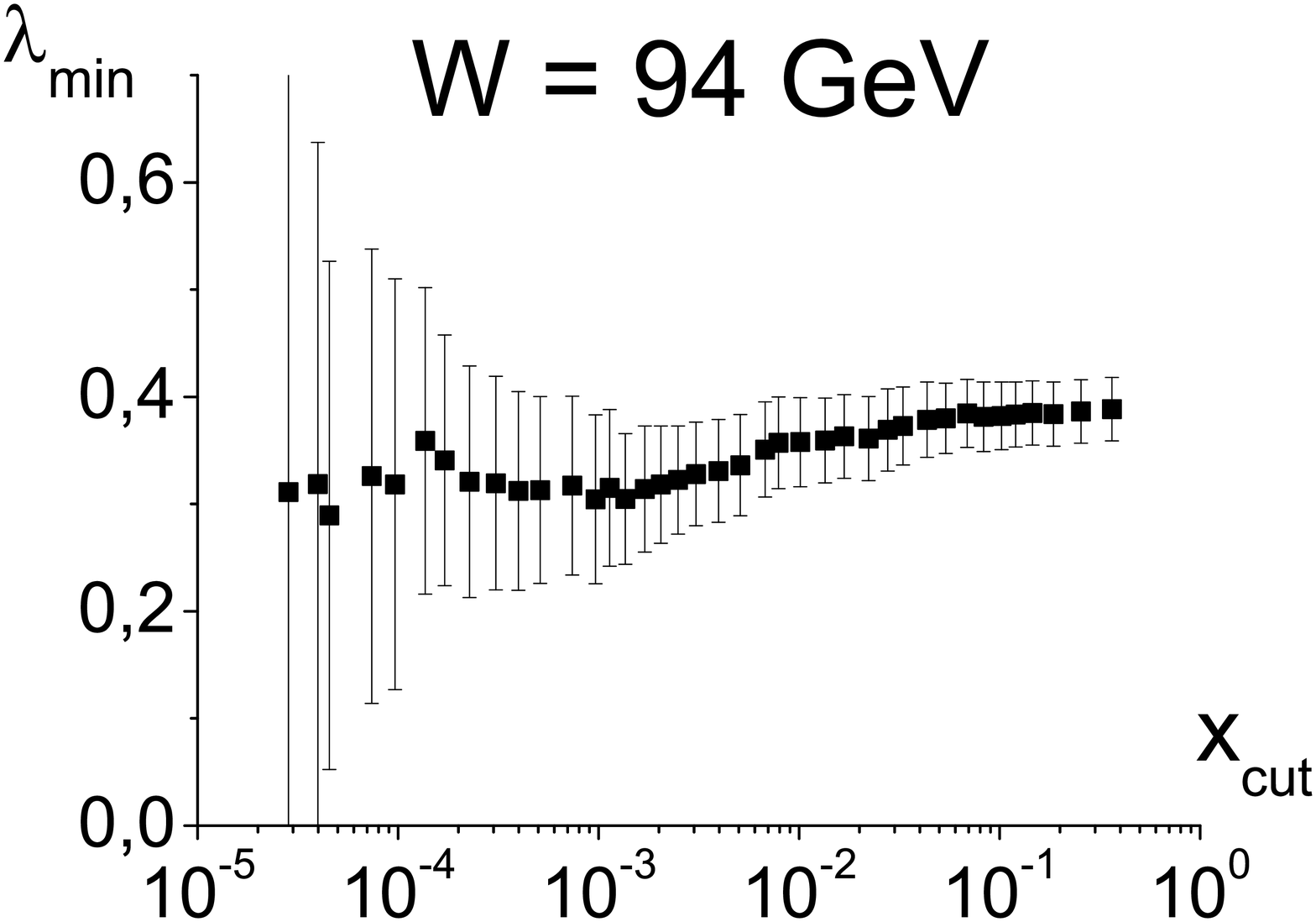}
\includegraphics[width=4.5cm,angle=0]{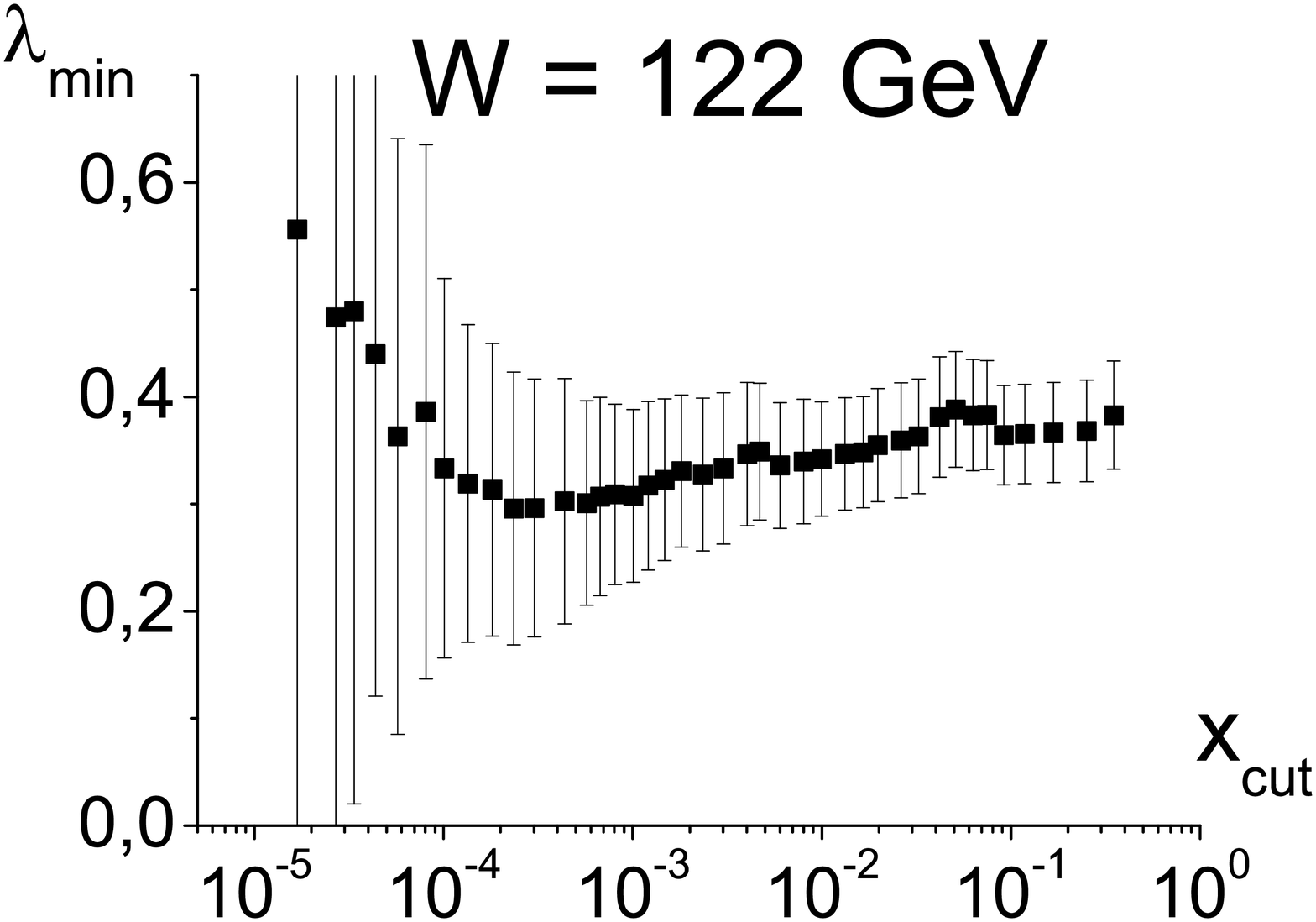}
\includegraphics[width=4.5cm,angle=0]{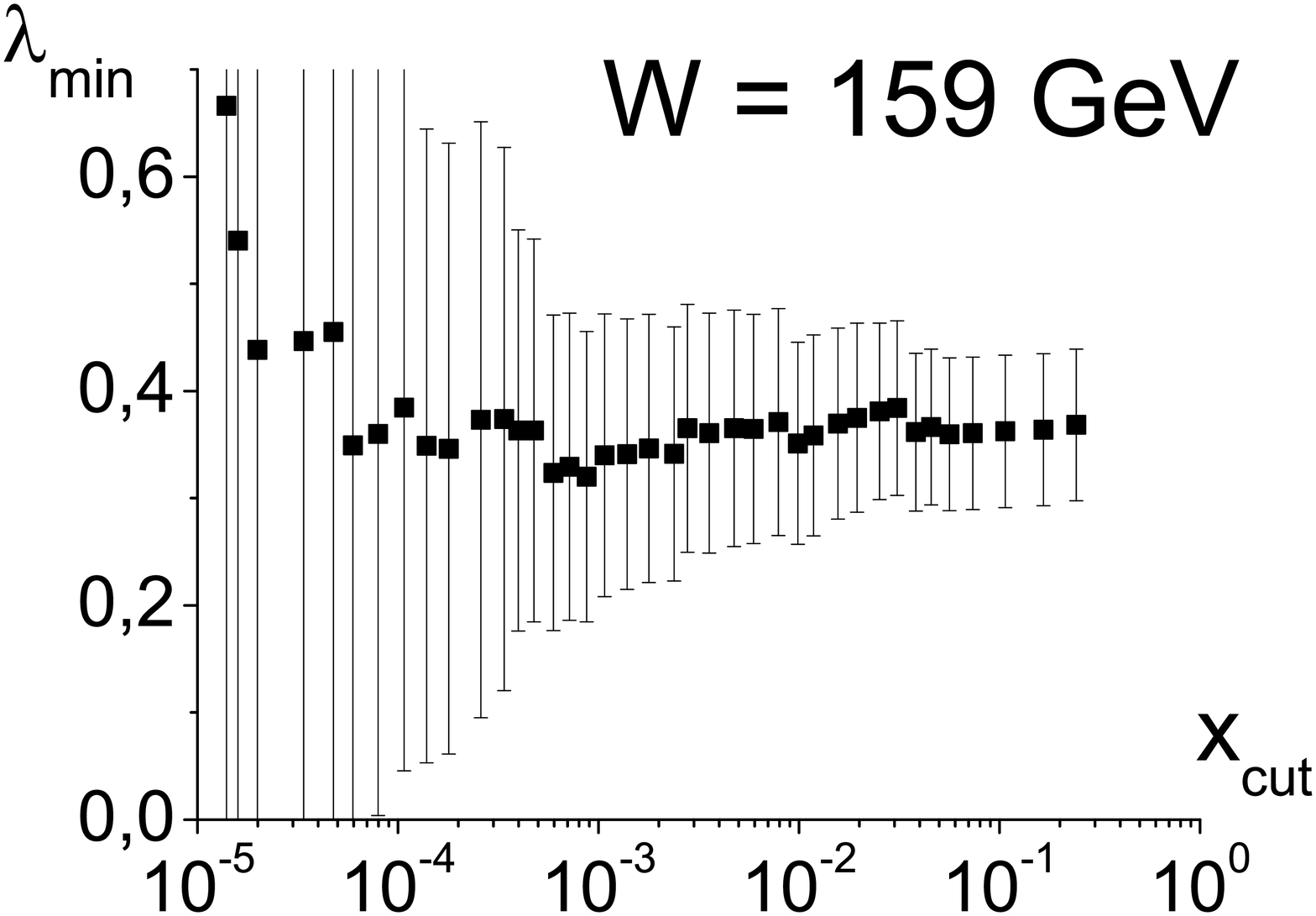}
\includegraphics[width=4.5cm,angle=0]{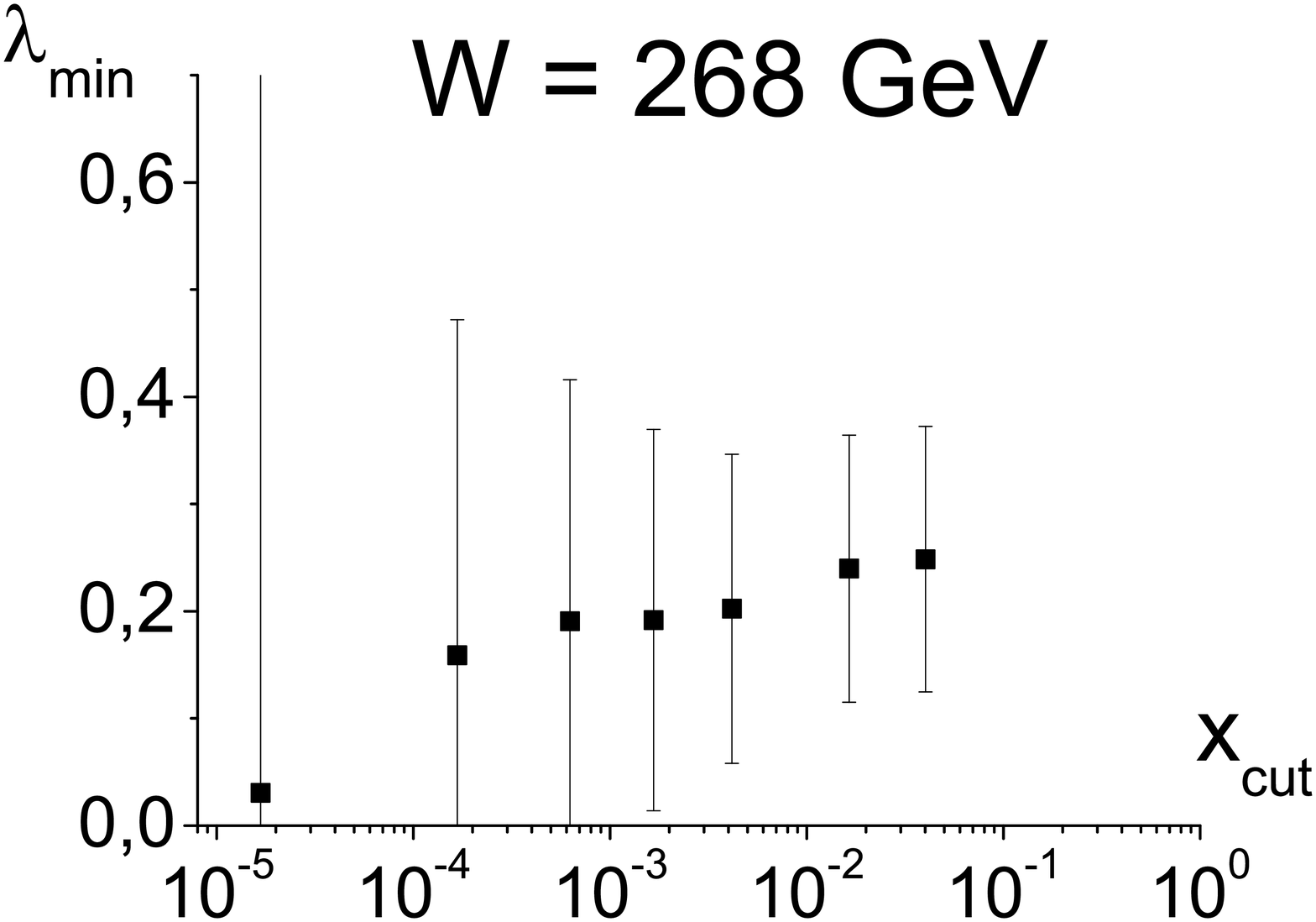}
\caption{$\lambda_{\rm{min}}$ as a function of $x_{\rm{cut}}$ for all energies $W\neq W_{\rm{ref}}=206$GeV}
\label{zesWyk5}
\end{figure}

In Fig. \ref{zesWyk5} we present plots of $\lambda_{\rm{min}}(x_{\rm{cut}})$. A given point on a plot represents $\lambda_{\rm{min}}$ calculated when we take into account only points with $x\leq x_{\rm{cut}}$.

It can be seen that for high $x_{\rm{cut}}$ uncertainties are much lower than for lower values. It is due to the bigger number of terms in the sum (\ref{defchi2}): the more terms we add the smaller deviation from $\lambda_{\rm{min}}$ is needed to change $\chi^2$ by 1. For energies $W$ which are close to $W_{\rm{ref}}$ uncertainties are larger. It is because for given $Q^2$ values of $x=\frac{Q^2}{Q^2+W^2-M^2}$ for $W$ and $W_{\rm{ref}}$ are similar, so values of scaling variable $\tau=Q^2 x^{\lambda}$ are also similar for wide range of $\lambda$ (this implies that $\chi^2(\lambda)$ is flat).  

\begin{figure}
\includegraphics[width=7cm,angle=0]{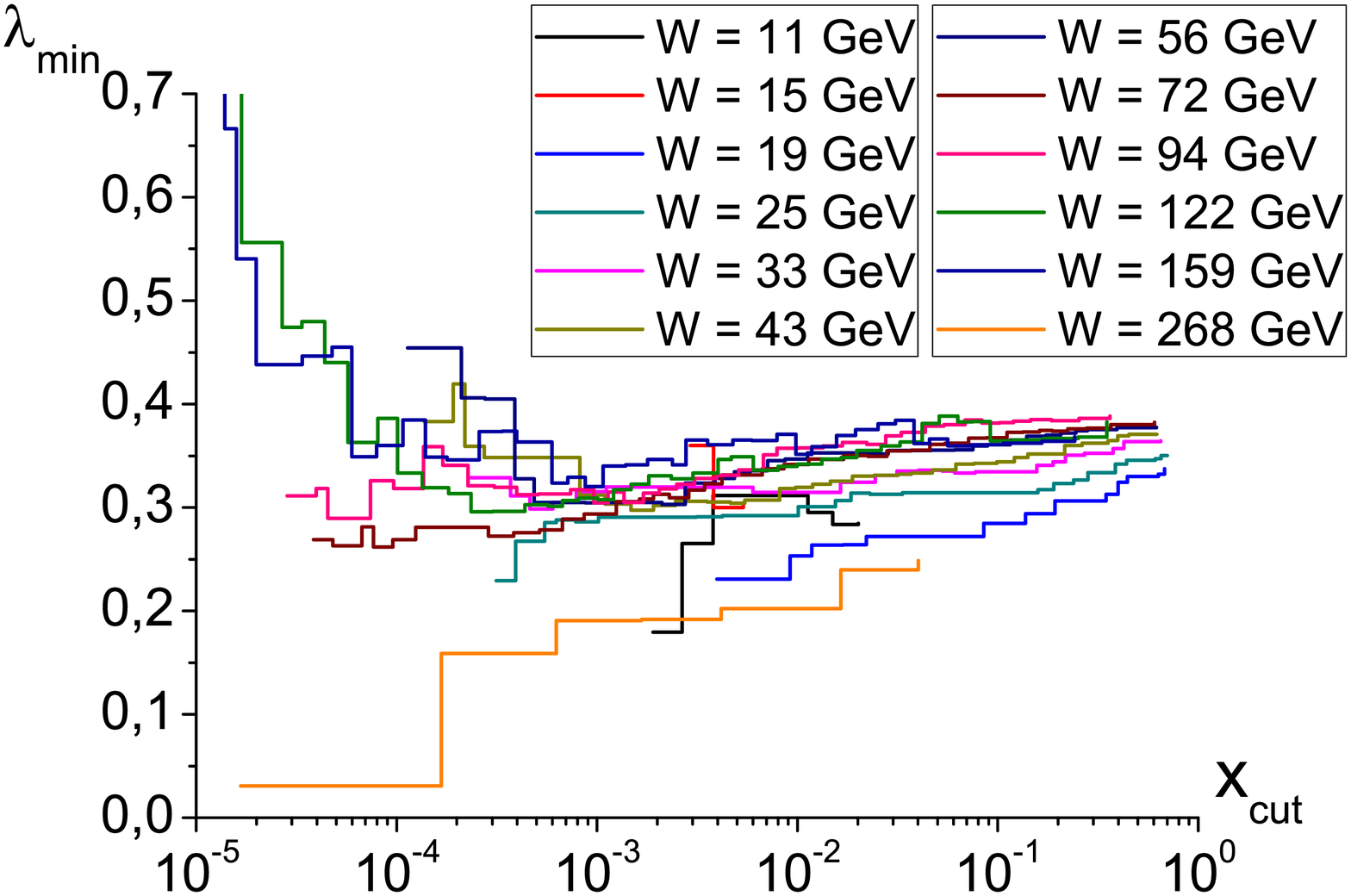}
\includegraphics[width=7cm,angle=0]{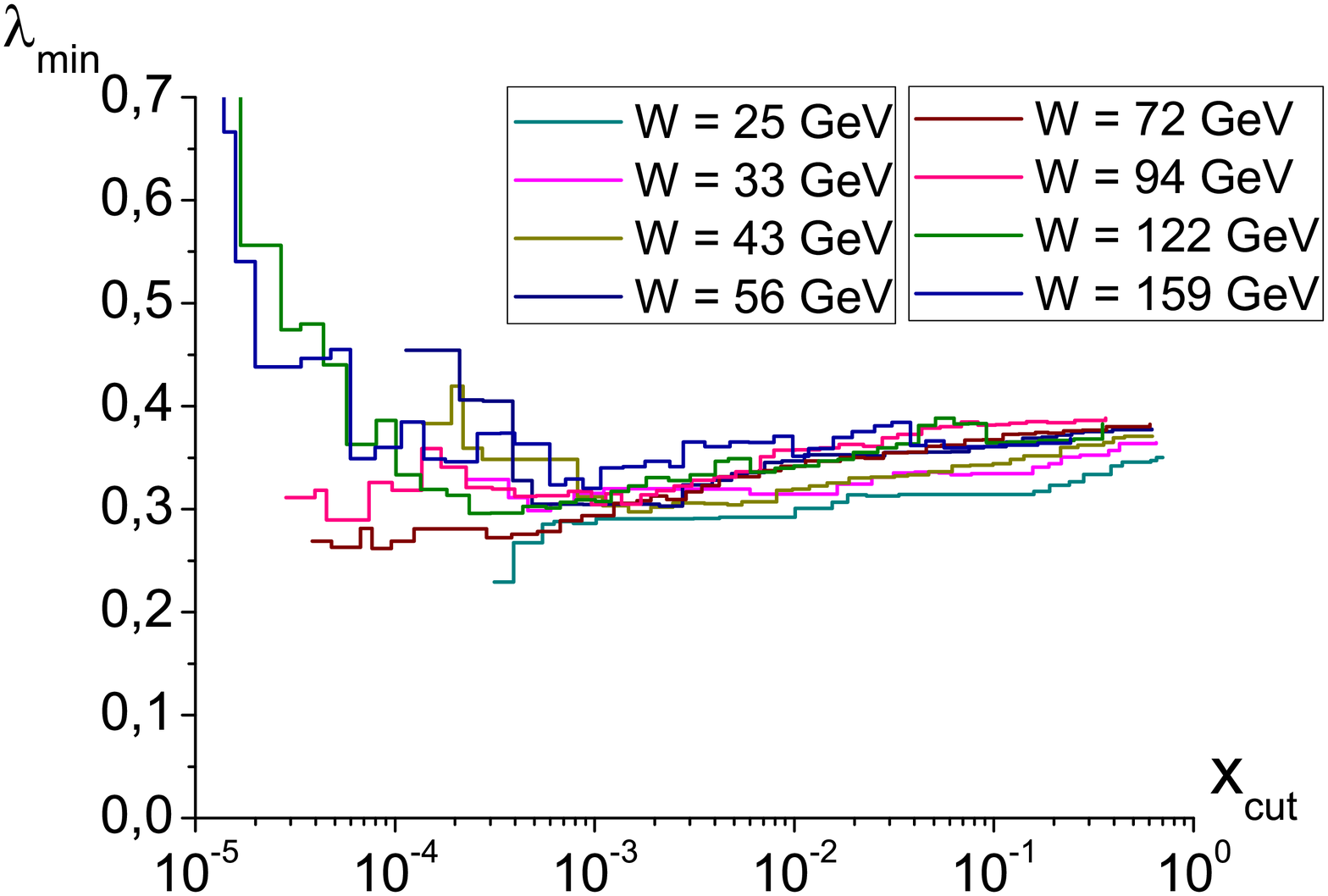}
\caption{$\lambda_{\rm{min}}$ as a function of $x_{\rm{cut}}$. Left plot presents all energies $W\neq W_{\rm{ref}}$, right one presents only energies $25$ GeV $\leq W \leq 159$ GeV. For clarity we have omitted uncertainties and joined points.}
\label{zesWyk6}
\end{figure}

In Fig. \ref{zesWyk6} merged plots from Figs. \ref{zesWyk5} without uncertainties are shown. Functions $\lambda_{\rm{min}}(x_{\rm{cut}})$ for $W$ = 12 , 15, 19 are separated from others. We will exclude them from our further analysis. We exclude also $W$ = 268 GeV because there are only few points for this energy and $\lambda_{\rm{min}}$ has very big uncertainties, moreover, for this energy ratios are smaller than 1 (see discussion in section \ref{metanalysB-x}).

We can see that for $x_{\rm{cut}}\leq 0.001$ functions $\lambda_{\rm{min}}$ for different $W$ diverge. In this region, however, uncertainties (caused by small number of points) are too large to compare these functions (see subsection \ref{sectlamdepW} where we compare $\lambda_{\rm{min}}$ for different $W$). To perform more precise analysis one should have better data (more points) in this region.
         
\subsection{$\chi^2_{\rm{nor}}(x_{\rm{cut}})$ for given $W$}
\label{sectchi^2nor}

$\chi^2$ defined in (\ref{defchi2}) gives us information how far points with energy $W$ lie from the curve $f_{\lambda}^{\rm{ref}}$ determined by points with energy $ W_{\rm{ref}}$. It rises with number of terms in the sum. Now we define "normalized" $\chi^2$ function:   
\begin{eqnarray}
\chi^2_{\rm{nor}}(W,x_{\rm{cut}};\lambda):=\frac{1}{N(W,x_{\rm{cut}})} \chi^2\left(W,x_{\rm{cut}};\lambda\right)= \nonumber \\
=\frac{1}{N(W,x_{\rm{cut}})} \sum\limits_{i\in W; x \leq x_{\rm{cut}}} \frac{\left[R^W_{i}\left(\lambda\right)-1\right]^2}{\left[\Delta R^W_{i}\left(\lambda \right)\right]^2},
\label{chinordef}
\end{eqnarray}
where $i\in W; x \leq x_{\rm{cut}}$ means that we sum over points with energy $W$ and values of $x$ are not larger than $x_{\rm{cut}}$; $N(W,x_{\rm{cut}})$ is the number of such points. 

\begin{figure}
\includegraphics[width=7cm,angle=0]{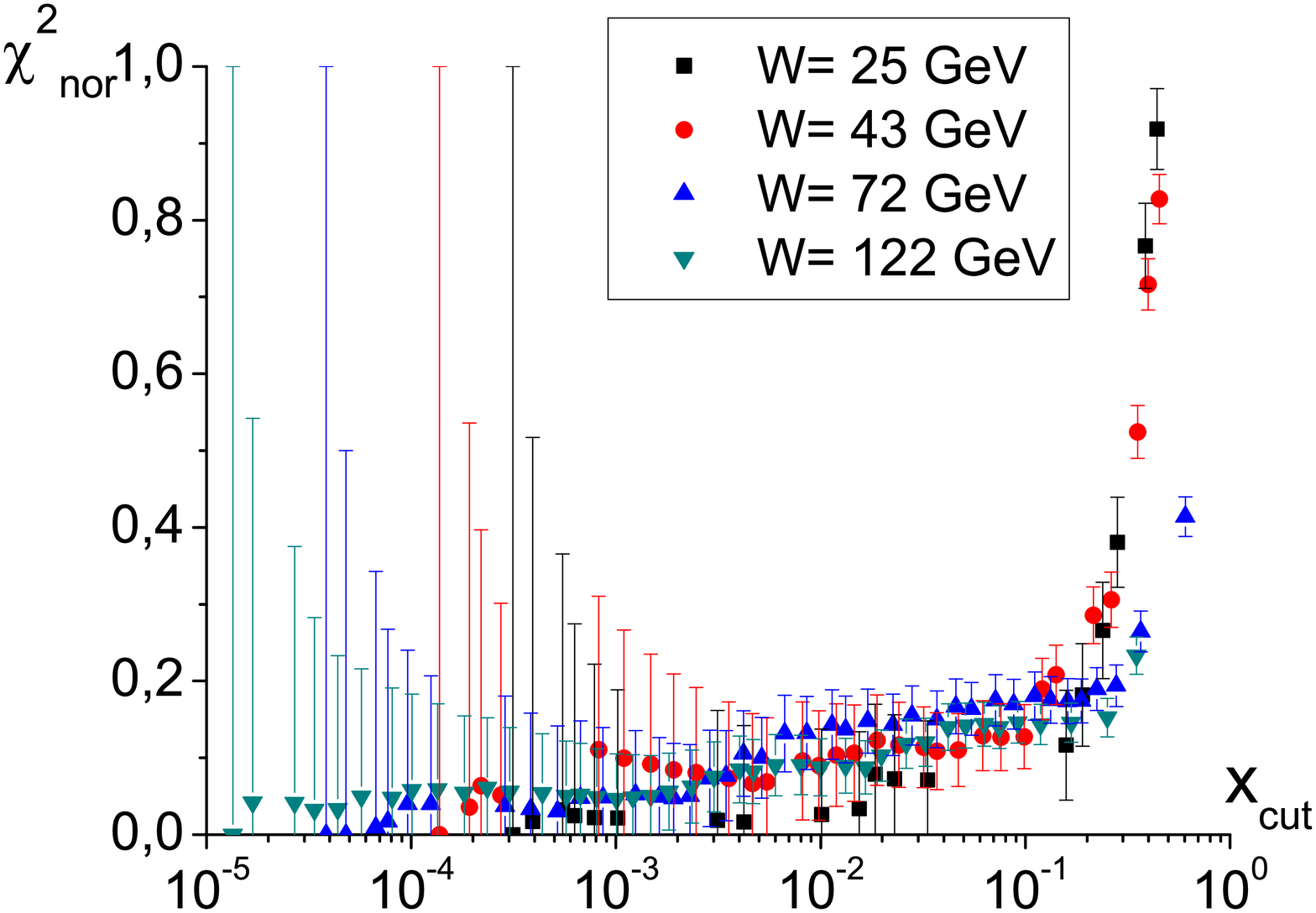}
\includegraphics[width=7cm,angle=0]{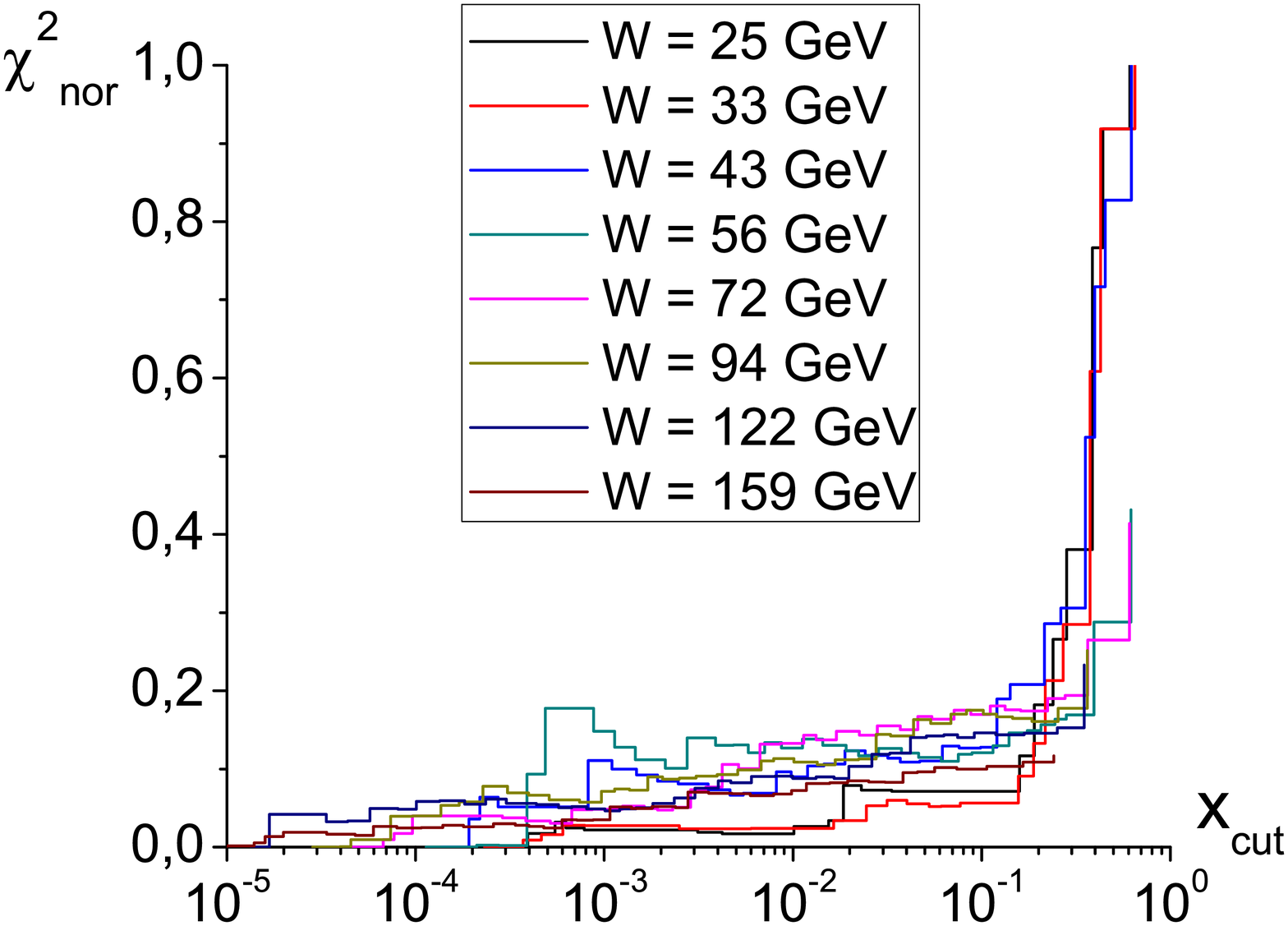}
\caption{$\chi^2_{\rm{nor}}(\lambda_{\rm{min}})$ as a function of $x_{\rm{cut}}$ for various $W$: left plot - for several energies, with uncertainties; right plot - for all energies $25$ GeV $\leq W \leq 159$ GeV, without uncertainties.} 
\label{zesWyk7}
\end{figure}

$\chi^2_{\rm{nor}}$ measures average deviation of points from the curve $f_{\lambda}^{\rm{ref}}$. $\lambda_{\rm{min}}(W,x_{\rm{cut}})$ was defined as a minimum of $\chi^2(W,x_{\rm{cut}};\lambda)$ so it is also a minimum of $\chi^2_{\rm{nor}}(W,x_{\rm{cut}};\lambda)$. We will concentrate on minimal values $\chi^2_{\rm{nor}}(W,x_{\rm{cut}}; \lambda_{\rm{min}}(W,x_{\rm{cut}}))$ and compare its values for various $x_{\rm{cut}}$ and $W$ (see Fig. \ref{zesWyk7}). We also define uncertainty:
\begin{equation}
\Delta \chi^2_{\rm{nor}}(\lambda_{\rm{min}}):=\chi^2_{\rm{nor}}(\lambda_{\rm{min}}+\Delta^{+} \lambda_{\rm{min}})-\chi^2_{\rm{nor}}(\lambda_{\rm{min}})= \frac{1}{N(W,x_{\rm{cut}})}.
\label{niepchinordef}
\end{equation}
Second equality comes from relation (\ref{defdeltalamb}).

In Fig. \ref{zesWyk7} we can see that one can distinguish two regions of $x_{\rm{cut}}$:
\begin{itemize}
\item $x_{\rm{cut}}<0.2$: values of $\chi^2_{\rm{nor}}(\lambda_{\rm{min}})$ are small and they rise slowly with $x_{\rm{cut}}$. In this region 
GS is present. For very small $x_{\rm{cut}}$ our analysis breaks down because of small number of points.
\item $x_{\rm{cut}}>0.2$: $\chi^2_{\rm{nor}}(\lambda_{\rm{min}})$ rise rapidly with $x_{\rm{cut}}$. GS is violated.
\end{itemize}

\subsection{$W$ dependence of $\lambda_{\rm{min}}$}
\label{sectlamdepW}

\begin{figure}
\includegraphics[width=7cm,angle=0]{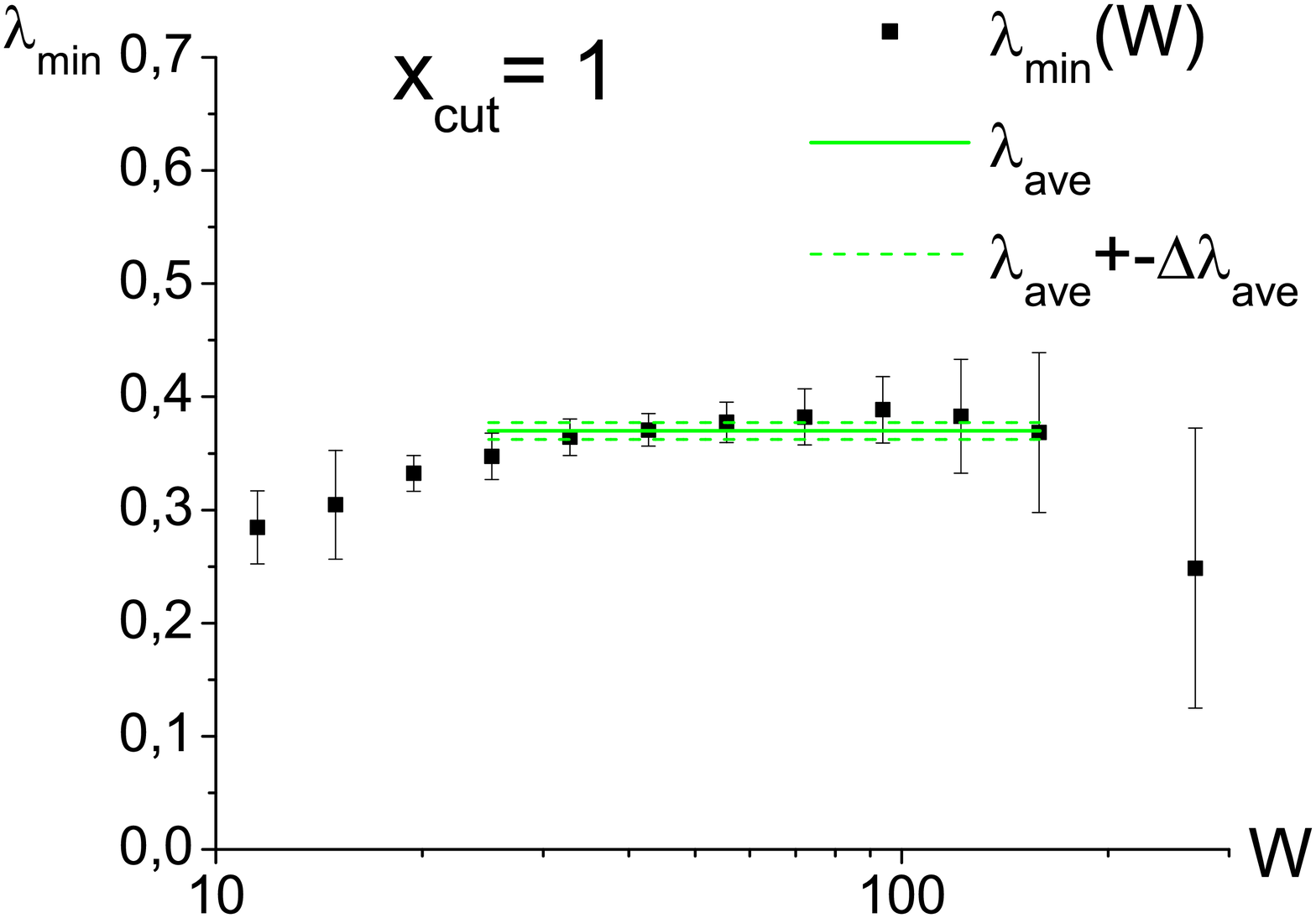}
\includegraphics[width=7cm,angle=0]{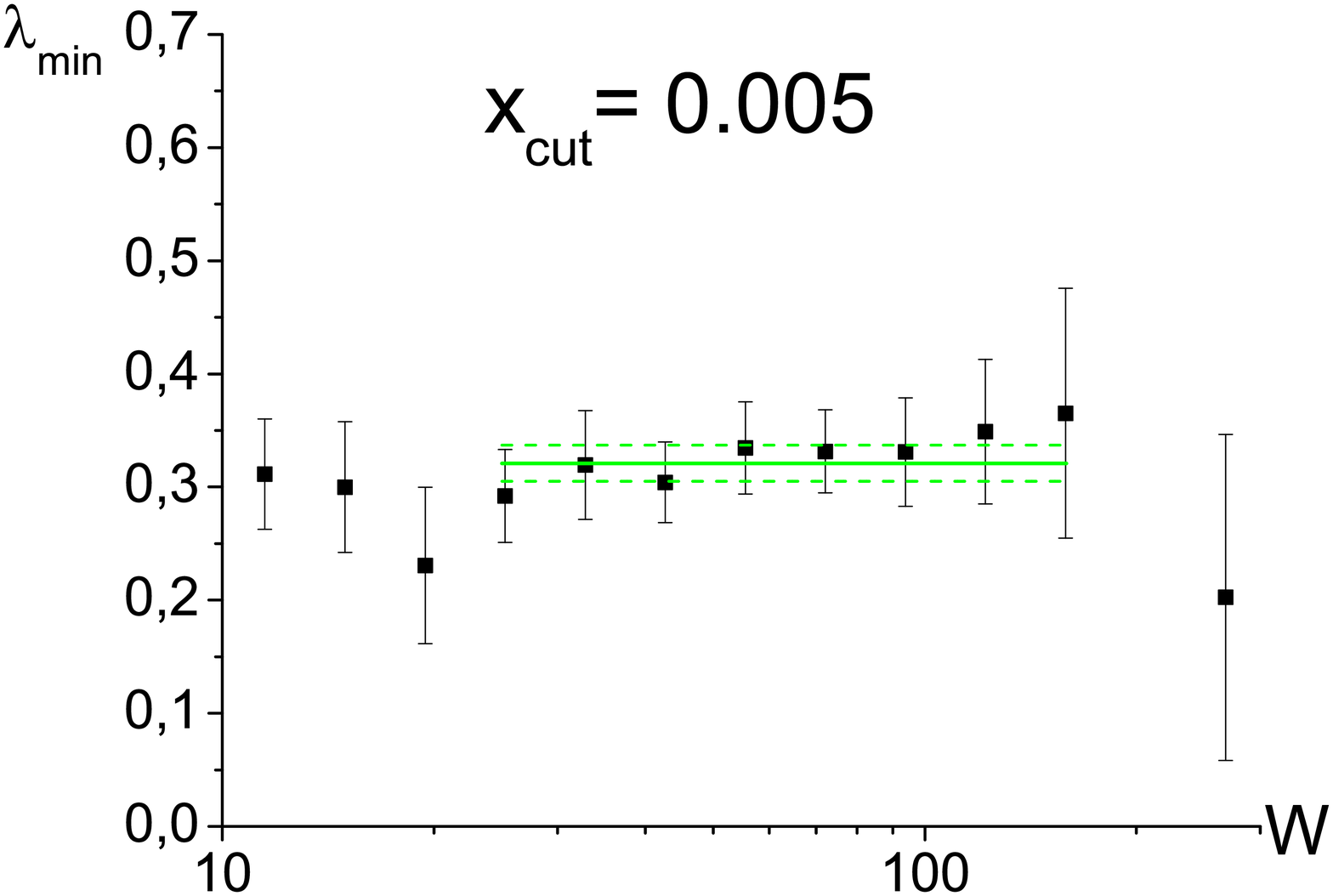}
\includegraphics[width=7cm,angle=0]{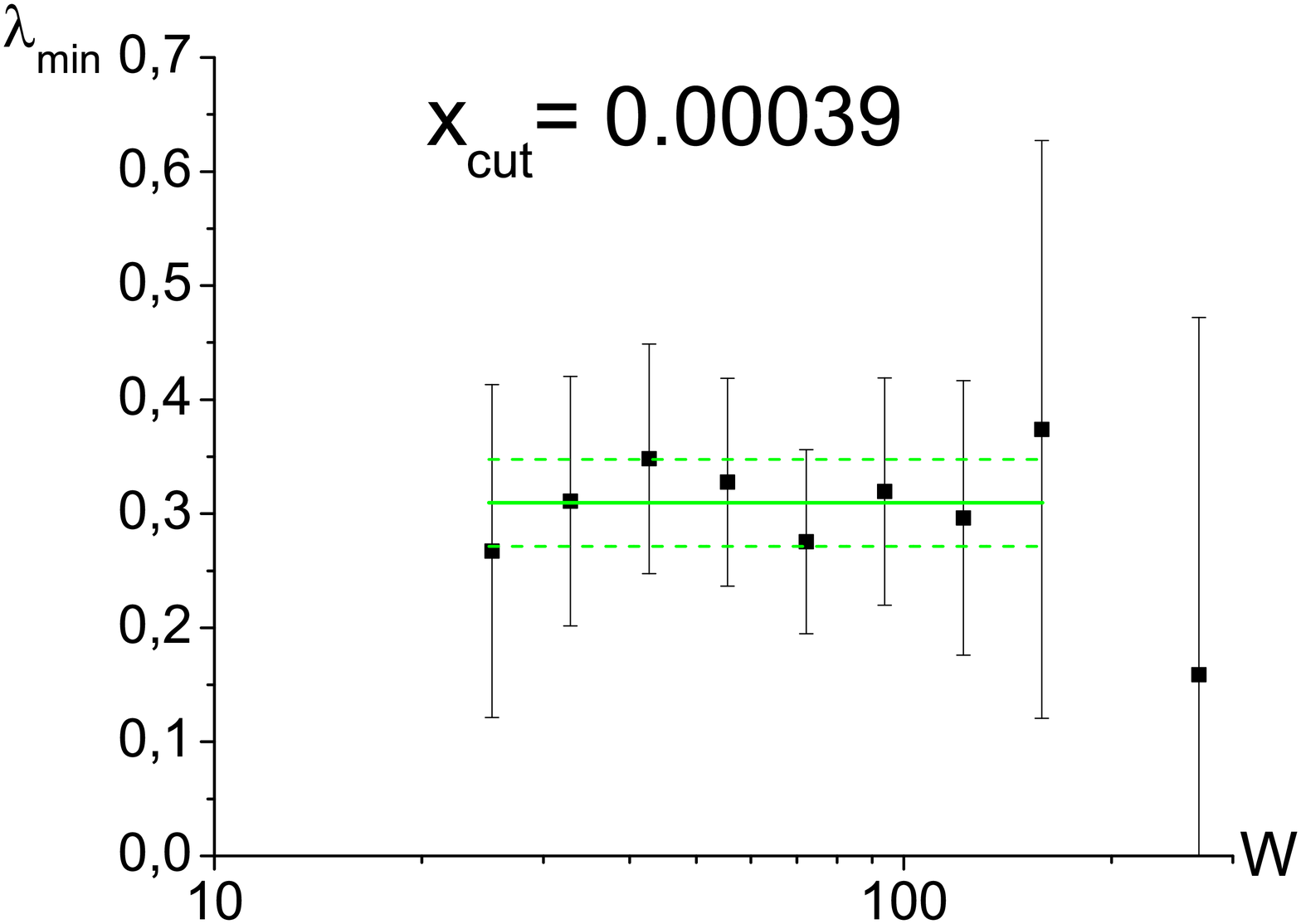}
\includegraphics[width=7cm,angle=0]{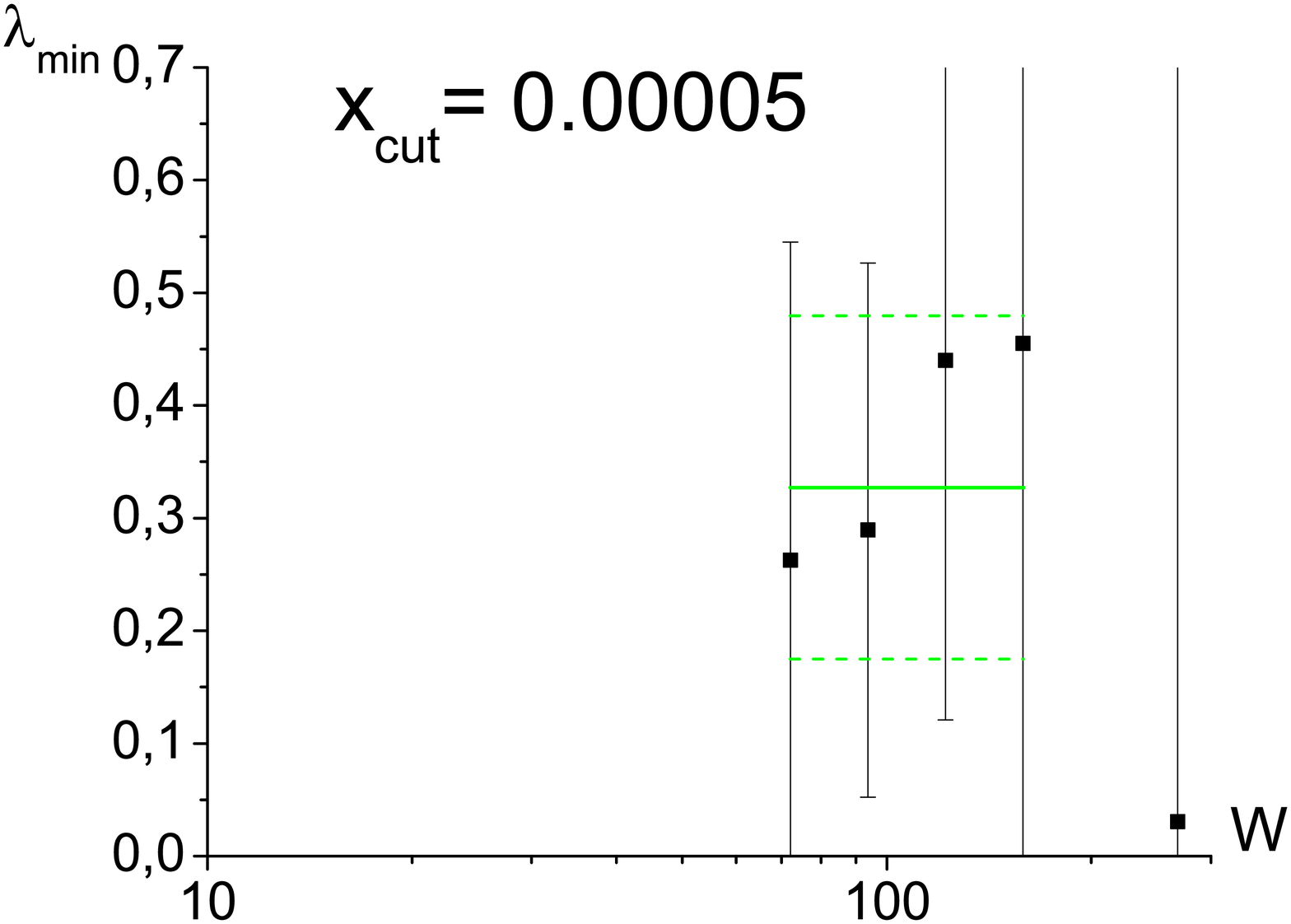}
\caption{$\lambda_{\rm{min}}$ as a function of $W$ for several values of $x_{\rm{cut}}$. Horizontal lines represents $\lambda_{\rm{ave}}$ and its uncertainties, they are plotted for energies which were used to calculate them ($25$GeV$\leq W \leq 159$GeV if all present).}
\label{zesWyk8}
\end{figure}

In Fig. \ref{zesWyk8} we show $\lambda_{\rm{min}}$ as a function of $W$ for several values of $x_{\rm{cut}}$ (black points). When we consider very small $x_{\rm{cut}}$ there are no points for some energies so one cannot find $\lambda_{\rm{min}}$.

For large $x_{\rm{cut}}$ one can see small $W$ dependence: $\lambda_{\rm{min}}$ rises with energy for $W < 19$ GeV and is constant for $ 25 \leq W \leq 159$ GeV (averaged value $\lambda_{\rm{ave}}$ and its uncertainties are shown by horizontal lines - see next subsection for more details). For small $x_{\rm{cut}}$ there is no $W$ dependence. This suggests that when GS is present $\lambda$ exponent does not depend on energy.
  
\subsection{Average over $W$: $\lambda_{\rm{ave}}(x_{\rm{cut}})$}
\label{sectlamave}

\begin{figure}
\centering
\includegraphics[width=12cm,angle=0]{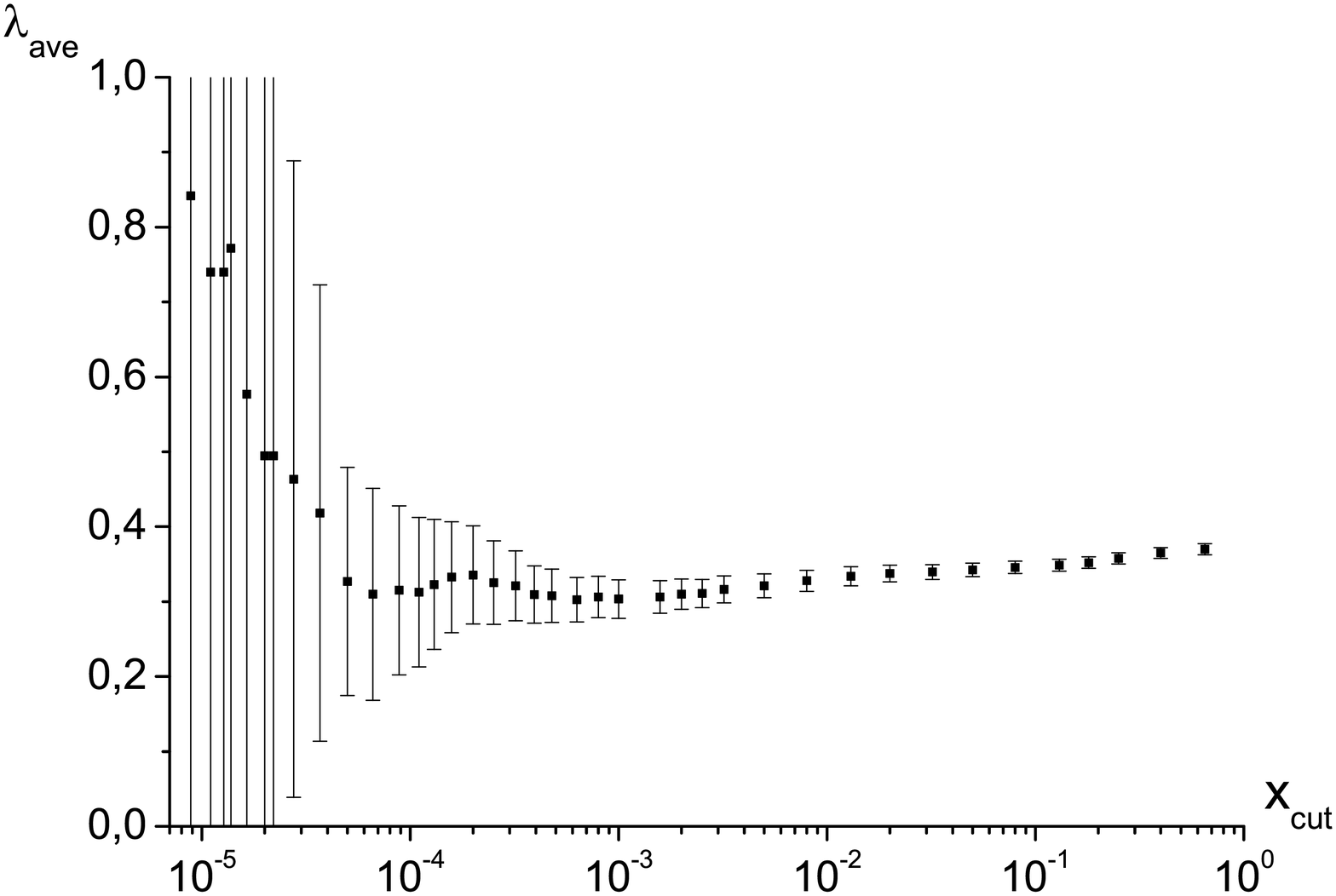}
\caption{Exponent $\lambda_{\rm{min}}$ averaged over energies (denoted as $\lambda_{\rm{ave}}$) as a function of $x_{\rm{cut}}$. The minimal value of $\lambda_{\rm{ave}}$ is $0.30\pm0.03$ for $x_{\rm{cut}}=0.0006$ and maximal is $0.370\pm0.008$ for $x_{\rm{cut}}=0.65$ (we do not consider values for very small $x_{\rm{cut}}$ because of theirs big uncertainties).}
\label{zesWyk9}
\end{figure}

For a given $x_{\rm{cut}}$ we can find value of $\lambda$ averaged over all energies - $\lambda_{\rm{ave}}$. Here we use only energies $25$ GeV $\leq W \leq 159$ GeV (see discussion in \ref{deplammin} and Fig. \ref{zesWyk8}). We define auxiliary function $\tilde{\chi}^2$ which takes into account uncertainties of $\lambda_{\rm{min}}$: 
\begin{equation}
\tilde{\chi}^2(x_{\rm{cut}};\lambda)=\sum\limits_{W\in \left\{25,\ldots,159\right\}} \frac{\left(\lambda_{\rm{min}}(W,x_{\rm{cut}})-\lambda\right)^2}{\left(\Delta \lambda_{\rm{min}}(W,x_{\rm{cut}})\right)^2},
\end{equation}
where we assume that $\Delta \lambda_{\rm{min}} = \frac{\Delta^- \lambda_{\rm{min}}+\Delta^+ \lambda_{\rm{min}}}{2}$ ($\Delta^\pm \lambda_{\rm{min}}$ are defined by formula (\ref{defdeltalamb})).

$\lambda_{\rm{ave}}$ is just a minimum of $\tilde{\chi}^2$ with respect to $\lambda$. We can also find uncertainty of $\lambda_{\rm{ave}}$ using condition (similar to formula (\ref{defdeltalamb})):
\begin{equation}
\tilde{\chi}^2(\lambda_{\rm{ave}}+ \Delta \lambda_{\rm{ave}})-\tilde{\chi}^2(\lambda_{\rm{ave}})=1.
\end{equation}

In Fig. \ref{zesWyk9} we show $\lambda_{\rm{ave}}$ as a function of $x_{\rm{cut}}$.

\section{Bjorken-$x$ binning} 
Now we will present results of $e^{+}p$ data analysis which is based on subsection \ref{metanalysB-x}. All quantities are denoted with prime which means that we are using Bjorken-$x$ binning. 

\subsection{$\lambda'_{\rm{min}}(x')$ for given $x'_{\rm{ref}}$}
\label{sectlam'}

\begin{figure}
\includegraphics[width=7cm,angle=0]{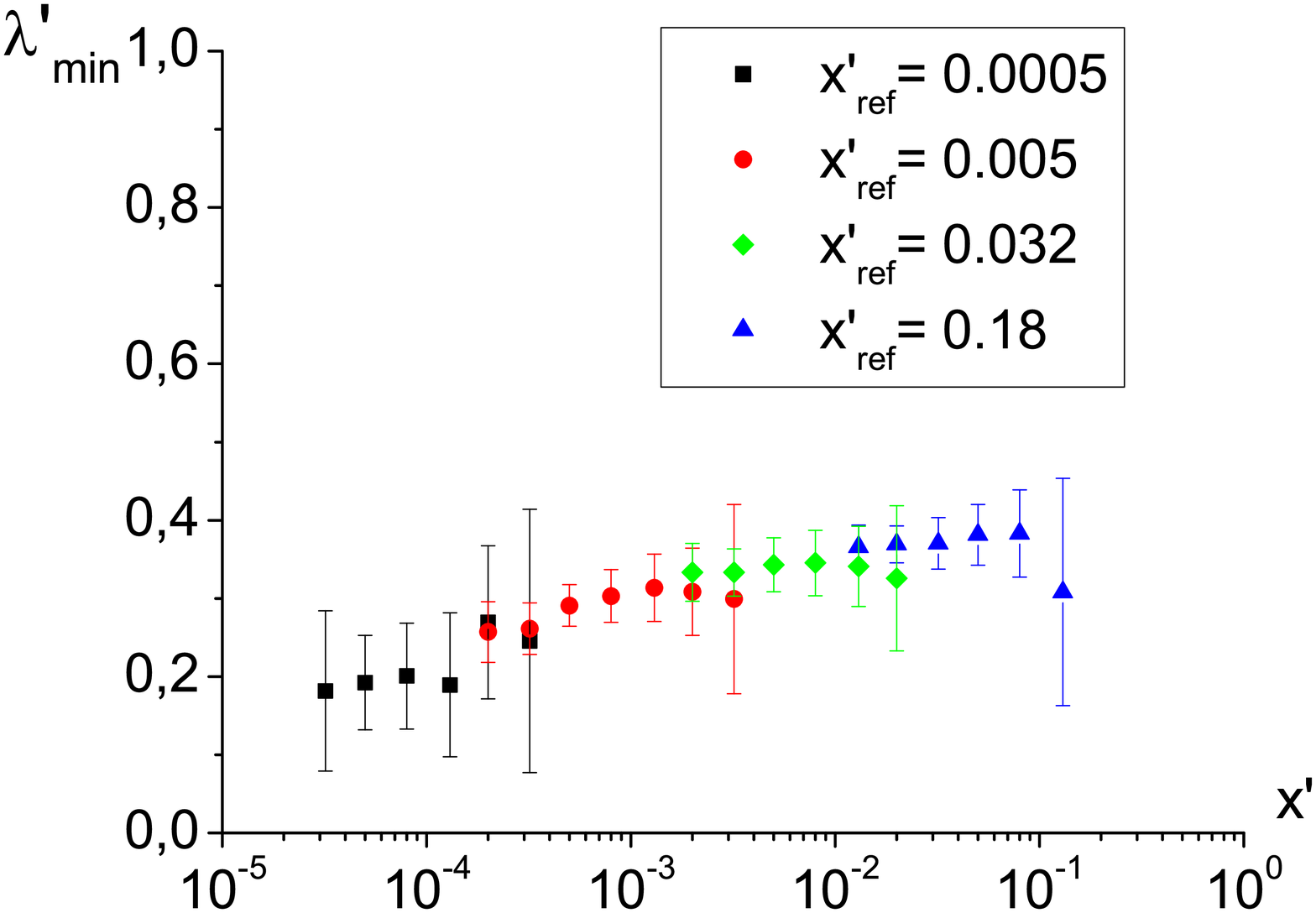}
\includegraphics[width=7cm,angle=0]{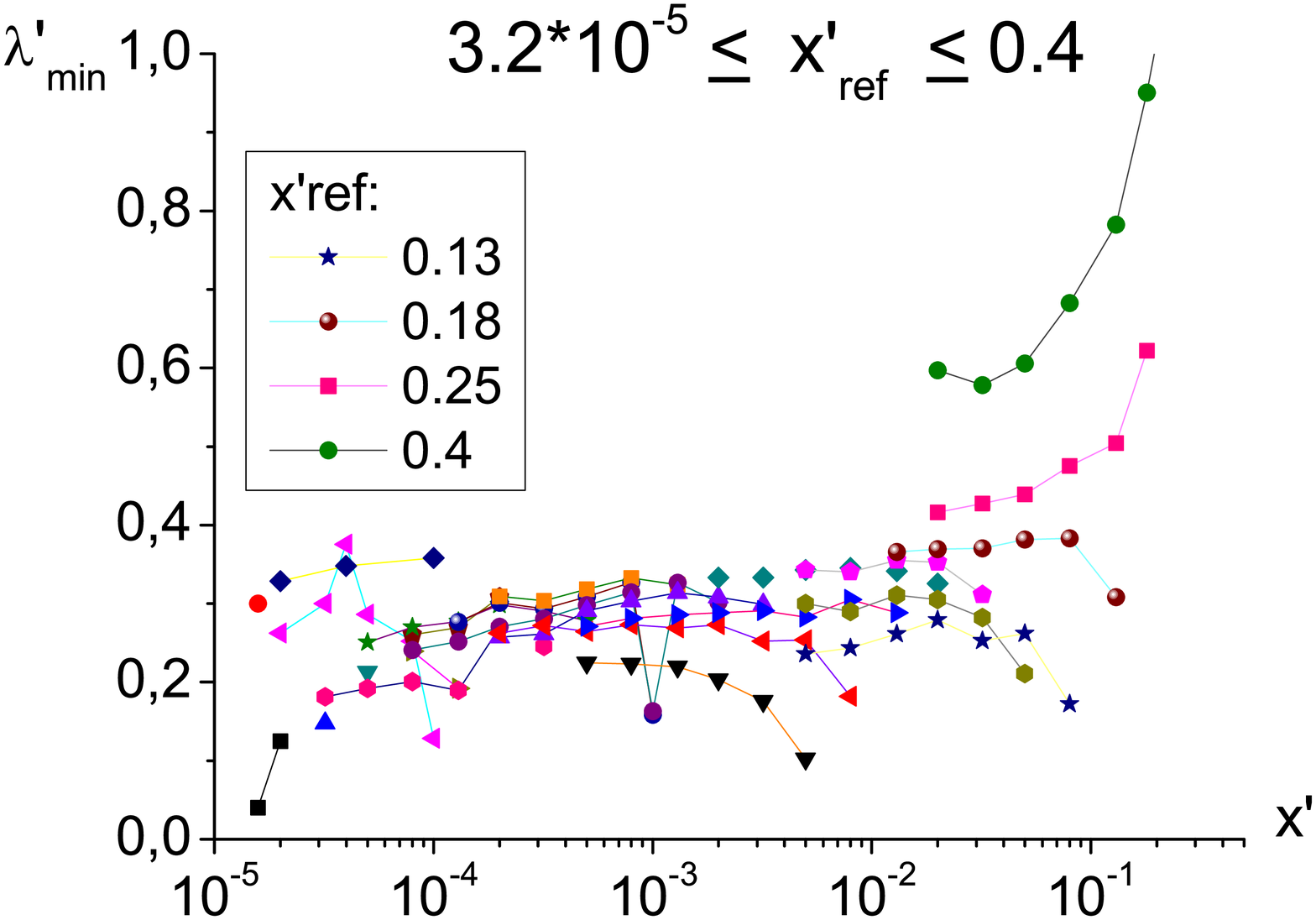}
\caption{Functions $\lambda'_{\rm{min}}(x')$: left plot - with uncertainties, for four $x'_{\rm{ref}}$; right plot - without uncertainties, for all $3.2 \cdot 10^{-5} \leq x'_{\rm{ref}} \leq 0.4$ (legend contains only several greatest $x'_{\rm{ref}}$). }
\label{zesWyk29}
\end{figure}

As we mentioned in \ref{metanalysB-x} one cannot use one $x'_{\rm{ref}}$ for all $x'$. However, choosing some $x'_{\rm{ref}}$ we can find $\lambda'_{\rm{min}}$ for several $x'$ ($x'$ must be smaller than $x'_{\rm{ref}}$ to ensure that ratios $R'_i$ are grater than 1). In left plot of Fig. \ref{zesWyk29} we show $\lambda'_{\rm{min}}$ with uncertainties as a function of $x'$ for four $x'_{\rm{ref}}$. 

Not all $x'$ from Table \ref{tableXbjor+} can be used as $x'_{\rm{ref}}$, if one chooses too small $x'_{\rm{ref}}$ then there is no $x'$ which has at least two points within $f^{\rm{ref'}}_{\lambda}$ domain. It turns out that we can use only $x'_{\rm{ref}} \geq 3.2 \cdot 10^{-5}$. Right plot of Fig. \ref{zesWyk29} shows $\lambda'_{\rm{min}}(x')$ for all such $x'_{\rm{ref}}$ except for 0.65 because it gives much greater $\lambda'_{\rm{min}}$ values. One can see that functions for $x'_{\rm{ref}}>0.2$ have bigger values, this is due to GS violation. 

\subsection{Ratios after minimization and $\chi'^2(\lambda')$}

\begin{figure}
\includegraphics[width=7cm,angle=0]{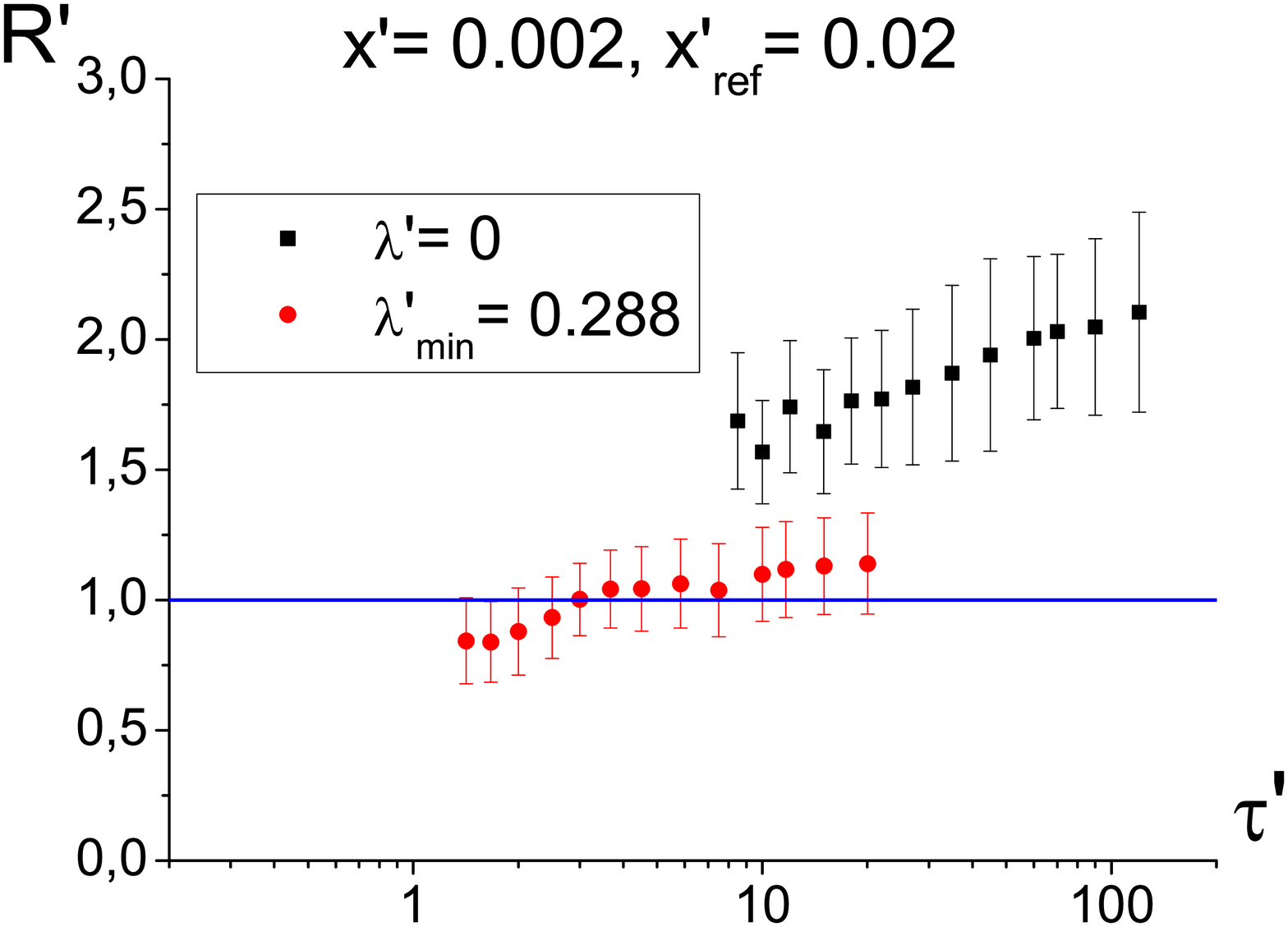}
\includegraphics[width=7cm,angle=0]{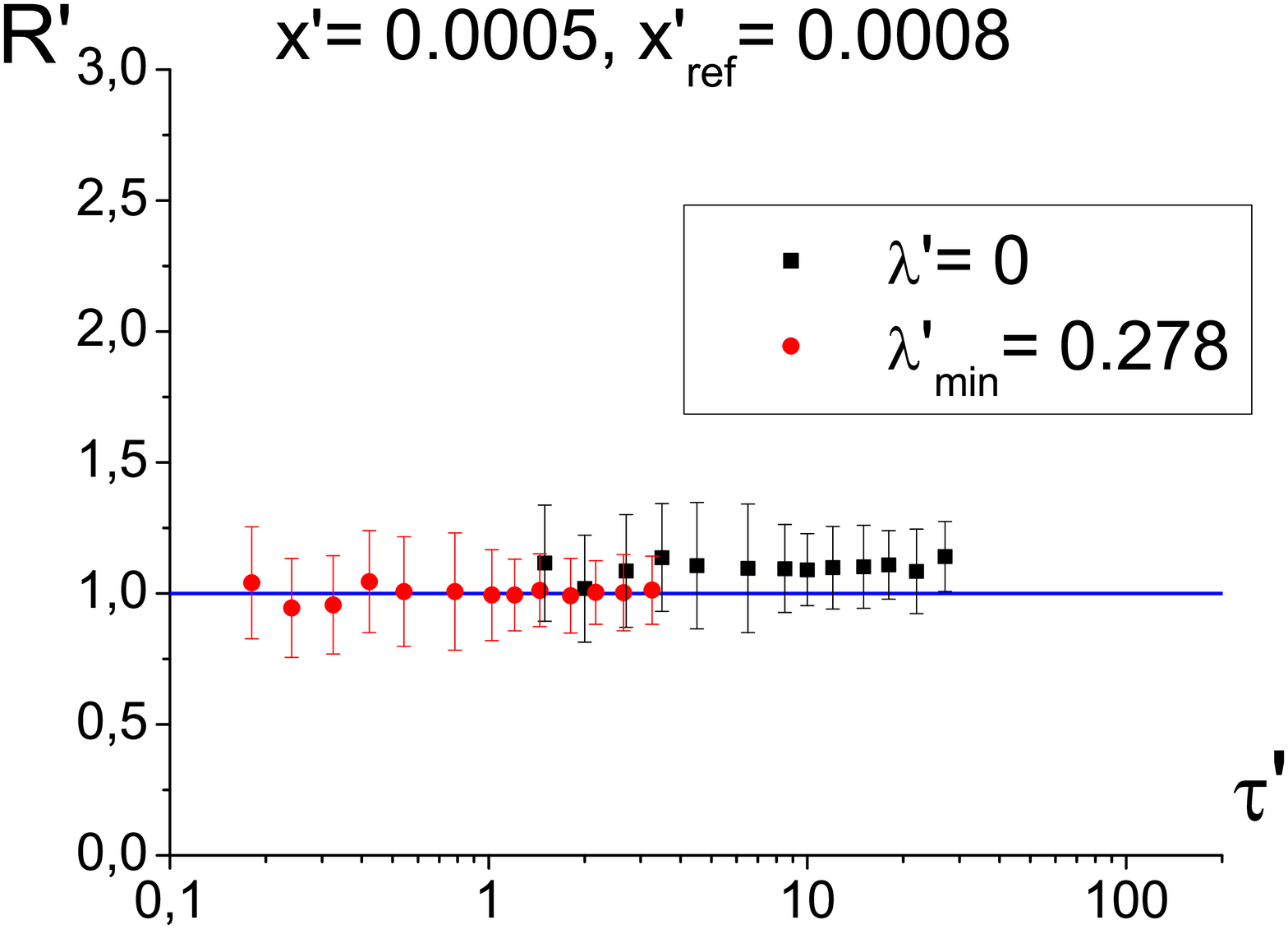}
\includegraphics[width=7cm,angle=0]{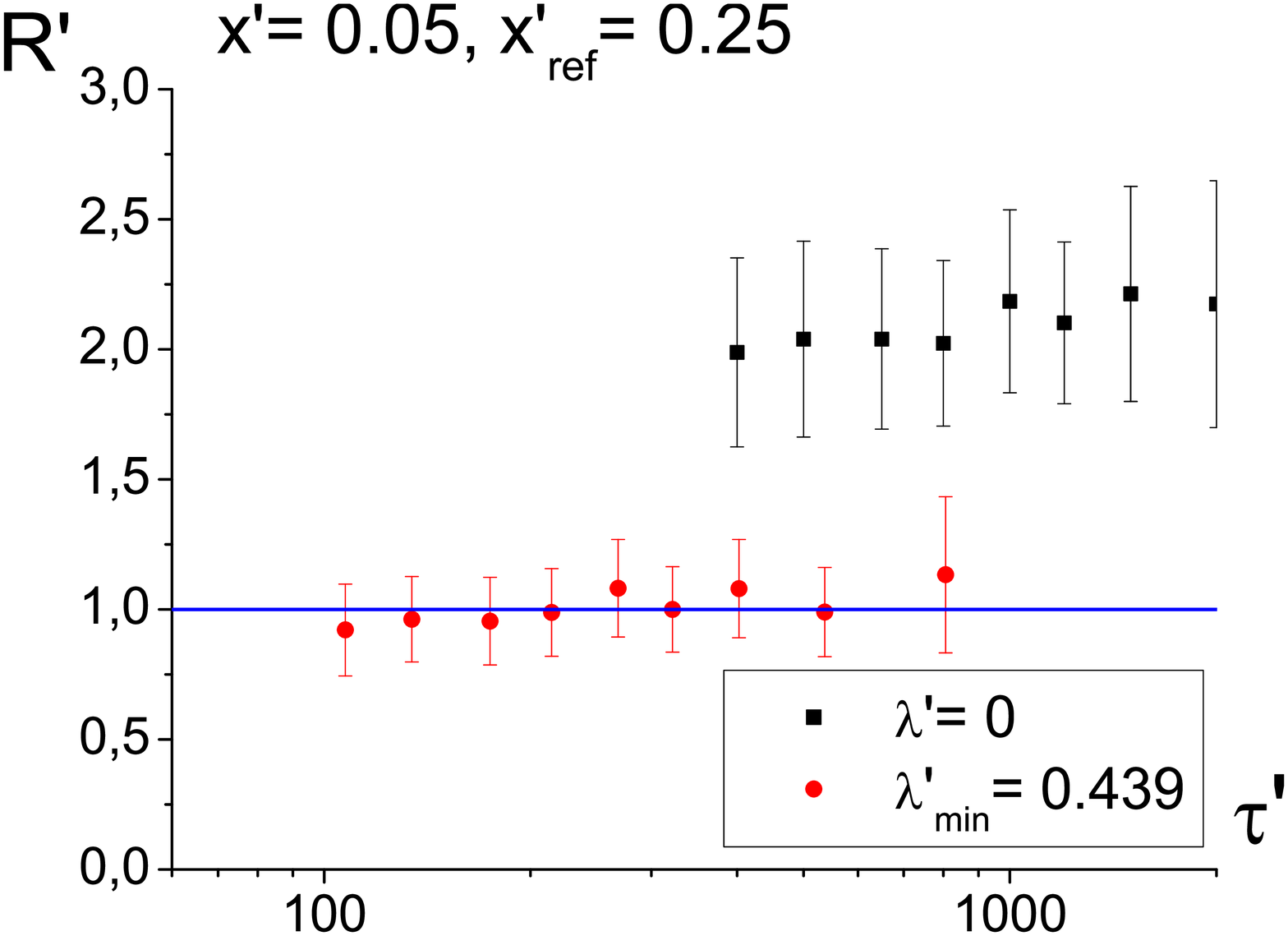}
\includegraphics[width=7cm,angle=0]{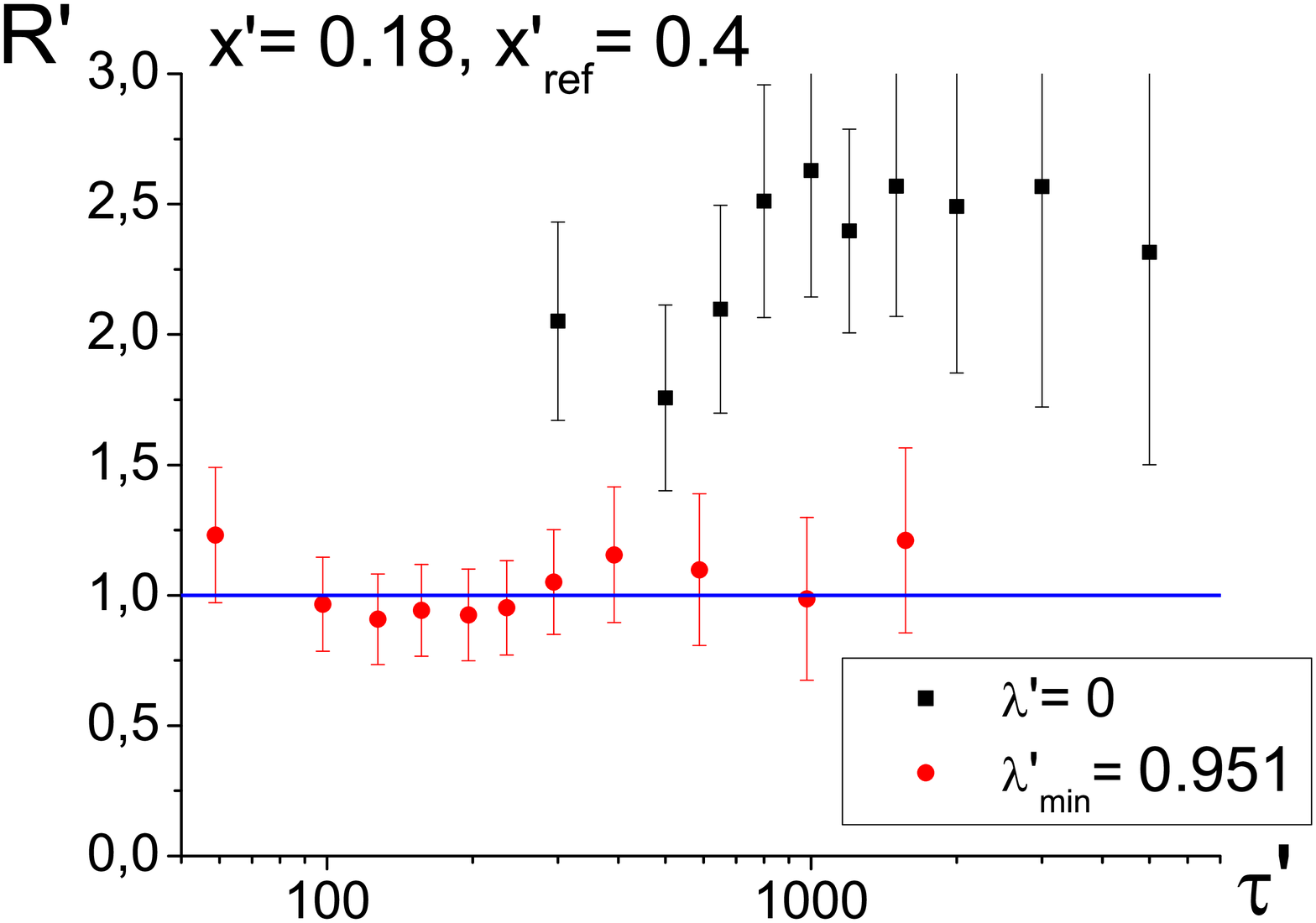}
\caption{Ratios $R'_{i}(\lambda)$ for $\lambda'=0$ and $\lambda'=\lambda'_{\rm{min}}$ for several combinations of $x'$ and $x'_{\rm{ref}}$.}
\label{zesWyk31}
\end{figure}

\begin{figure}
\includegraphics[width=7cm,angle=0]{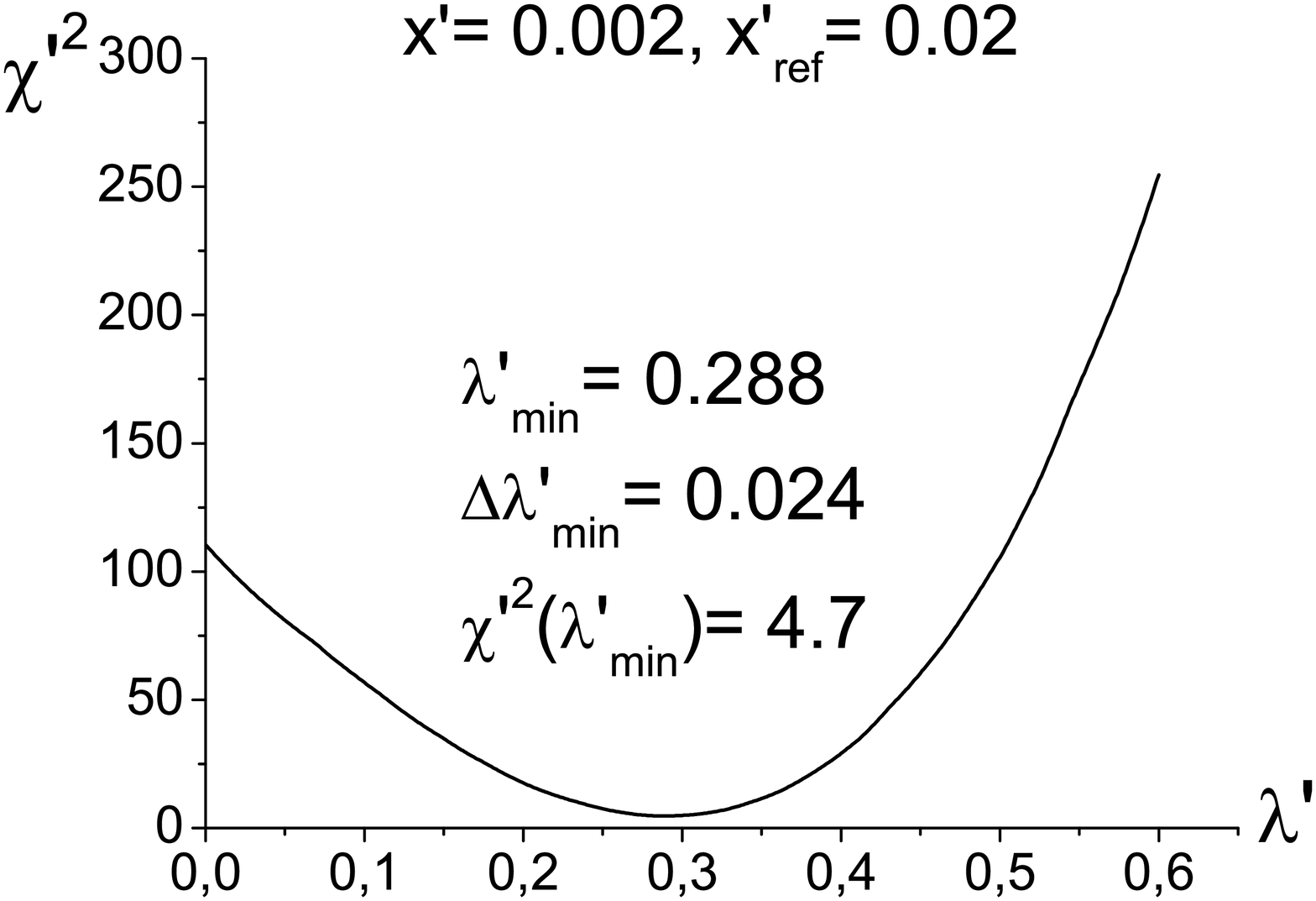}
\includegraphics[width=7cm,angle=0]{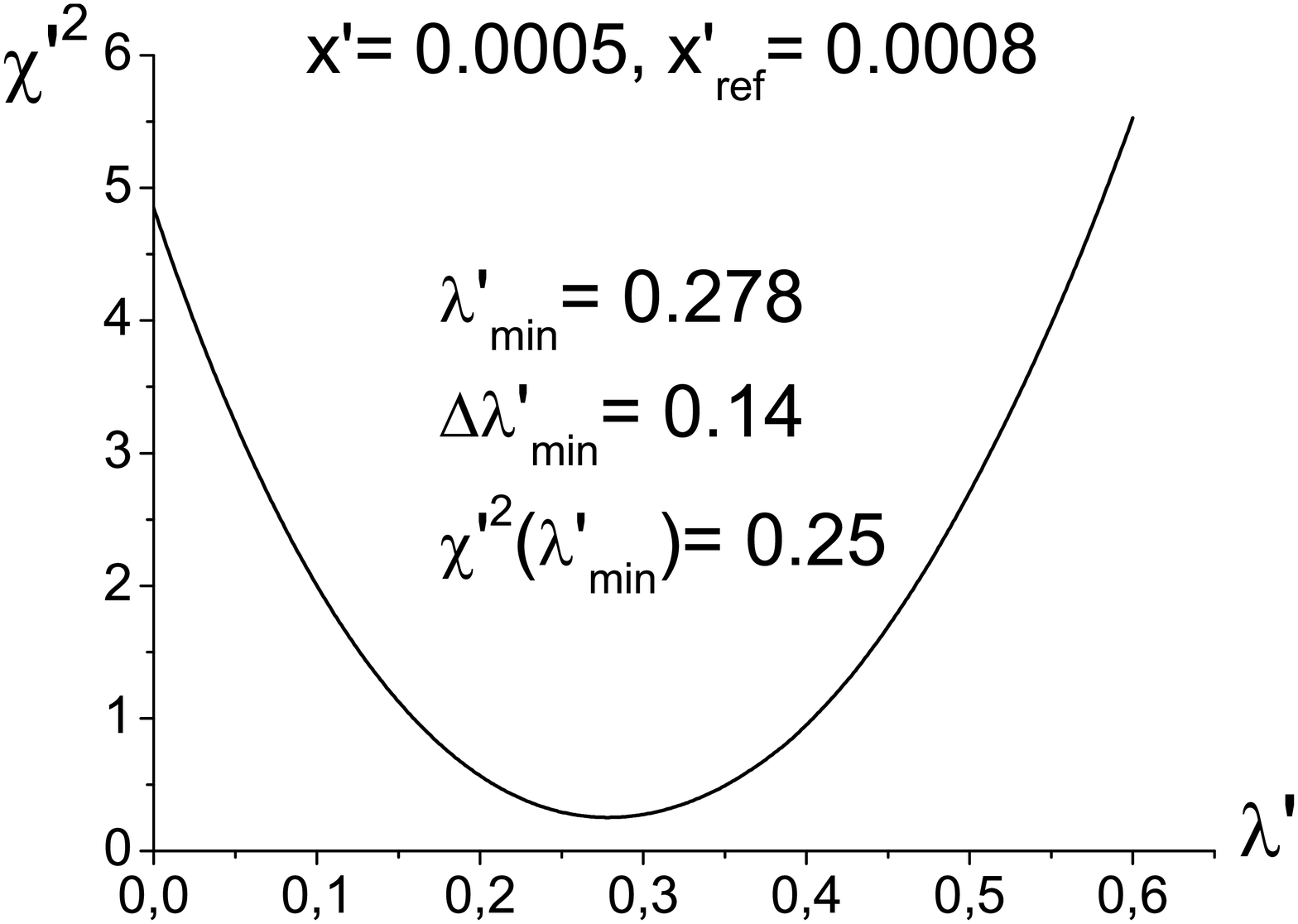}
\caption{Functions $\chi'^2(\lambda')$ for two combinations of $x'$ and $x'_{\rm{ref}}$. We show also values of $\lambda'_{\rm{min}}$, $\Delta \lambda'_{\rm{min}}$ and minimal value of $\chi'^2$.}
\label{zesWyk32}
\end{figure}

In Fig. \ref{zesWyk31} we show ratios (defined by formula (\ref{ratioXdef})) for $\lambda'=0$ and $\lambda'=\lambda'_{\rm{min}}$ for several combinations of $x'$ and $x'_{\rm{ref}}$. We can see that if $x'$ and $x'_{\rm{ref}}$ are similar (second plot) then these ratios are almost 1 even for $\lambda'=0$, this means that $\chi'^2(\lambda')$ is flat (see right plot in Fig. \ref{zesWyk32}) and error of $\lambda'_{\rm{min}}$ is large. Last plot in Fig. \ref{zesWyk31} shows ratios for $x'_{\rm{ref}}=0.4$ \textit{i.e.} value for which GS is not present. We can, however, make this ratios equal 1 but it is possible only for very large $\lambda'_{\rm{min}}$. 

In Fig. \ref{zesWyk32} we show functions $\chi'^2(\lambda')$ for first two combinations ($x'$, $x'_{\rm{ref}}$). We can see that values of $\chi'^2$ decrease several dozen times when we minimize them.

\subsection{$\chi_{\rm{nor}}'^2$ for given $x'_{\rm{ref}}$}
\label{sectchi'nor}

Similarly like in subsection \ref{sectchi^2nor} we introduce $\chi_{\rm{nor}}'^2(x',x'_{\rm{ref}};\lambda')$ which measures average deviation of points from the curve $f_{\lambda}^{\rm{ref'}}$: 
\begin{eqnarray}
\chi_{\rm{nor}}'^2(x'_{\rm{ref}},x';\lambda'):=\frac{1}{N(x',x'_{\rm{ref}})} \chi'^2(x'_{\rm{ref}},x';\lambda')= \nonumber \\
=\frac{1}{N(x',x'_{\rm{ref}})} \sum_{i=1}^{N(x',x'_{\rm{ref}})} \frac{(R'_{i}(\lambda')-1)^2}{(\Delta R'_{i}(\lambda'))^2}.
\label{chi'nordef}
\end{eqnarray}

The most important for us is a minimal value $\chi'^2_{\rm{nor}}(\lambda'_{\rm{min}})$. Its uncertainty is given by:
\begin{equation}
\Delta \chi'^2_{\rm{nor}}(\lambda'_{\rm{min}}):=\chi'^2_{\rm{nor}}(\lambda'_{\rm{min}}+\Delta^{+} \lambda'_{\rm{min}})-\chi'^2_{\rm{nor}}(\lambda'_{\rm{min}})=\frac{1}{N(x',x'_{\rm{ref}})}
\label{niepchi'nordef}.
\end{equation}

\begin{figure}
\includegraphics[width=7cm,angle=0]{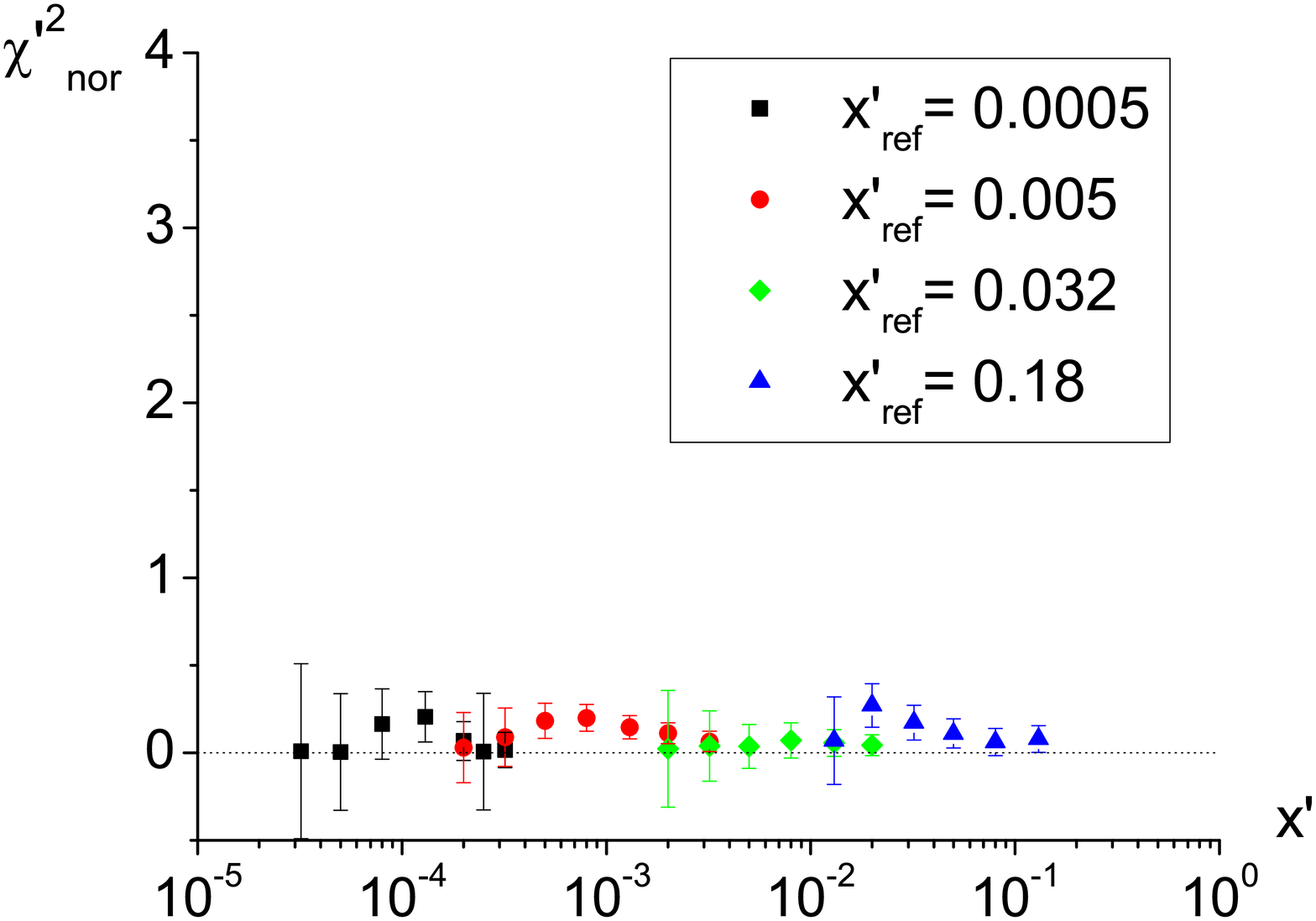}
\includegraphics[width=7cm,angle=0]{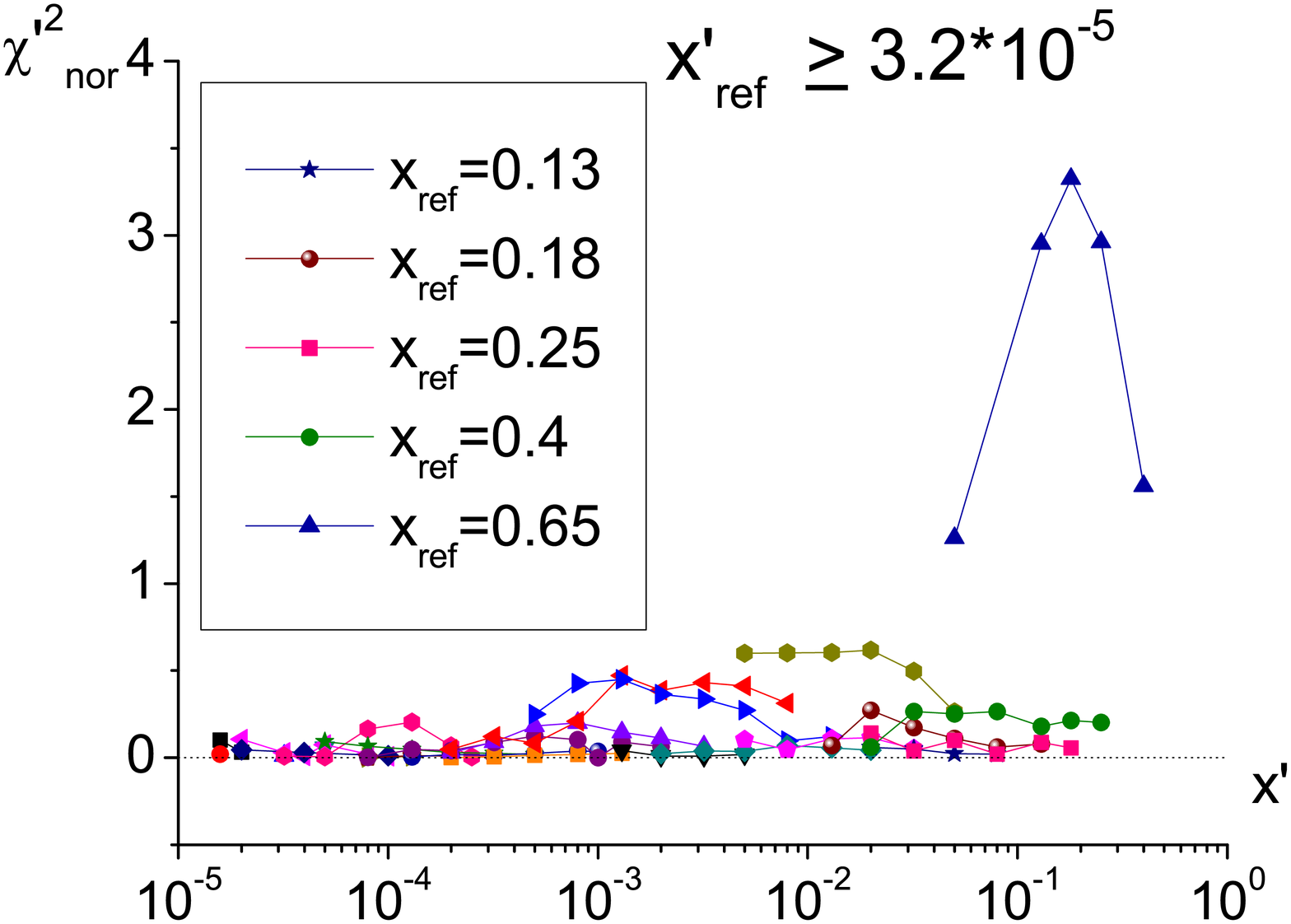}
\caption{$\chi'^2_{\rm{nor}}(\lambda'_{\rm{min}})$ as a function of $x'$: left plot - with uncertainties, for four $x'_{\rm{ref}}$; right plot - without uncertainties, for all $x'_{\rm{ref}} \geq 3.2 \cdot 10^{-5}$ (legend contains only several greatest $x'_{\rm{ref}}$).}
\label{zesWyk33}
\end{figure}

In Fig. \ref{zesWyk33} we show $\chi'^2_{\rm{nor}}(\lambda'_{\rm{min}})$ as a function of $x'$ for various $x'_{\rm{ref}}$. We can see that for most $x'_{\rm{ref}}$ values of $\chi'^2_{\rm{nor}}(\lambda'_{\rm{min}})$ are small, only for $x'_{\rm{ref}}=0.65$ we cannot minimize $\chi'^2_{\rm{nor}}$ with sufficient precision.

\subsection{Average over $x'_{\rm{ref}}$}
\label{sectchi'norave}

\begin{figure}
\includegraphics[width=7cm,angle=0]{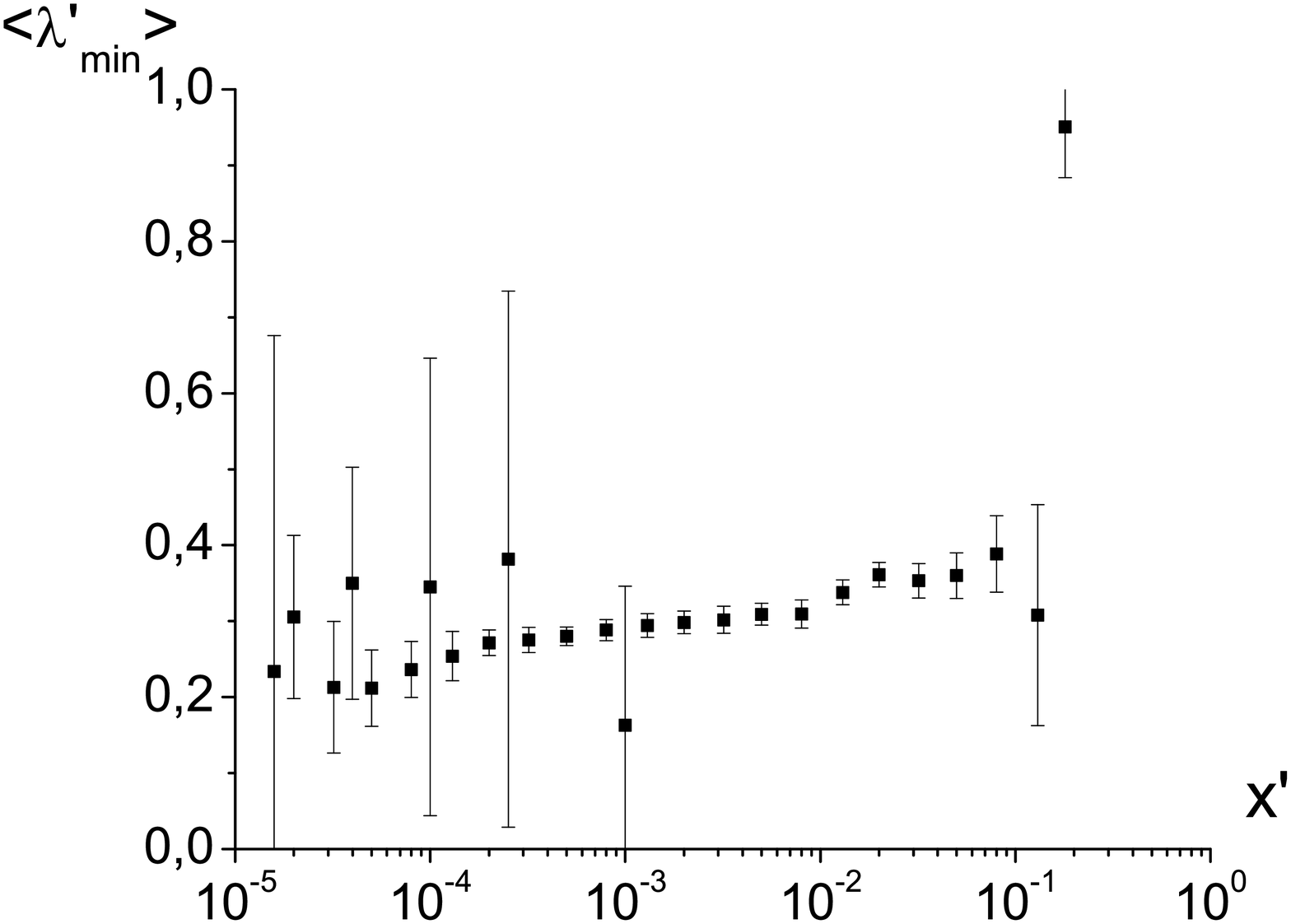}
\includegraphics[width=7cm,angle=0]{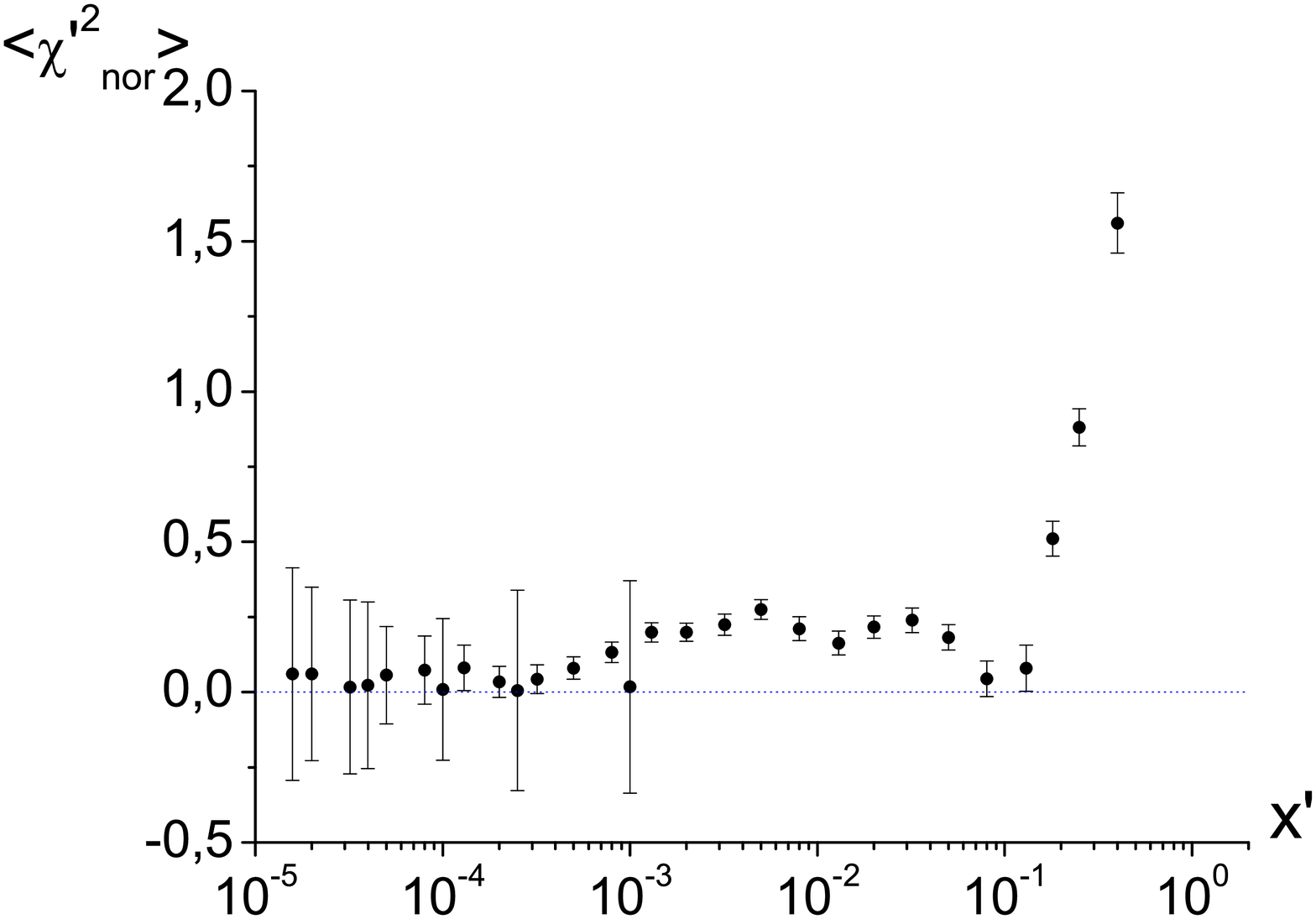}
\caption{Left plot: $\left\langle  \lambda'_{\rm{min}} \right\rangle$ as a function of $x'$. The minimal value is $0.21\pm0.05$ for $x'=0.00005$ while maximal is $0.39\pm0.05$ for $x'=0.08$. Right plot: $\left\langle  \chi'^2_{\rm{nor}}(\lambda_{\rm{min}}) \right\rangle$ as a function of $x'$.}
\label{zesWyk34}
\end{figure}

In subsections \ref{sectlam'} and \ref{sectchi'nor} we showed $\lambda'_{\rm{min}}$ and $\chi'^2_{\rm{nor}}(\lambda'_{\rm{min}})$ as functions of $x'$ for given $x'_{\rm{ref}}$. To eliminate this $x'_{\rm{ref}}$-dependence we will average these quantities over $x'_{\rm{ref}}$ (averages will be denoted using bracket $\left\langle \right\rangle$). Procedure of averaging is similar to that which we use in subsection \ref{sectlamave}: to get an average of a quantity $T(x',x'_{\rm{ref}})$ over $x'_{\rm{ref}}$ we define an auxiliary function:
\begin{equation}
\tilde{\chi}_T^2\Big( x';t \Big):=\sum\limits_{x'_{\rm{ref}}} \frac{\left(T(x',x'_{\rm{ref}})-t \right)^2}{\left(\Delta T(x',x'_{\rm{ref}})\right)^2}.
\end{equation}
The summation in above formula is defined in the following way: if $x' < 0.18$ than we sum over all $x'_{\rm{ref}}$ such that $T(x',x'_{\rm{ref}})$ exists and $x'_{\rm{ref}} \leq 0.18$; if $x' \geq 0.18$ we sum over all $x'_{\rm{ref}} > x'$ - when we want to find $\left\langle T(x')\right\rangle $ for $x'\lesssim 0.18$ we should average only over $x'_{\rm{ref}}<0.18$, otherwise GS violation for $x'_{\rm{ref}}>0.18$ will bias results (for $x'\ll 0.18$ all $x'_{\rm{ref}}$ are smaller than $0.18$ so constraint $x'_{\rm{ref}}<0.18$ is not important). For $x' \geq 0.18$ we have at our disposal only $x'_{\rm{ref}}>0.18$.

$\left\langle T(x')\right\rangle $ is defined as a minimum of $\tilde{\chi}'^2\Big( x';t \Big)$, its uncertainty $\Delta \left\langle T(x')\right\rangle $ is found using equation:
\begin{equation}
\tilde{\chi}_T^2 \Big(x'; \left\langle T(x')\right\rangle + \Delta \left\langle T(x')\right\rangle \Big)-\tilde{\chi}_T^2 \Big(x'; \left\langle T(x')\right\rangle \Big)=1.
\end{equation}

In left plot of Fig. \ref{zesWyk34} we show function $\left\langle  \lambda'_{\rm{min}}(x') \right\rangle$. 
For some $x'$ these averaged quantities have very large uncertainties, this is because of small numbers of  $x'_{\rm{ref}}$.

Right plot of Fig. \ref{zesWyk34} shows function $\left\langle \chi'^2_{\rm{nor}}(x';\lambda'_{\rm{min}}) \right\rangle$. We see that for $x>0.13$ function $\left\langle \chi'^2_{\rm{nor}}(x';\lambda'_{\rm{min}}) \right\rangle$ rises rapidly (this is because for $x\geq0.18$ we averaged over $x'_{\rm{ref}}>0.2$ where GS is not present).

\subsection{Average over $x'$ and $x'_{\rm{ref}}$}
\label{sectchi'noraveave}

\begin{figure}
\includegraphics[width=7cm,angle=0]{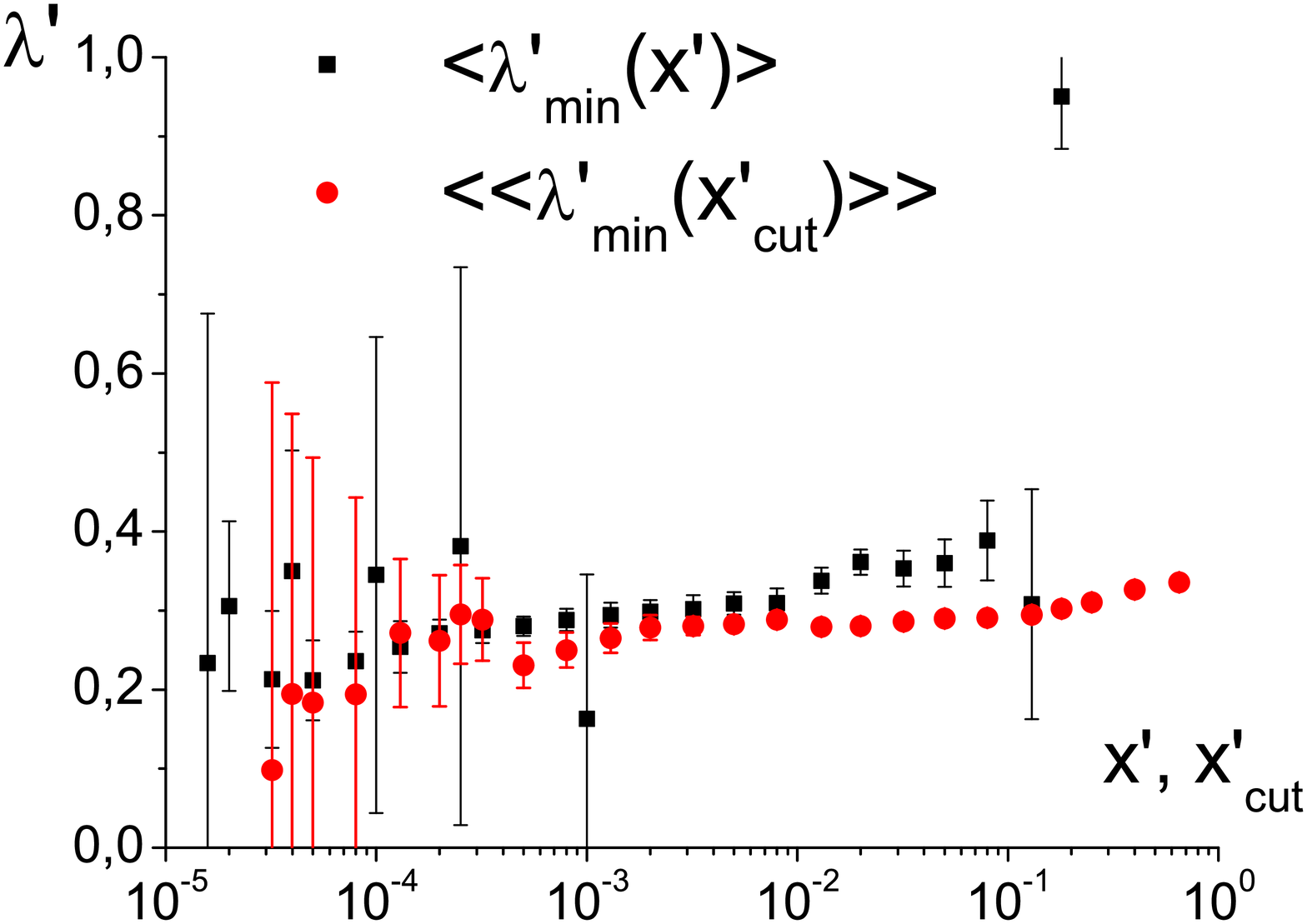}
\includegraphics[width=7cm,angle=0]{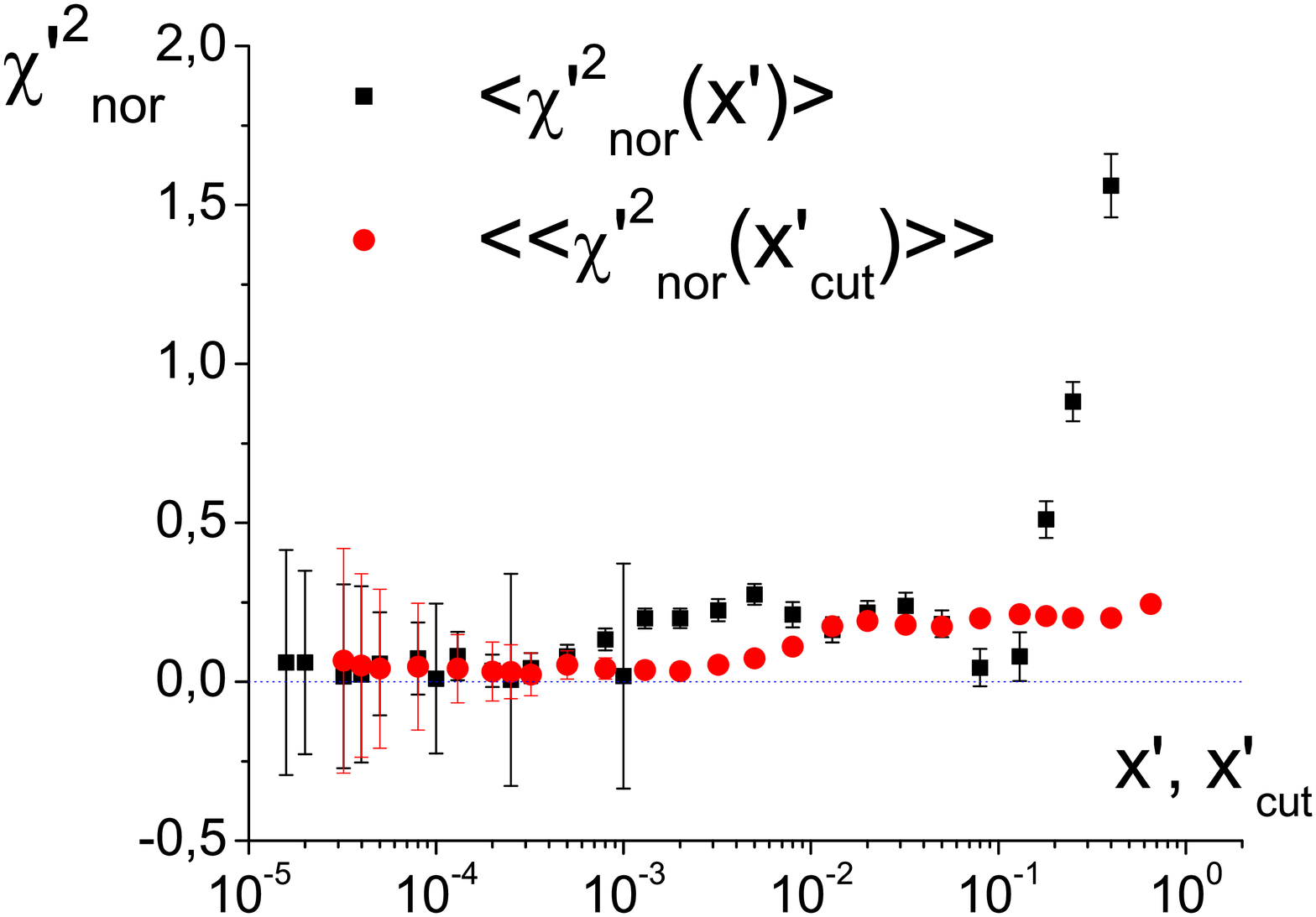}
\caption{$\left\langle  \lambda'_{\rm{min}} \right\rangle$ and $\left\langle  \chi'^2_{\rm{nor}}(\lambda'_{\rm{min}}) \right\rangle$ as functions of
$x'$ (black points) compared with $\left\langle \left\langle \lambda'_{\rm{min}} \right\rangle \right\rangle$ and $\left\langle \left\langle
\chi'^2_{\rm{nor}}(\lambda'_{\rm{min}}) \right\rangle \right\rangle$ as functions of $x'_{\rm{cut}}$ (red points). 
Fact that $x'$ and $x'_{\rm{cut}}$ are different arguments was indicated on horizontal axes.}
\label{zesWyk39}
\end{figure}

In previous subsection we described averaging of quantity $T(x',x'_{\rm{ref}})$ over $x'_{\rm{ref}}$, now we will also average it over $x'$.

Similarly like in the case of energy binning we introduce parameter $x'_{\rm{cut}}$ such that we consider only data with $x'\leq x'_{\rm{cut}}$. As previously we define an auxiliary function:

$$\widetilde{\widetilde{\chi}}_T^2\Big( x'_{\rm{cut}};t \Big):=\sum\limits_{x'_{\rm{ref}} \leq x'_{\rm{cut}}} \sum\limits_{x'< x'_{\rm{ref}}} \frac{\left(T(x',x'_{\rm{ref}})-t \right)^2}{\left(\Delta T(x',x'_{\rm{ref}})\right)^2}$$

$\left\langle \left\langle T(x'_{\rm{cut}})\right\rangle \right\rangle$ is a minimum of $\widetilde{\widetilde{\chi}}_T^2$ with respect to $t$.

In that way we can find $\left\langle \left\langle  \lambda'_{\rm{min}}(x'_{\rm{cut}}) \right\rangle \right\rangle$ and $\left\langle \left\langle  \chi'^2_{\rm{nor}}(x'_{\rm{cut}};\lambda'_{\rm{min}}) \right\rangle \right\rangle$. In Fig. \ref{zesWyk39} we show these functions (red points); we compare them with quantities averaged only over $x'_{\rm{ref}}$ (black points). We should be careful comparing both results because arguments of functions have different meaning: $x'$ means that we calculate $\left\langle \lambda'_{\rm{min}}\right\rangle$ using only points with this value of Bjorkjen variable while $x'_{\rm{cut}}$ means that we calculate $\left\langle \left\langle  \lambda'_{\rm{min}}(x'_{\rm{cut}}) \right\rangle \right\rangle$ using also points with $x'<x'_{\rm{cut}}$. This is the reason why we have difference between $\left\langle \lambda'_{\rm{min}}(x')\right\rangle$ and $\left\langle \left\langle  \lambda'_{\rm{min}}(x'_{\rm{cut}}) \right\rangle \right\rangle$ for large $x'$: dominant contribution to $\left\langle \left\langle  \lambda'_{\rm{min}}(x'_{\rm{cut}}) \right\rangle \right\rangle$ comes from points with smaller $x'$.

\subsection{Comparison of energy and Bjorken-$x$ binnings}

\begin{figure}
\centering
\includegraphics[width=9cm,angle=0]{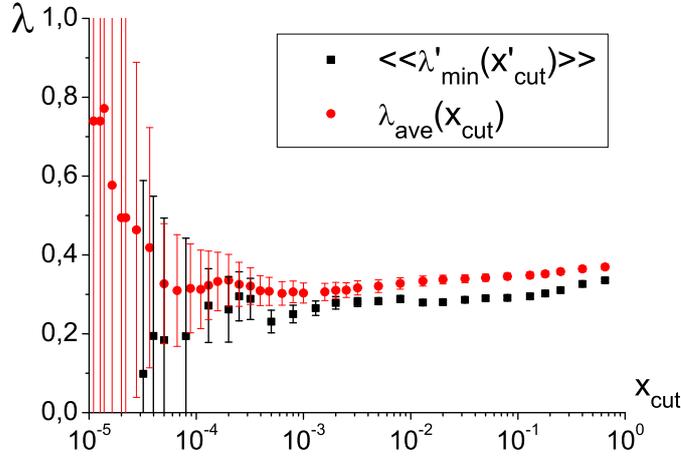}
\caption{Comparison of $\left\langle \left\langle  \lambda'_{\rm{min}}(x'_{\rm{cut}}) \right\rangle \right\rangle$ with $\lambda_{\rm{ave}}(x_{\rm{cut}})$. $0.3\leq \lambda_{\rm{ave}} \leq 0.37$ and 0.1$<\left\langle \left\langle \lambda'_{\rm{min}} \right\rangle \right\rangle<0.336$}
\label{zesWyk30}
\end{figure}

In Fig. \ref{zesWyk30} we compare $\left\langle \left\langle  \lambda'_{\rm{min}}(x'_{\rm{cut}}) \right\rangle \right\rangle$ (defined for Bjorken-$x$ binning in subsection \ref{sectchi'noraveave}) and function $\lambda_{\rm{ave}}(x_{\rm{cut}})$ (defined for energy binning in subsection \ref{sectlamave}). One can see that these functions differ strongly for small $x_{\rm{cut}}$, however uncertainties are too big to make some clear statement. This difference influences also values for larger $x'_{\rm{cut}}$. Moreover, one should remember that when we calculated $\lambda_{\rm{ave}}$ (see section \ref{sectlamave}) we omitted energies $W<25$ GeV and $W=268$ GeV which for large $x_{\rm{cut}}$ have smaller values of $\lambda_{\rm{min}}$ (see Fig. \ref{zesWyk8}). 

In previous sections we found that GS is present for $x<0.2$ (for both binnings). As a optimal value of parameter $\lambda$ we adopt $\left\langle \left\langle  \lambda'_{\rm{min}}(0.2) \right\rangle \right\rangle \equiv \lambda_{\rm{Bj}}$ for Bjorken-$x$ binning and $\lambda_{\rm{ave}}(0.2)\equiv \lambda_{\rm{En}}$ for energy binning \textit{i.e.} values obtained by averaging results of $\lambda_{\rm{min}}$ for all $x\leq0.2$:
\begin{eqnarray}
\lambda_{\rm{Bj}} & = 0.302 \: \pm  \:0.004 \\
\lambda_{\rm{En}} & = 0.352 \: \pm  \: 0.008 .
\end{eqnarray}

\section{$Q^2$ dependence of $\lambda$ exponent}
\label{Q^2dependence}

\subsubsection{HERA results for $\lambda(Q^2)$}

We know from experiment that for small $x$ proton structure function $F_2$ behaves like:
\begin{equation}
F_2 \sim x^{-\lambda_{F}(Q^2)},
\end{equation} 
where $Q^2$ dependence of $\lambda_F$ can be extracted from the HERA data. We use data points from \cite{lamodQ2Kowalski} (see also \cite{lamodQ2}) where values of $\lambda$ were given for $4.5 \textrm{ GeV}^2 \leq Q^2 \leq 150 \textrm{ GeV}^2$. Unfortunately, we do not know for what $x$-range $\lambda_{F}(Q^2)$ was found.
In \cite{imprGS} an eyeball fit to these points was given:
\begin{equation}
\lambda_{\rm{eff}}(Q^2)=0.13+0.1 \left(\frac{Q^2}{10}\right)^{0.35}.
\label{eyeballfit}
\end{equation}

We will check now whether we can find some $Q^2$ dependence of $\lambda$ using our analysis and compare it to $\lambda_F(Q^2)$ values and $\lambda_{\rm{eff}}(Q^2)$. Since we have no $\lambda_{F}(Q^2)$ values for $Q^2 < 4.5 \textrm{ GeV}^2$ and $Q^2 > 150 \textrm{ GeV}^2$ we use in these regions extrapolation of $\lambda_{\rm{eff}}$ function. In our analysis of geometrical scaling we use points with $0.1 \textrm{ GeV}^2 \leq Q^2 \leq 20000 \textrm{ GeV}^2$ so for most of this range $\lambda_{\rm{eff}}$ function cannot be treated as a good extrapolation.   
 
\subsection{Energy binning}
\subsubsection{Method of analysis}

We will use energy binning first because in \cite{imprGS} it was argued that in pp collisions some $Q^2$ dependence of $\lambda$ can be seen and is similar to $\lambda_{\rm{eff}}(Q^2)$ (for pp collisions data energy binning is used).

We introduce $Q^2$ dependence of $\lambda$ by generalization of formula (\ref{deftau}):
\begin{equation}
\tau = Q^2 x^{\lambda(Q^2)}, 
\end{equation}
where $\lambda(Q^2)$ is unknown function which we want to find.

We use similar method like previously (see section \ref{sectMethodoffinding}), but instead of one parameter $\lambda$ we have a set of parameters
$\left(\lambda_1, \lambda_2,\ldots,\lambda_N \right)\equiv \vec{\lambda}$. For every $Q^2$ value we assign one coordinate $\lambda_i$: some coordinates correspond to points with energy $W$, some to those with energy $W_{\rm{ref}}$. If $Q^2$ values are present both for $W$ and $W_{\rm{ref}}$ we assign them the same coordinate of $\vec{\lambda}$ (\textit{i.e.} we assume that function $\lambda(Q^2)$ is the same for $W$ and $W_{\rm{ref}}$). It turns out that for $W_{\rm{ref}}=206$ GeV almost all $Q^2$ values present for $W$ are also present for $W_{\rm{ref}}$ thus the number of parameters is equal to the number of points with energy $W_{\rm{ref}}=206$ GeV.

The reference curve $f_{\vec{\lambda}}^{\rm{ref}}(\tau)$ joins points with energy $W_{\rm{ref}}$ \textit{i.e.} ($Q_k^2 x_k^{\lambda_k}$,$\tilde{\sigma}_k$), where $\lambda_k$ is coordinate of $\vec{\lambda}$ associated with $Q^2_k$. In the same way we can generalize other quantities from section \ref{sectMethodoffinding}:  $R^W_{i}(\vec{\lambda})$, $\Delta R^W_{i}(\vec{\lambda})$, $\chi^2(W,x_{\rm{cut}};\vec{\lambda})$. 

$\vec{\lambda}_{m}=\left(\lambda^{m}_1, \lambda^{m}_2,\ldots,\lambda^{m}_N \right)$ is by definition a vector which minimizes $\chi^2$ for given $W$ and $x_{\rm{cut}}$. To find $\lambda^m_{i}$ uncertainties we use similar conditions to (\ref{defdeltalamb}):
\begin{equation}
\chi^2 \left(\lambda^{m}_1, \lambda^{m}_2,\ldots,\lambda^{m}_i \pm \Delta^\pm \lambda^{m}_i,\ldots,\lambda^{m}_N \right)-\chi^2 \left(\lambda^{m}_1, \lambda^{m}_2,\ldots,\lambda^{m}_N \right)=1.
\label{defnieplamvec}
\end{equation}
In that way we are able to find uncertainties for every coordinate.   
There is, however, a subtlety in such definition - it gives true uncertainties if all $\lambda_i$ are independent variables. It is probably not true in our case. To find correct values of uncertainties one should use covariance matrix, but because it would complicate calculations enormously and would not change our conclusions we decided to use simpler estimate of error given by eq. (\ref{defnieplamvec}). However, we should treat our results for uncertainties as an approximation and be aware that they are probably underestimated.

To perform numerical calculations we need to start from some initial point $\vec{\lambda}_{0}$. We use results of previous section where it was assumed that $\lambda$ does not depend on $Q^2$ - in section \ref{deplammin} we found $\lambda_{\rm{min}}$ values (see for example Fig. \ref{zesWyk5}). It is natural to take $\vec{\lambda}_{0}=\left(\lambda_{\rm{min}}, \lambda_{\rm{min}},\ldots,\lambda_{\rm{min}} \right)$ as a starting point.

\subsubsection{Results}

\begin{figure}
\includegraphics[width=7cm,angle=0]{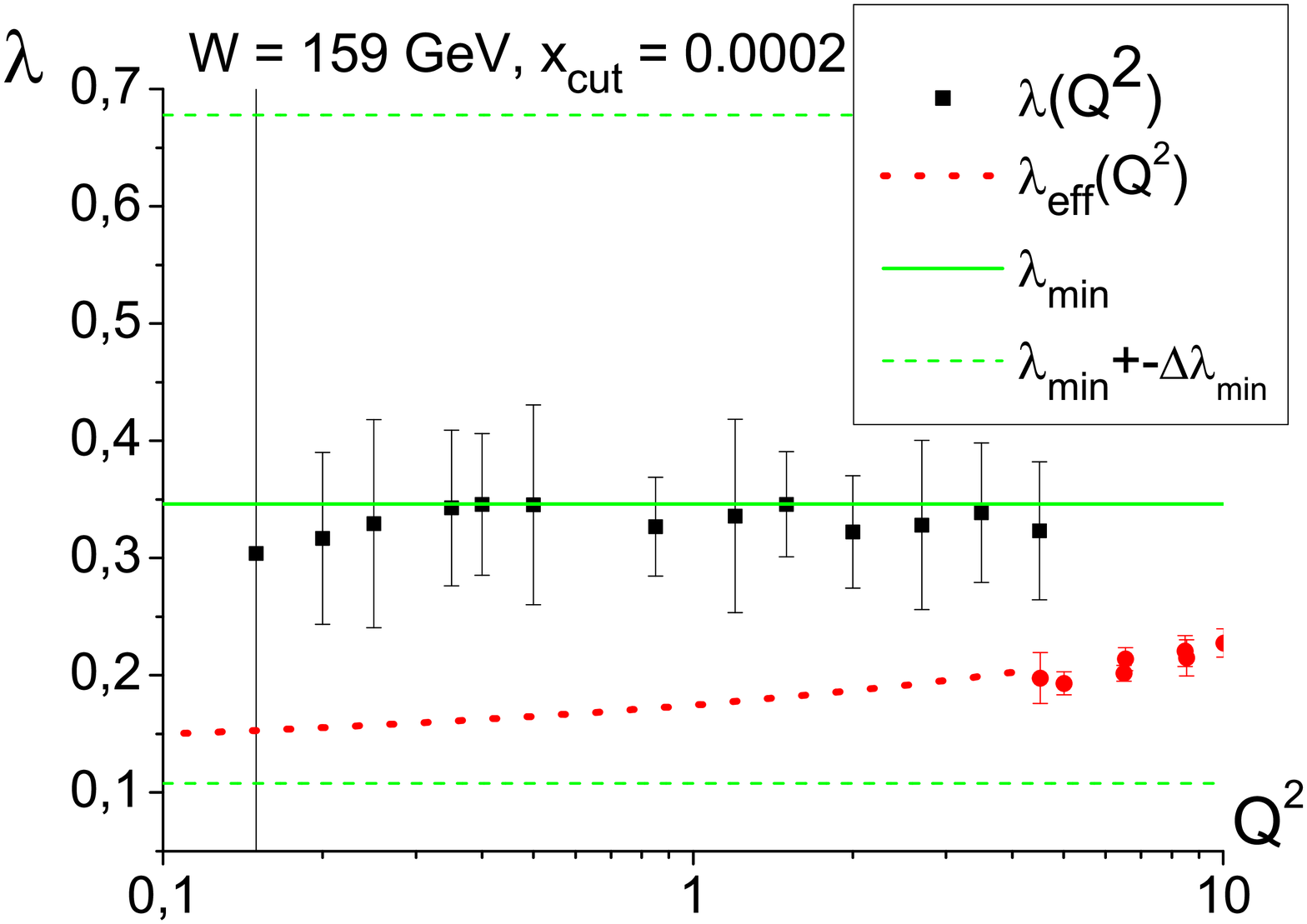}
\includegraphics[width=7cm,angle=0]{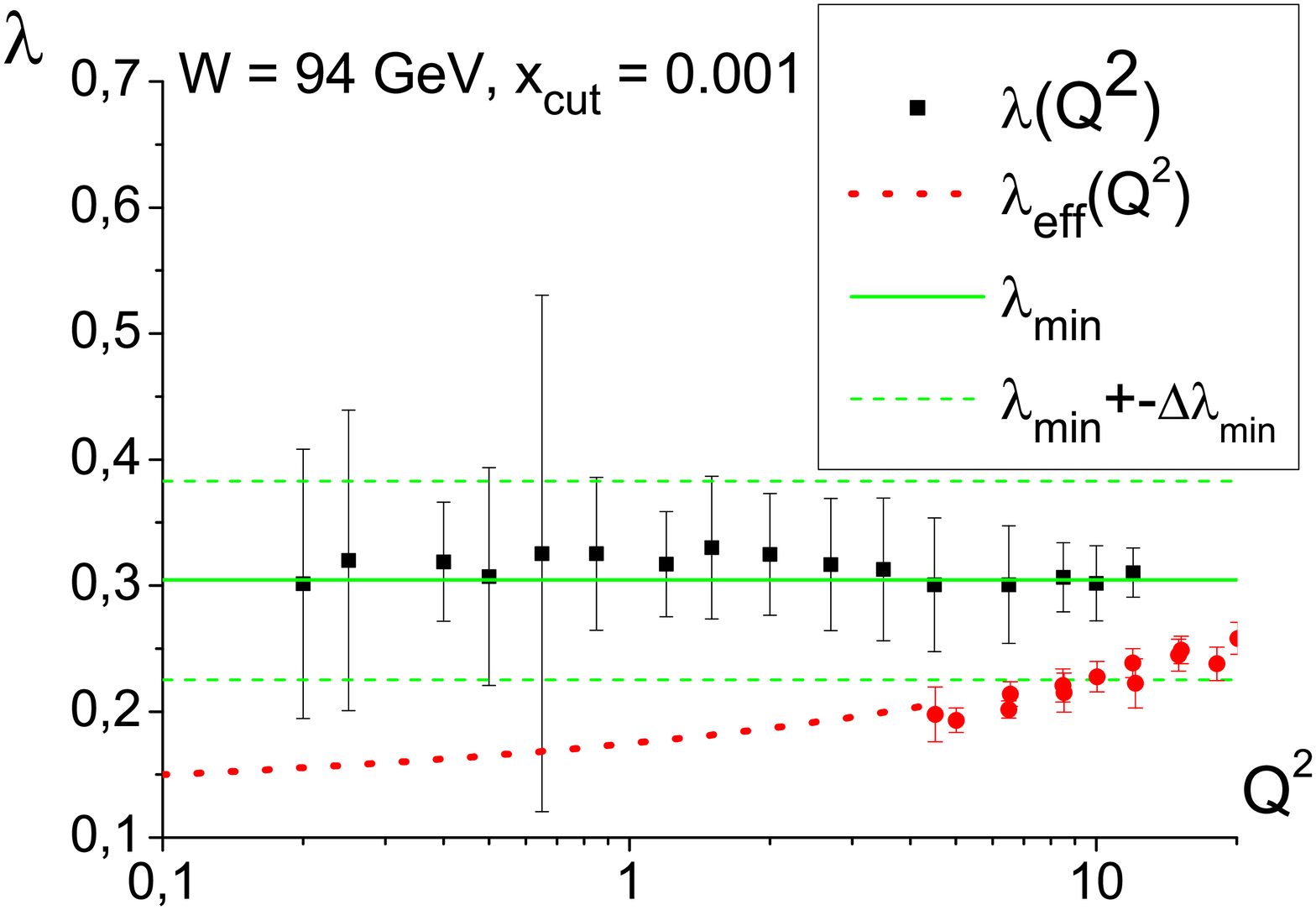}
\includegraphics[width=7cm,angle=0]{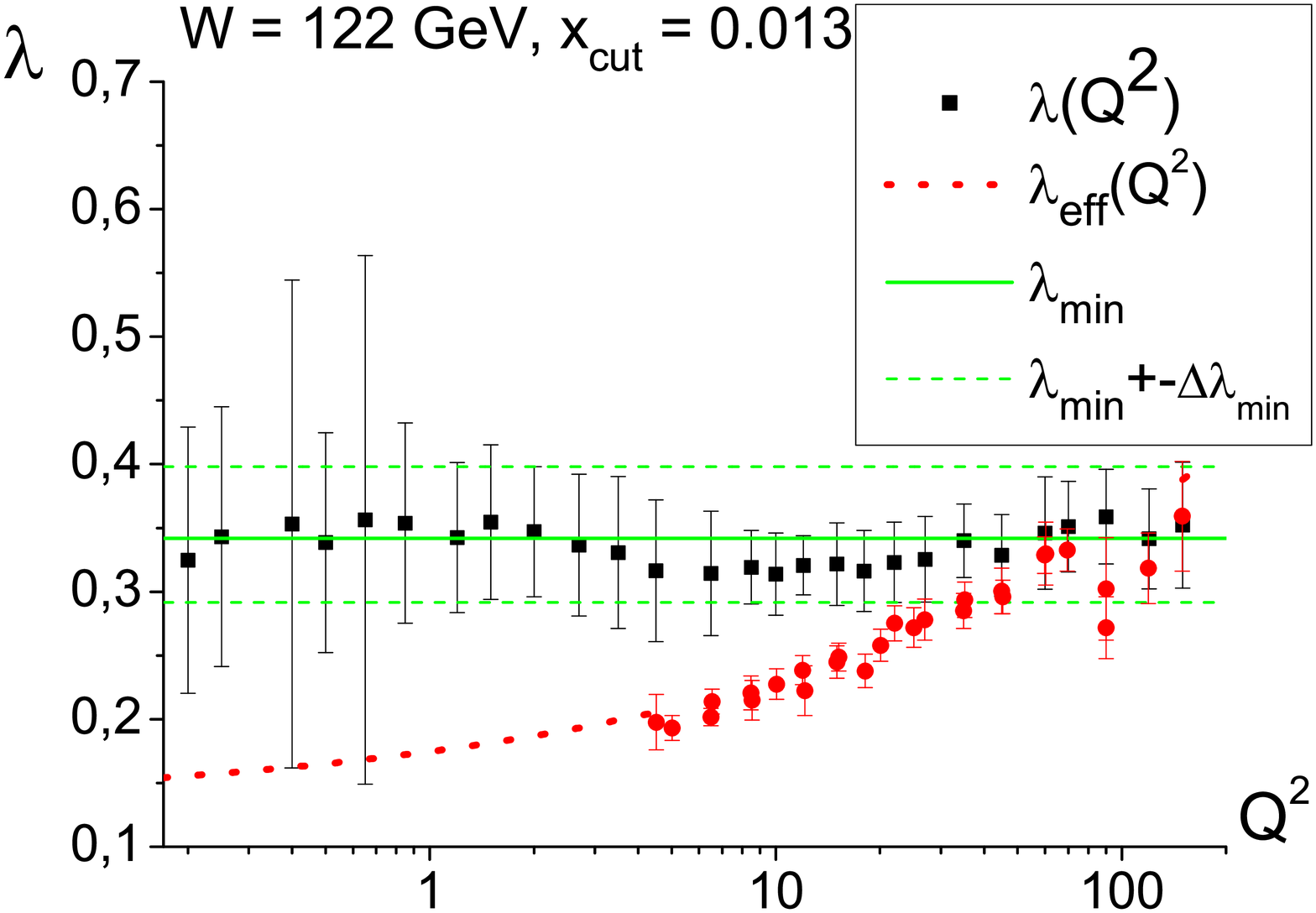}
\includegraphics[width=7cm,angle=0]{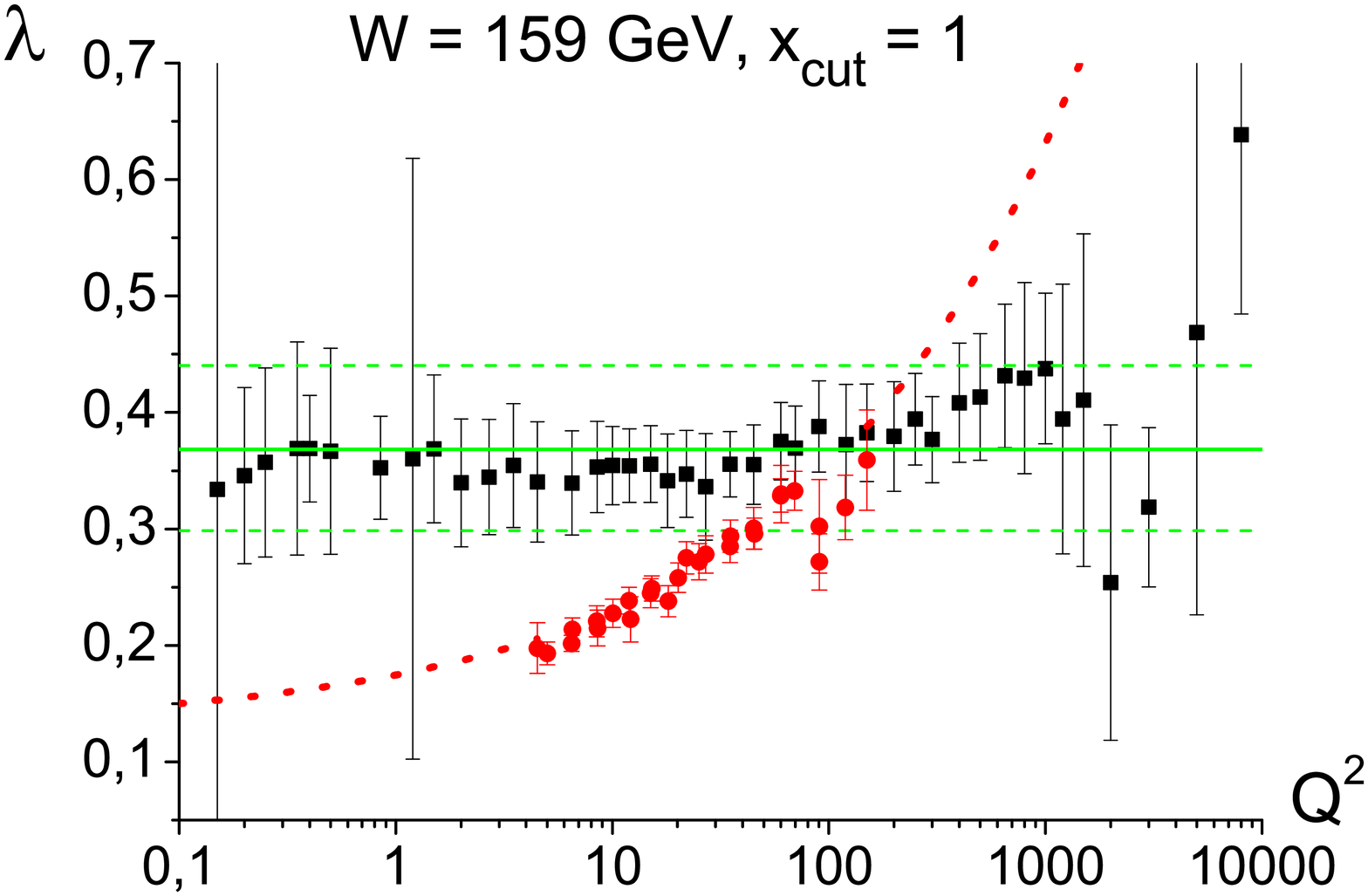}
\caption{$\lambda(Q^2)$ (\textit{i.e.} coordinates of $\vec{\lambda}_{m}$ plotted in terms of $Q^2$) for four ($x_{\rm{cut}}$,$W$) combinations. $\lambda_F(Q^2)$ values are shown as red points; red, dotted line represents $\lambda_{\rm{eff}}(Q^2)$ plotted for $Q^2 < 4.5 \textrm{ GeV}^2$ and $Q^2 > 150 \textrm{ GeV}^2$.}
\label{zesWyk10}
\end{figure}

In Fig. \ref{zesWyk10} we show four examples of functions $\lambda(Q^2)$, they are represented by black points $(Q^2_i, \lambda^{m}_i)$ (we plotted only coordinates of $\vec{\lambda}_{m}$ which correspond to points with energy $W$). As we can see these functions are very similar to $\lambda_{\rm{min}}$ (green, solid line).

Because for every $Q^2$ we have one parameter we can minimize $\chi^2$ almost to 0. Table \ref{table2} shows comparison of minimized values of $\chi^2$ for $\lambda=\textrm{const}.$ (denoted as $\chi^2(\lambda_{\rm{min}})$), $\lambda=\lambda(Q^2)$ (denoted as $\chi^2(\vec{\lambda}_m)$). Last column shows values of $\chi^2$ calculated for $\lambda_{\rm{eff}}(Q^2)$. We can see that $\chi^2$ for $\lambda_{\rm{eff}}(Q^2)$ are several times greater than for $\lambda_{\rm{min}}$.   

\begin{table}
\centering
\begin{tabular}{|c|c|c|c|} \hline
Energy and $x$ value & $\chi^2(\lambda_{\rm{min}})$ & $\chi^2(\vec{\lambda}_m)$ & $\chi^2\left(\lambda_{\rm{eff}}\right)$ \\ \hline
$W=159$GeV, $x_{\rm{cut}}=0.0002$ & 0.34 & $8.8\times 10^{-13}$ & 0.65 \\
$W=94$GeV, $x_{\rm{cut}}=0.001$ & 0.93 & $ 2.4 \times10^{-12}$ & 2.09 \\
$W=122$GeV, $x_{\rm{cut}}=0.013$ & 2.37 & $3\times 10^{-10}$ & 10.2 \\
$W=159$GeV, $x_{\rm{cut}}=1$ & 4.9 & 0.022 & 46 \\ \hline
\end{tabular}
\caption{Values of $\chi^2$: minimized using one parameter $\lambda$ (second column), minimized using many parameters $\left(\lambda_1, \lambda_2,\ldots,\lambda_N \right)$ (third column) and calculated for $\lambda_{\rm{eff}}(Q^2)$ (last column).}
\label{table2}
\end{table}

Value $\chi^2(\vec{\lambda}_m) \approx 10^{-10}$ means that ratios are equal 1 with a precision of the order $10^{-7}$. On the other hand uncertainties of those ratios calculated from formula (\ref{uncRat}) are of order $10^{-1}$ (see Fig. \ref{zesWyk4}). This shows that in fact we do not obtain results which improve geometrical scaling, $Q^2$ dependence of $\lambda$ which we have found is caused by fluctuations, it can be also seen in Fig. \ref{zesWyk10}. 

Using this analysis we do not obtain any nontrivial $Q^2$ dependence of $\lambda$ regardless of $x_{\rm{cut}}$ or $W$. In particular $\lambda_{\rm{eff}}(Q^2)$ (found using HERA data) do not improve geometrical scaling but even makes it even worse comparing to $\lambda=\textrm{const.}$ (we should remember, however, that eyeball fit (\ref{eyeballfit}) is good only for quite narrow range of $Q^2$).

One should note that original data are divided into Bjorken-$x$ bins. In subsection \ref{sectDivisionofdata} we changed binning but after this procedure we have obtained data of worse quality (\textit{i.e.} for given $W$ data points are not smooth - see \ref{sectXoryg}). When we perform analysis for $\lambda=\textrm{const.}$ data obtained in that way are sufficient, however, $Q^2$ dependence is expected to be quite mild and quality of our data is probably not sufficient to analyze it. Indeed, in the next section we will show that for Bjorken-$x$ binning (where data have better quality) $Q^2$ dependence occurs for some $x'$ and $x'_{\rm{ref}}$.

\subsection{Bjorken-$x$ binning}

Now we will check if $Q^2$ dependence can be found for Bjorken-$x$ binning. Quantities are denoted with prime which means that we use Bjorken-$x$ binning.

We will search for $\lambda(Q^2)$ for given $x'$ and $x'_{\rm{ref}}$ (see subsection \ref{metanalysB-x}). The method is similar like for energy binning: we assign to every $Q^2$ one parameter $\lambda'_i$ (coordinate of $\vec{\lambda}'$) - some coordinates correspond to $x'$ points, some to $x'_{\rm{ref}}$.

$\vec{\lambda}'_{m}=\left(\lambda'^{m}_1, \lambda'^{m}_2,\ldots,\lambda'^{m}_N \right)$ is by definition a vector which minimizes $\chi'^2$ for given $x'$ and $x'_{\rm{ref}}$.

\subsubsection{Results}

\begin{table}
\centering
\begin{tabular}{|c|c|c|c|} \hline
 & $\chi'^2(\lambda'_{\rm{min}})$ & $\chi'^2(\vec{\lambda}'_m)$ & $\chi'^2\left(\lambda_{\rm{eff}}\right)$ \\ \hline
$x'_{\rm{ref}}=0.0005$, $x'=0.00032$  & 0.16 & $2.2\times 10^{-10}$ & 0.09 \\
$x'_{\rm{ref}}=0.002$, $x'=0.0005$ & 0.12 & $ 2.3 \times10^{-12}$ & 0.54 \\
$x'_{\rm{ref}}=0.0013$, $x'=0.0008$ & 0.55 & $7.1\times 10^{-9}$ & 0.43 \\
$x'_{\rm{ref}}=0.032$, $x'=0.008$  & 0.70 & $1.7\times 10^{-12}$ & 19.3 \\
$x'_{\rm{ref}}=0.08$, $x'=0.02$  & 8.0 & $1.5\times 10^{-11}$ & 2.5 \\
$x'_{\rm{ref}}=0.18$, $x'=0.08$  & 0.80 & $1.5\times 10^{-12}$ & 45.9 \\ \hline
\end{tabular}
\caption{Values of $\chi'^2$: minimized using one parameter $\lambda'$ (second column), minimized using many parameters $\left(\lambda'_1, \lambda'_2,\ldots,\lambda'_N \right)$ (third column) and calculated for $\lambda_{\rm{eff}}(Q^2)$ (last column).}
\label{tablechiminimal}
\end{table}

\begin{figure}
\includegraphics[width=7cm,angle=0]{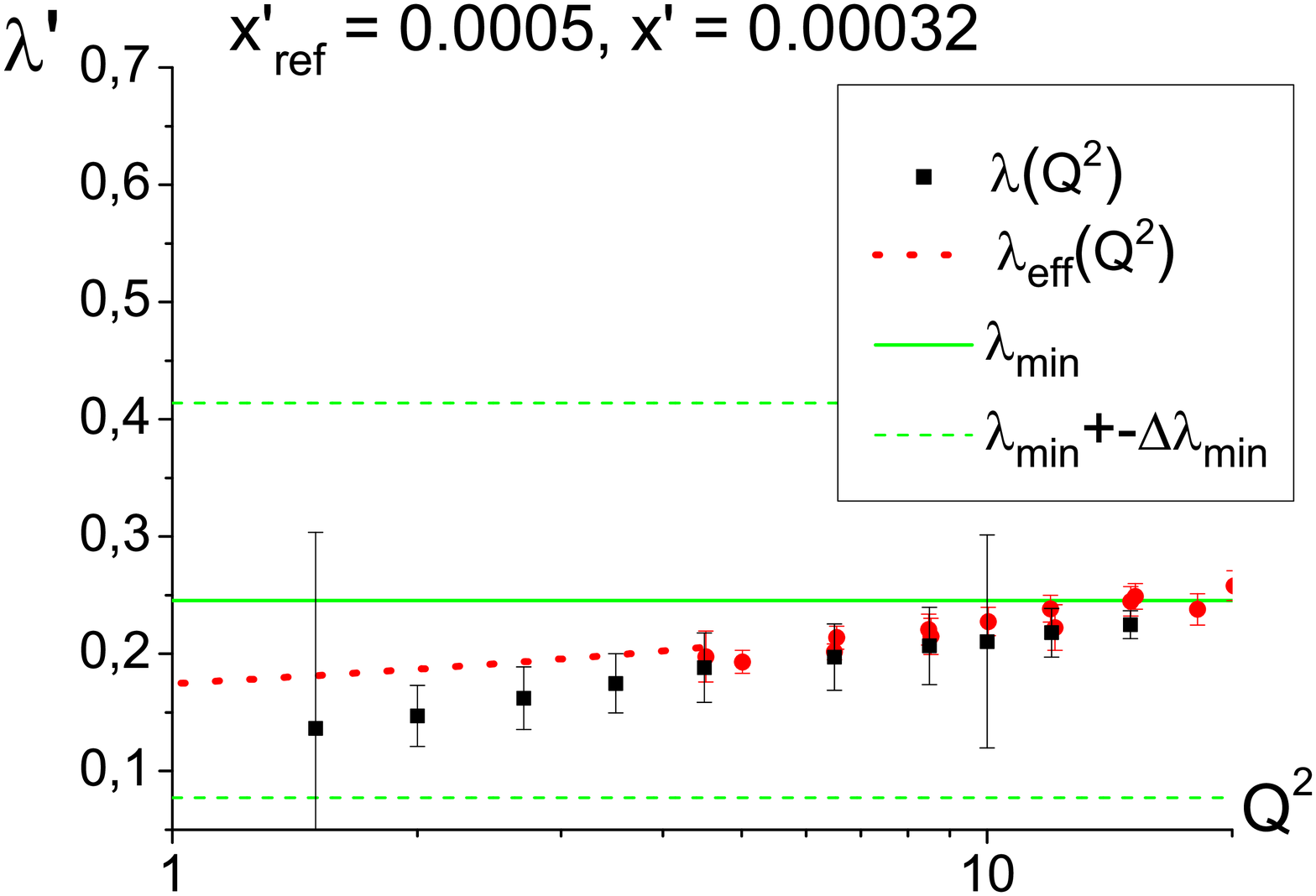}
\includegraphics[width=7cm,angle=0]{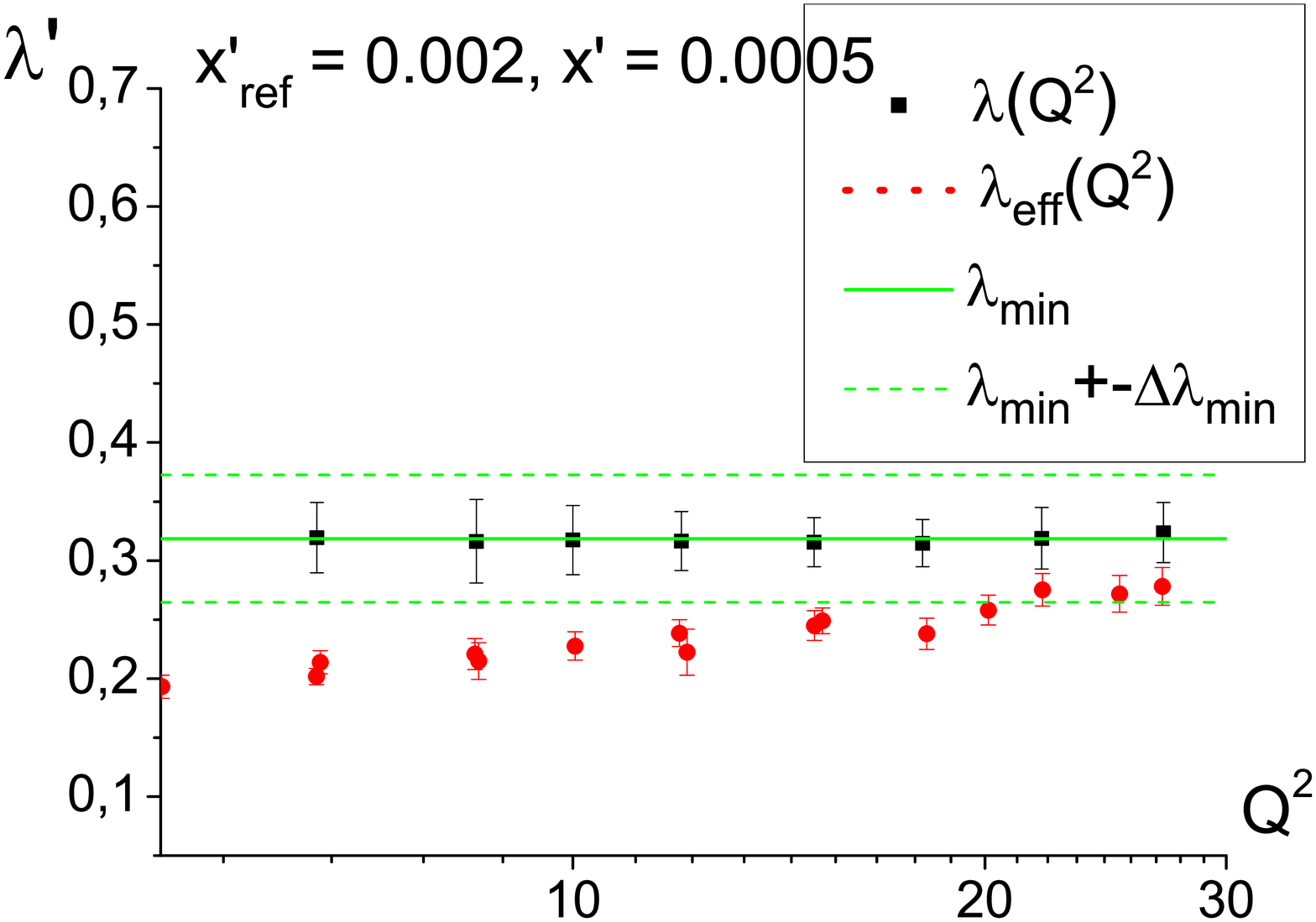}
\includegraphics[width=7cm,angle=0]{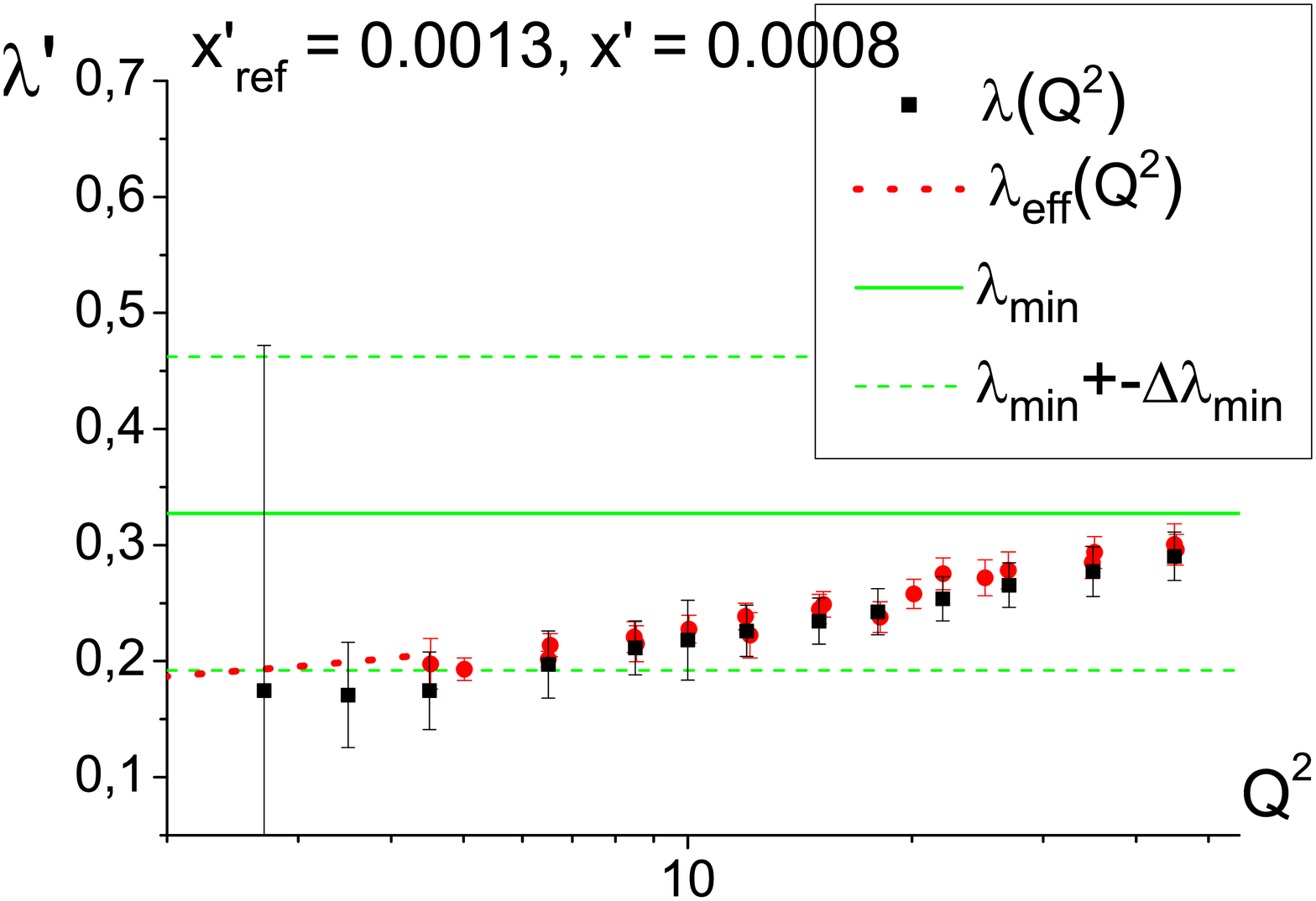}
\includegraphics[width=7cm,angle=0]{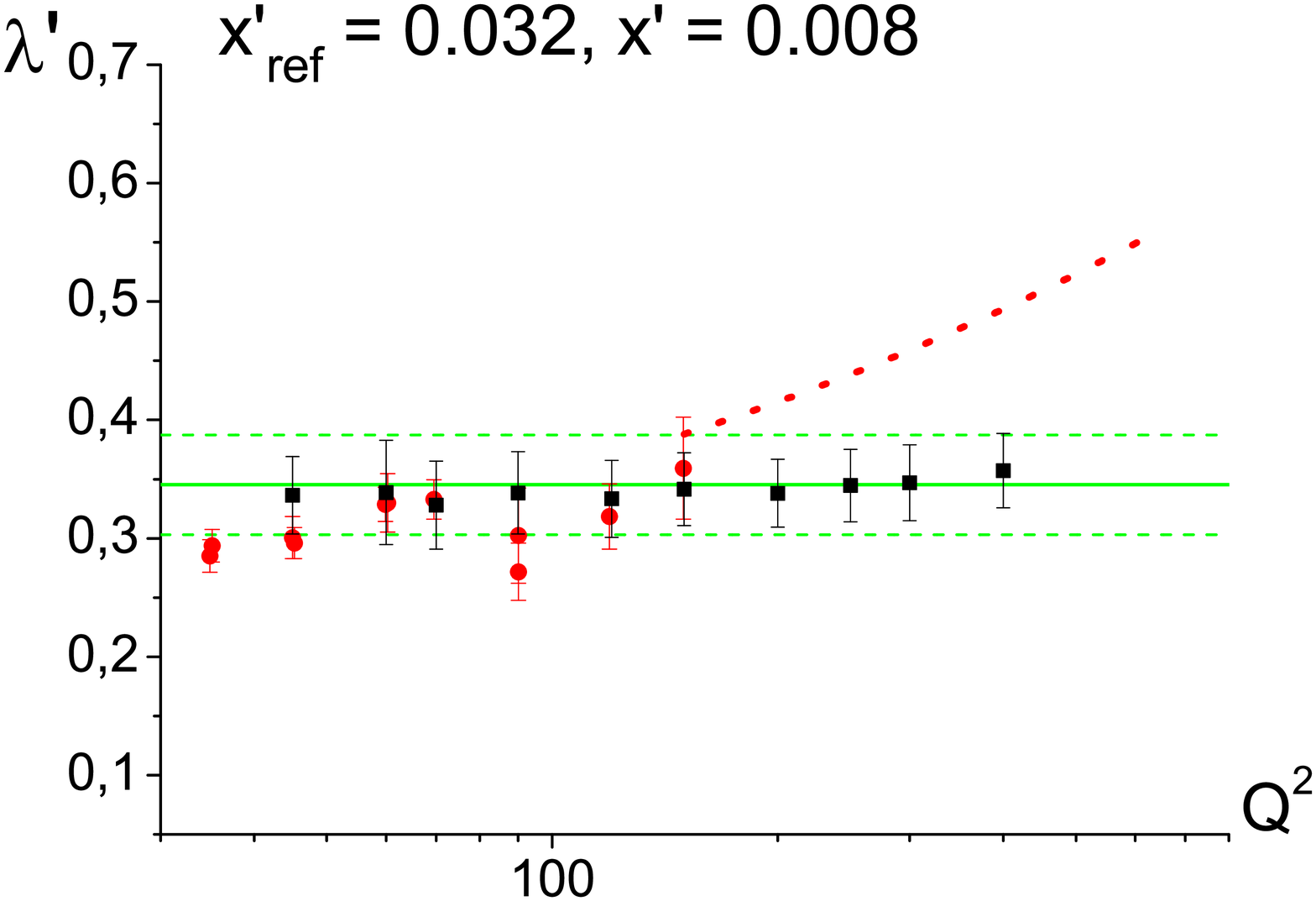}
\includegraphics[width=7cm,angle=0]{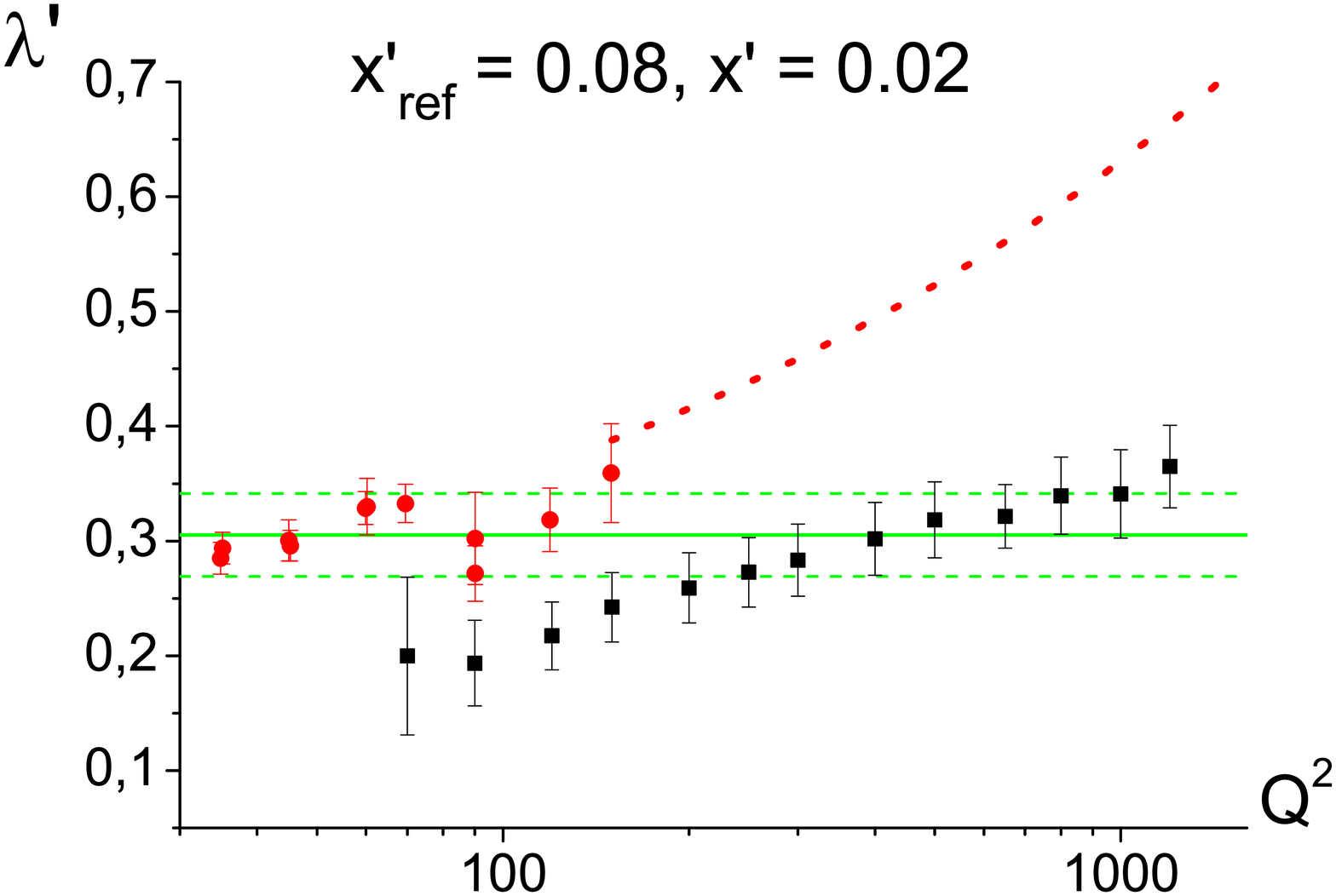}
\includegraphics[width=7cm,angle=0]{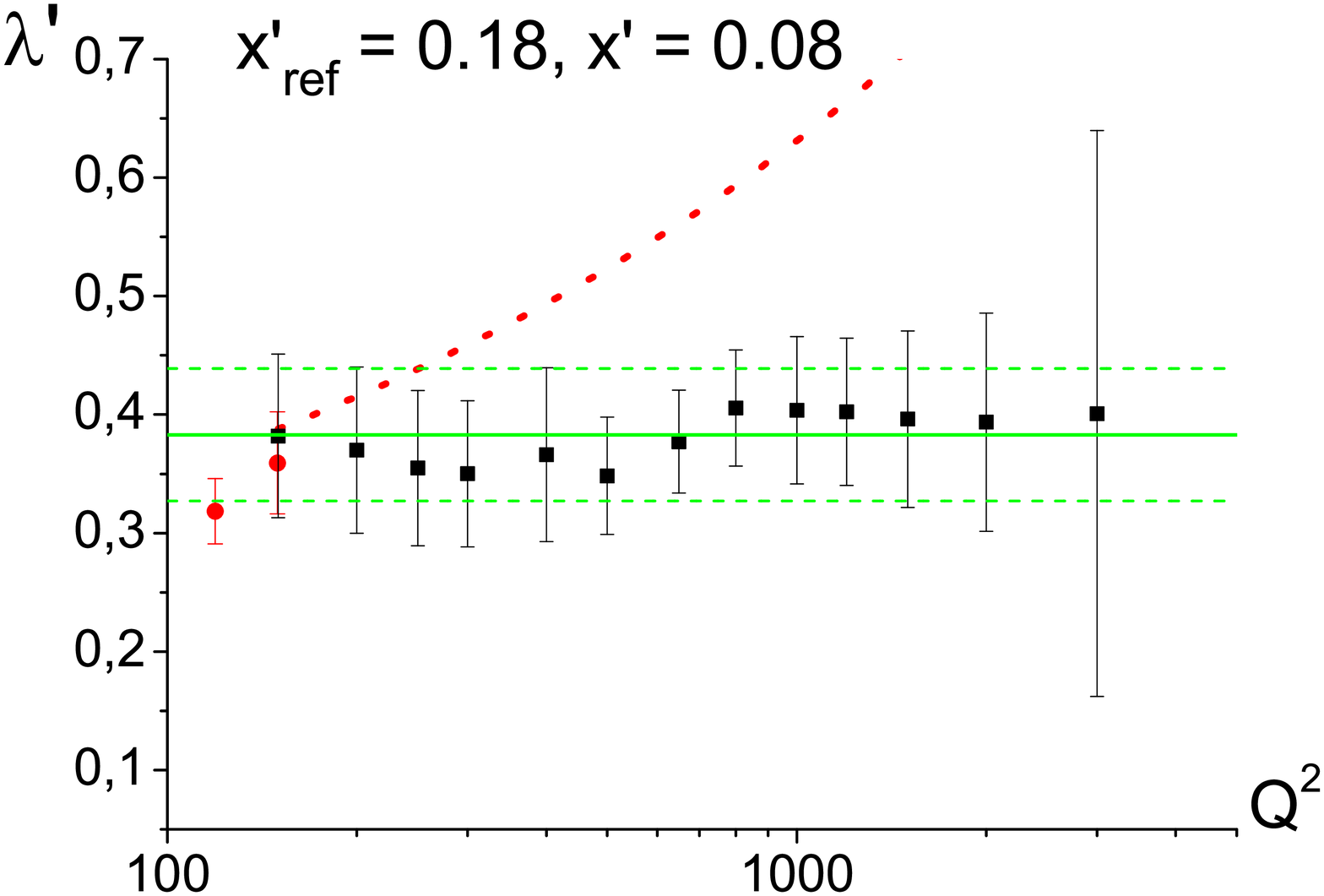}
\caption{$\lambda'(Q^2)$ (\textit{i.e.} coordinates of $\vec{\lambda}'_{m}$ which correspond to $x'$, plotted in terms of $Q^2$) for six ($x'_{\rm{ref}}$,$x'$) combinations. $\lambda_F(Q^2)$ values are shown as red points; red, dotted line represents $\lambda_{\rm{eff}}(Q^2)$ plotted for $Q^2 < 4.5 \textrm{ GeV}^2$ and $Q^2 > 150 \textrm{ GeV}^2$.}
\label{zesWyk40}
\end{figure}

In Fig. \ref{zesWyk40} we show six examples of functions $\lambda'(Q^2)$ (black points). One can see that for some combinations of $x'_{\rm{ref}}$ and $x'$ we can find $Q^2$ dependence of $\lambda'$. Moreover, for some pairs ($x'_{\rm{ref}}$, $x'$) this dependence is quite similar to $\lambda_{\rm{eff}}(Q^2)$. There are, however, pairs ($x'_{\rm{ref}}$, $x'$) for which there is no $Q^2$ dependence or it is different than $\lambda_{\rm{eff}}(Q^2)$. In fact we have not found any rule which describes for which ($x'_{\rm{ref}}$, $x'$) $Q^2$ dependence is present. 

In Table \ref{tablechiminimal} we show values of $\chi'^2$ for pairs ($x'_{\rm{ref}}$, $x'$) used in Fig. \ref{zesWyk40}. We can compare minimized values for $\lambda'=\textrm{const.}$ (second column) and $\lambda'=\lambda'(Q^2)$ (third column); in fourth column we show $\chi'^2$ calculated for $\lambda_{\rm{eff}}(Q^2)$:
\begin{itemize} 
\item similarly like for energy binning $\chi'^2$ can be minimized to 0 with great precision (third column) when we use independent parameters assigned to different $Q^2$.
\item for most ($x'_{\rm{ref}}$, $x'$) using $\lambda'=\textrm{const.}$ one can minimize $\chi'^2$ to quite small values (i.e $\chi'^2\lesssim 1$). For these pairs also $\chi'^2(\lambda_{\rm{eff}})$ are small if we use it only for proper range of $Q^2$ (\textit{i.e.} we do not extrapolate $\lambda_{\rm{eff}}$ outside $4.5 \textrm{ GeV}^2 \leq Q^2 \leq 150 \textrm{ GeV}^2$).
\end{itemize}

To perform more precise analysis one should have values of $\lambda_{F}$ for wider range of $Q^2$. Also, as we mentioned at the beginning one should know for what $x$-range values $\lambda_{F}(Q^2)$ were found.

\chapter{Results for $e^{-}p$ data}

In this Chapter we will present results for $\lambda$ obtained for $e^{-}p$ collisions and compare them with $e^{+}p$ results. $e^{-}p$ data are taken from Tables 14-16 \cite{H1ZueusWork}. To simplify notation we denote cross section $F_2/Q^2$ in $e^{+}p$ collisions by $\tilde{\sigma}^{+}$ and cross section $F_2/Q^2$ in $e^{-}p$ collisions by $\tilde{\sigma}^{-}$.

Data for $e^{-}p$ are poorer than for $e^{+}p$: range of $Q^2$ values are limited to $90\textrm{ GeV}^2$ $\leq Q^2 \leq 20000\textrm{ GeV}^2$ whereas in $e^{+}p$ case it was $0.1 \textrm{ GeV}^2$ $\leq Q^2 \leq 20000 \textrm{ GeV}^2$. To compare results we exclude from $e^{+}p$ data points with $Q^2<90\rm{GeV}^2$ (this means also exclusion low $x$ values due to relation $x=\frac{Q^2}{Q^2+W^2-M^2}$).

\section{Energy binning}

\subsection{Data}

We use the same binnig as for $e^{+} p$ \textit{i.e.} logarithmic binning with step 1.3 (see Table \ref{tablebinningeMinp}). We choose $W_{\rm{ref}}=206$ GeV (like previously).

\begin{table}
\begin{tabular}{|c|c|c|c|c|c|c|c|} \hline
$W'_{\rm{min}}\rm{ [GeV]}$ & 10 & 13 & 16.9 & 22 & 28.6 & 37.1 & 48.3 \\ \hline
$W'_{\rm{max}}\rm{ [GeV]}$ & 13 & 16.9 & 22 & 28.6 & 37.1 & 48.3 & 62.7 \\ \hline
$W\rm{ [GeV]}$ & 11.5 & 15 & 19.4 & 25.3 & 32.8 & 42.7 & 55.5 \\ \hline
Number of $e^{+}p$ points & 0 & 0 & 6 & 9 & 8 & 12 & 14 \\ \hline
Number of $e^{-}p$ points & 0 & 1 & 3 & 4 & 6 & 9 & 11 \\ \hline \hline
$W'_{\rm{min}}\rm{ [GeV]}$ & 62.7 & 81.6 & 106 & 137.9 & 179.2 & 233 &  \\ \hline
$W'_{\rm{max}}[\rm{ [GeV]}]$ & 81.6 & 106 & 137.9 & 179.2 & 233 & 302.9 &  \\ \hline
$W\rm{ [GeV]}$ & 72.2 & 93.8 & 122 & 158.5 & 206.1 & 267.9 &  \\ \hline
Number of $e^{+}p$ points & 17 & 17 & 18 & 18 & 18 & 4 &  \\ \hline
Number of $e^{-}p$ points & 13 & 14 & 16 & 16 & 18 & 4 &  \\ \hline
\end{tabular}
\caption{Energy bins and energies assigned to them. Number of points which are present in different bins are also written.}
\label{tablebinningeMinp}
\end{table}

\begin{figure}
\includegraphics[width=7cm,angle=0]{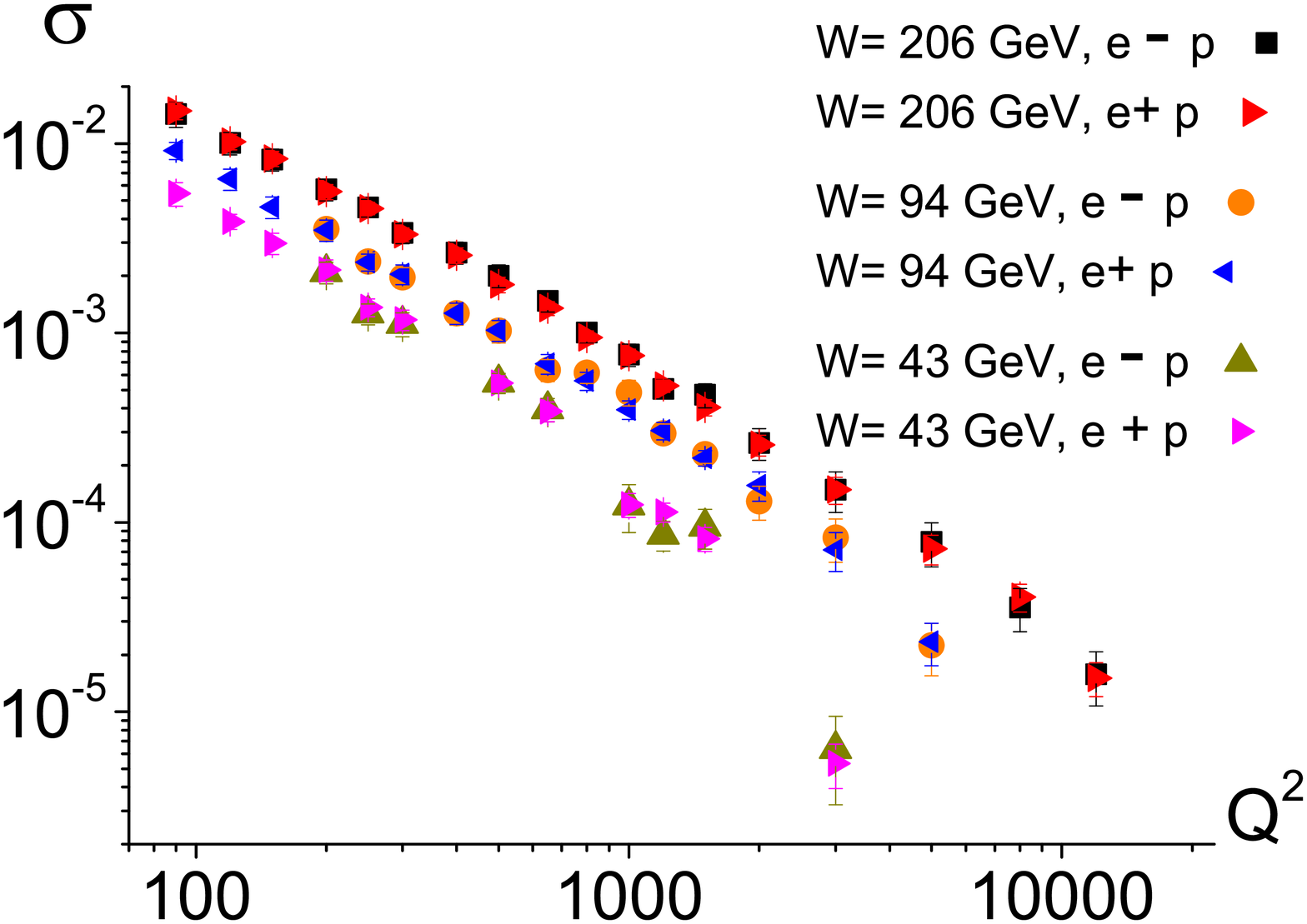}
\includegraphics[width=7cm,angle=0]{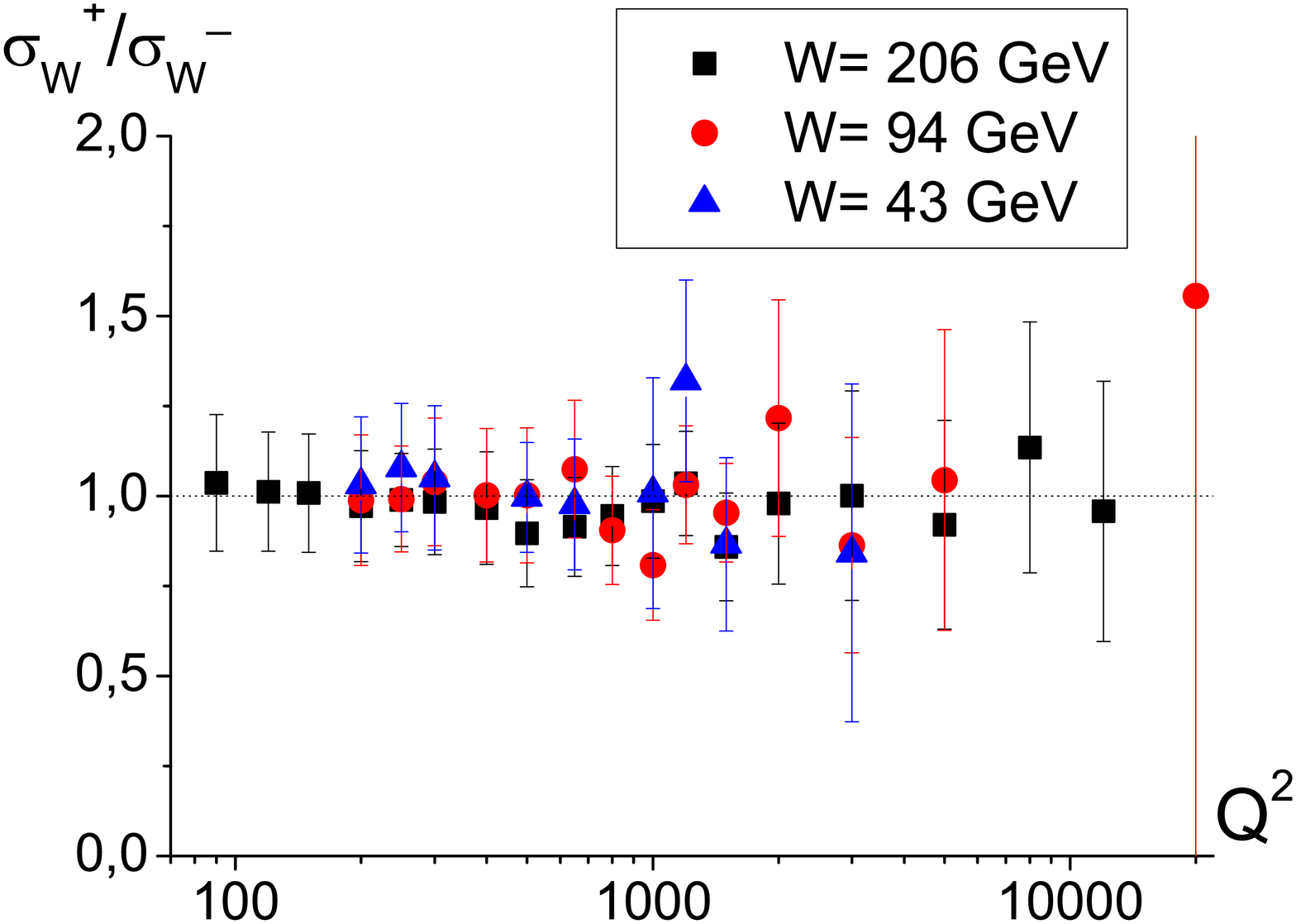}
\caption{Left: functions $\tilde{\sigma}^{-}_W(Q^2)$ and $\tilde{\sigma}^{+}_W(Q^2)$ for three $W$. Right: ratios $\tilde{\sigma}^{+}_W/\tilde{\sigma}^{-}_W$ for these energies.}
\label{zesWyk36}
\end{figure}

In left Fig. \ref{zesWyk36} we show functions $\tilde{\sigma}^{-}_W(Q^2)$ and $\tilde{\sigma}^{+}_W(Q^2)$ for three energies. Right plot presents ratios $\tilde{\sigma}^{+}_W(Q^2)/\tilde{\sigma}^{-}_W(Q^2)$ for these energies. It is easy to see, that values of cross section for $e^{+}p$ and $e^{-}p$ are similar.

\subsection{$\lambda_{\rm{min}}(x_{\rm{cut}})$ for given $W$}
\begin{figure}
\includegraphics[width=4.5cm,angle=0]{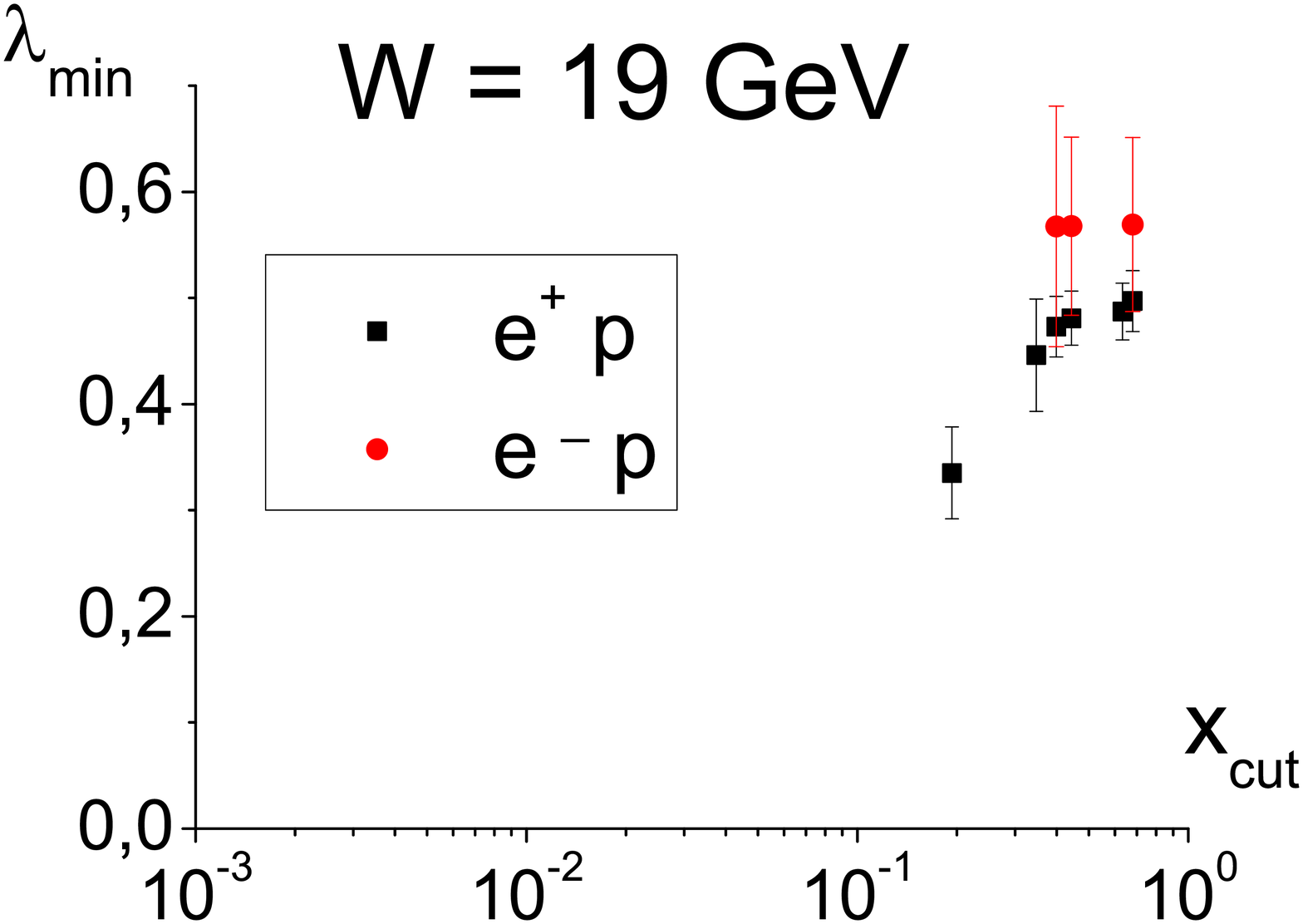}
\includegraphics[width=4.5cm,angle=0]{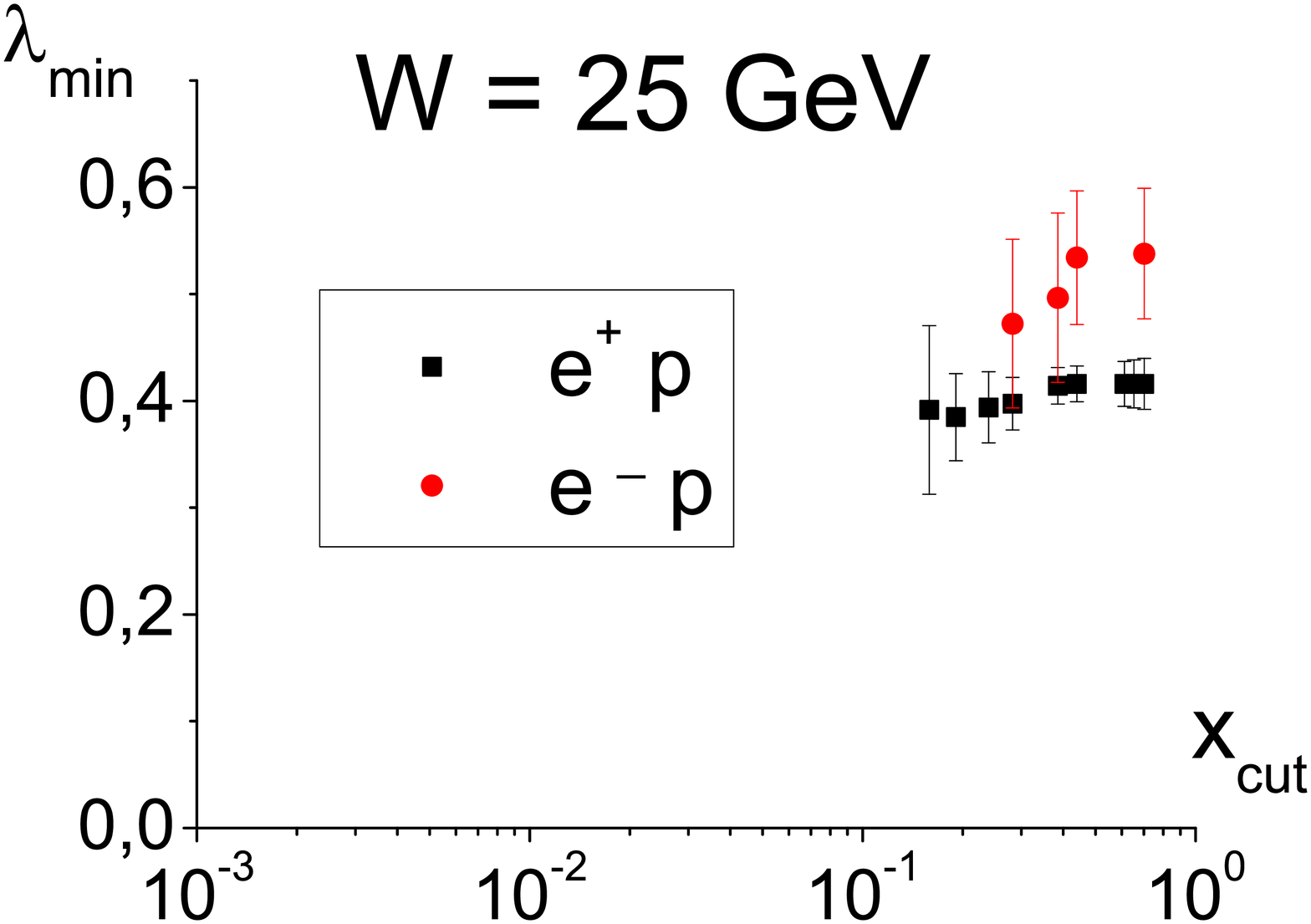}
\includegraphics[width=4.5cm,angle=0]{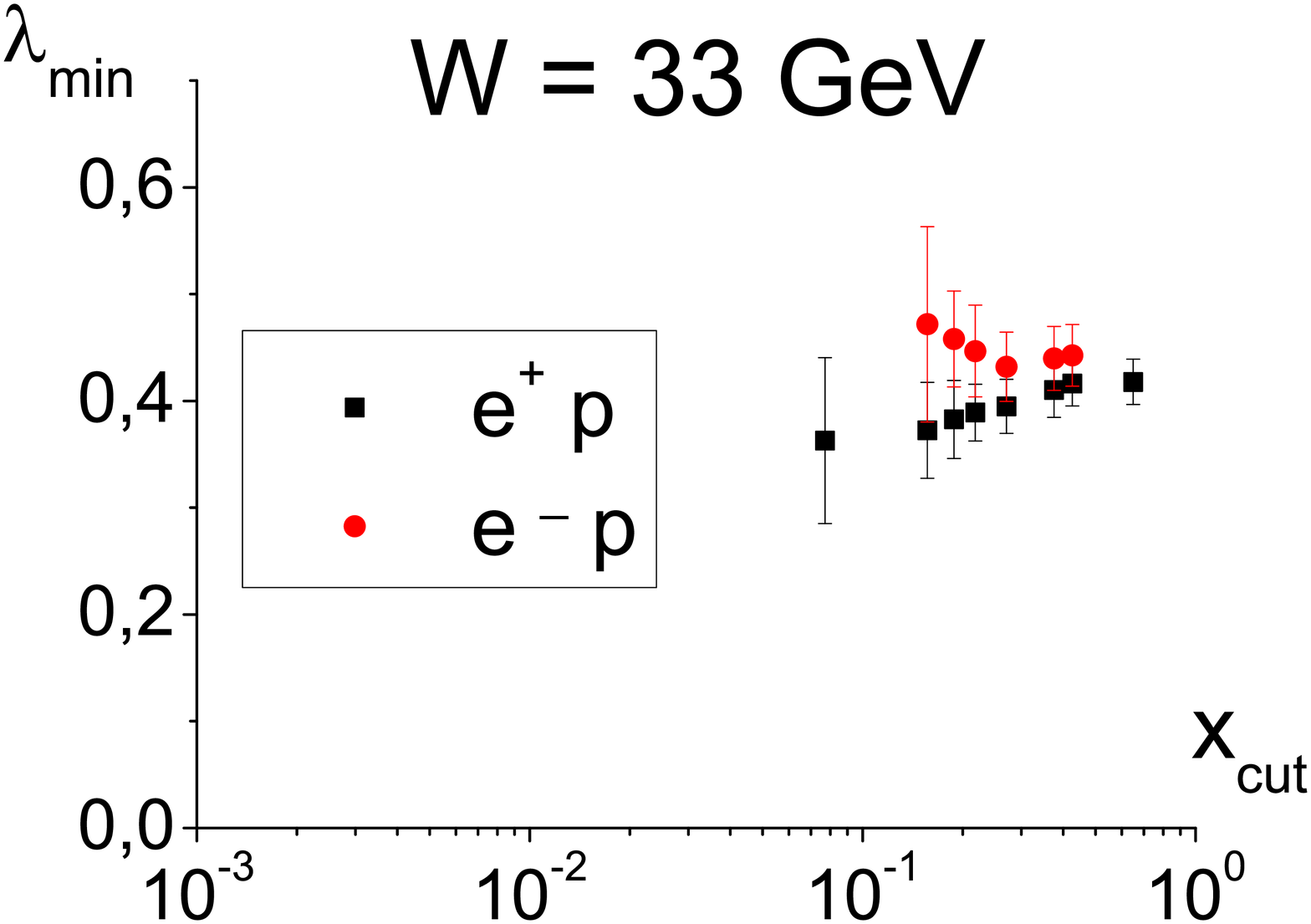}
\includegraphics[width=4.5cm,angle=0]{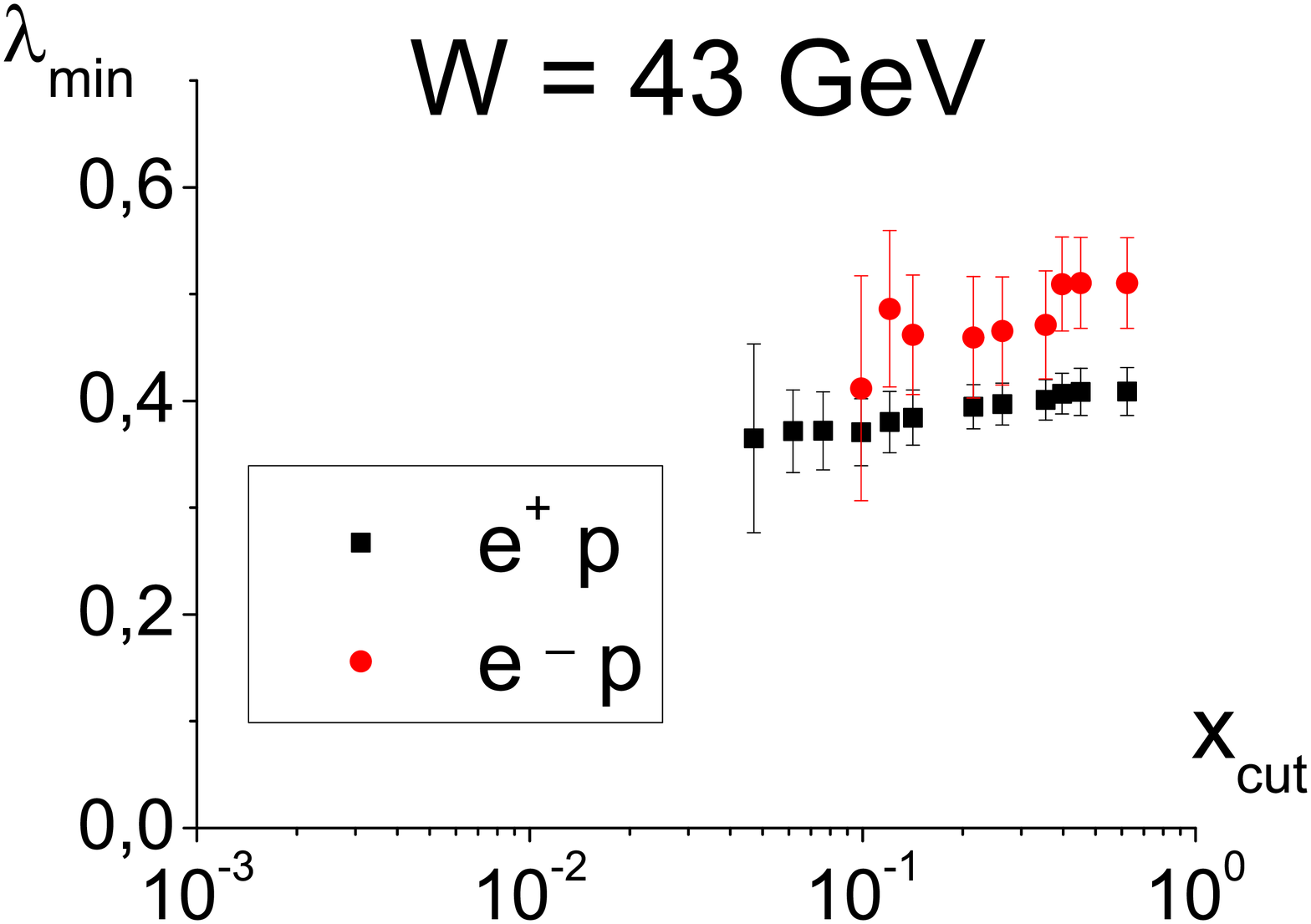}
\includegraphics[width=4.5cm,angle=0]{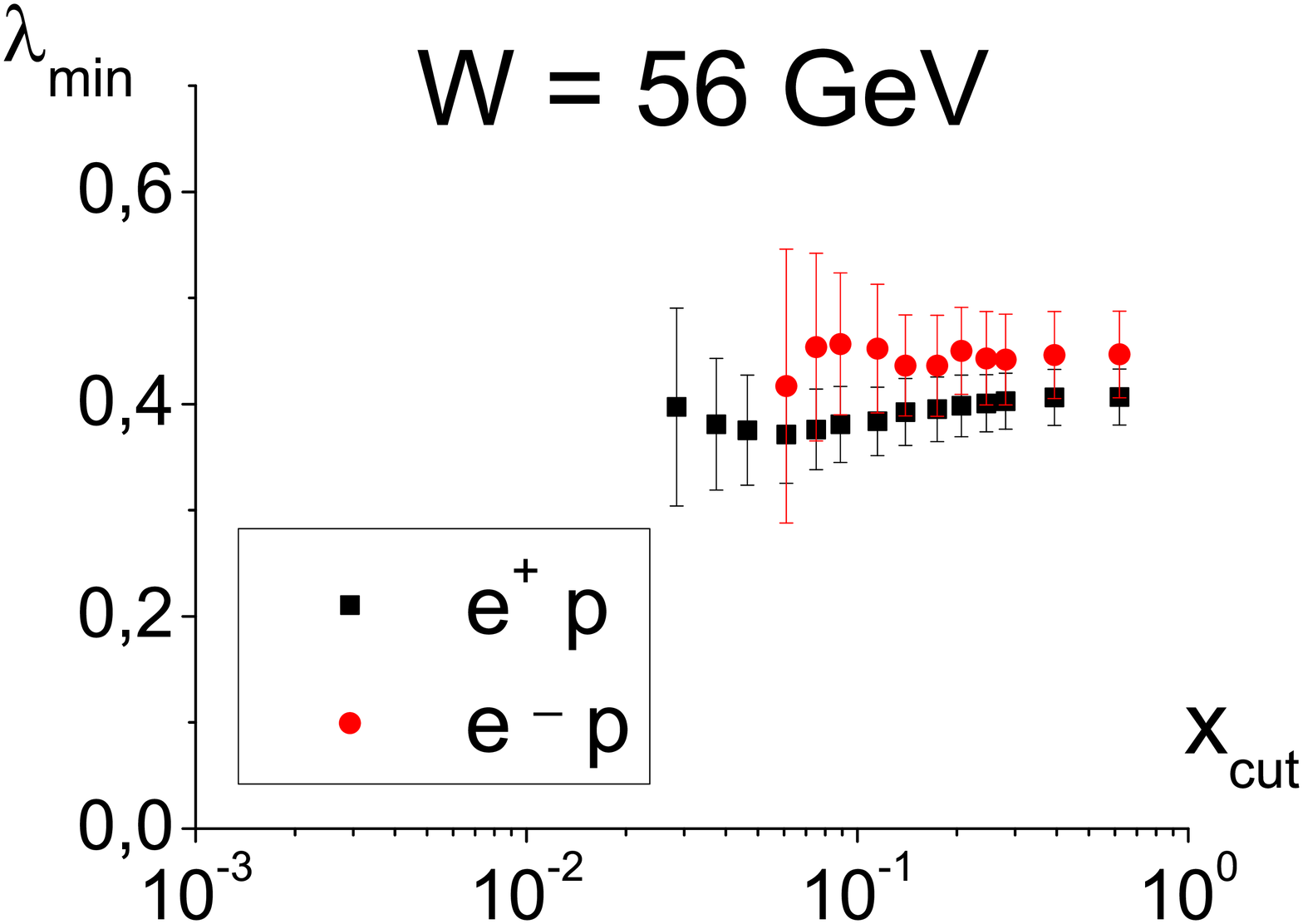}
\includegraphics[width=4.5cm,angle=0]{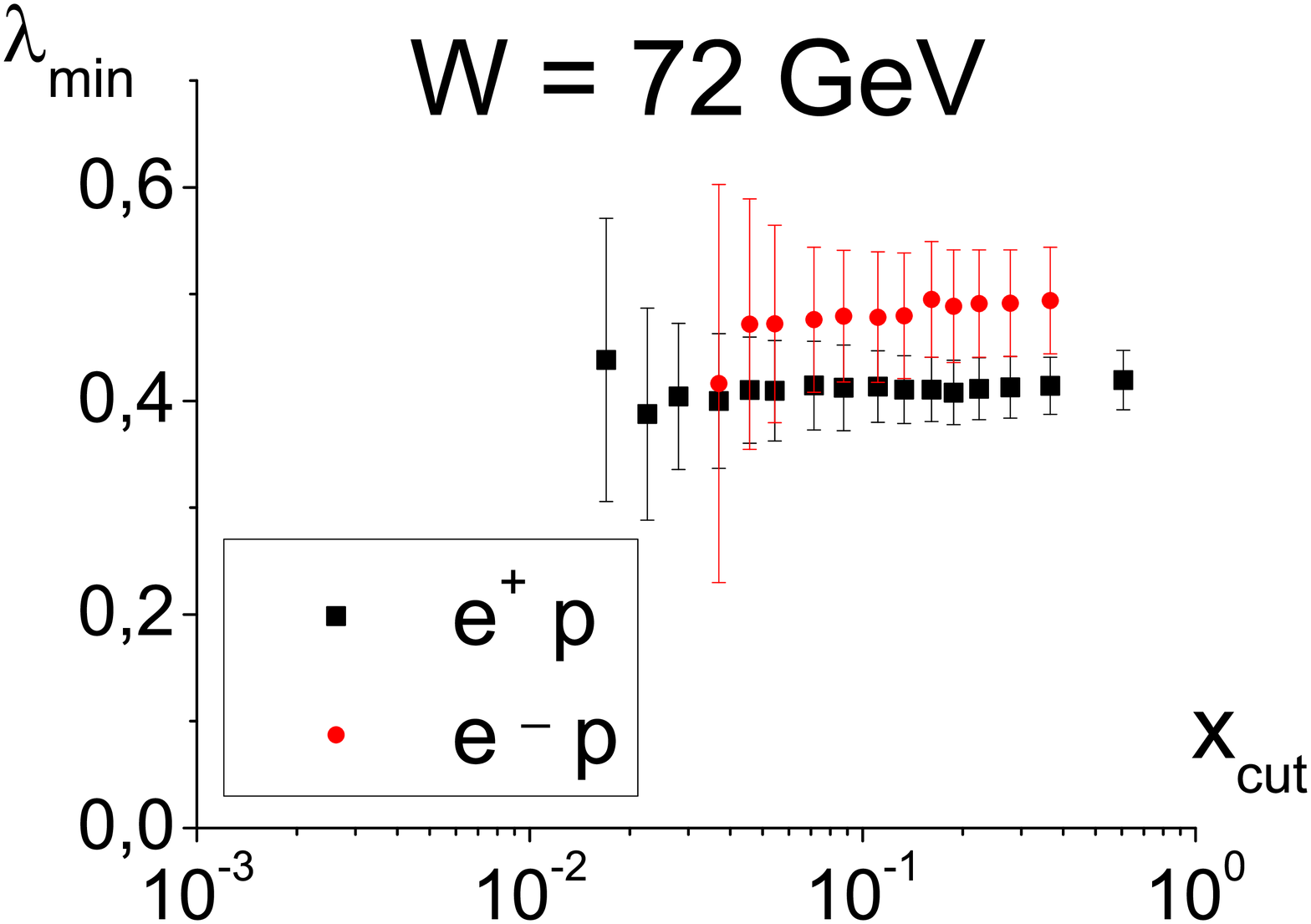}
\includegraphics[width=4.5cm,angle=0]{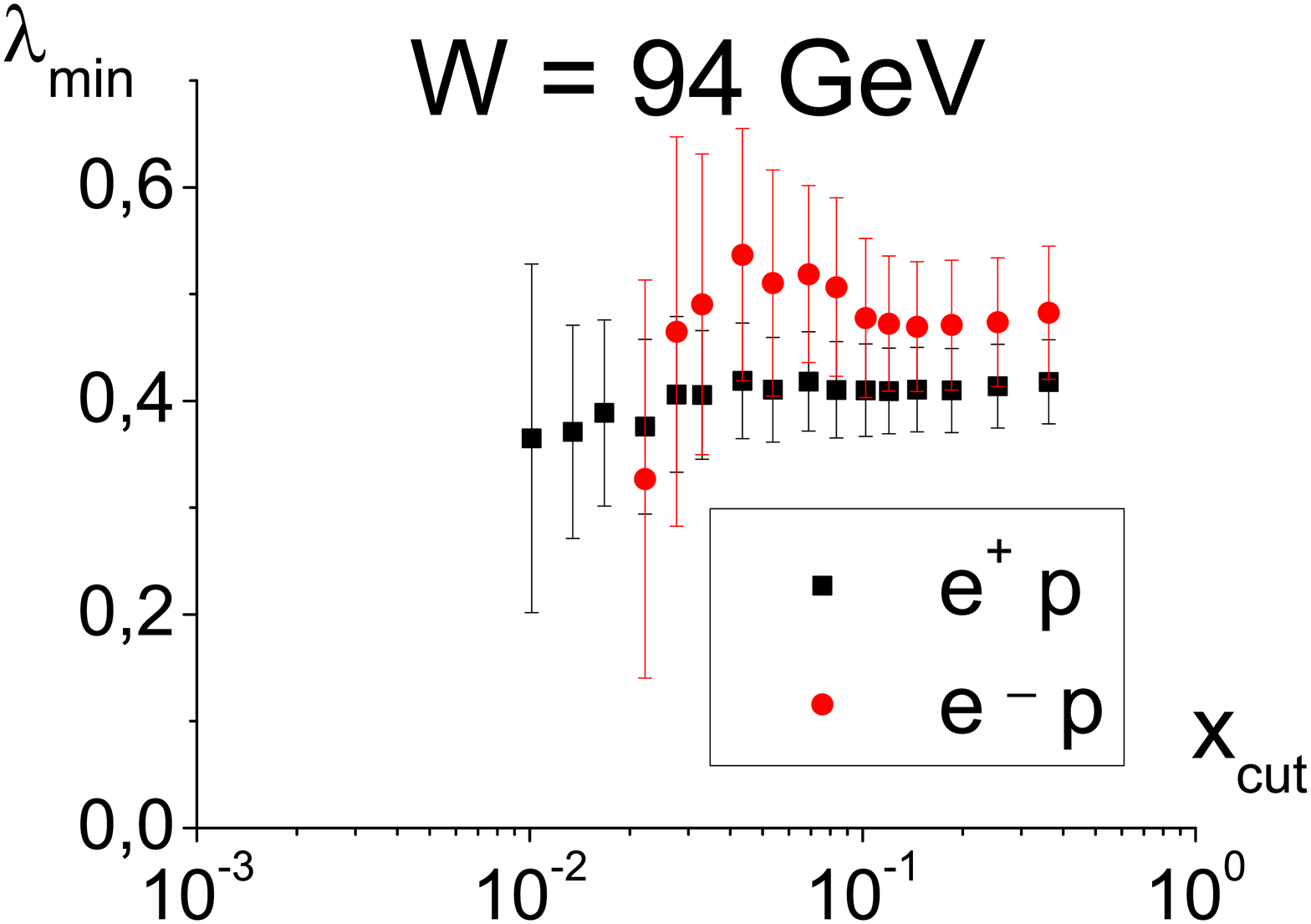}
\includegraphics[width=4.5cm,angle=0]{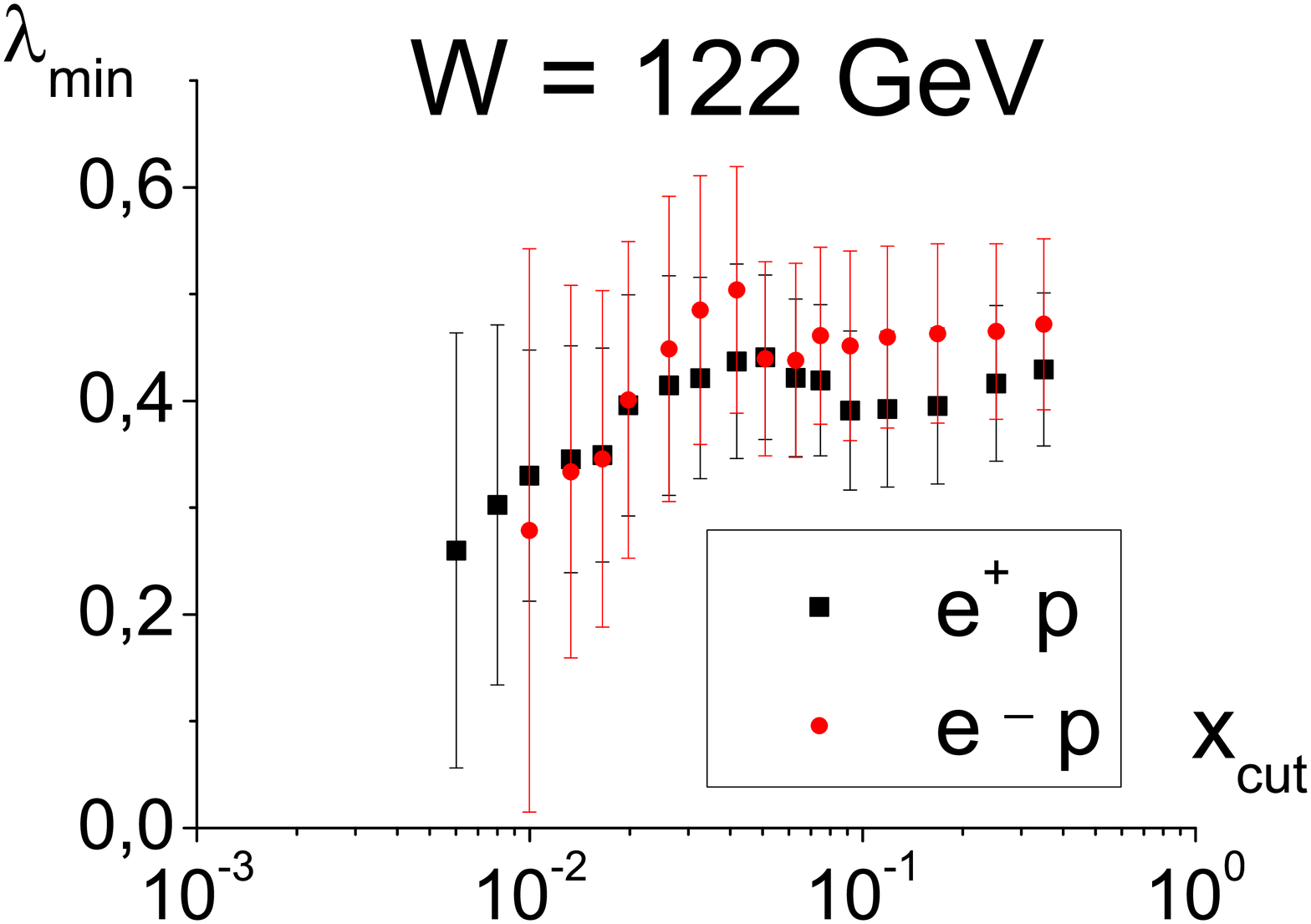}
\includegraphics[width=4.5cm,angle=0]{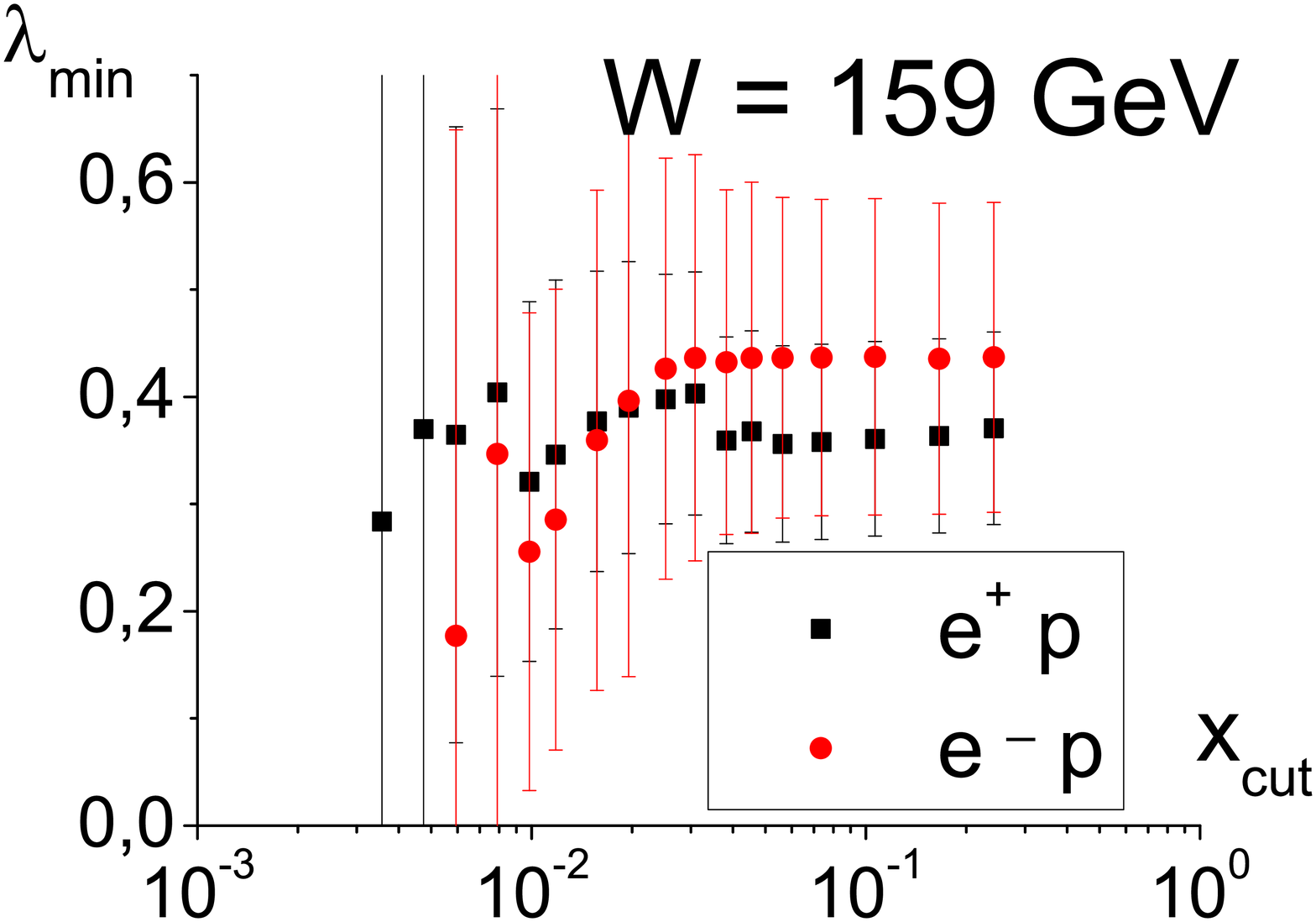}
\caption{Functions $\lambda_{\rm{min}}(x_{\rm{cut}})$ for $e^{-}p$ and $e^{+}p$.}
\label{zesWyk13}
\end{figure}

In Fig. \ref{zesWyk13} we present plots $\lambda_{\rm{min}}(x_{\rm{cut}})$ for $e^{-}p$ and $e^{+}p$. We can see that values of $\lambda_{\rm{min}}$ for $e^{-}p$ are little bit bigger than for $e^{+}p$. It is quite surprising since cross sections for $e^{+}p$ and $e^{-}p$ are similar with good precision (see Fig. \ref{zesWyk36}). This is probably caused by low quality of data after energy binning (see \ref{sectXoryg}), indeed as we will see in next section, for Bjorken-$x$ binning difference between $e^{-}p$ and $e^{+}p$ results is smaller.

\subsection{$\chi^2_{\rm{nor}}(x_{\rm{cut}})$ for given $W$}

\begin{figure}
\includegraphics[width=7cm,angle=0]{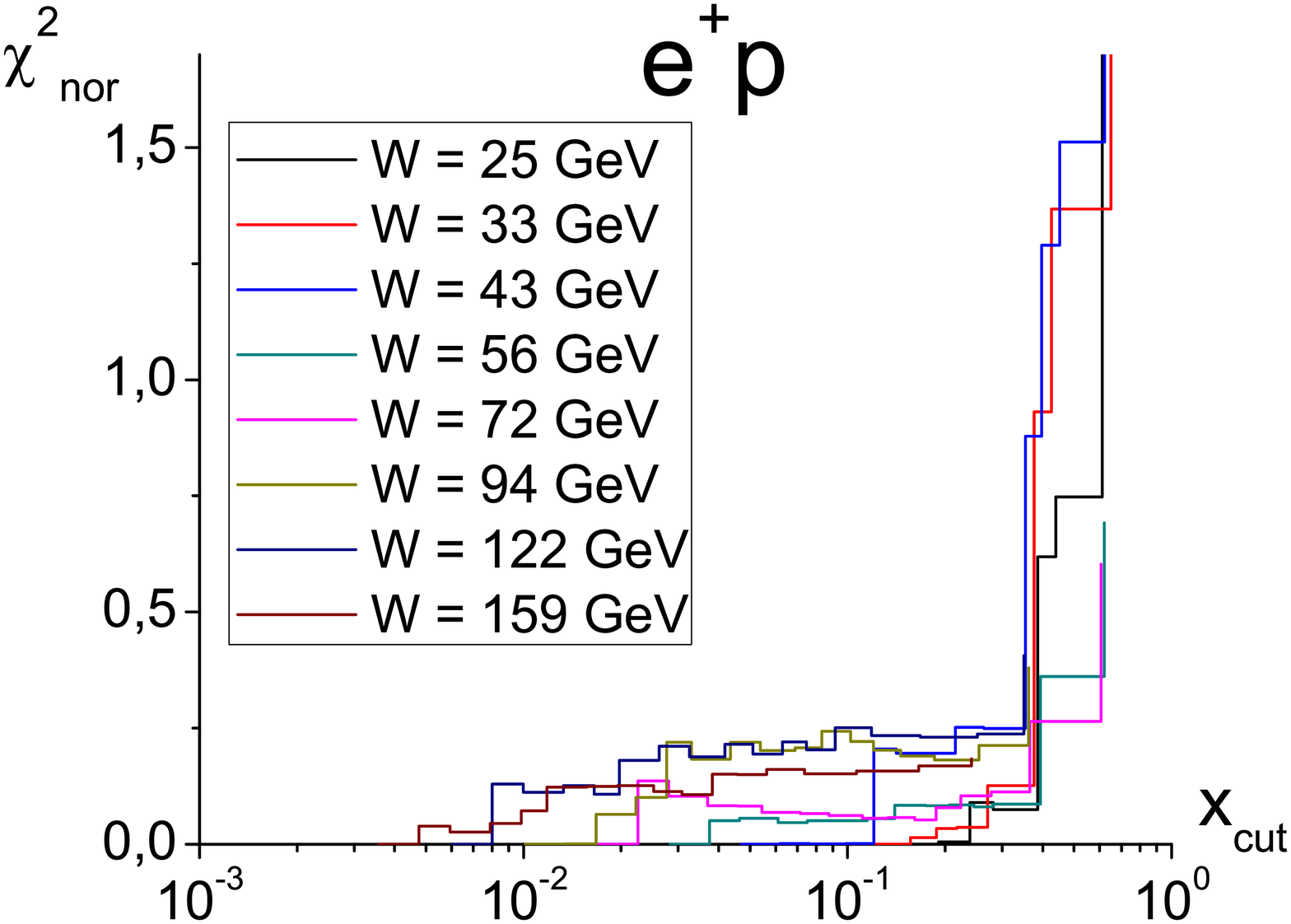}
\includegraphics[width=7cm,angle=0]{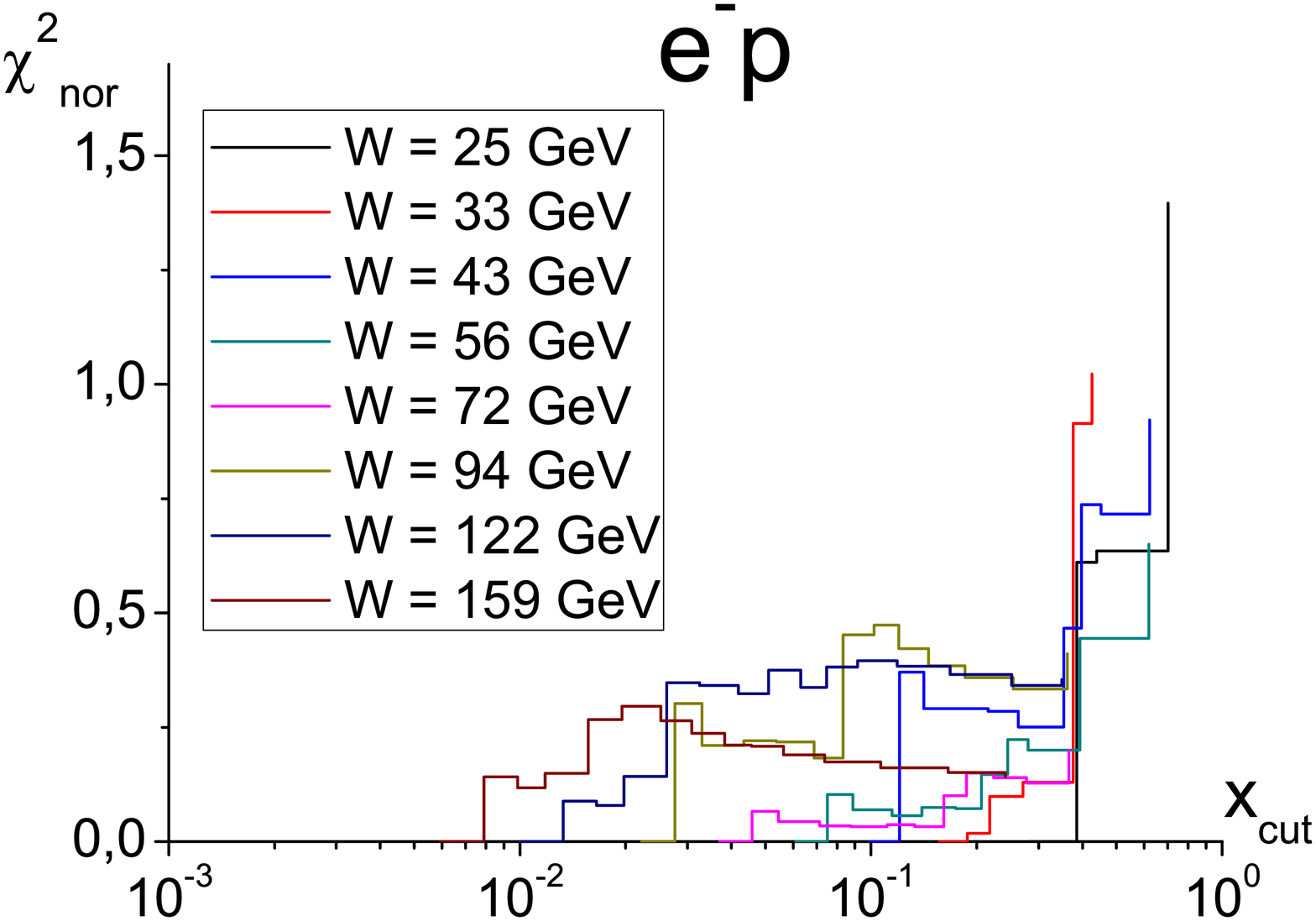}
\caption{$\chi^2_{\rm{nor}}(\lambda_{\rm{min}})$ as a function of $x_{\rm{cut}}$ for $25$ GeV $\leq W \leq 159$ GeV: for $e^{+}p$ (left plot) and for $e^{-}p$ (right plot).} 
\label{zesWyk37}
\end{figure}

In Fig. \ref{zesWyk37} we compare functions $\chi^2_{\rm{nor}}(x_{\rm{cut}})$ for $e^{-}p$ and $e^{+}p$: we can see that $\chi^2$ can be better minimized for $e^{+}p$. In both cases $x_{\rm{cut}}>0.2$ is a region where GS disappears (the same result we obtain in section \ref{sectchi^2nor}).  

\subsection{$\lambda(W)$ for given $x_{\rm{cut}}$}

\begin{figure}[h]
\includegraphics[width=7cm,angle=0]{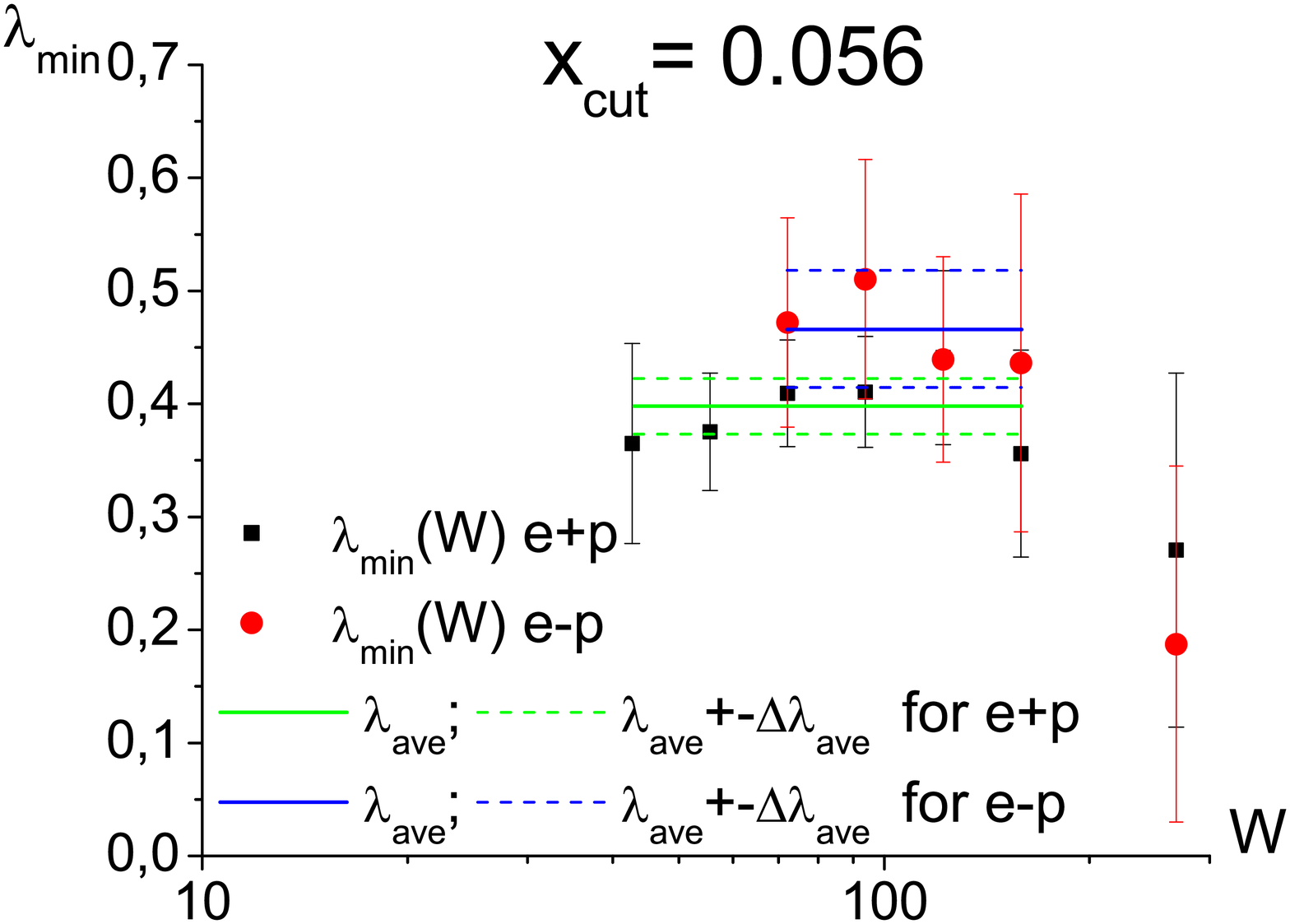}
\includegraphics[width=7cm,angle=0]{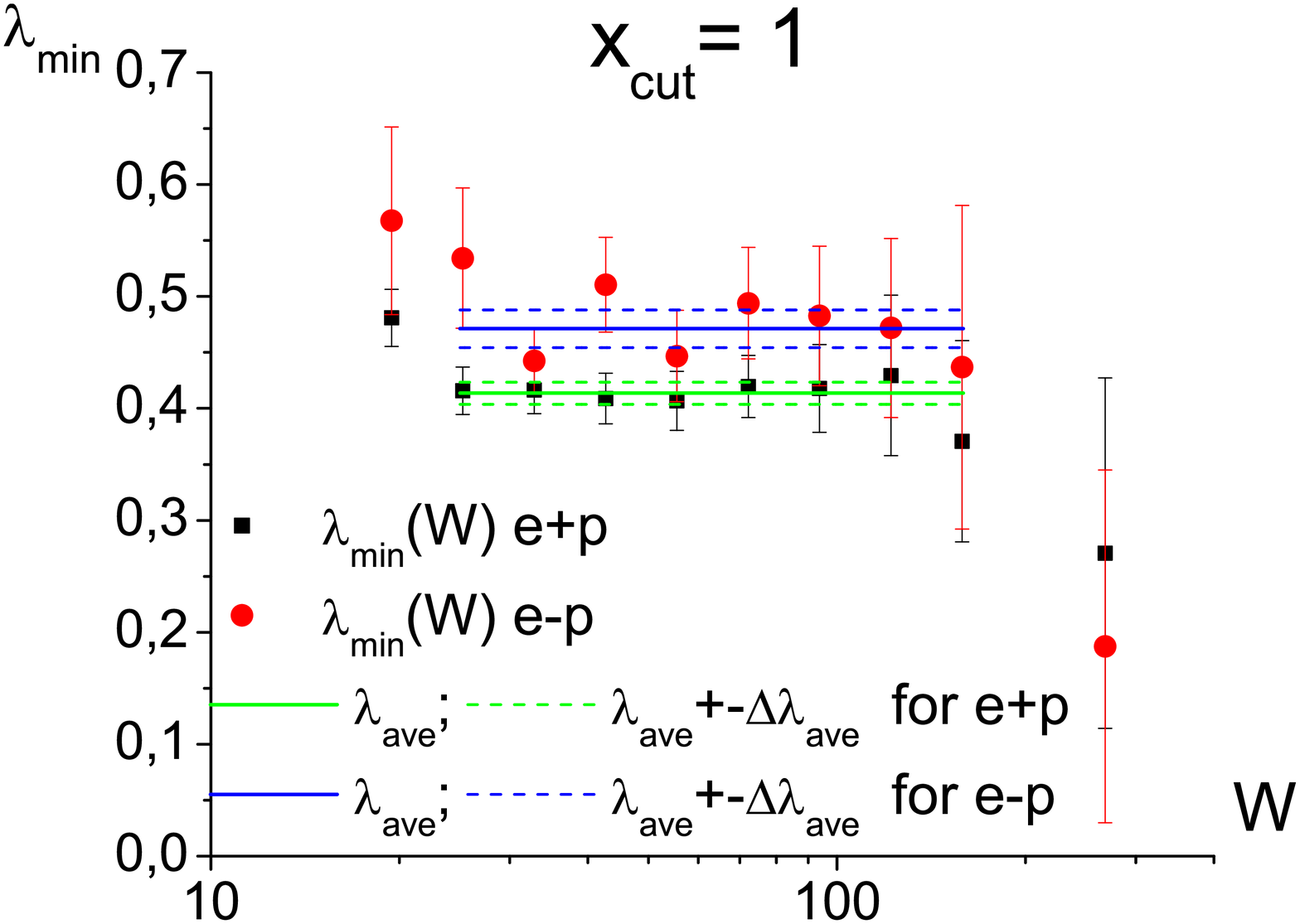}
\caption{$\lambda_{\rm{min}}(W)$ for two $x_{\rm{cut}}$ for $e^{+}p$ and $e^{-}p$. Horizontal lines represents $\lambda_{\rm{ave}}$ and its uncertainties (green for $e^{+}p$ and blue for $e^{-}p$), they are plotted for energies which were used to calculate them ($25$ GeV$\leq W \leq 159$ GeV if all energies are present).}
\label{zesWyk14}
\end{figure}

In Fig. \ref{zesWyk14} we show $\lambda_{\rm{min}}(W)$ comparison between $e^{+}p$ and $e^{-}p$ for two values of $x_{\rm{cut}}$. Horizontal lines are $\lambda_{\rm{ave}}$ and theirs uncertainty (see section \ref{sectlamave} for definition of $\lambda_{\rm{ave}}$).

\subsection{$\lambda_{\rm{ave}}(x_{\rm{cut}})$}

\begin{figure}[h]
\centering
\includegraphics[width=9cm,angle=0]{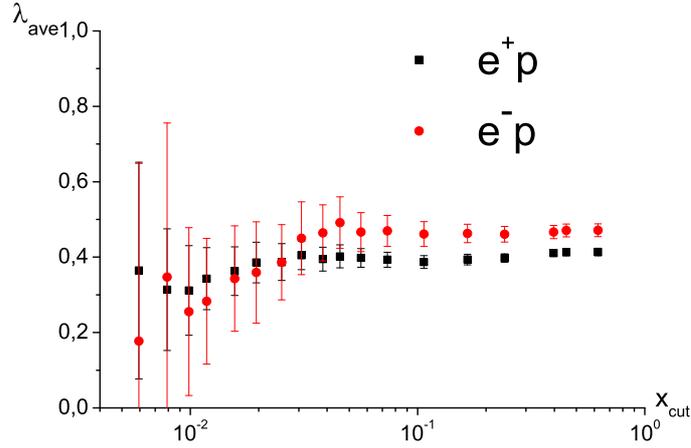}
\caption{$\lambda_{\rm{ave}}(x_{\rm{cut}})$ for $e^{+}p$ and $e^{-}p$.}
\label{zesWyk15}
\end{figure}

Fig. \ref{zesWyk15} shows $\lambda_{\rm{ave}}$ as a function of $x_{\rm{cut}}$ (red points), it is compared with $e^{+}p$ results (black points).

\section{Bjorken-$x$ binning}

\subsection{Data}

\begin{table}[b]
\begin{tabular}{|c|c|c|c|c|c|c|c|} \hline
$x'$ & 0.002 & 0.0032 & 0.005	&	0.008	&	0.013	&	0.02 & 0.032 \\ \hline
Number of $e^{+}p$ points &	2	&	3	&	6	&	7	&	9	&	12	&	13 \\ \hline
Number of $e^{-}p$ points &	2	&	2	&	4	&	5	&	6	&	9	&	10 \\ \hline \hline
$x'$ &	0.05	&	0.08	&	0.13	&	0.18	&	0.25	&	0.4	&	0.65 \\ \hline
Number of $e^{+}p$ points & 15	&	15	&	13	&	16	&	15	&	16	&	11 \\ \hline
Number of $e^{-}p$ points &	12	&	12	&	12	&	13	&	14	&	15	&	7 \\ \hline
\end{tabular}
\caption{$x'$ values used in analysis and numbers of points in bins for $e^{+}p$ and $e^{-}p$ data.}
\label{tableXbjor}
\end{table}

Structure of $e^{-}p$ data is similar to $e^{+}p$ \textit{i.e.} they are divided into bins with definite value of Bjorken-$x$ variable. As previously we denote those values by $x'$. In fact bins for $e^{-}p$ and $e^{+}p$ data are the same, in Table \ref{tableXbjor} we show values $x'$ which we will use and number of points in these bins.

\begin{figure}
\includegraphics[width=7cm,angle=0]{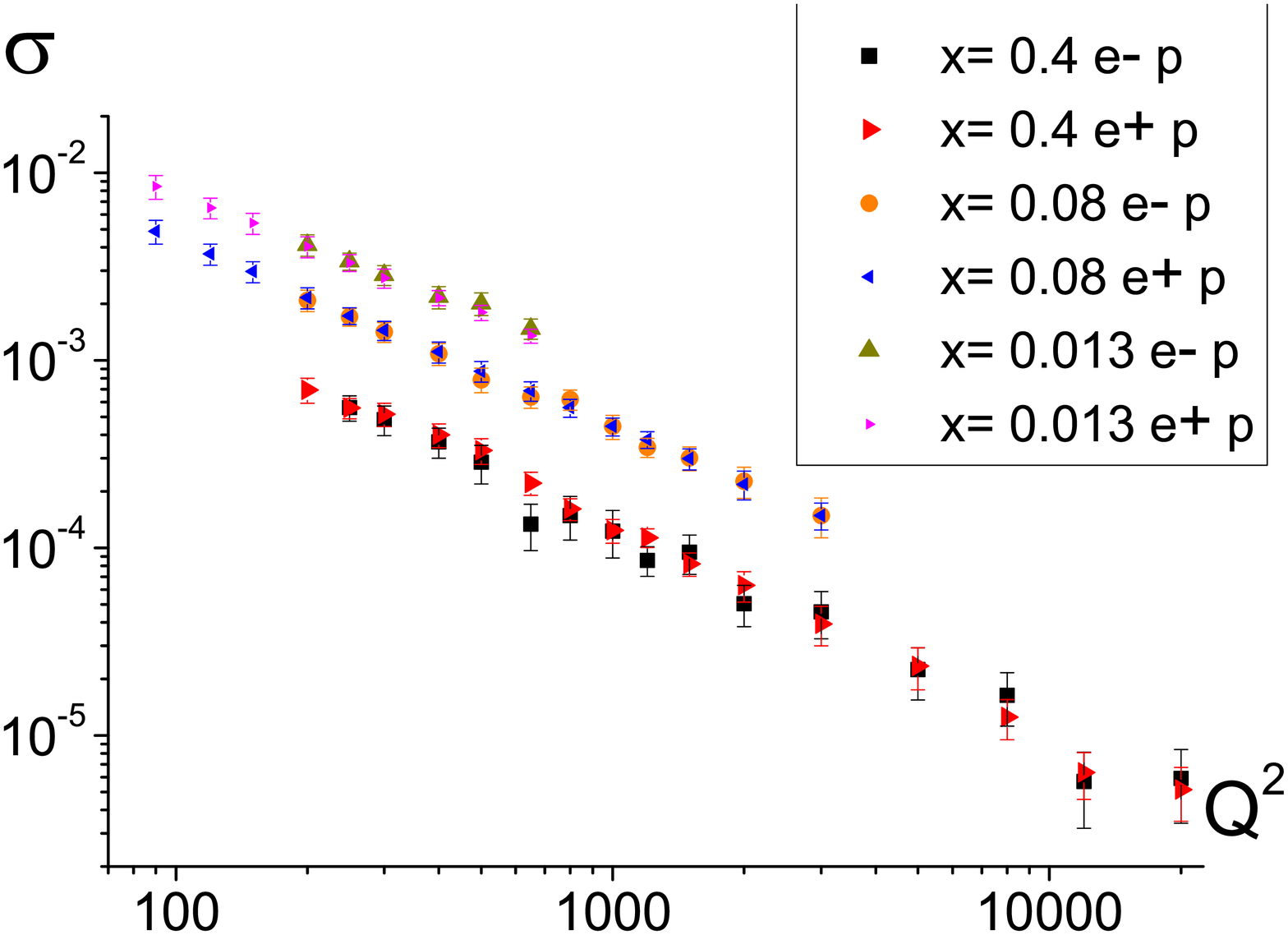}
\includegraphics[width=7cm,angle=0]{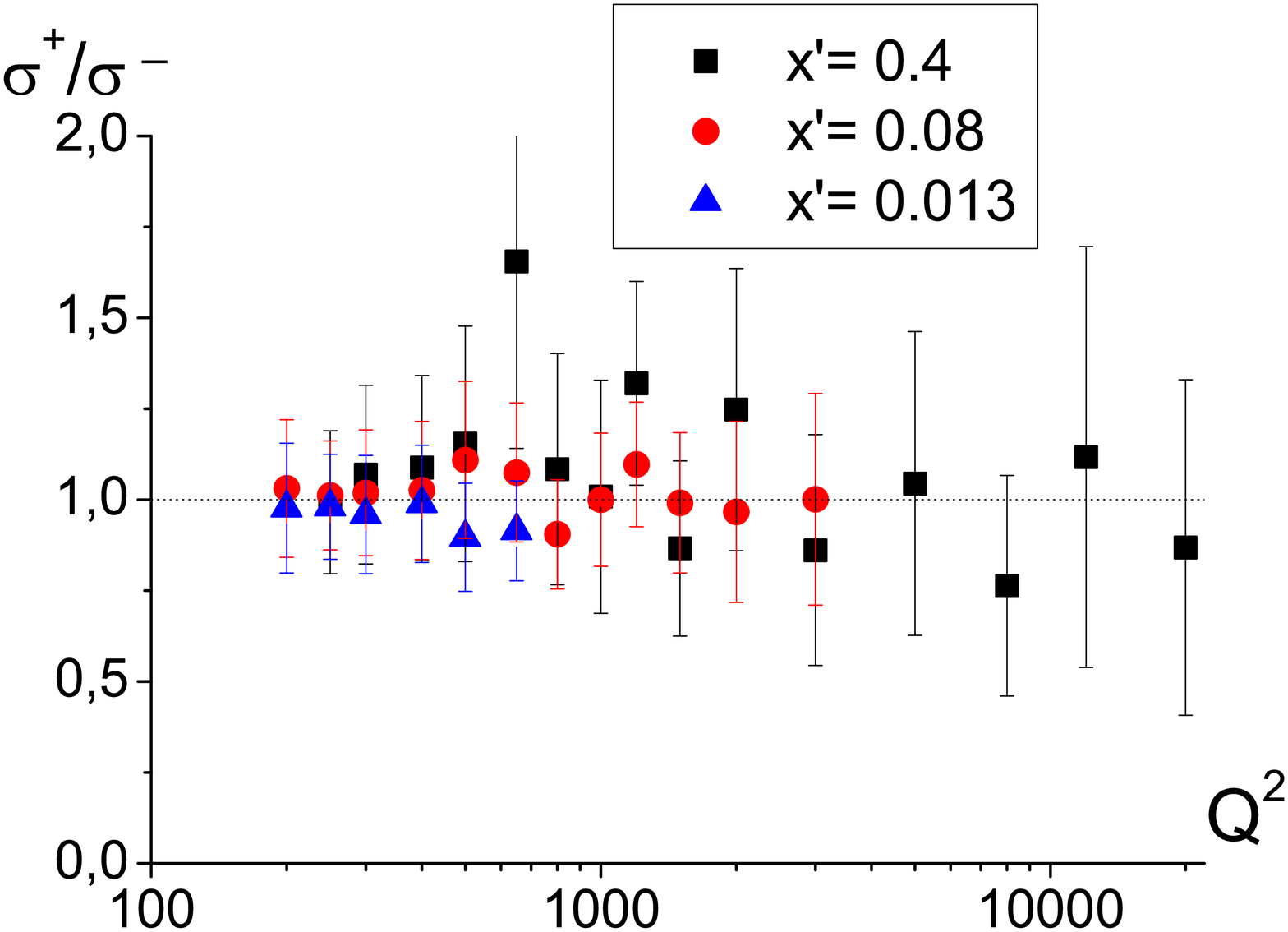}
\caption{Left: Functions $\tilde{\sigma}^{-}(Q^2)$ and $\tilde{\sigma}^{+}(Q^2)$ for three $x'$. Right: Ratios of $e^{+}p$ and $e^{-}p$ cross sections for these $x'$.}
\label{zesWyk35}
\end{figure}

In left plot of Fig. \ref{zesWyk35} we show functions $\tilde{\sigma}^{-}(Q^2)$ and $\tilde{\sigma}^{+}(Q^2)$ for three $x'$. Right plot shows ratio of these cross sections $\tilde{\sigma}^{+}(Q^2)/\tilde{\sigma}^{-}(Q^2)$. As we can see for given $x'$  $\tilde{\sigma}^{-}(Q^2)$ and $\tilde{\sigma}^{+}(Q^2)$ are almost the same.

\subsection{Results for $\left\langle  \lambda'_{\rm{min}}(x') \right\rangle$ and $\left\langle  \chi'^2_{\rm{nor}}(x';\lambda'_{\rm{min}}) \right\rangle$}

\begin{figure}[h]
\includegraphics[width=7cm,angle=0]{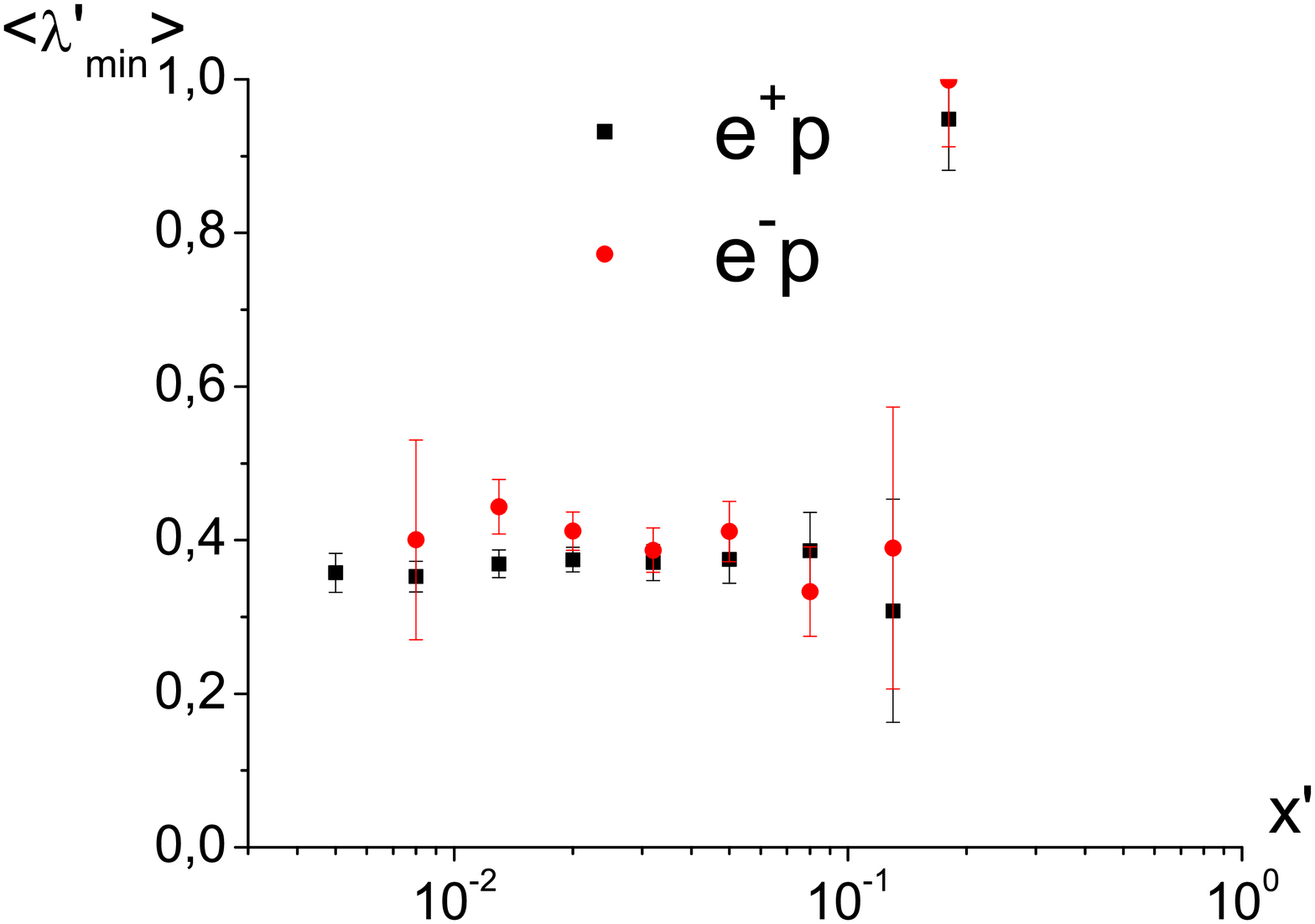}
\includegraphics[width=7cm,angle=0]{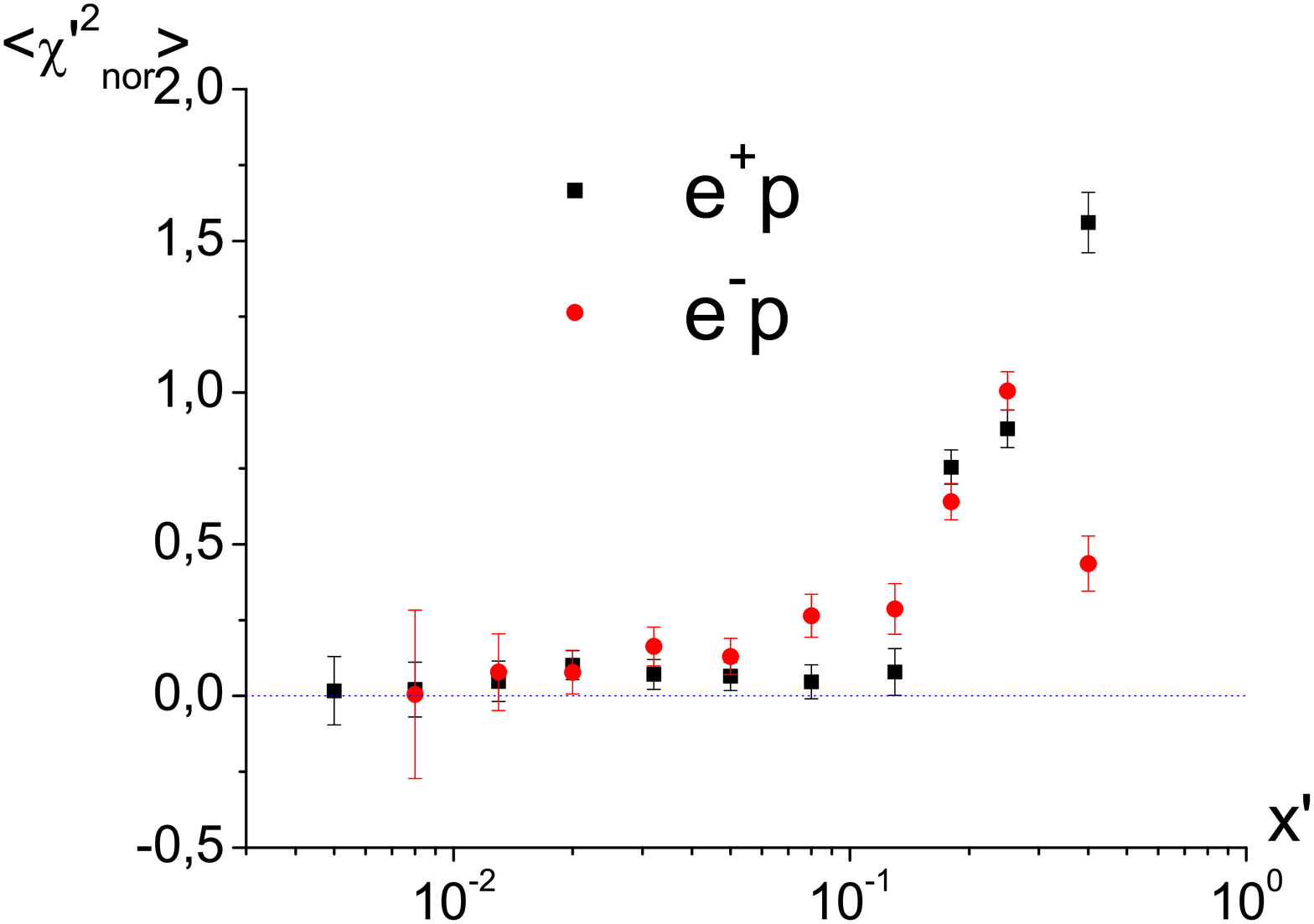}
\caption{Left plot: $\left\langle  \lambda'_{\rm{min}} \right\rangle$ as a function of $x'$ for $e^{+}p$ and $e^{-}p$. Right plot: $\left\langle  \chi'^2_{\rm{nor}}(\lambda_{\rm{min}}) \right\rangle$ as a function of $x'$ for $e^{+}p$ and $e^{-}p$.}
\label{zesWyk38}
\end{figure}

In subsection \ref{metanalysB-x} we said that we cannot use one $x'_{\rm{ref}}$ for all $x'$ (because there is no such $x'_{\rm{ref}}$ for which domain of $f^{\rm{ref'}}_{\lambda}$ covers range of $\tau$ for all other $x'$). For $e^-p$ we have the same problem so, as previously, we choose some $x'_{\rm{ref}}$ and find $\lambda'_{\rm{min}}$ and $\chi'^2_{\rm{nor}}(\lambda'_{\rm{min}})$ for all $x'<x'_{\rm{ref}}$ it is possible (condition $x'<x'_{\rm{ref}}$ guarantees that ratios are greater than 1 - see subsection \ref{metanalysB-x}). Furthermore, we can average these quantities over $x'_{\rm{ref}}$ obtaining $\left\langle  \lambda'_{\rm{min}}(x') \right\rangle$ and $\left\langle  \chi'^2_{\rm{nor}}(x';\lambda'_{\rm{min}}) \right\rangle$ (see subsection \ref{sectchi'norave} for details of averaging).

In left plot of Fig. \ref{zesWyk38} we show comparison between $\left\langle  \lambda'_{\rm{min}} \right\rangle$ for $e^{+}p$ and $e^{-}p$. Similarly as for energy binning values of $\left\langle  \lambda'_{\rm{min}} \right\rangle$ for $e^{-}p$ are a little bit bigger than for $e^{+}p$, here however this difference is smaller than uncertainties.
  
In right plot we compare $\left\langle  \chi'^2_{\rm{nor}}(x';\lambda'_{\rm{min}}) \right\rangle$: both functions are similar and for $x>0.2$ GS is violated (the same as for energy binning).

\chapter{Summary}

The main goal of this dissertation was an analysis of Geometrical Scaling in Deep Inelastic Scattering. It was performed using experimental data taken at HERA $ep$ collider. 

We have performed two different kinds of analysis: Bjorken-$x$ binning and energy binning. Data were originally prepared using Bjorken-$x$ binning so their quality is better.

In Chapter 3 we have presented a method of finding $\lambda$ (essentially the same for both binnings): it is based on observation that when GS is present all points form one curve. We have constructed ratios of cross sections for different energies $W$ (or $x$) and found $\lambda$ for which they are closest to 1. 

In Chapter 4 we have used this method to analyze $e^+ p$ data. For energy binning we have shown that values of $\lambda$ are almost the same for all energies $W\geq25$ GeV. We have shown also that $\lambda$ rises when we add to analysis points with greater $x$ (except for very small $x$ but there uncertainties are very large). This rise was also confirmed using Bjorken-$x$ binning. When we have compared $\lambda$ results for both binnings we have obtained some difference. In particular averaged values for $x<0.2$, which we adopt as the best estimation of $\lambda$, are different:
\begin{eqnarray*}
\lambda_{\rm{Bj}} & = 0.302 \: \pm  \: 0.004 \\
\lambda_{\rm{En}} & = 0.352 \: \pm  \: 0.008
\end{eqnarray*}
and $\lambda_{\rm{Bj}}$ is expected to be more reliable result due to better quality of data. 

In both binnings we obtain the same result for the range of $x$ where GS is present: 
\begin{equation}
x<0.2 .
\end{equation}
For very small $x$, however, the analysis breaks down due to the lack of points and one cannot draw firm conclusions concerning GS.

Last section of Chapter 4 was devoted to $Q^2$-dependence of $\lambda$. Some theoretical and experimental results indicate that indeed, $\lambda=\lambda(Q^2)$. To find such dependence we have searched for a function $\lambda(Q^2)$ which minimizes ratios better than $\lambda= \textrm{const.}$ For energy binning we have not found such dependence (probably due to unsatisfactory quality of data). For Bjorken-$x$ binning for some pairs of $x'$ we have found  $Q^2$-dependence but the results do not show enough regularity that would allow for clear interpretation.
  
In Chapter 5 we have analyzed $e^- p$ data. They are, however, much poorer than for $e^+ p$ so we have not used them as source of information about GS. We have just compared $e^- p$ and $e^+ p$ results. The main conclusion of this section is the following: values of $\lambda$ for $e^- p$ are similar to $e^+ p$ but they are a little bit bigger. Range of $x$ where GS is present is the same in both cases \textit{i.e.} $x<0.2$.

In Appendix C we have shown analysis for data obtained directly from experiment (\textit{i.e.} without structure functions $F_L$, $x \tilde{F}_3$, $F^{\gamma Z}_2$ parametrisations). We have obtained the same results as in Chapter 4.

The future work on GS should be concentrated on two aspects:

\begin{itemize}
\item improvement of data quality for DIS: one can use some numerical parameterizations of structure functions to generate data with better quality (especially for energy binning). In particular one can improve analysis for small $x$ where more points are needed. 
\item GS in hadrons collisions: LHC produces data from $pp$ and heavy ion collisions, it turns out that in such processes GS also can be found, more detailed analysis should be performed. Methods presented in this thesis with small modifications should be suitable for such purpose.
\end{itemize}

\appendix

\chapter{Analysis checking for energy binning}

During our analysis for energy binning we use some methods of dealing with data. Here we would check how results change if we modify them. Analysis is performed for $e^{+}p$ data because they are richer than data for $e^{-}p$. 

\section{Choice of binning}
\label{sectEnergybinning}
In section \ref{sectDivisionofdata} we divided data into sets with definite energies - logarithmic binning with step 1.3 (see point (a) below) was used. We can, however, consider different binnings: logarithmic or linear with various steps. Now we shall compare results which are obtained using the following binnings:

\begin{enumerate}[(a)]
\item Logarithmic binning with step 1.3 - every consecutive border is 1.3 times greater than preceding one (starting from 10 GeV). Value of energy $W$ is the mean of two limiting values $W'_{\rm{min}}$, $W'_{\rm{max}}$ between which it lies:

\begin{tabular}{|c|c|c|c|c|c|c|c|} \hline
$W'_{\rm{min}}\rm{ [GeV]}$ & 10 & 13 & 16.9 & 22 & 28.6 & 37.1 & 48.3 \\ \hline
$W'_{\rm{max}}\rm{ [GeV]}$ & 13 & 16.9 & 22 & 28.6 & 37.1 & 48.3 & 62.7 \\ \hline
$W\rm{ [GeV]}$ & 11.5 & 15 & 19.4 & 25.3 & 32.8 & 42.7 & 55.5 \\ \hline \hline
$W'_{\rm{min}}\rm{ [GeV]}$ & 62.7 & 81.6 & 106 & 137.9 & 179.2 & 233 & \\ \hline
$W'_{\rm{max}}[\rm{ [GeV]}]$ & 81.6 & 106 & 137.9 & 179.2 & 233 & 302.9 & \\ \hline
$W\rm{ [GeV]}$ & 72.2 & 93.8 & 122 & 158.5 & 206.1 & 267.9 & \\ \hline
\end{tabular}

\item Logarithmic binning with step 1.2 - every consecutive border is 1.2 times greater then preceding one (starting from 10 GeV). Value of energy $W$ is the mean of two limiting values $W'_{\rm{min}}$, $W'_{\rm{max}}$ between which it lies:
\begin{eqnarray*}
W \in \left\{\rm{11, 13.2, 15.8, 19, 22.8, 27.4, 32.8, 39.4, 47.3, 56.8,}\right. \\
\left. \rm{68.1, 81.7, 98.1, 117.7, 141.2, 169.5, 203.4, 244} \right\}.
\end{eqnarray*}

\item Logarithmic binning with step 1.4 - every consecutive border is 1.4 times greater than preceding one (starting from 10 GeV). Value of energy $W$ is the mean of two limiting values $W'_{\rm{min}}$, $W'_{\rm{max}}$ between which it lies:
\begin{eqnarray*}
W \in \left\{\rm{12., 16.8, 23.5, 32.9, 46.1, 64.5, 90.4, 126.5, 177.1, 247.9}\right\}.
\end{eqnarray*}
\item Linear binning with step 40 - all bins have the same width 40 GeV(starting from 0). Value of energy $W$ is the mean of two limiting values $W'_{\rm{min}}$, $W'_{\rm{max}}$ between which it lies: 

\begin{tabular}{|c|c|c|c|c|c|c|c|} \hline
$W'_{\rm{min}}\rm{ [GeV]}$ & 0 & 40 & 80 & 120 & 160 & 200 & 240 \\ \hline
$W'_{\rm{max}}\rm{ [GeV]}$ & 40 & 80 & 120 & 160 & 200 & 240 & 280 \\ \hline
$W\rm{ [GeV]}$ & 20 & 60 & 100 & 140 & 180 & 220 & 260 \\ \hline
\end{tabular}

\item Linear binning with step 30 - all bins have the same width 30 GeV (starting from 0). Value of energy $W$ is the mean of two limiting values $W'_{\rm{min}}$, $W'_{\rm{max}}$ between which it lies:
\begin{eqnarray*}
W \in \left\{\rm{0, 30, 60, 90, 120, 150, 180, 210, 240, 270}\right\}.
\end{eqnarray*}
\item Linear binning with step 14 - all bins have the same width 14 GeV (starting from 0). Value of energy $W$ is the mean of two limiting values $W'_{\rm{min}}$, $W'_{\rm{max}}$ between which it lies:
\begin{eqnarray*}
W \in \left\{\rm{7, 21, 35, 49, 63, 77, 91, 105, 119, 133,}\right. \\
\left. \rm{147, 161, 175, 189, 203, 217, 231, 245} \right\}.
\end{eqnarray*}

\end{enumerate}

Except for changing the binning, calculations are the same described in section \ref{sectMethodoffinding}. 

\begin{figure}
\includegraphics[width=7cm,angle=0]{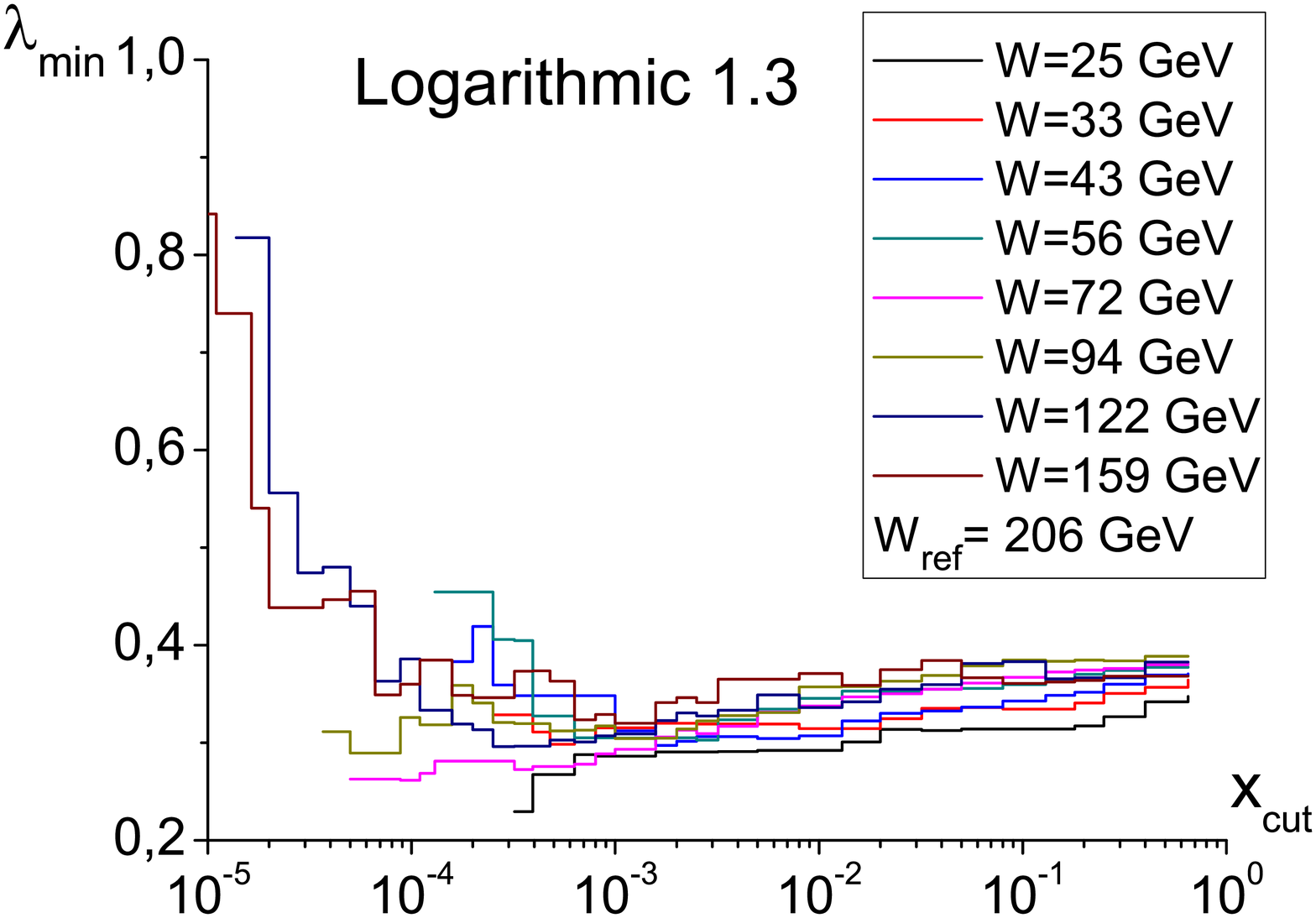}
\includegraphics[width=7cm,angle=0]{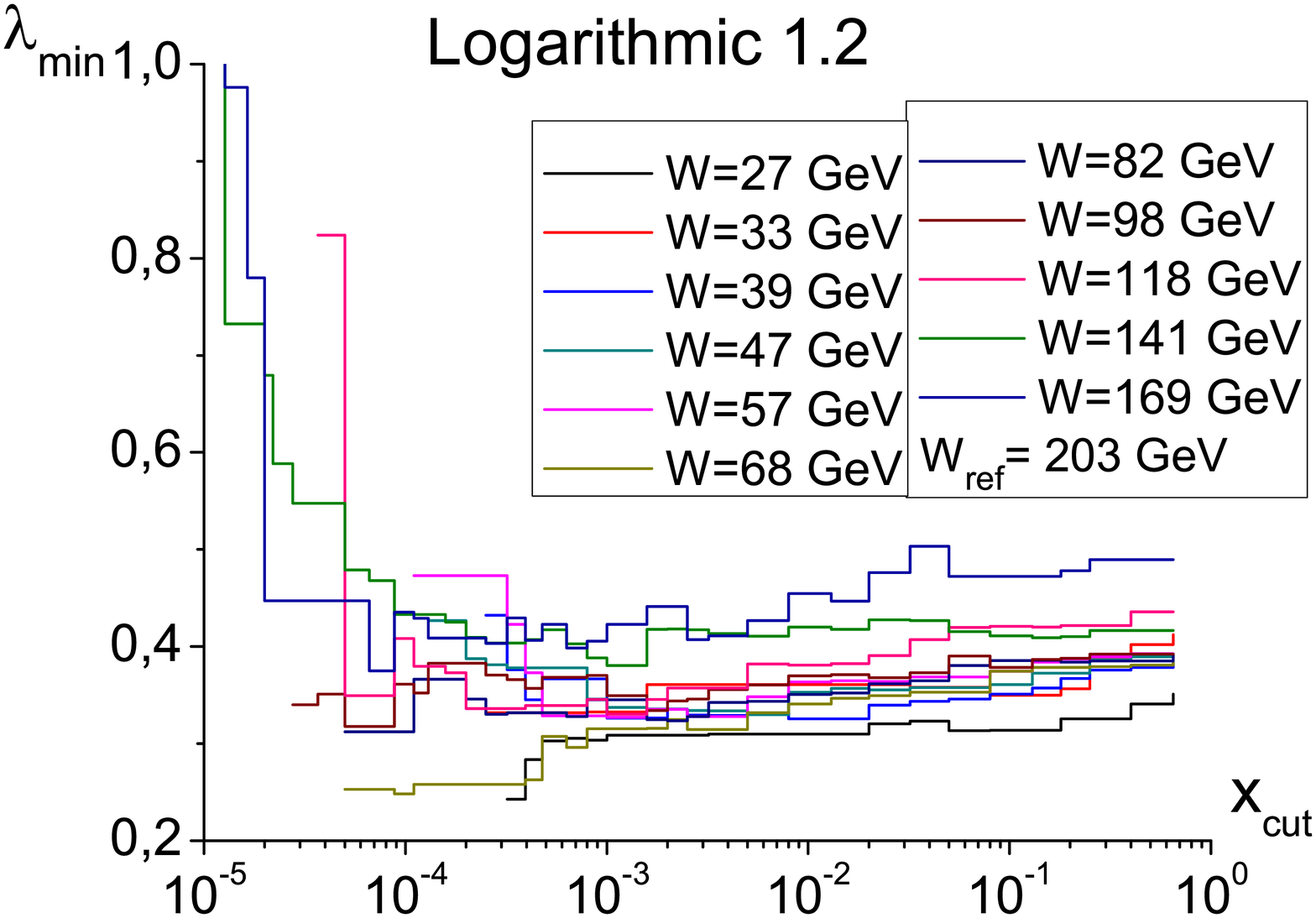}
\includegraphics[width=7cm,angle=0]{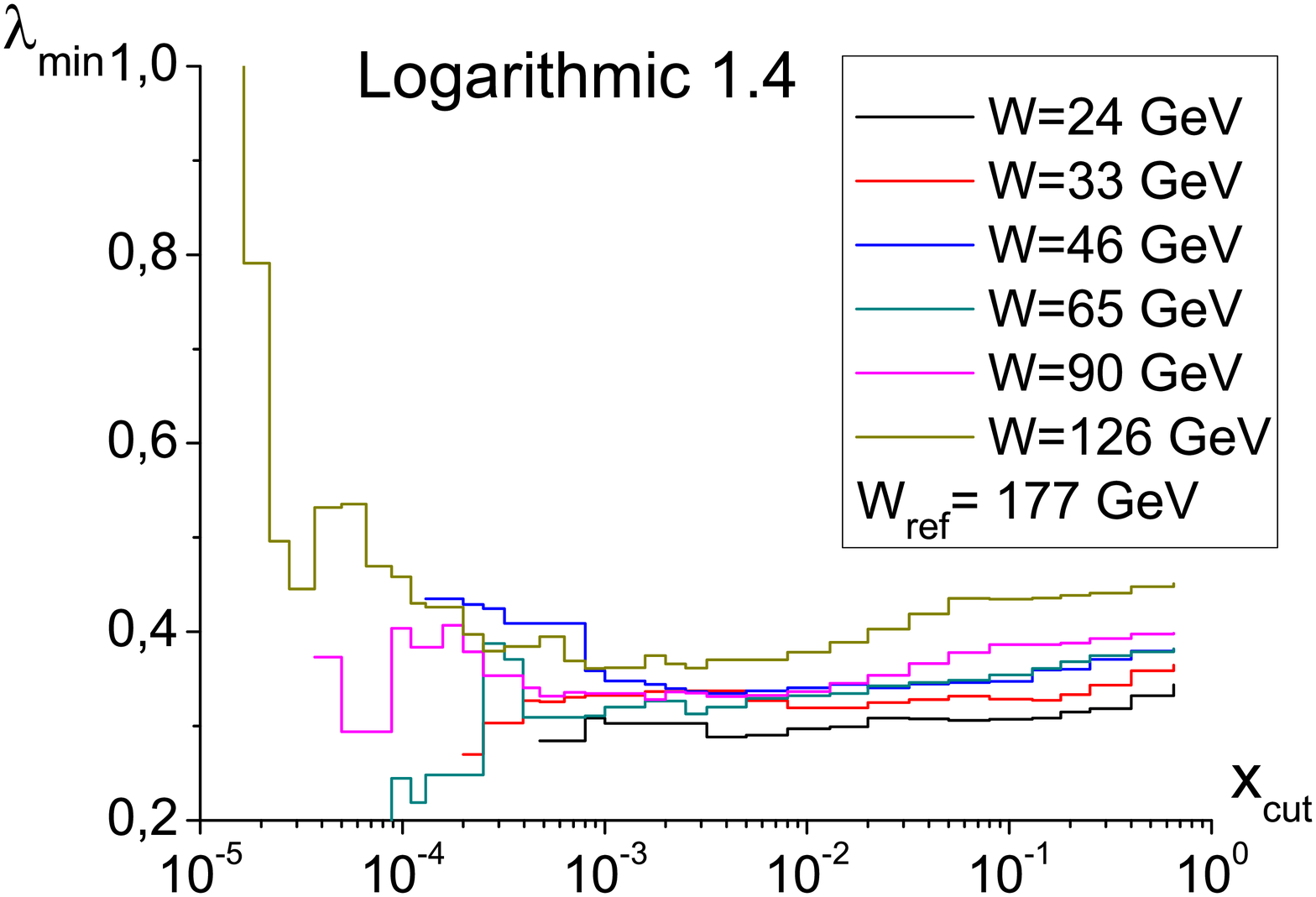}
\includegraphics[width=7cm,angle=0]{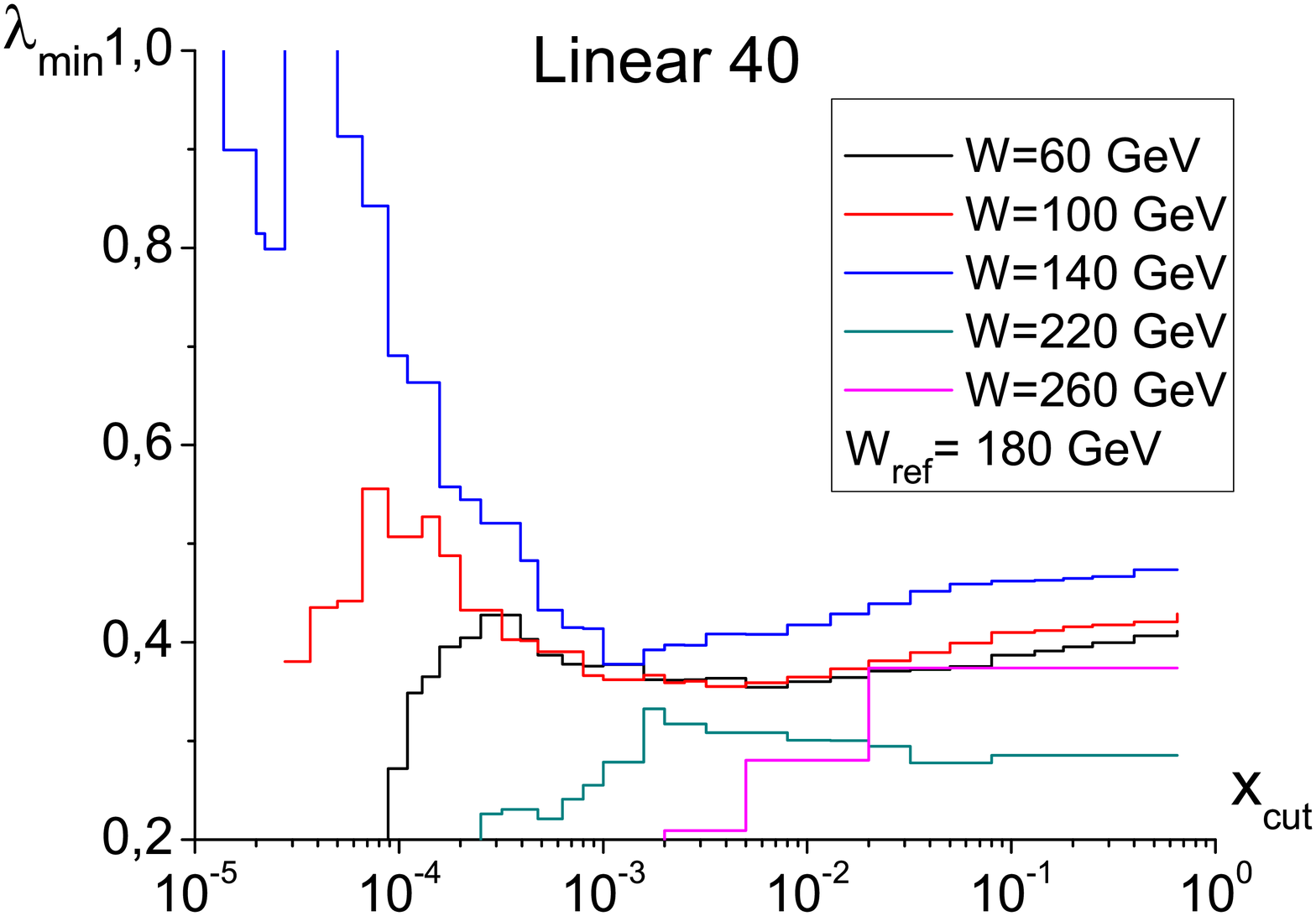}
\includegraphics[width=7cm,angle=0]{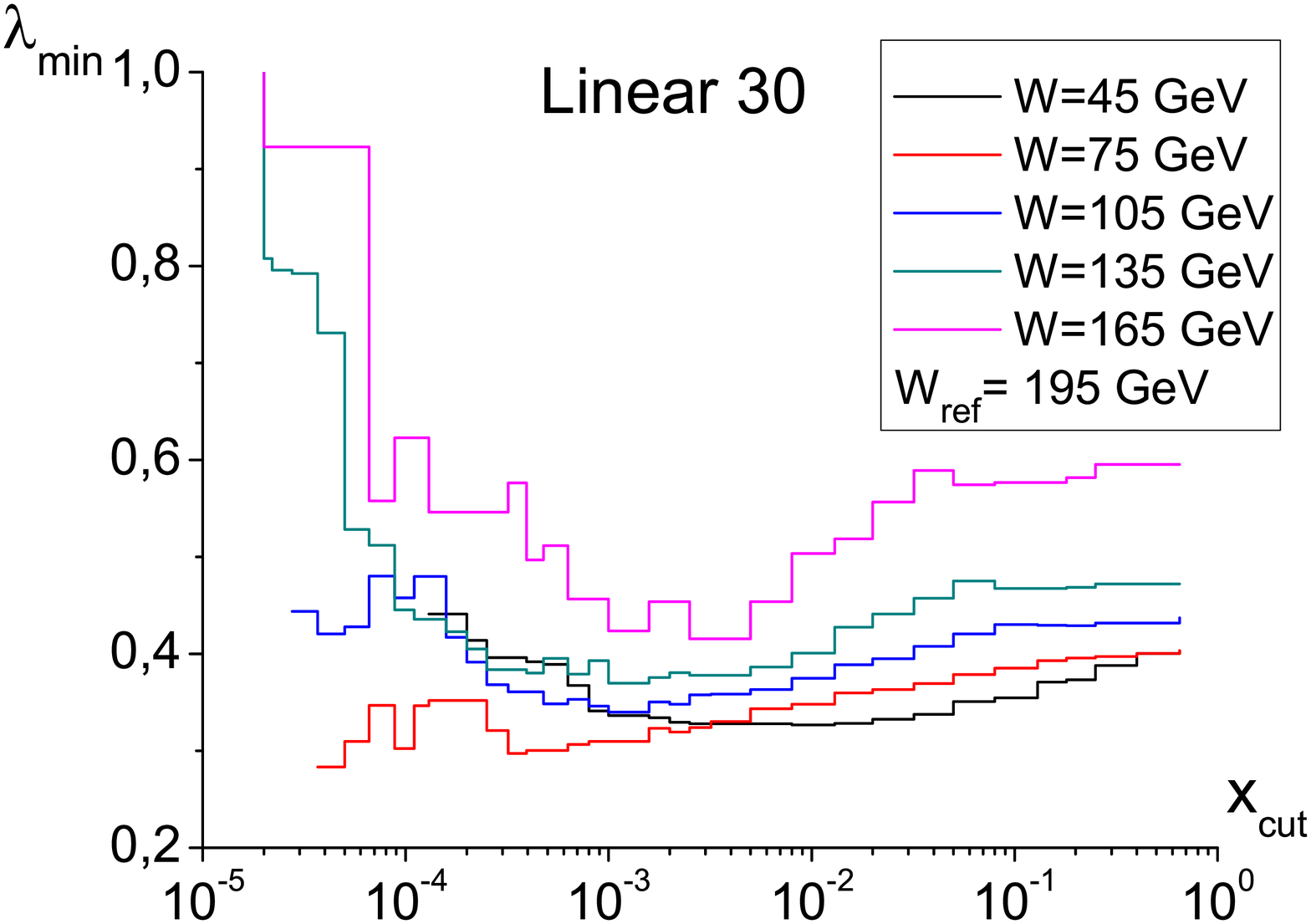}
\includegraphics[width=7cm,angle=0]{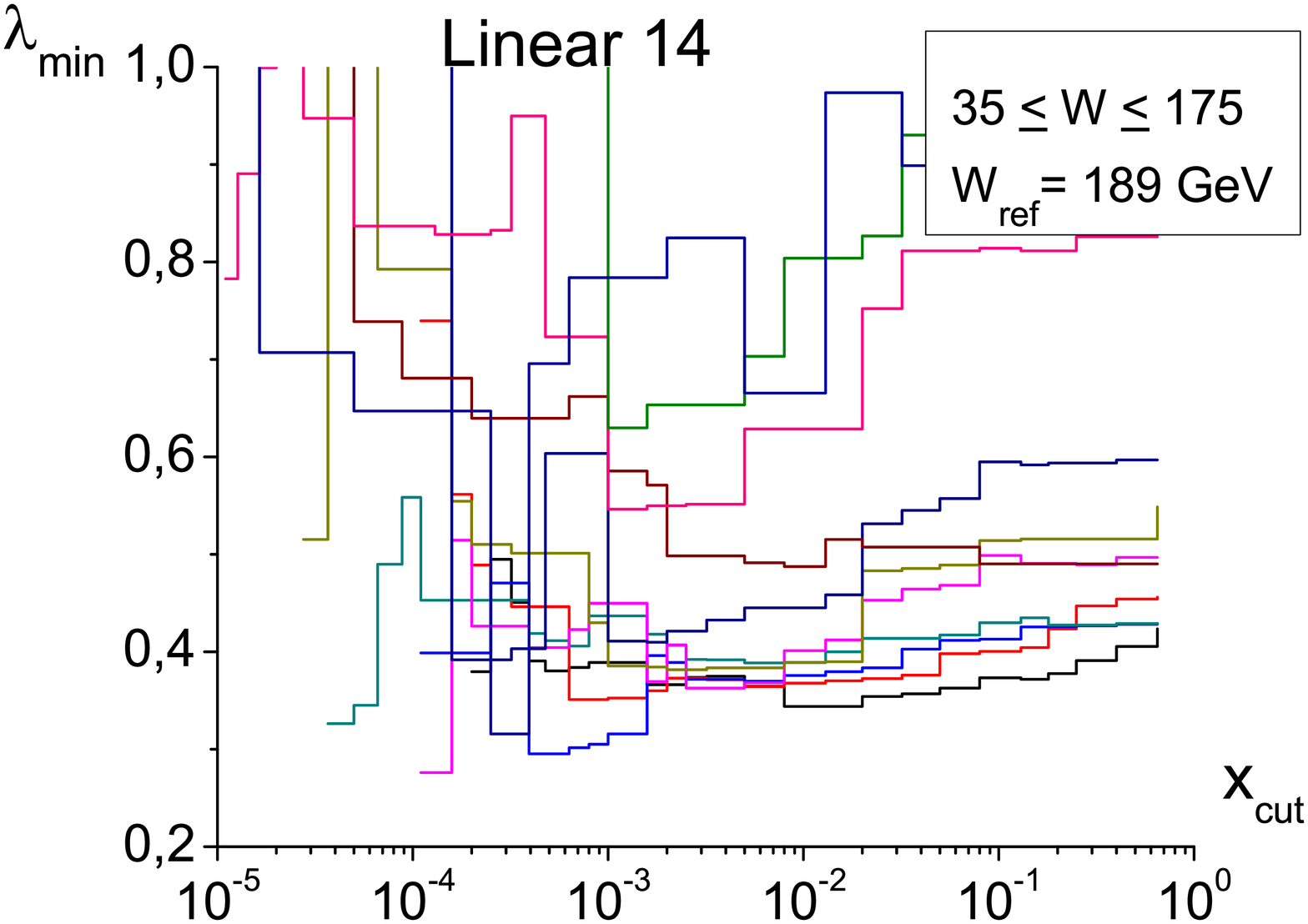}
\caption{Set of functions $\lambda_{\rm{min}} \left( x_{\rm{cut}}\right)$ for different binnings. Every plot shows one binning described in (a)-(f). We also write for which energies $\lambda_{\rm{min}} \left( x_{\rm{cut}}\right)$ were plotted and which energy was chosen as $W_{\rm{ref}}$.}
\label{zesWyk17}
\end{figure}

In Fig. \ref{zesWyk17} we show how set of functions $\lambda_{\rm{min}} \left( x_{\rm{cut}}\right)$ (for energies $20 \textrm{ GeV}<W< W_{\rm{ref}}$) look like for different binnings.

\begin{figure}
\centering
\includegraphics[width=9cm,angle=0]{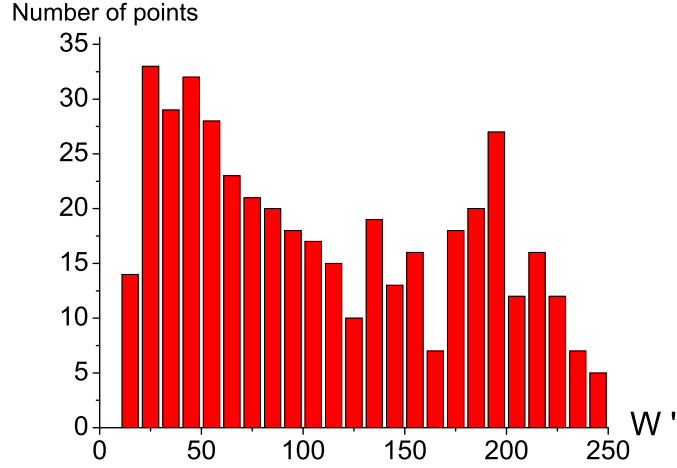}
\caption{Number of points for different energies $W'$. $W'$ is calculated from equation (\ref{w'def}) \textit{i.e.} it is an energy for Bjorken-$x$ binning.}
\label{zesWyk18}
\end{figure}

We can see that logarithmic binnings are more appropriate than the linear ones. This is due to the fact that most of data have small $W'$ values (see Fig. \ref{zesWyk18}) so for linear binning high energies have small number of points. This is a disadvantage because we choose $W_{\rm{ref}}$ to be large (for high energies range of $Q^2$ values is the widest). Logarithmic bins for high energies are relatively wide so there are more points.

For logarithmic binning with step 1.3 we obtain functions $\lambda_{\rm{min}} \left( x_{\rm{cut}}\right)$ for different $W$'s which form the narrowest band. This is the reason why we have used it in Chapters 4 and 5 (we have not presented here $\chi^2_{\rm{nor}}(x_{\rm{cut}})$ funtions because it turns out that they are similar for different binnings and comparison of $\lambda_{\rm{min}} \left( x_{\rm{cut}}\right)$ functions is a better criterion).

\section{Points with the same $Q^2$. Original values of Bjorken variable}
\label{sectXoryg}
As previously we use notation (introduced in section \ref{sectBinning}): $x'$ - value of Bjorken variable in Bjorken-$x$ binning; $x$ - value of Bjorken variable in energy binning (calculated from relation $x=\frac{Q^2}{Q^2+W^2-M^2}$, where $W$ is energy of point in energy binning). We also define two scaling variables: $\tau=Q^2 x^\lambda$ and $\tau'=Q^2 x'^\lambda$. If two points have the same $W$ and $Q^2$ then their $\tau$ are also the same (this is not true for $\tau'$). 
 
\begin{figure}
\centering
\includegraphics[width=7cm,angle=0]{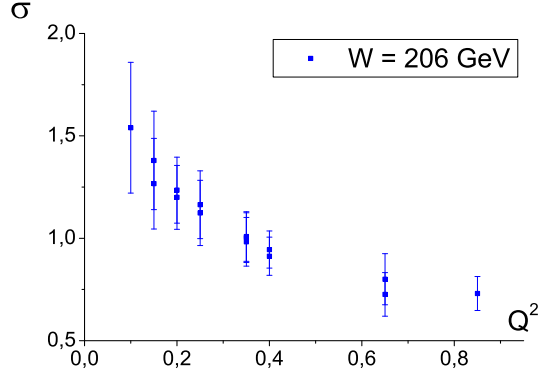}
\caption{Cross section $\tilde{\sigma}^{206}(Q^2)$ for $Q^2< 1$ when points with the same $Q^2$ are not averaged.}
\label{zesWyk19}
\end{figure}

In section \ref{sectDivisionofdata} we said that in original data for a given energy $W$ several points can have the same $Q^2$ values and different $F_2$ values. This results in an ambiguity of the value of function $\tilde{\sigma}^W(Q^2)$ (see Fig. \ref{zesWyk19}). This ambiguity is also present when we consider functions $\tilde{\sigma}^W(\tau)$. 

\begin{figure}[h]
\includegraphics[width=7cm,angle=0]{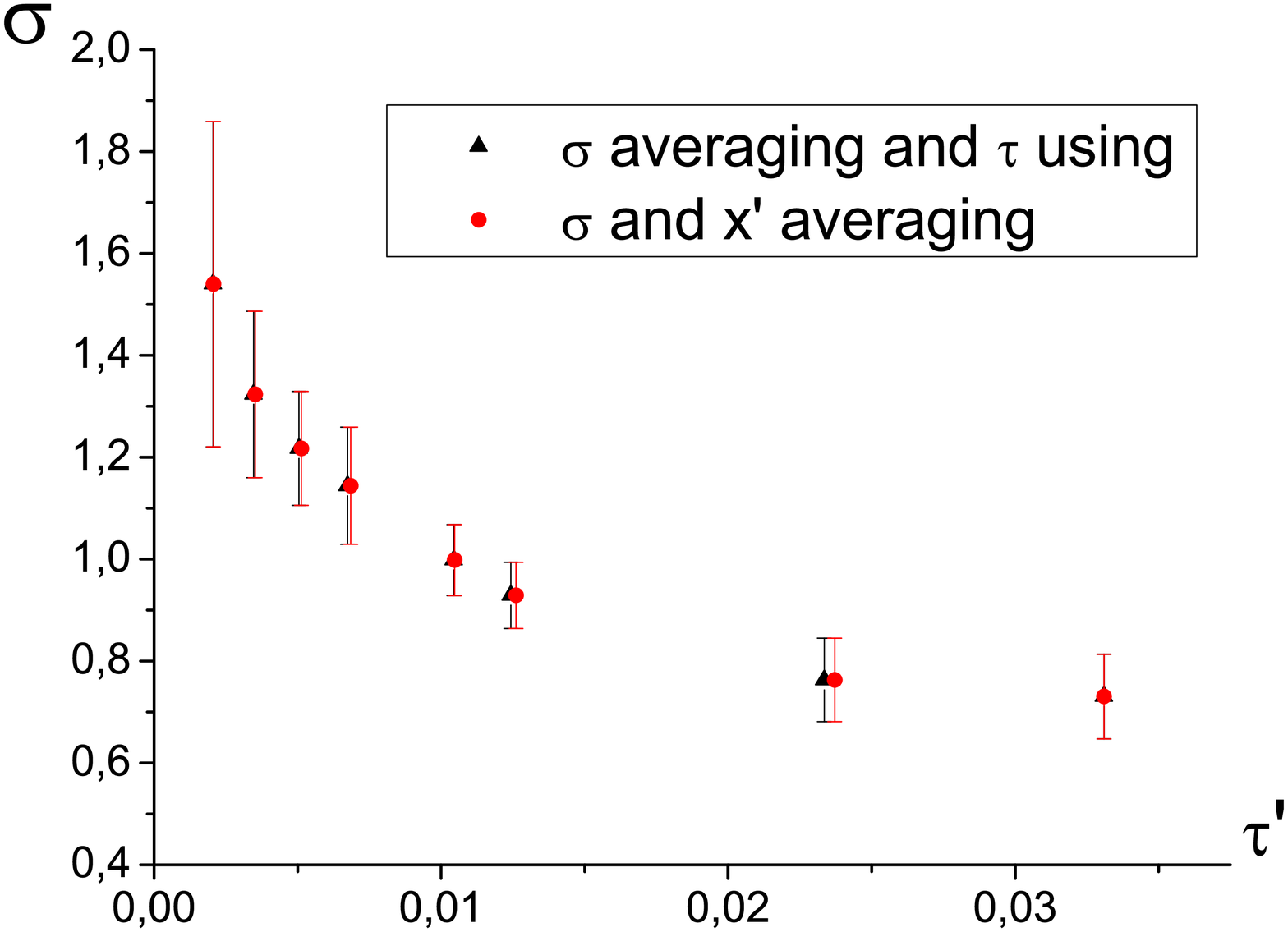}
\includegraphics[width=7cm,angle=0]{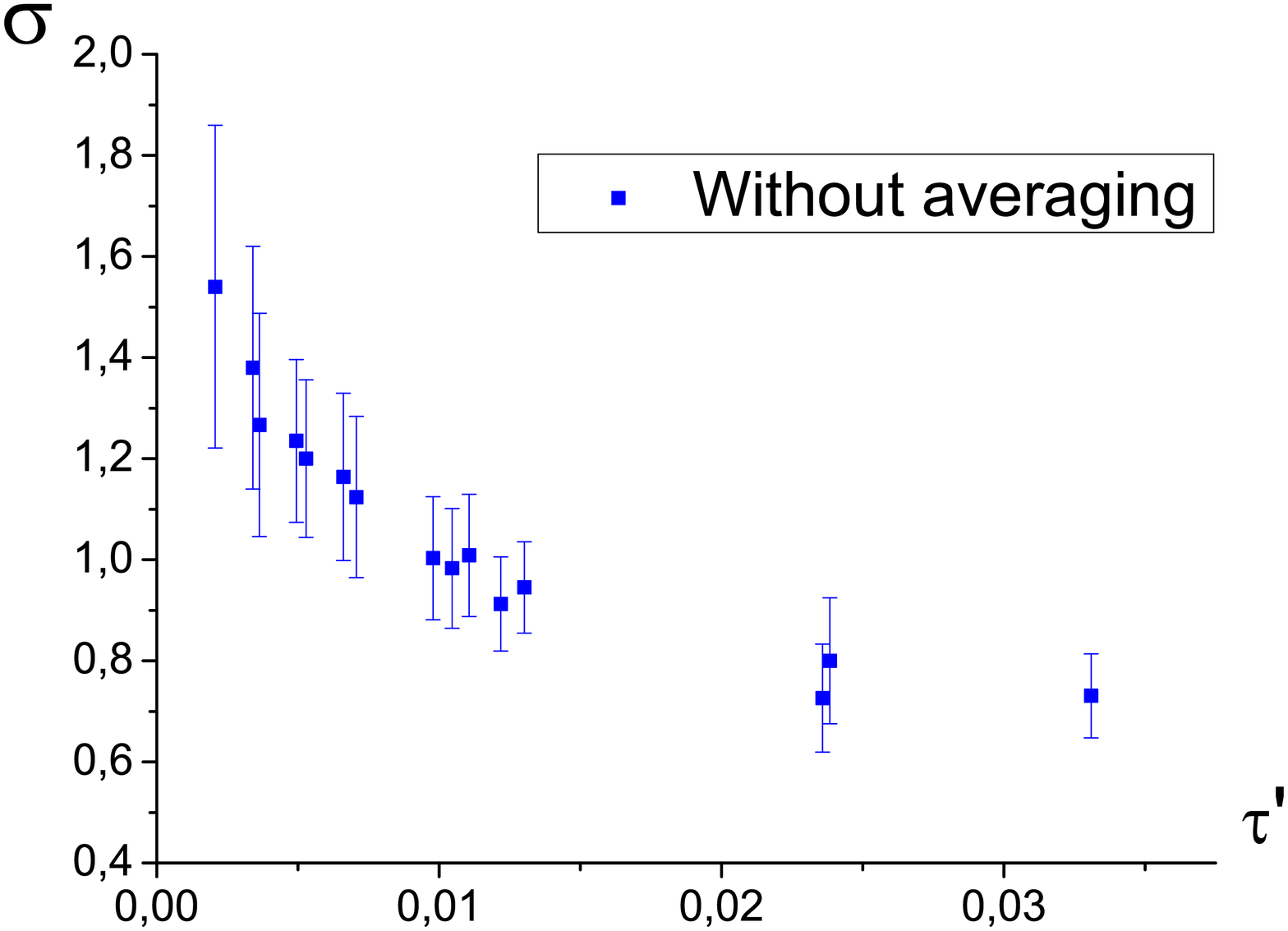}
\caption{$\tilde{\sigma}^{206}$ as a function of scaling variable with $\lambda=0.3$, only points with $Q^2< 1$ are shown. Three methods are demonstrated: (a) and (b) in the left plot, (c) in the right.}
\label{zesWyk20}
\end{figure}

We will discuss the following methods of elimination ambiguity in $\tilde{\sigma}^W$:
\begin{enumerate}[(a)]
\item $\tilde{\sigma}$ averaging and $\tau$ using: if two (or more points) have the same $Q^2$ (so also the same $\tau$) but different values of cross section $\tilde{\sigma}^W_1$, $\tilde{\sigma}^W_2$ we replace them by one point with $\tilde{\sigma}^W=\left(\tilde{\sigma}^W_1+\tilde{\sigma}^W_1\right)/2$ and $\Delta \tilde{\sigma}^W=\frac{1}{2} \sqrt{\left(\Delta \tilde{\sigma}^W_1\right)^2+\left(\Delta \tilde{\sigma}^W_2\right)^2}$. As a scaling variable we use $\tau$. It is a "full" energy binning (all quantities are determined in energy binning). This method was used in Chapters 4 and 5.

\item $\tilde{\sigma}$ and $x'$ averaging: if two (or more points) have the same $Q^2$ we replace them by one point with averaged values of $\tilde{\sigma}^W$ and $x'$. As a scaling variable we use $\tau'$ with averaged $x'$. This means that we divide data into energy bins but also we use values of Bjorken variable determined in Bjorken-$x$ binning (in some sense we "mix" binnings). 

\item Without averaging: even if two points have the same value of $Q^2$ they in general have different $x'$ so they split when we plot function $\tilde{\sigma}^W(\tau')$ (see right plot in Fig. \ref{zesWyk20}). However, this splitting is quite small and functions $\tilde{\sigma}^W(\tau')$ are not smooth (see left plot in \ref{zesWyk20}). Similarly like in (b) we "mix" binnings.
\end{enumerate}

In Fig. \ref{zesWyk20} we show plots $\tilde{\sigma}^W$ as a function of scaling variable using these three methods. All functions are plotted for $W=206$ GeV and $\lambda=0.3$. Methods (a) and (b) give very similar functions $\tilde{\sigma}^{206}$ so we compare them on the same plot.
\newline

\subsubsection{Functions $\lambda_{\rm{min}}(x_{\rm{min}})$}

\begin{figure}[ht]
\includegraphics[width=7cm,angle=0]{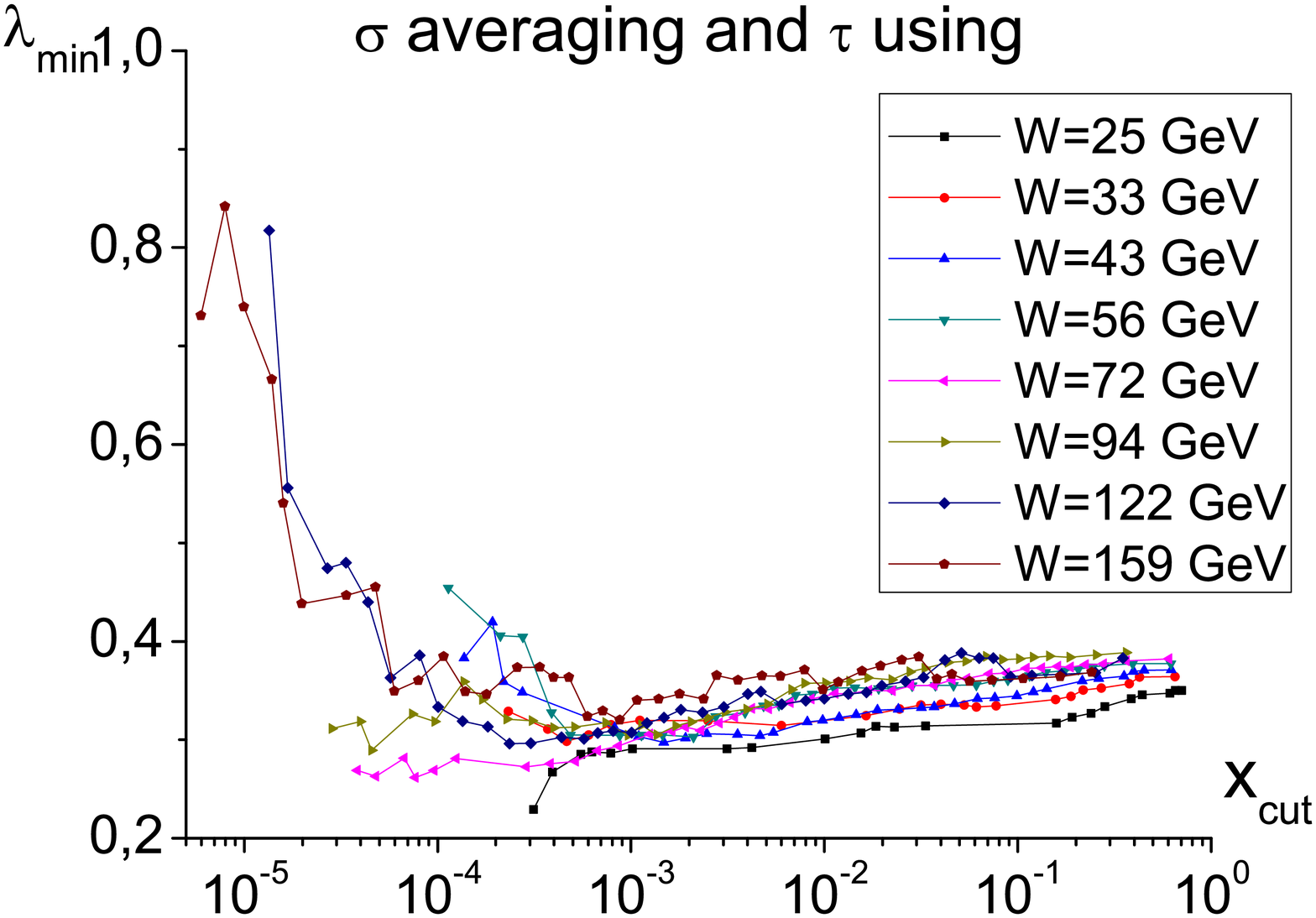}
\includegraphics[width=7cm,angle=0]{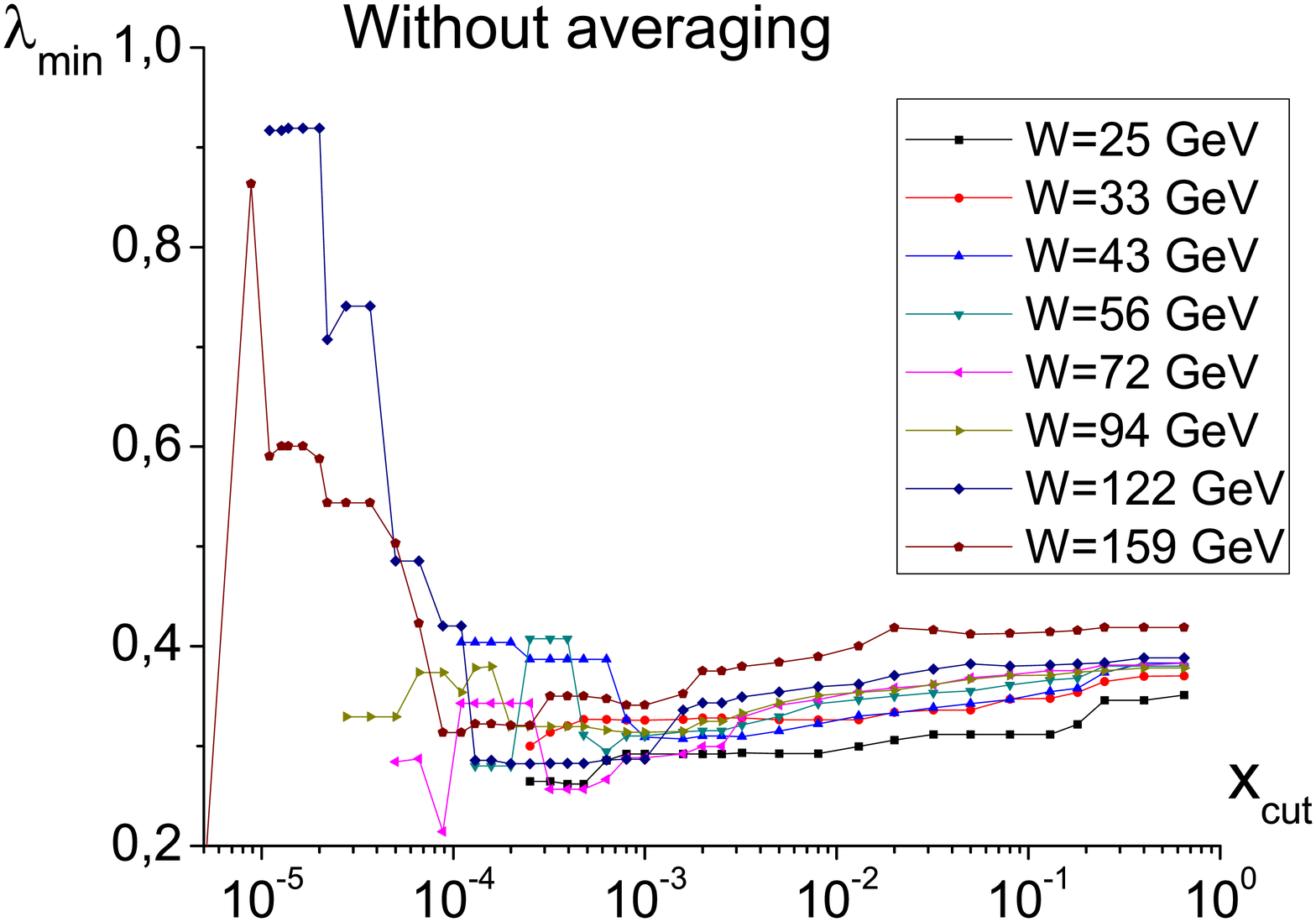}
\includegraphics[width=7cm,angle=0]{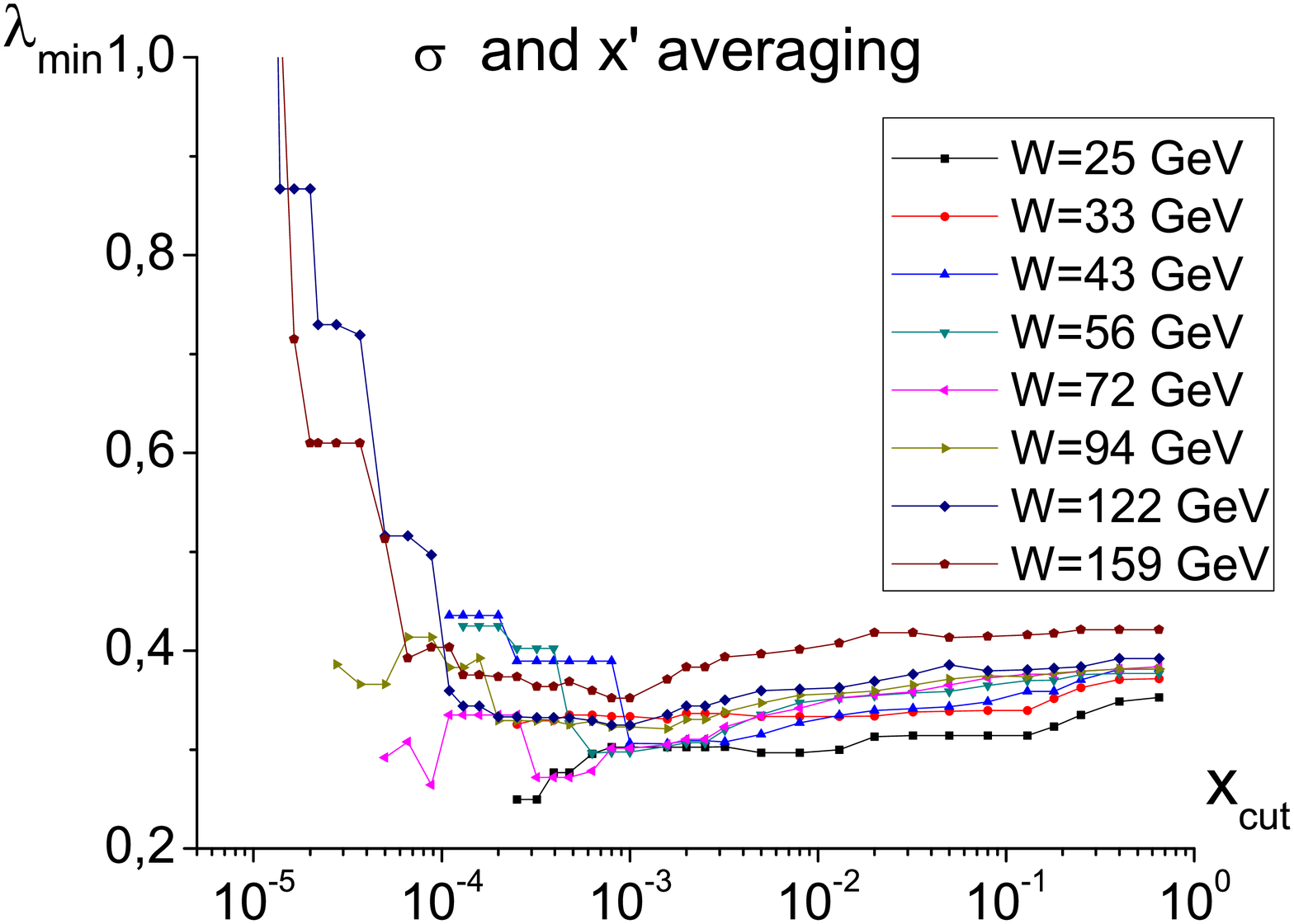}
\caption{Set of functions $\lambda_{\rm{min}} \left( x_{\rm{cut}}\right)$ for different methods of $\tilde{\sigma}^W$-ambiguity elimination. Every plot shows one method described in (a)-(c).}
\label{zesWyk21}
\end{figure}

In the Fig. \ref{zesWyk21} we show how functions $\lambda_{\rm{min}}(x_{\rm{min}})$ look like for different methods described in (a)-(c). We can see that for (a) functions $\lambda_{\rm{min}}(x_{\rm{min}})$ for various energies form the narrowest band. 
This means that using "full" energy binning we get slightly better results than for "mixed" binning.

\newpage

\subsubsection{Roughness of data}
\begin{figure}
\includegraphics[width=7cm,angle=0]{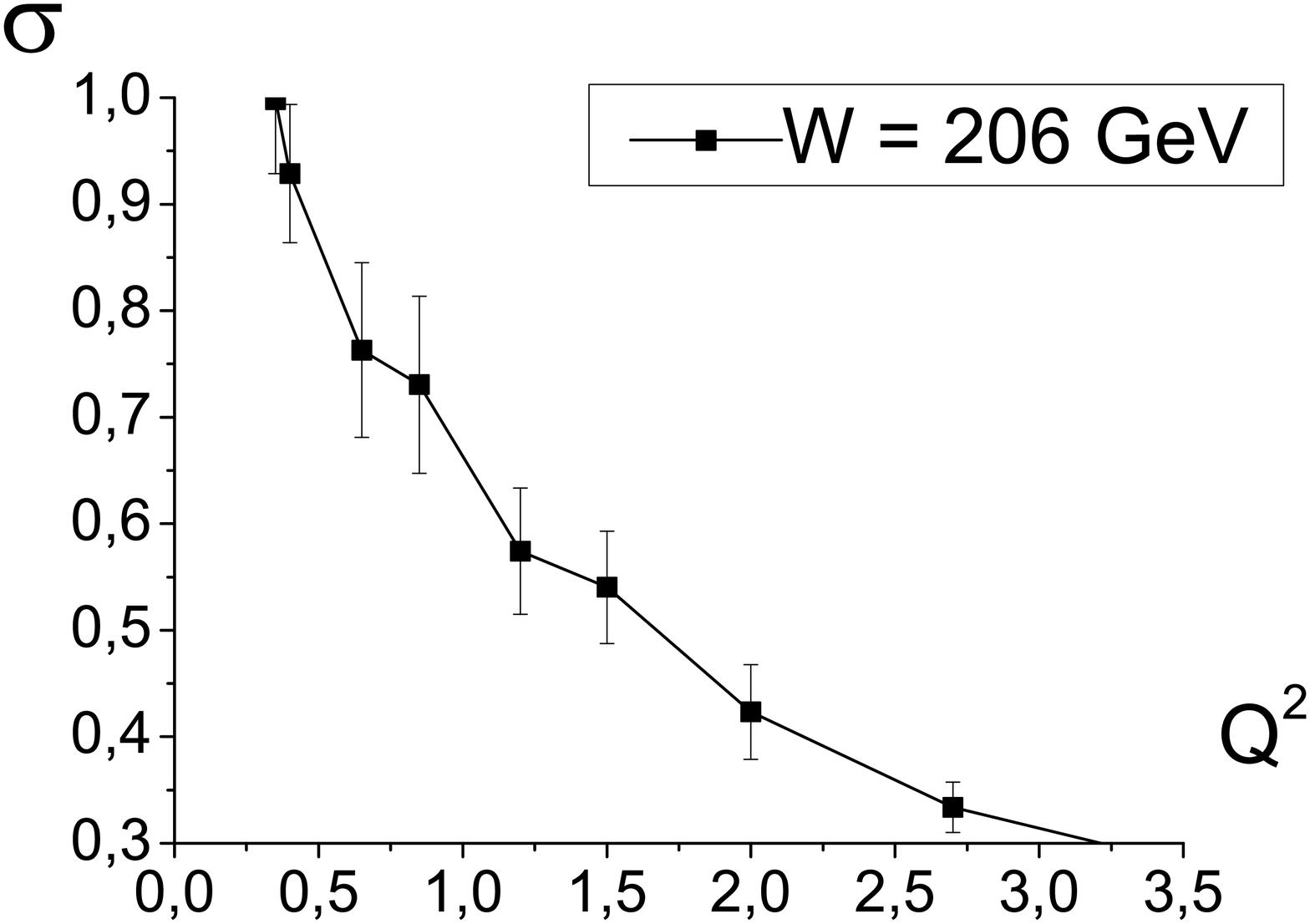}
\includegraphics[width=7cm,angle=0]{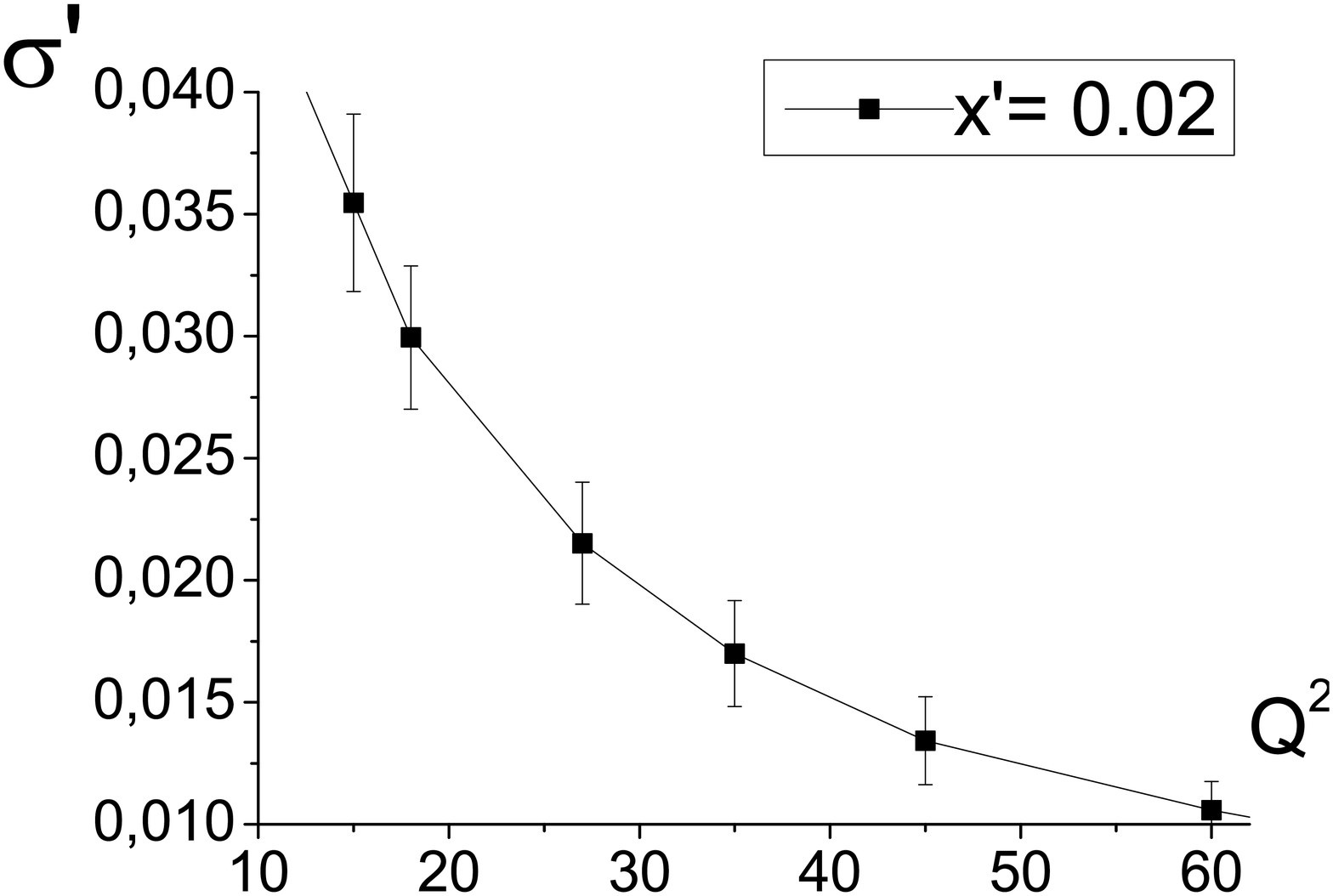}
\caption{Example of cross section $\tilde{\sigma}(Q^2)$ for energy binning (left plot) and Bjorken-$x$ binning (right plot). In both cases only a fragment of function was shown.}
\label{zesWyk41}
\end{figure}
In previous chapters we used method (a) to eliminate ambiguity in $\tilde{\sigma}^W$. However, it does not smoothen data completely and some roughness remains: in left plot of Fig. \ref{zesWyk41} we show a fragment of plot $\tilde{\sigma}^W(Q^2)$ for $W=206$ GeV (we used this energy as $W_{\rm{ref}}$ so black line  is a  $f_{\lambda}^{\rm{ref}}$ for $\lambda=0$). 

In right plot of Fig. \ref{zesWyk41} we can see example of $\tilde{\sigma}'(Q^2)$ for Bjorken-$x$ binning: data are more smooth than for energy binning. This is because original data taken from \cite{H1ZueusWork} are divided using Bjorken-$x$ binning. After binning changing (see subsection \ref{sectDivisionofdata}) data have worse quality.

\section{Choice of $W_{\rm{ref}}$}
\label{sectDiffrentchoiceofW}

\begin{figure}[ht]
\includegraphics[width=7cm,angle=0]{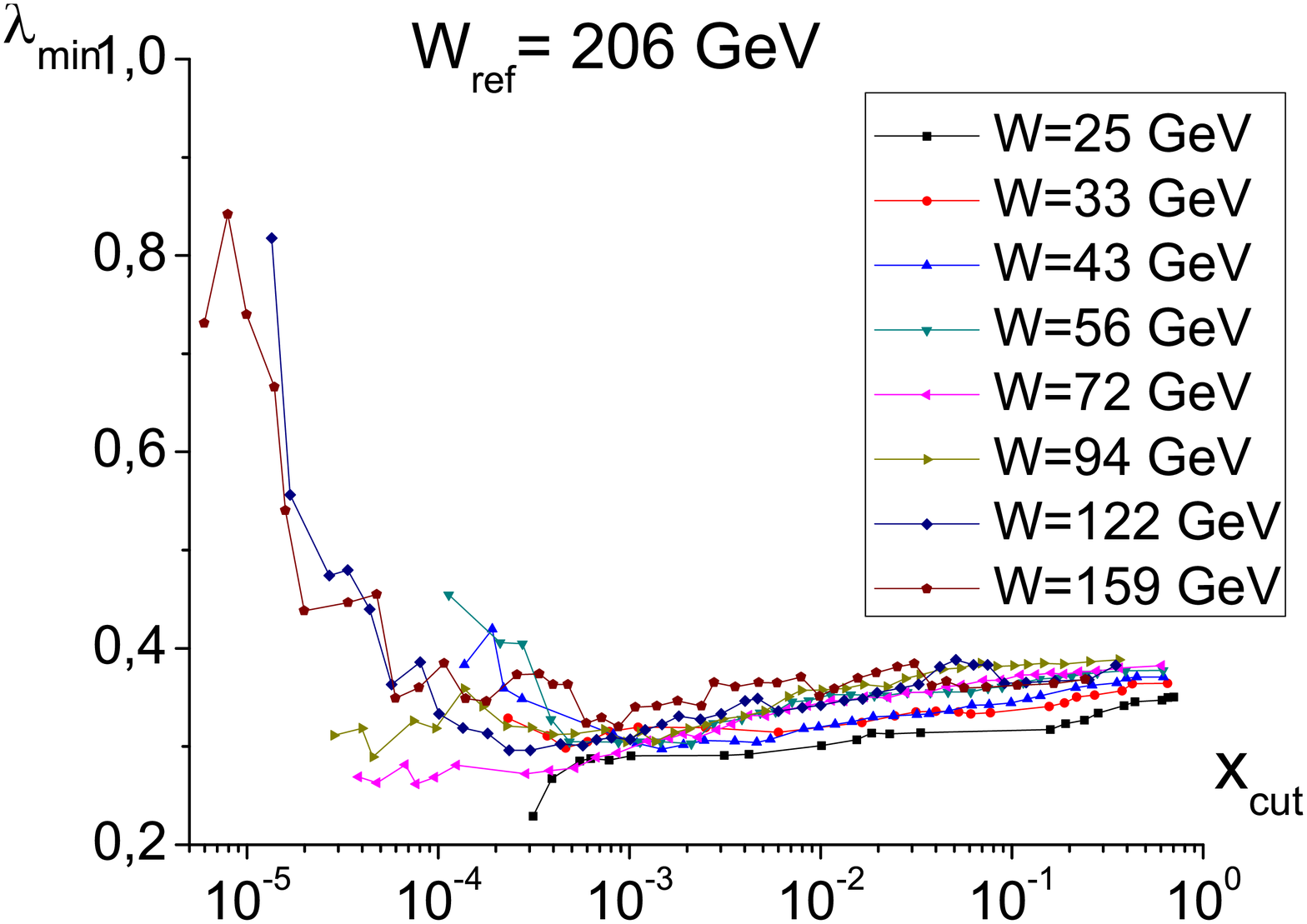}
\includegraphics[width=7cm,angle=0]{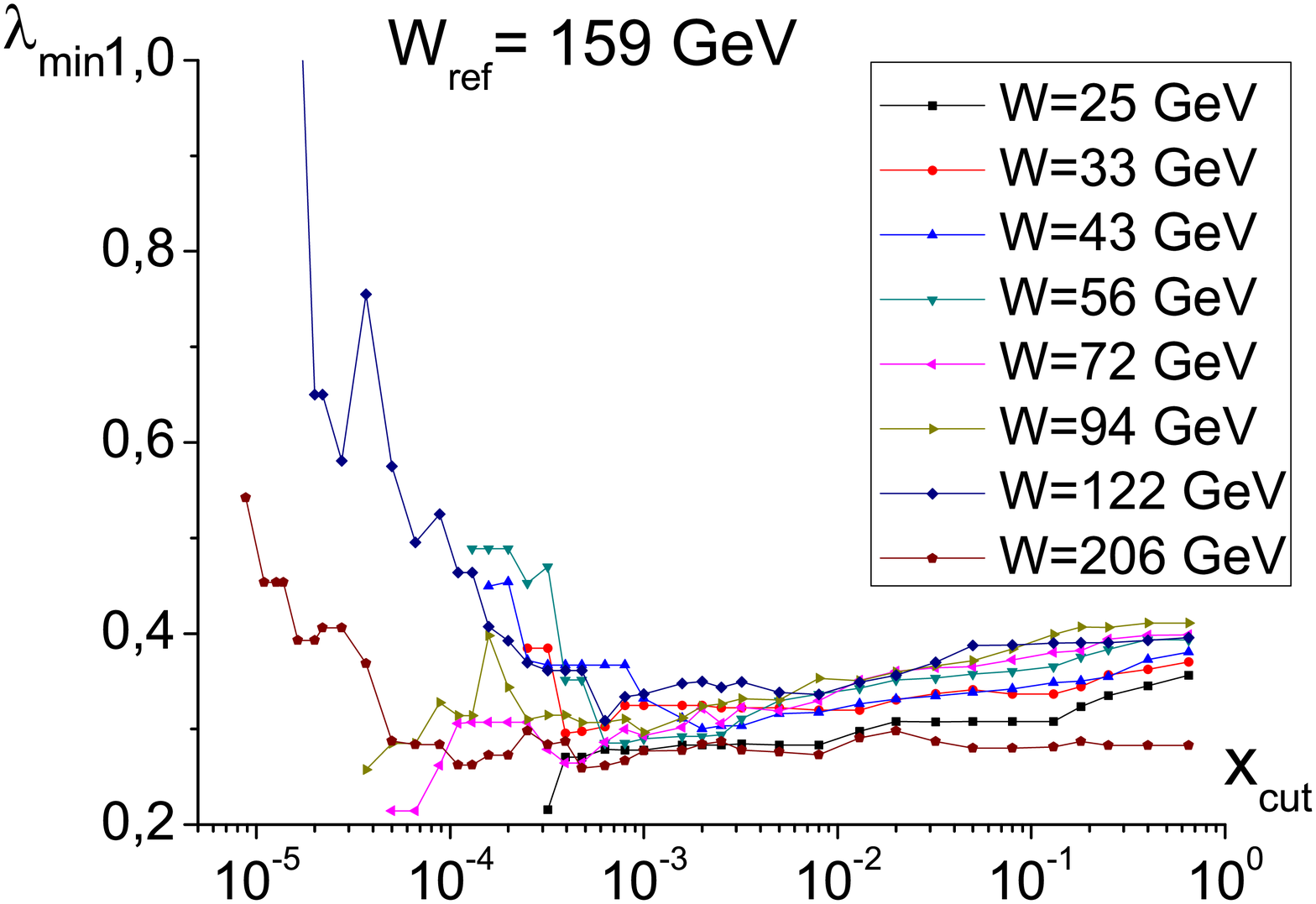}
\includegraphics[width=7cm,angle=0]{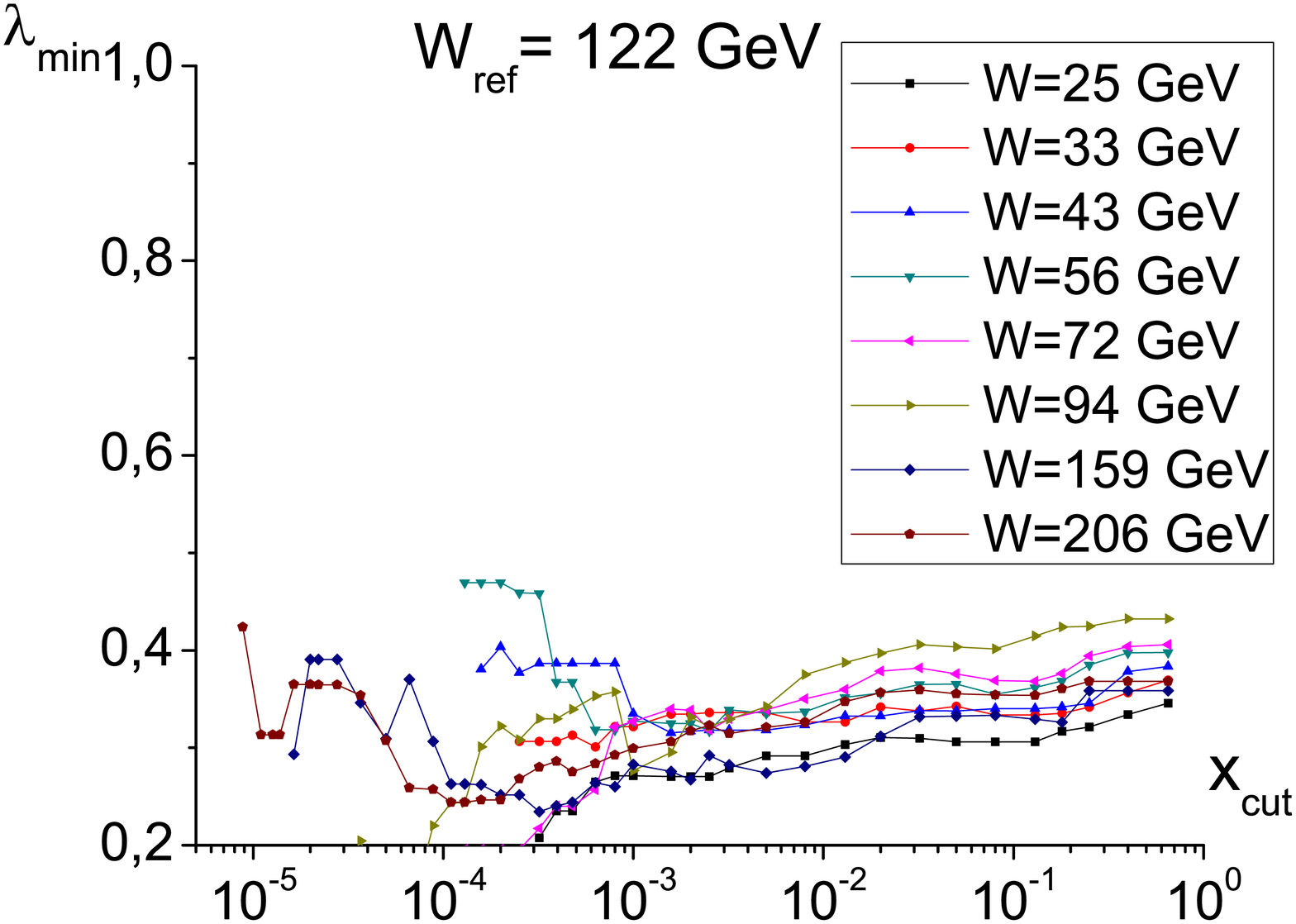}
\includegraphics[width=7cm,angle=0]{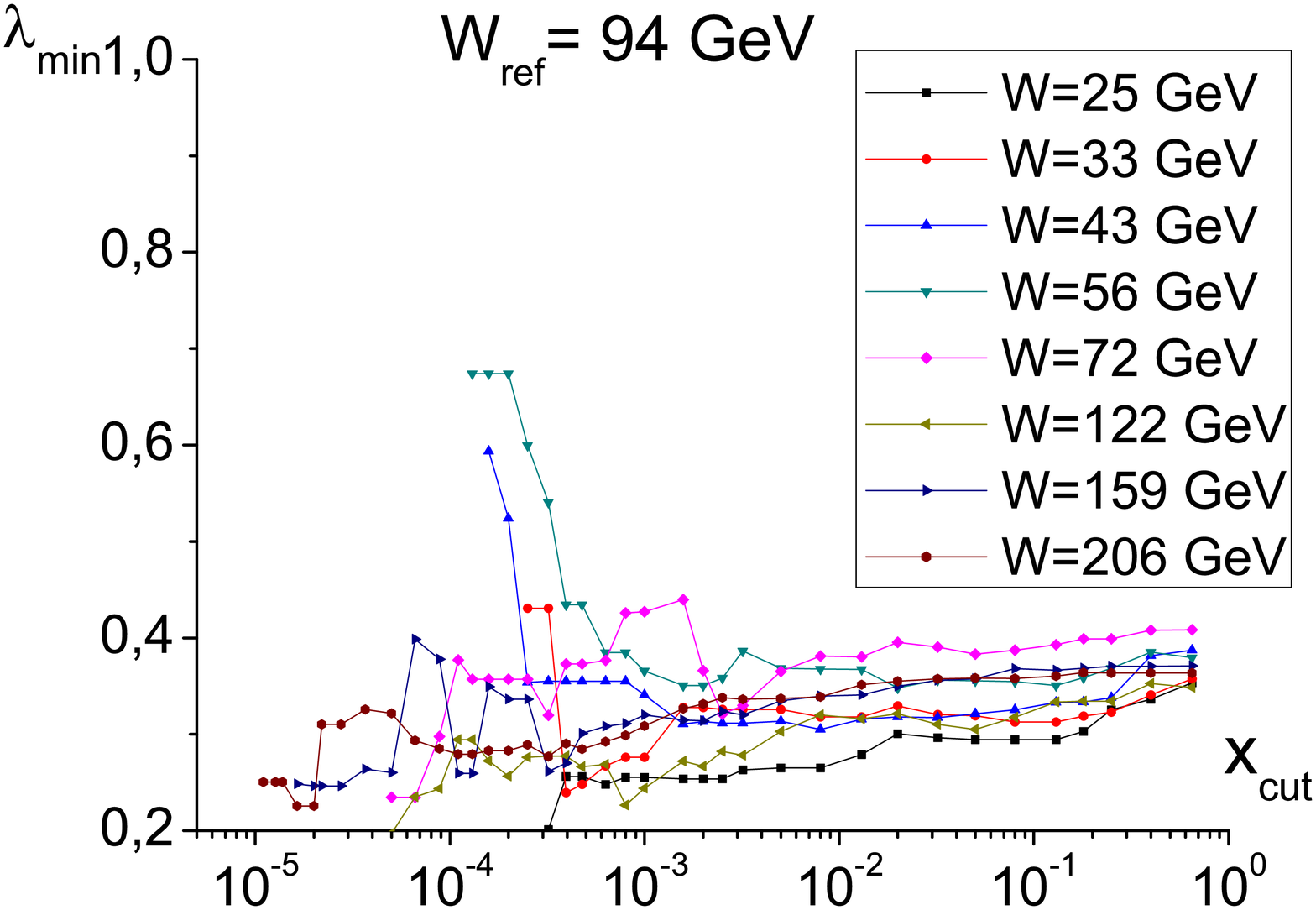}
\caption{Set of functions $\lambda_{\rm{min}} \left( x_{\rm{cut}}\right)$ for different choice of $W_{\rm{ref}}$.}
\label{zesWyk22}
\end{figure}

In section \ref{sectMethodoffinding} we chose $W_{\rm{ref}}=206$ GeV. The reason was that it gives us the widest range of $\tau$ values and is one of the biggest energy we have (energy 268 GeV, which is the biggest, has only a few points - see table in section \ref{sectDivisionofdata}). 

In the Fig. \ref{zesWyk22} we show how set of functions $\lambda_{\rm{min}} \left( x_{\rm{cut}}\right)$ looks like for different choices of reference energy. We can see that for higher $W_{\rm{ref}}$ functions $\lambda_{\rm{min}}(x_{\rm{min}})$ form narrower band.

\newpage

\section{$\lambda_{\rm{ave}}$ calculated without energy binning}

\begin{figure}
\centering
\includegraphics[width=12cm,angle=0]{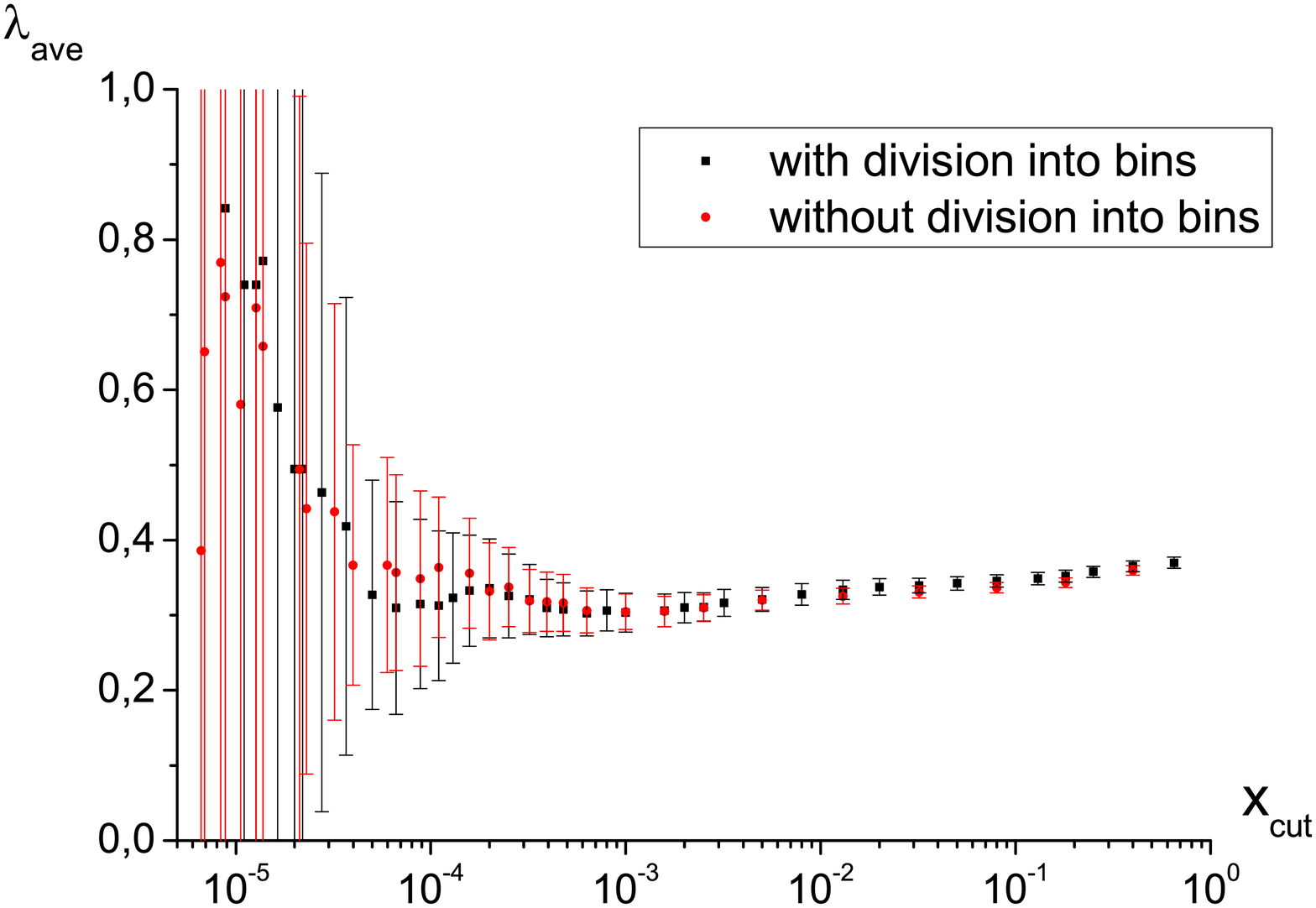}
\caption{$\lambda_{\rm{ave}}$ as a function of $x_{\rm{cut}}$.}
\label{zesWyk23}
\end{figure}

In section \ref{sectlamave} we calculated value of $\lambda$ averaged over energies (denoted by $\lambda_{\rm{ave}}$). To do this we used $\lambda_{\rm{min}}$ calculated for definite energies $25$ GeV $\leq W \leq 159$ GeV. 

Here we will calculate $\lambda_{\rm{ave}}$ in the easier way - without energy binning. We need, however, binning to determine $W_{\rm{ref}}$, and we use binning from section \ref{sectDivisionofdata}: we choose points with $W'\in$(179.2 GeV, 233 GeV) and assign them energy 206 GeV (see Table \ref{table1}). These points outline reference curve $f^{\rm{ref}}_{\lambda}$ and we can perform analysis similar to that described in section \ref{sectMethodoffinding} but now we use all points without energy binning. We define $\chi^2$ function analogously to (\ref{defchi2}): 
\begin{equation}
\chi^2(x_{\rm{cut}};\lambda):=\sum\limits_{i: x \leq x_{\rm{cut}}} \frac{(R_{i}(\lambda)-1)^2}{(\Delta R_{i}(\lambda))^2},
\end{equation}
where we sum over all points which have $x \leq x_{\rm{cut}}$, ratios $R_{i}$ and theirs uncertainties are defined in (\ref{ratiodef}) and (\ref{uncRat}). In this sum we did not exclude points assigned to $W_{\rm{ref}}$ because for them $R_{i}\equiv 1$.  

$\lambda_{\rm{ave}}$ for given $x_{\rm{cut}}$ is found as a minimum of $\chi^2$. Similarly to (\ref{defdeltalamb}) we estimate uncertainty by formula $\chi^2(\lambda_{\rm{ave}}\pm \Delta^\pm \lambda_{\rm{ave}})-\chi^2(\lambda_{\rm{ave}})=1$.

In the Fig. \ref{zesWyk23} we compare functions $\lambda_{\rm{ave}}$ obtained using division into bins (\textit{i.e.} Fig. \ref{zesWyk9}) and without it. As we can see both methods give the same results. This holds true even though in first method we do not take into account points with  $W< 25$ GeV or $W =268$ GeV.

\chapter{Analysis without $Q^2$ uncertainties}

In section \ref{sectdata} we estimated uncertainties of $Q^2$ using some assumption about binning of data. Here we will omit these uncertainties and compare obtained results with results from Chapter 4 (we analyze $e^+ p$ data using energy binning).

If we assume that $\Delta Q^2=0$ uncertainty of $\tilde{\sigma}=\frac{F_{2}}{Q^2}$ is given by:
\begin{equation}
\Delta_p \tilde{\sigma}=\frac{\Delta F_{2}}{Q^2}.
\end{equation}
We denoted it with subscript $p$ to emphasize fact that we do not take into account $\Delta Q^2$. Now all quantities are directly provided by experiment. 

Full uncertainty we denote without subscript: 
\begin{equation}
\Delta \tilde{\sigma}=\sqrt{\left(\frac{\Delta F_{2}}{Q^2}\right)^2+\left(\frac{F_{2}}{(Q^2)^2} \Delta Q^2\right)^2}.
\end{equation}

\begin{figure}
\centering
\includegraphics[width=10cm,angle=0]{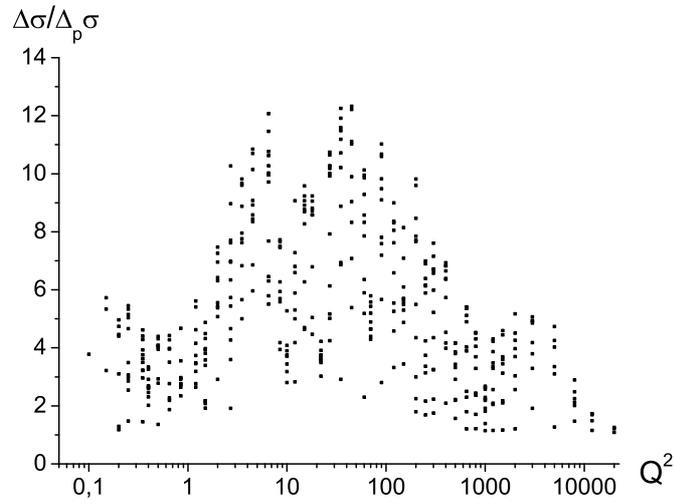}
\caption{Ratios of full uncertainty of $\tilde{\sigma}$ to uncertainty without $\Delta Q^2$ for all experimental points.}
\label{zesWyk24}
\end{figure}

In Fig. \ref{zesWyk24} we show ratios $\Delta \tilde{\sigma}/\Delta_p \tilde{\sigma}$ for all experimental points plotted as a function of $Q^2$. We can see that for most of the points this ratio is substantially greater than 1 \textit{i.e.} most of uncertainty $\Delta \tilde{\sigma}$ comes from $\Delta Q^2$. This means that using $\Delta_p \tilde{\sigma}$ we underestimate uncertainties very strongly.

\begin{figure}
\includegraphics[width=7cm,angle=0]{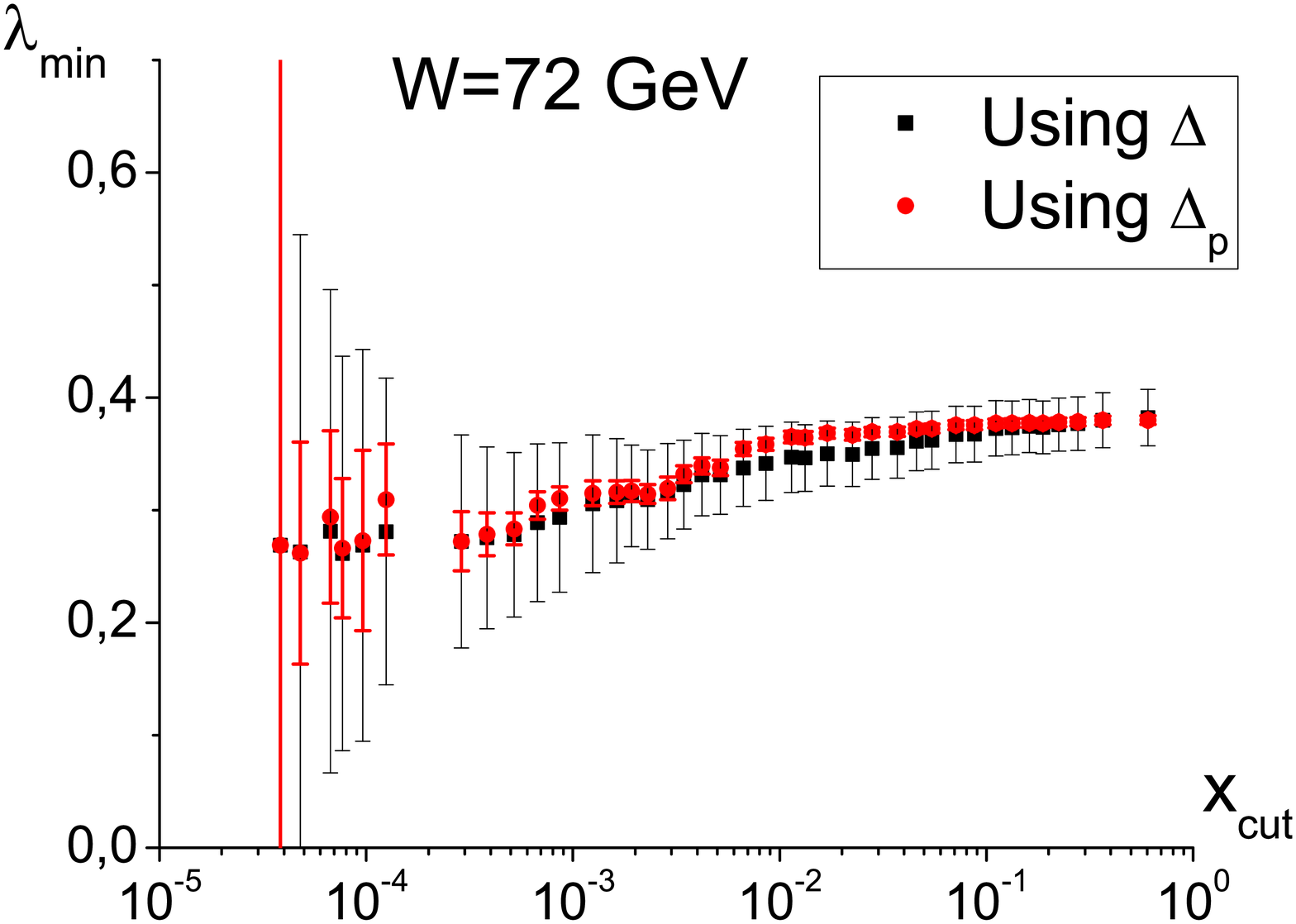}
\includegraphics[width=7cm,angle=0]{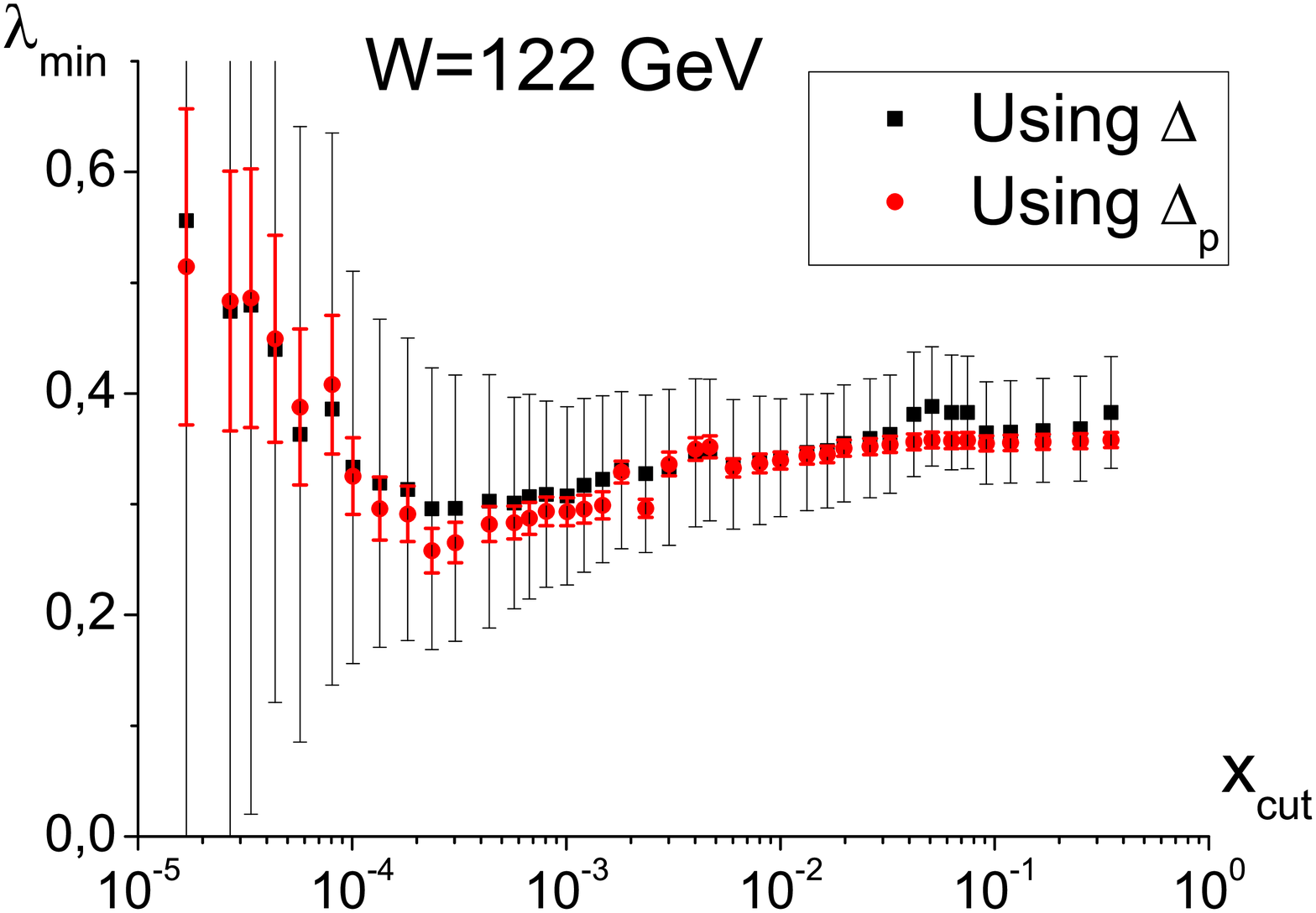}
\caption{Functions $\lambda_{\rm{min}}(x_{\rm{cut}})$ for energies 72 GeV and 122 GeV. They were obtained using $\Delta \tilde{\sigma}$ or $\Delta_p \tilde{\sigma}$}
\label{zesWyk26}
\end{figure}

\begin{figure}
\includegraphics[width=7cm,angle=0]{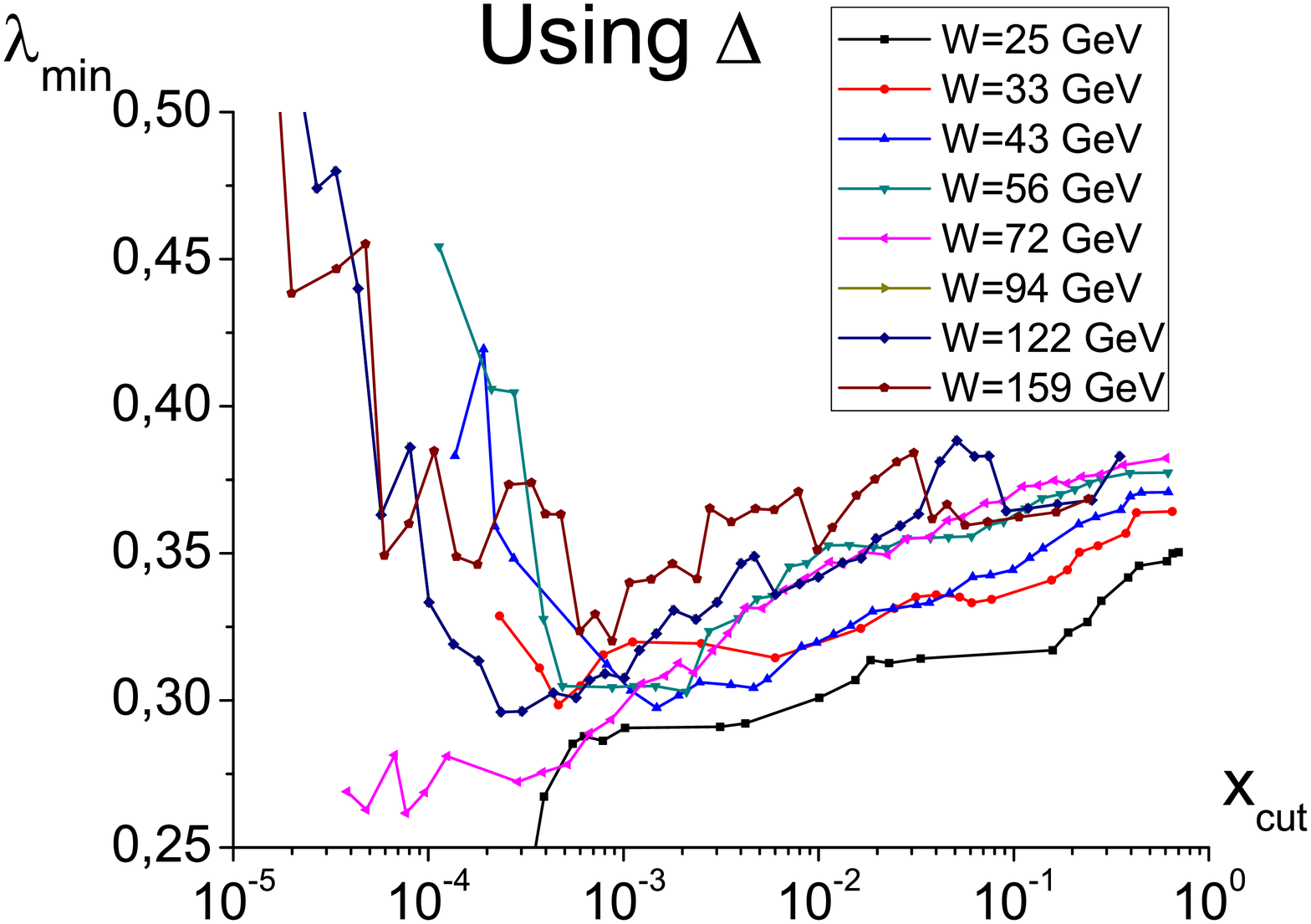}
\includegraphics[width=7cm,angle=0]{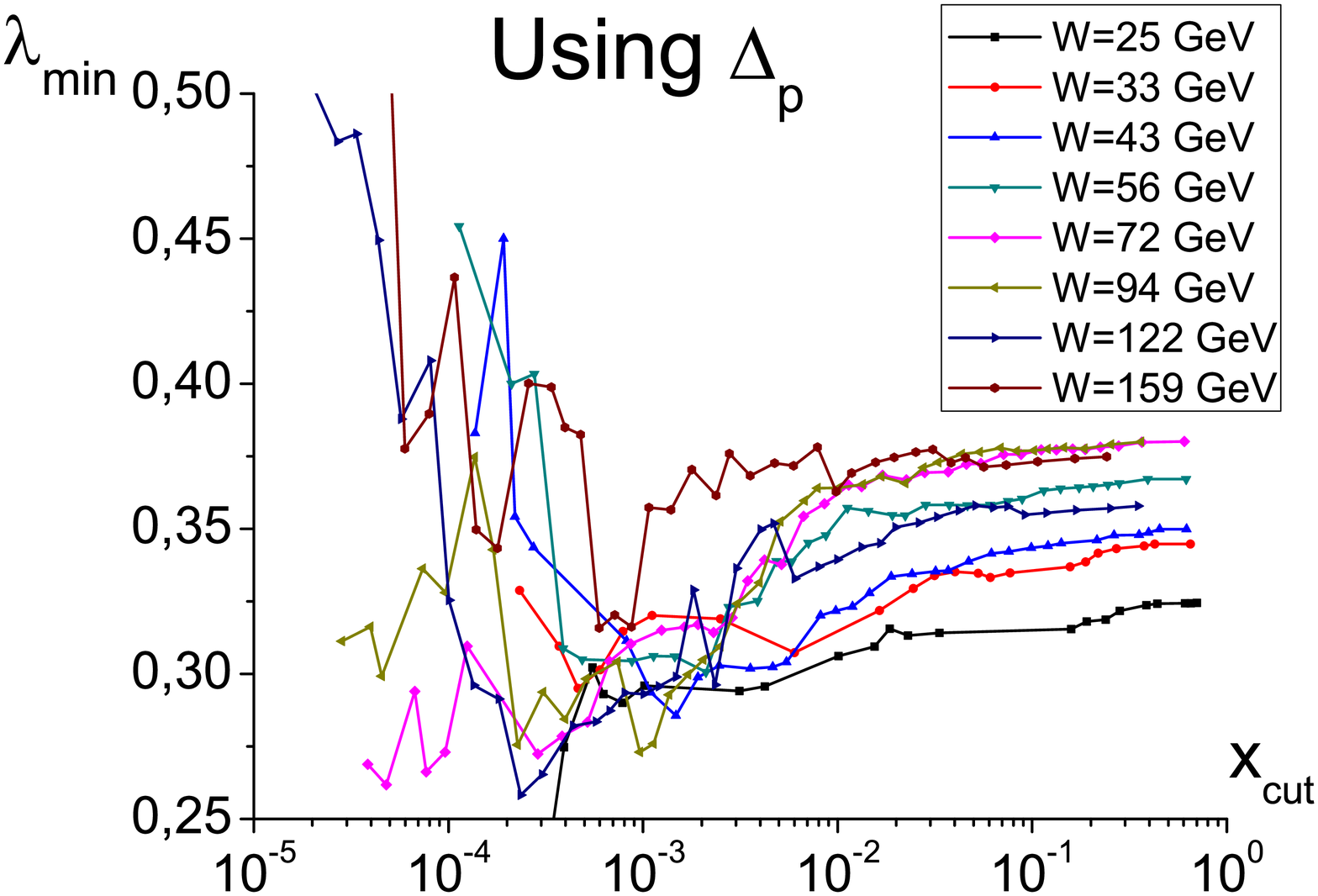}
\caption{Set of functions $\lambda_{\rm{min}}(x_{\rm{cut}})$ for $25$ GeV $\leq W \leq 159$ GeV obtained using $\Delta \tilde{\sigma}$ or $\Delta_p \tilde{\sigma}$}
\label{zesWyk25}
\end{figure}

\begin{figure}
\centering
\includegraphics[width=10cm,angle=0]{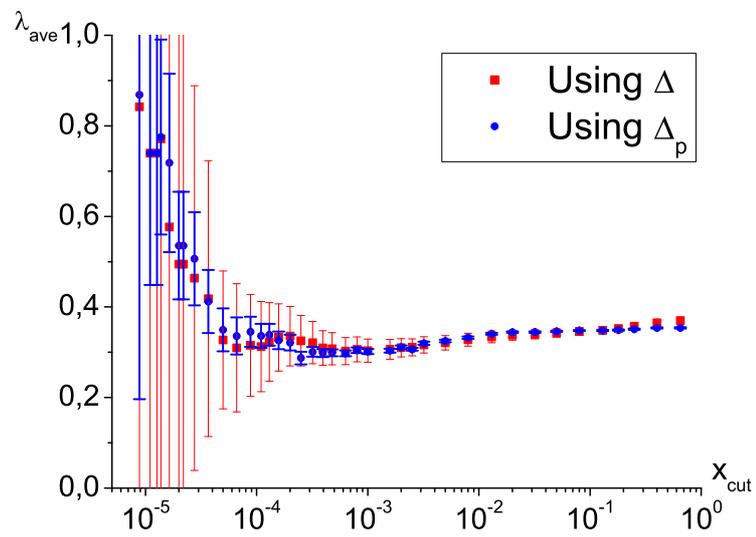}
\caption{$\lambda_{\rm{ave}}(x_{\rm{cut}})$ obtained using $\Delta \tilde{\sigma}$ or $\Delta_p \tilde{\sigma}$}
\label{zesWyk27}
\end{figure}

In Figs. \ref{zesWyk26}, \ref{zesWyk25} and \ref{zesWyk27} we show some quantities which were introduced in Chapter 4. We compare results of two method of calculations: when we use $\Delta \tilde{\sigma}$ (\textit{i.e.} we take into account $\Delta Q^2$) and when we use $\Delta_p \tilde{\sigma}$ (we omit $\Delta Q^2$). As we expect results for $\Delta_p \tilde{\sigma}$ have much smaller uncertainties but values obtained using both methods are practically the same.

\chapter{Reduced cross section}

Structure function $F_2$ can be decomposed in the following way:
\begin{equation}
F_2=\sigma_{r} + f(F_L,x \tilde{F_3}, F^{\gamma Z}_2),
\end{equation}
where $\sigma_r$ we call reduced cross section (like in \cite{H1ZueusWork}) and $f(F_L,x \tilde{F_3}, F^{\gamma Z}_2)\geq0$ is a combination of structure functions.

In \cite{H1ZueusWork} values of structure function $F_2$ was found by measurements of $\sigma_r$ in experiment and calculations of $f(F_L,x \tilde{F_3}, F^{\gamma Z}_2)$ using structure functions parameterization. 

It turns out that in most cases $f(F_L,x \tilde{F_3}, F^{\gamma Z}_2) \ll \sigma_r$ so values of $\sigma_r$ are similar to $F_2$. We can however check how results of our analysis change when we use $\sigma_r$ instead of $F_2$. We define:
\begin{equation}
\tilde{\sigma}_r:=\frac{\sigma_r}{Q^2},
\end{equation}
where we added subscript $r$ to distinguish it from $\tilde{\sigma}$ defined in (\ref{defsigmtylde}).

We will use exactly the same method like in section \ref{ePlresEB} (comparison will be made using $e^+ p$ data and energy binning), in particular we choose the same $W_{\rm{ref}}$ \textit{i.e.} 206 GeV. We denote $\lambda_{\rm{min}}$ values of exponent calculated using $F_2$ and $\lambda^r_{\rm{min}}$ values calculated using $\sigma_r$.

It turns out that $f(F_L,x \tilde{F_3}, F^{\gamma Z}_2)$ was not calculated for all points for which $\sigma_r$ was measured. This means that using $\sigma_r$ instead of $F_2$ we have more points at our disposal. It tourns out that the vast majority of these additional points has energy $W=268$ GeV. 

\begin{figure}
\includegraphics[width=4.5cm,angle=0]{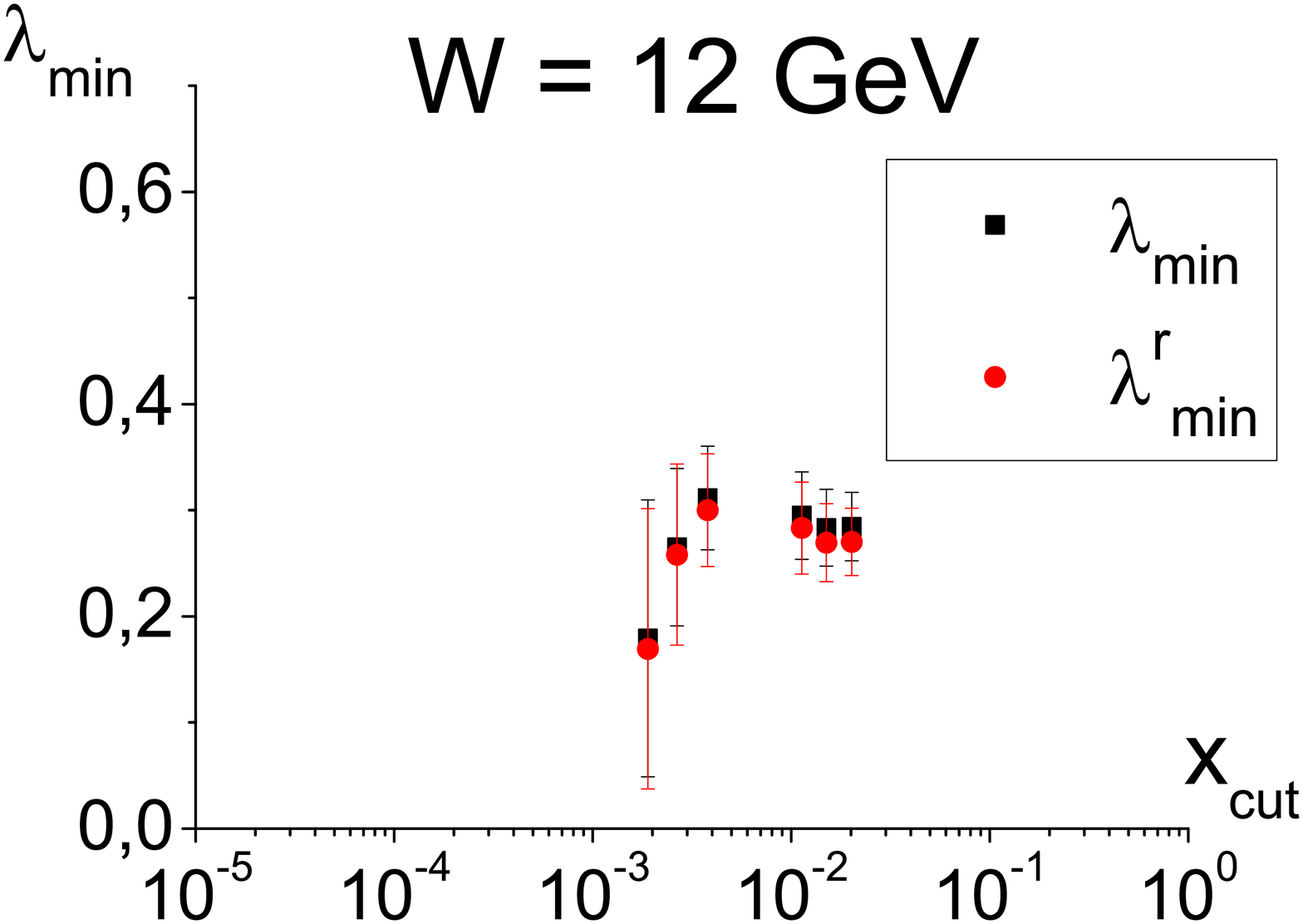}
\includegraphics[width=4.5cm,angle=0]{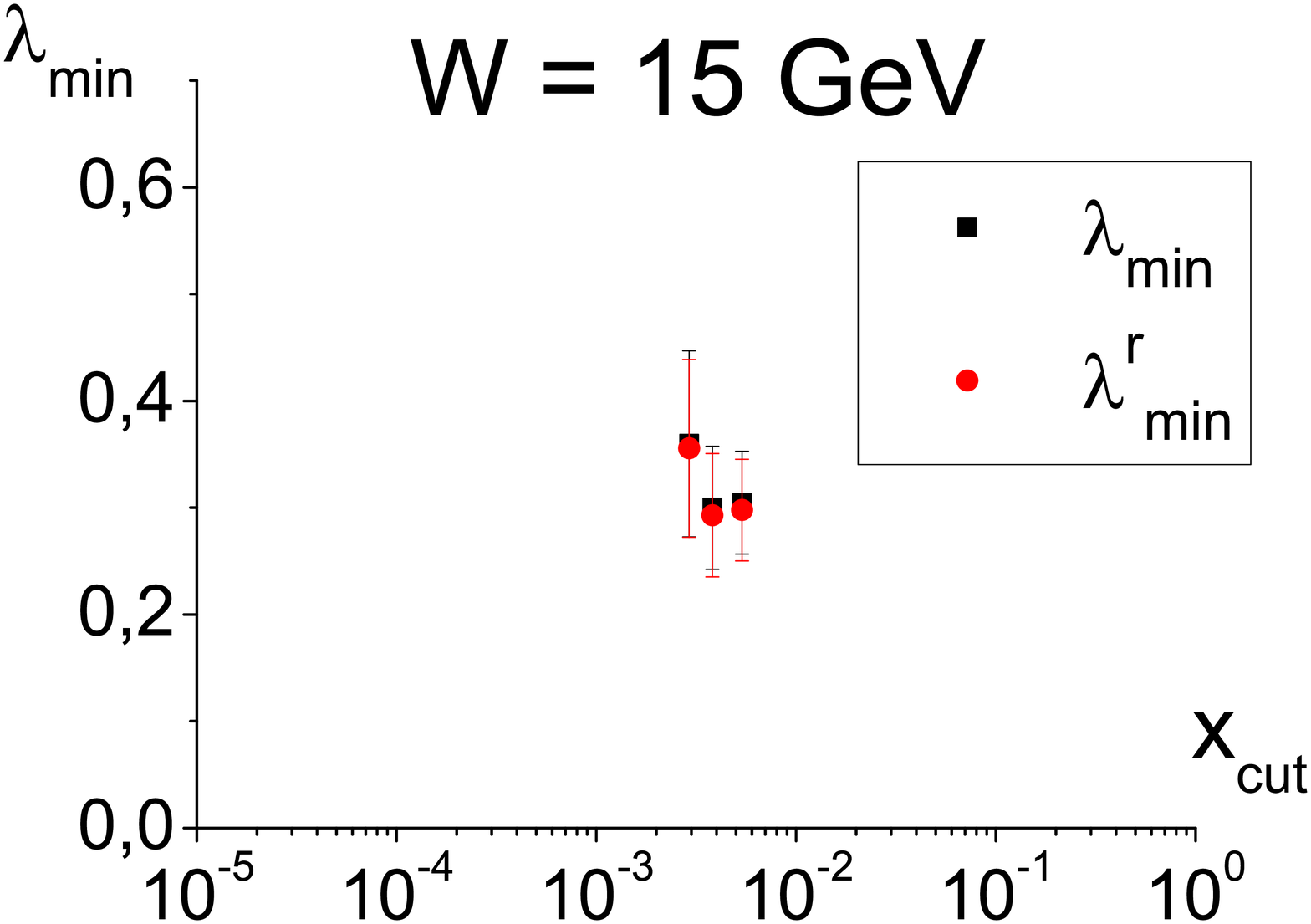}
\includegraphics[width=4.5cm,angle=0]{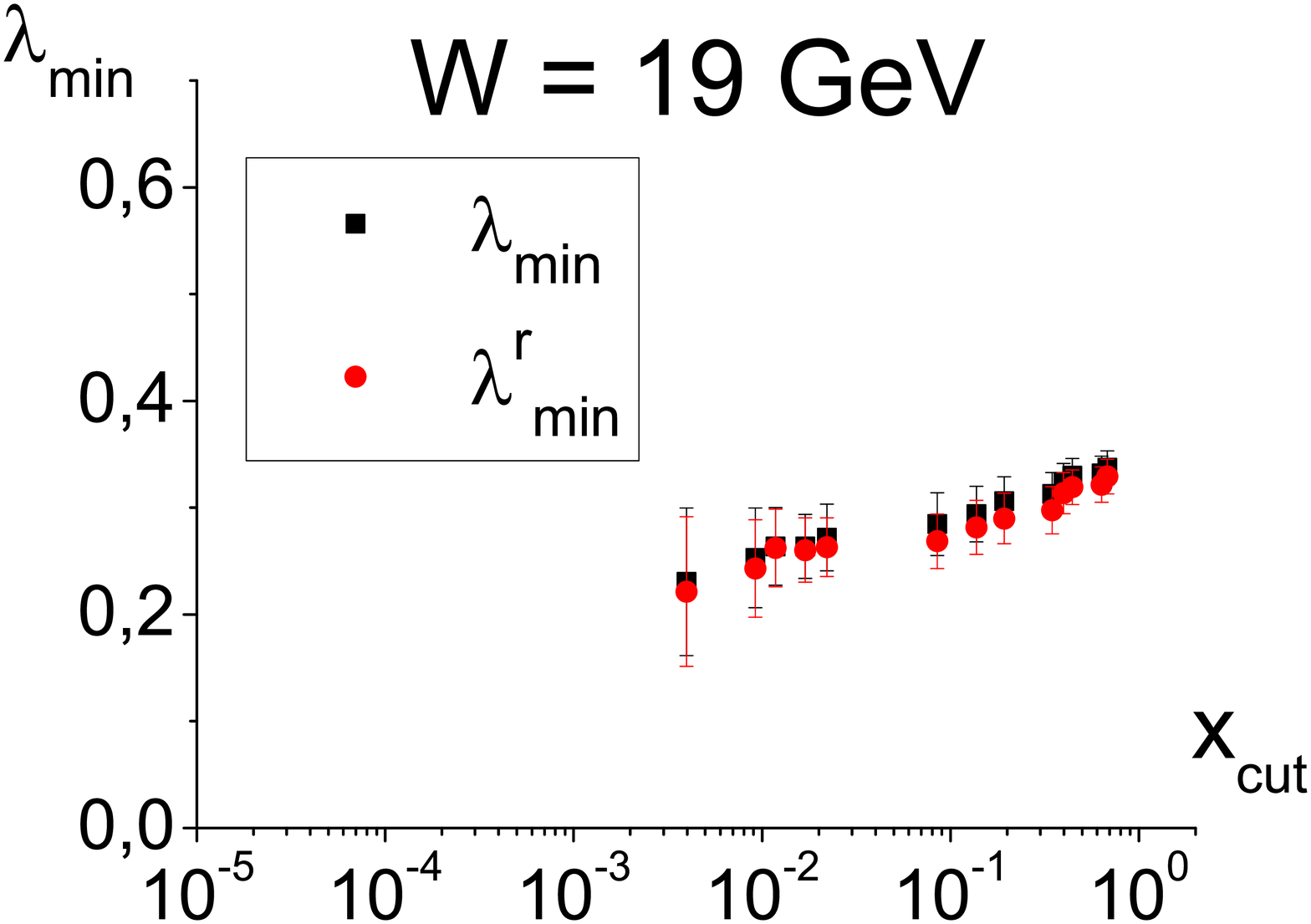}
\includegraphics[width=4.5cm,angle=0]{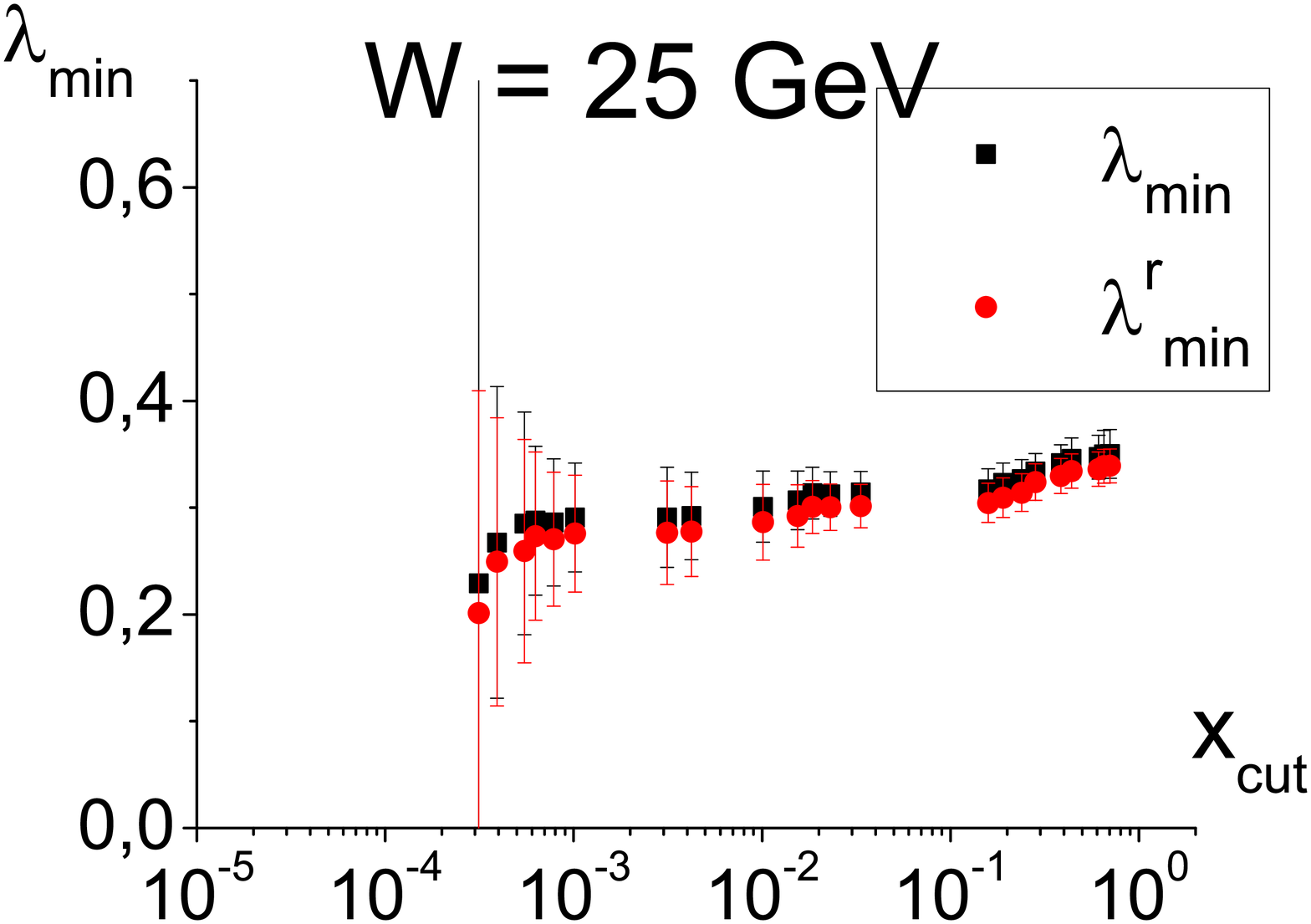}
\includegraphics[width=4.5cm,angle=0]{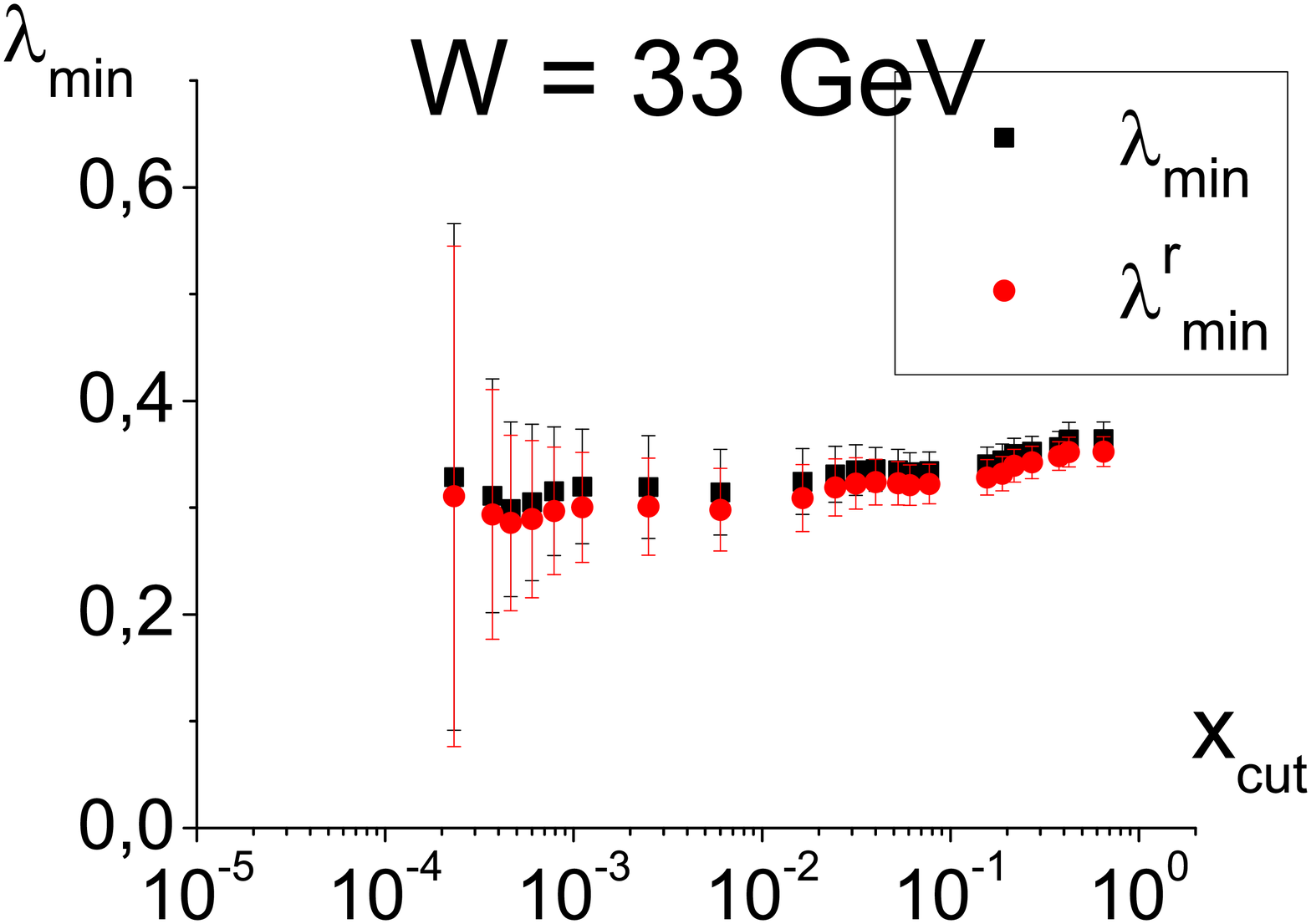}
\includegraphics[width=4.5cm,angle=0]{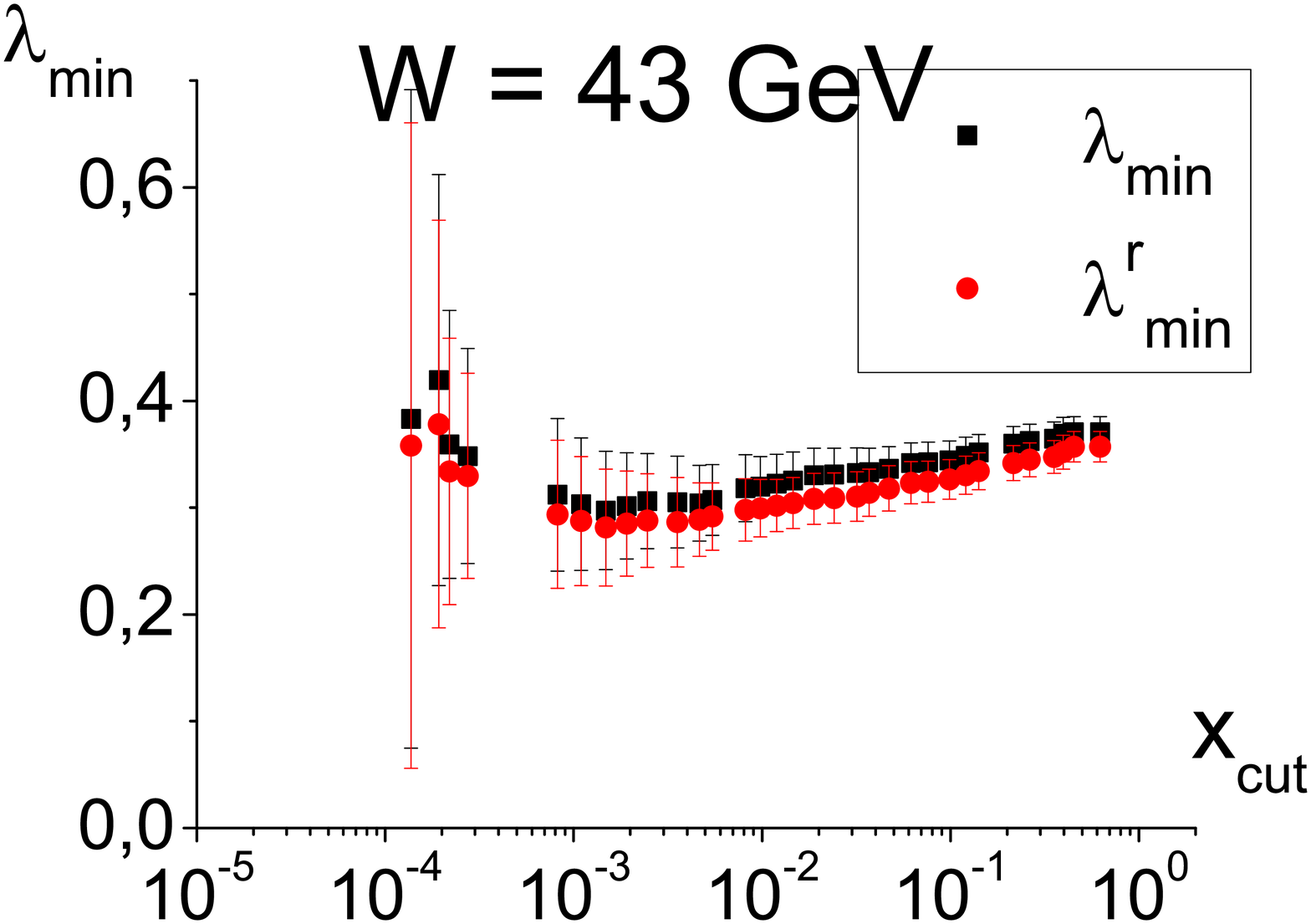}
\includegraphics[width=4.5cm,angle=0]{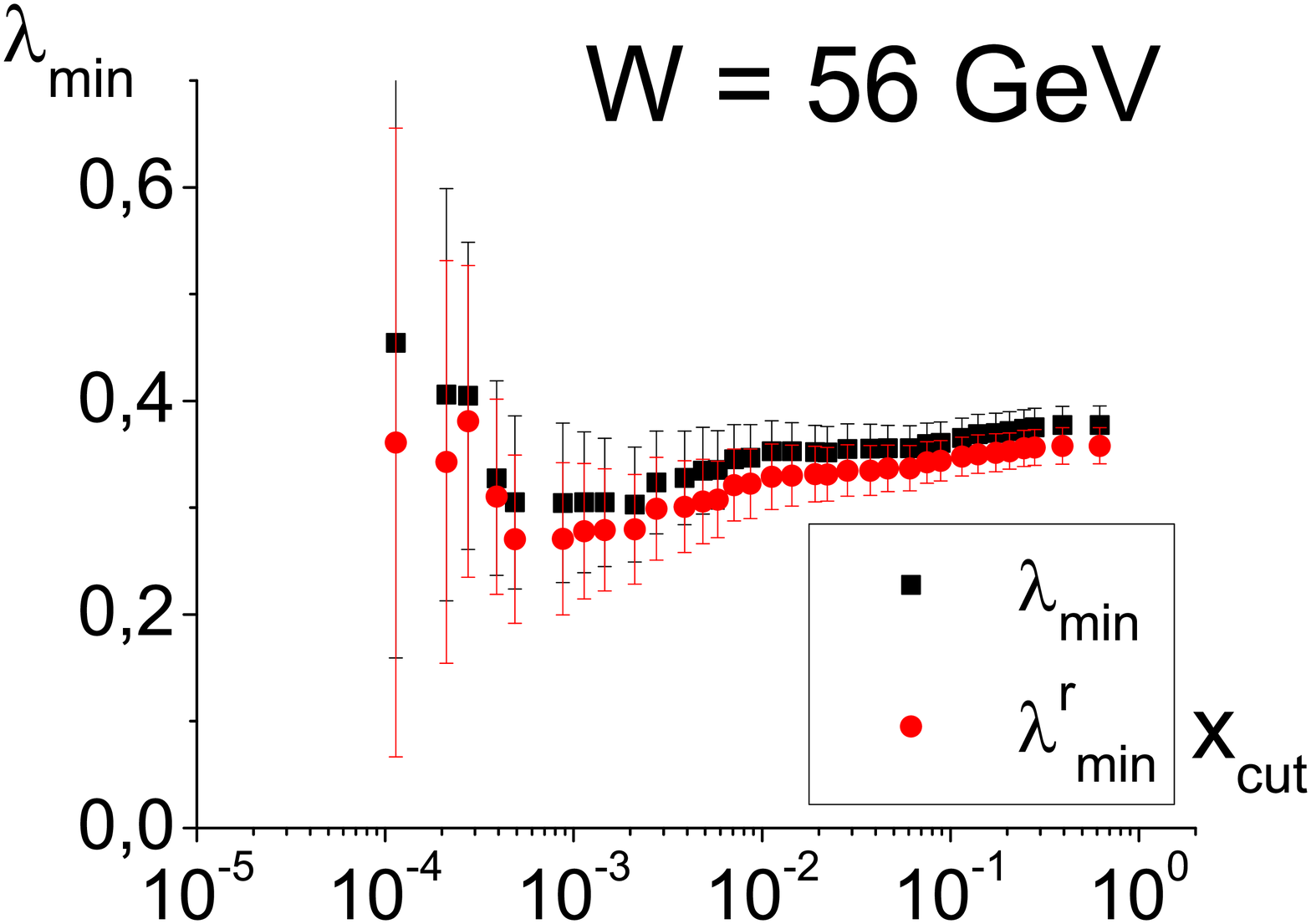}
\includegraphics[width=4.5cm,angle=0]{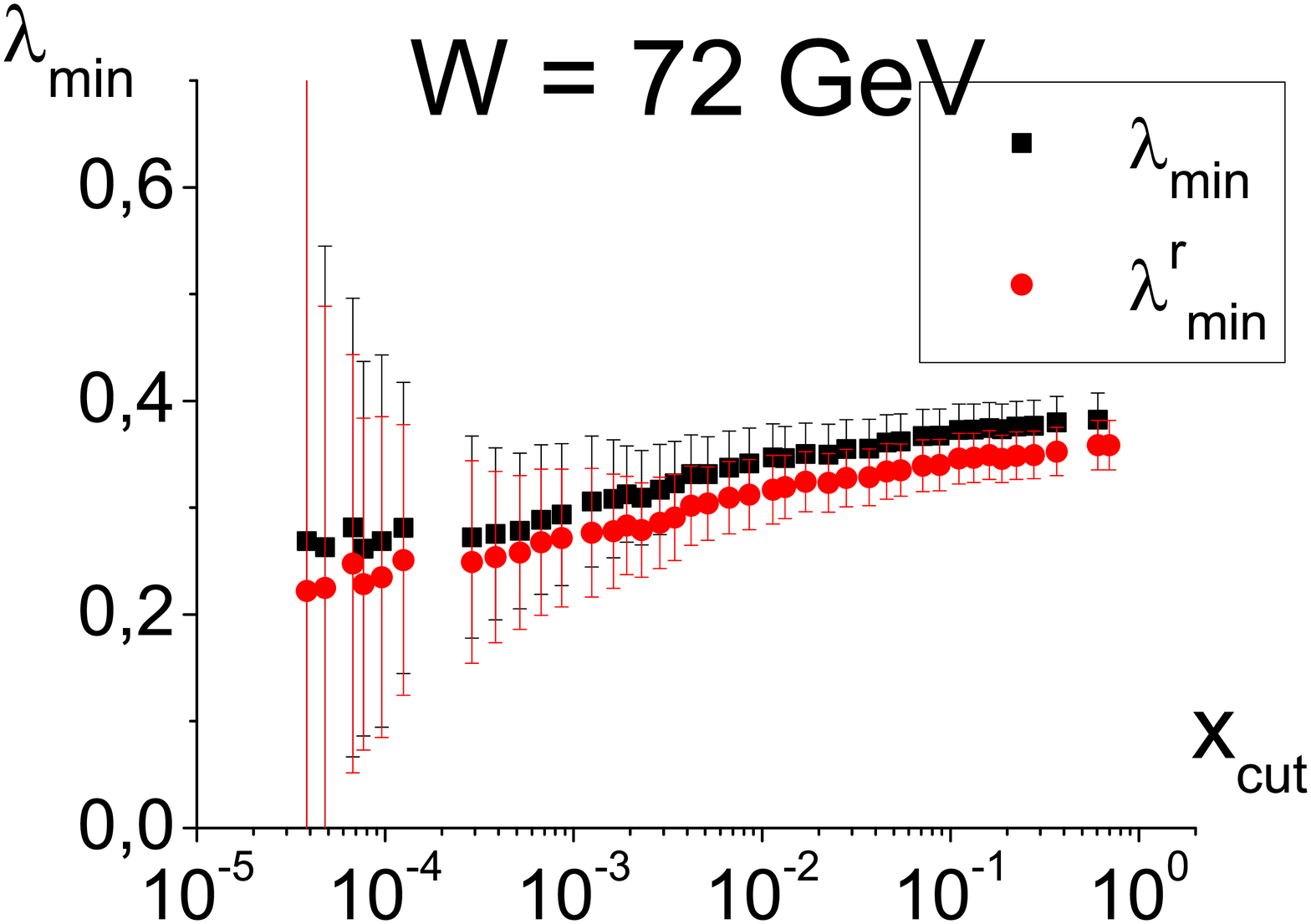}
\includegraphics[width=4.5cm,angle=0]{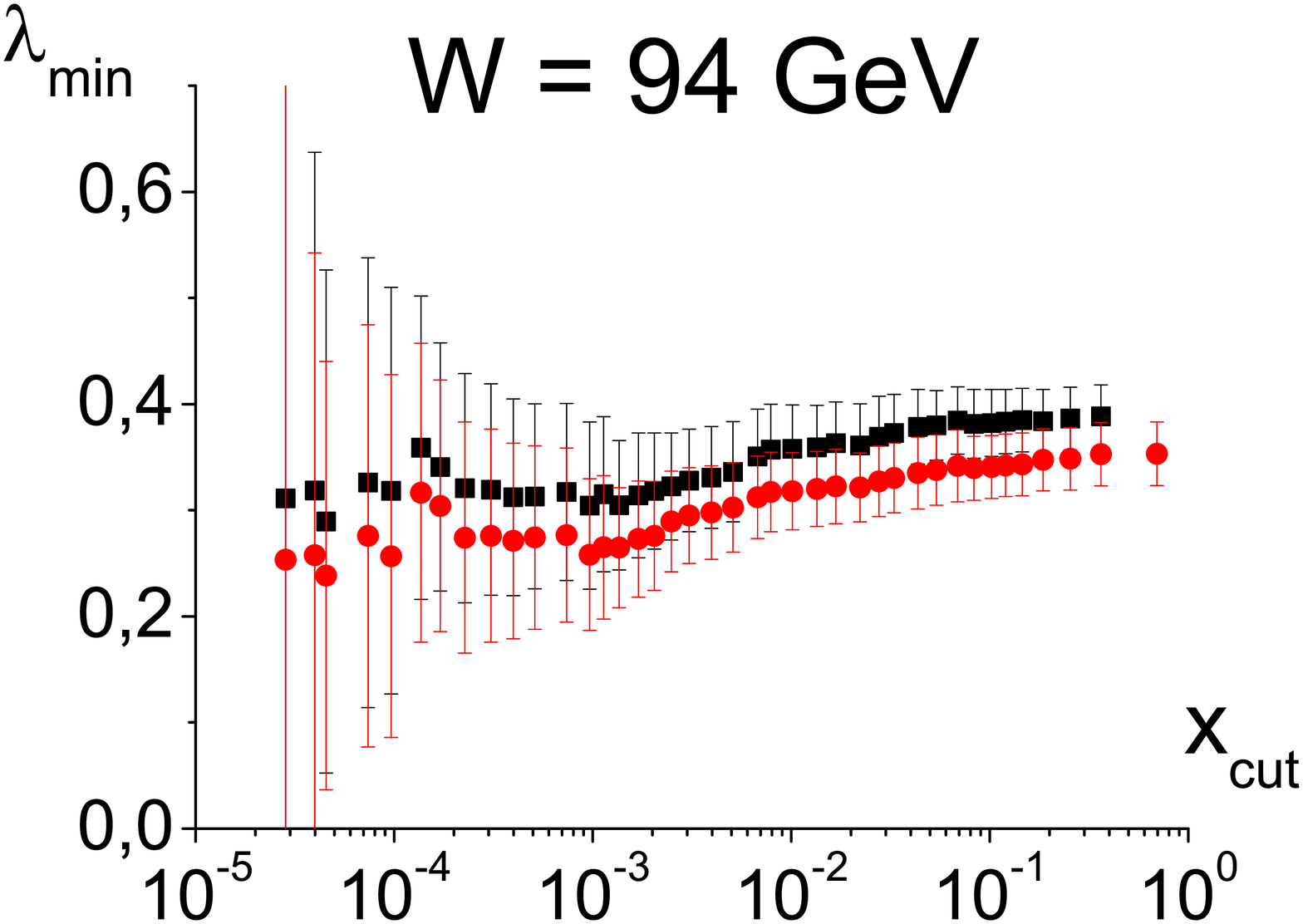}
\includegraphics[width=4.5cm,angle=0]{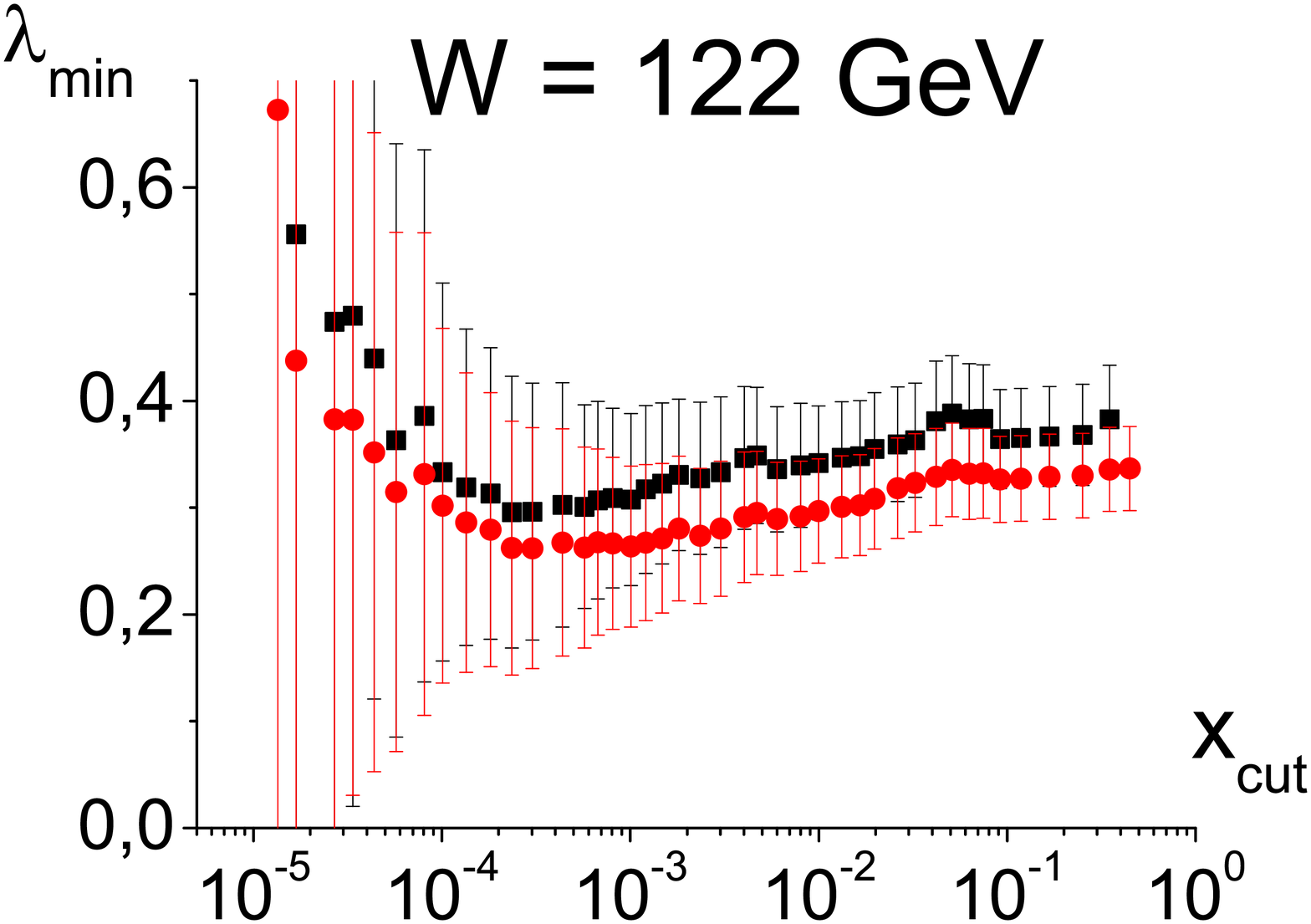}
\includegraphics[width=4.5cm,angle=0]{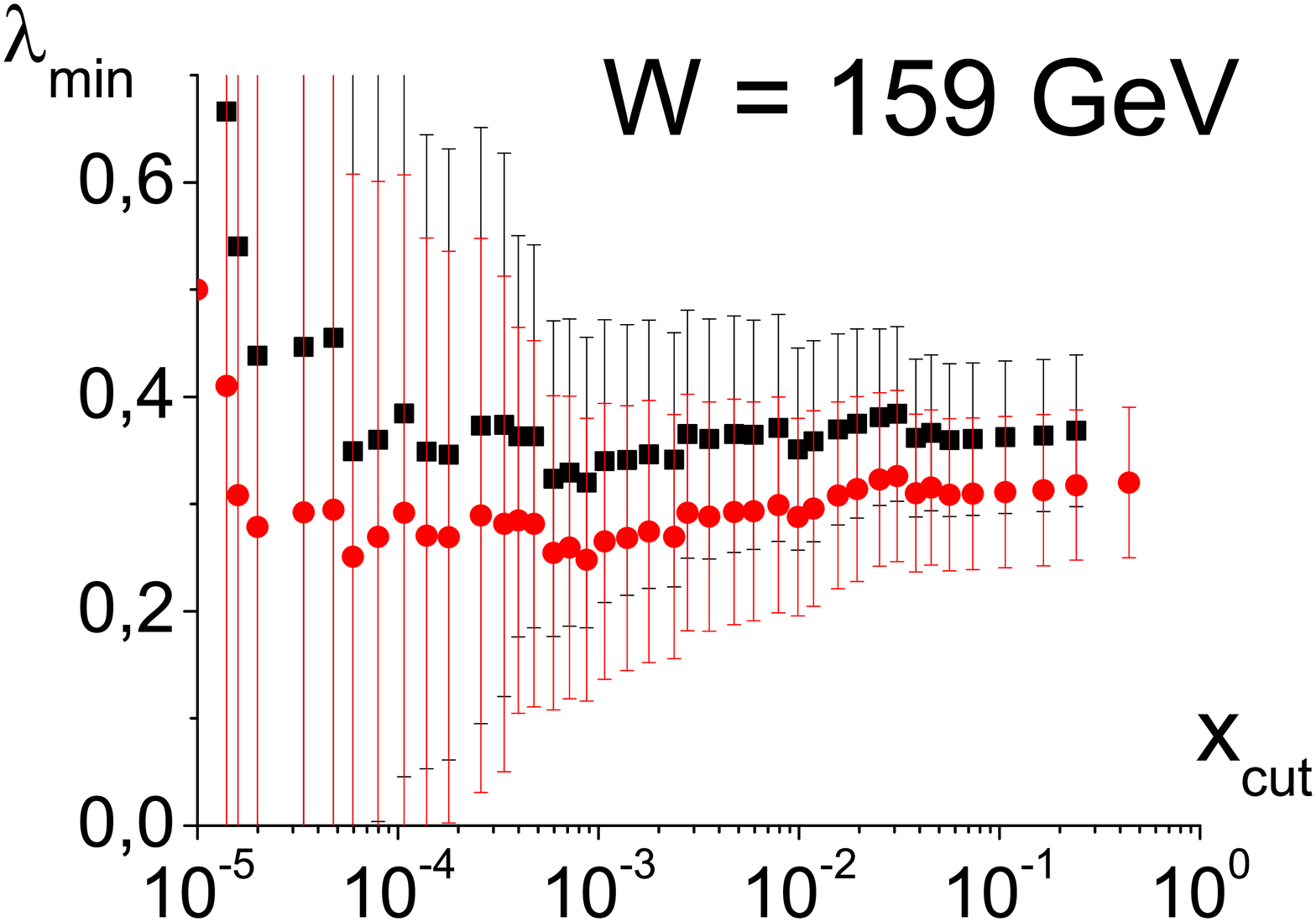}
\includegraphics[width=4.5cm,angle=0]{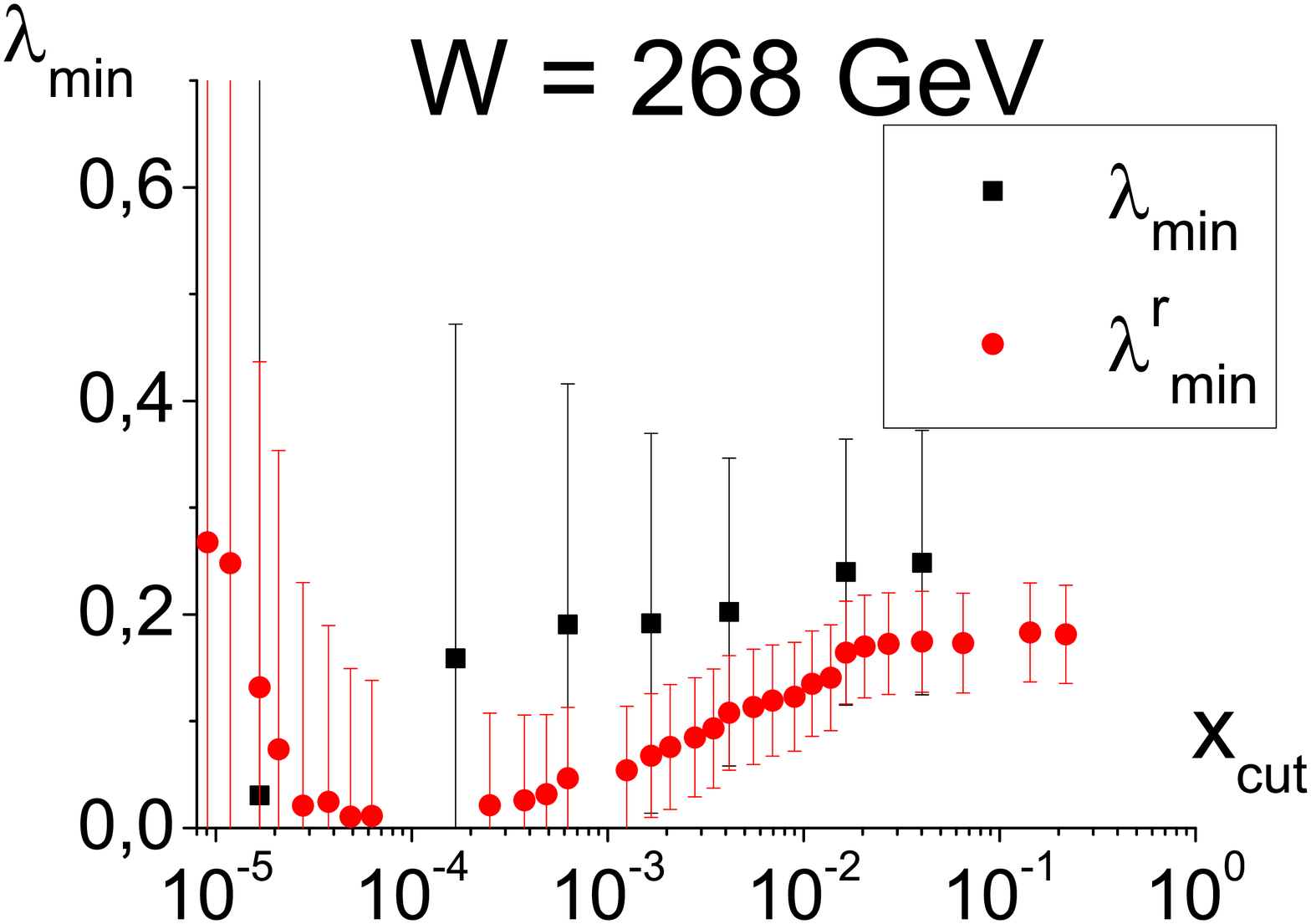}
\caption{$\lambda_{\rm{min}}$ and $\lambda^r_{\rm{min}}$ as a functions of $x_{\rm{cut}}$ for all energies $W\neq W_{\rm{ref}}=206.1$GeV}
\label{zesWyk42}
\end{figure}

In Fig. \ref{zesWyk42} we present plots with comparison between $\lambda_{\rm{min}}(x_{\rm{cut}})$ and $\lambda^r_{\rm{min}}(x_{\rm{cut}})$. We can see that for all energies these functions are similar (except for $W$=268 GeV where much more points are present). 

\newpage
\thispagestyle{empty}

\begin{flushleft}
This work was supported in part by the Polish NCN grant 2011/01/B/ST2/00492.
\end{flushleft}

\end{document}